\documentclass[twocolumn,aps,pre,reprint,superscriptaddress,showpacs,11pt,notitlepage, amsmath, a4paper,nofootinbib]{revtex4-2}

\usepackage{amsmath}
\usepackage{amsfonts}
\usepackage{amssymb}
\usepackage{bbold}
\usepackage{float}
\usepackage{todonotes}
\usepackage{graphicx}
\usepackage{verbatim}
\usepackage{color}
\usepackage{hyperref}
\usepackage{caption}
\usepackage{epigraph}

\usepackage{rotating}
\usepackage{multirow}
\usepackage{amsmath}
\usepackage{datetime}

\usepackage{empheq}
\usepackage[most]{tcolorbox}

\usepackage{chngcntr}

\newtcbox{\mymath}[1][]{
	nobeforeafter,
	math upper,
	tcbox raise base,
	enhanced,
	colframe=black!40,
	colback=black!8,
	boxrule=0.8pt,
	arc=2pt,
	#1
}

\newcommand{\refquote}[1]{\begin{quote} {\textsf{\emph{#1}}} \end{quote}}

\newcommand{\id}{\mathbb{1}}

\makeatletter 

\renewcommand\onecolumngrid{
	\do@columngrid{one}{\@ne}%
	\def\set@footnotewidth{\onecolumngrid}
	\def\footnoterule{\kern-6pt\hrule width 1.5in\kern6pt}%
}

\renewcommand\twocolumngrid{
	\def\footnoterule{
		\dimen@\skip\footins\divide\dimen@\thr@@
		\kern-\dimen@\hrule width.5in\kern\dimen@}
	\do@columngrid{mlt}{\tw@}
}%

\makeatother    

\begin{document}
	
\title{%
	\LARGE Random matrix theory: the results, the applications, and the analytical tools \\
	\large A `Taught Course Centre' lecture series  \\
	at the University of Bath, UK}
\author{Joseph W. Baron}
\email{jwb96@bath.ac.uk}
\affiliation{Department of Mathematical Sciences, University of Bath, Claverton Down, Bath, BA2 7AY}

\begin{abstract}
	Random matrix theory has established itself as a theoretical cornerstone of the mathematical sciences over the past century. It has undeniable utility in areas of research as diverse as nuclear physics, finance, ecology and disordered systems. The purpose of these notes is twofold. First, the most famous and widely used classic results are derived in a pedagogical manner, mostly using the comparatively elementary and transparent cavity method. The significance of each result is then demonstrated in the context of a particular application. There are also some select exercises at the end of each section. In the second part of these notes, a reference guide of analytical techniques for the random-matrix/disordered-systems practitioner is provided. Introducing the diagrammatic, replica, path-integral, and supersymmetric formalisms from first principles, we rederive some of the aforementioned classic results, particularly focussing on the simplest one -- the semicircle law. Innovations such as the population dynamics method and the tools of free probability theory are also included. We discuss the merits of each analytical approach, and we highlight the contexts in which each becomes particularly useful. 
\end{abstract}

\maketitle

\onecolumngrid

\newpage

\tableofcontents

\newpage

\section{Introduction}
\subsection{What is random matrix theory?}
At its core, random matrix theory (RMT) poses a very simple question, one that can be understood by any student on the first day of their Physics or Mathematics degree. It is as follows. Of course, we can (normally) calculate the eigenvalues and eigenvectors of a square matrix by simply applying the usual tools of linear algebra. When the matrix dimension $N$ is 1, 2 or, 3, (maybe 4, if you are lucky), you can do it with a pen and paper. For larger $N$, a computer can often do it for you. But what can we say about the eigenvalues and eigenvectors of large (possibly infinitely large) matrices? 

In attempting to answer this, we may take inspiration from statistical mechanics. To understand the behaviour of large collections of microscopic particles on a macroscopic level, one need not concern oneself with the behaviour of each constituent particle individually. Instead, one considers the \textit{statistical} properties of the ensemble. Likewise, in random matrix theory, we content ourselves to be agnostic of the specifics of the particular matrix entries. We modify the above question: Given some summary statistics of the matrix elements, what then can we say about the eigenvalues and eigenvectors of a large matrix?

\begin{figure}[h]
	\centering 
	\includegraphics[scale = 0.5]{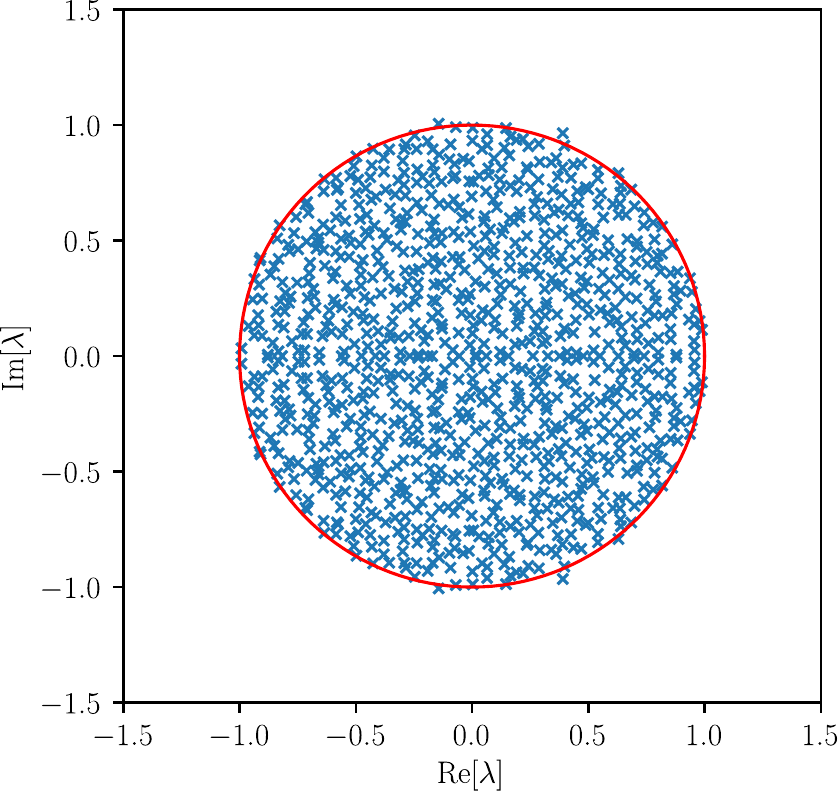} 
	\captionsetup{justification=raggedright,singlelinecheck=false}
	\caption{The eigenvalues of a square matrix of size $N = 1000$ whose elements are all independent centred Gaussian random variables with variance $1/N$. Eigenvalues are indicated with blue crosses. The unit circle is indicated with a solid red line. }\label{fig:circlelaw}
\end{figure}

The results that we obtain from these considerations often turn out to have a wonderful simplicity. Let us take a classic example. If one fills a large matrix of dimension $N$ with independent real random entries with variance $1/N$ (drawn, for instance, from a Gaussian distribution), the eigenvalues all fall within the unit circle in the complex plane, and are uniformly distributed therein, when $N \to \infty$ (see Fig. \ref{fig:circlelaw}). Such beautiful, and perhaps unexpected, simply stated truths are not only of intrinsic interest to the theorist/mathematician, they also possess tremendous applicability, as we shall see. This owes to the fundamental nature of the random matrix theory question and the ready-to-go, user-friendly quality of the results.

\subsection{Why is RMT so useful?}
Many of the problems that are at the current forefront of science relate to complexity. This complexity may arise from the large data sets made available to us by modern technology, the deliberate intricacy of the artificial neural networks used in machine learning, or the intrinsic complexity of ecological, financial or biological systems. The tools that are brought to bear in RMT, which muster some order from an initially complicated problem, do not merely offer some hope of application to the aforementioned research areas by vague analogy, but can in fact be applied directly.

As we will discuss, random matrix theory is able to predict with astonishing reliability the statistics of energy-level separations in large nuclei \cite{mehta2004random, haq1982fluctuation, weidenmuller2009random}. It also lies at the heart of the debate on the causes of stability/instability in complex ecosystems \cite{may1972will, allesina2015stability}. Remarkably, RMT also allows us to understand under what circumstances we are able to make reliable inferences from noisy data \cite{baik2005phase} (which is centrally important in finance \cite{laloux2000random, potters2005financial}, for example). Elsewhere in Physics, RMT makes an appearance in topics as wide-ranging as spin-glasses \cite{mezard1987spin}, the metal-insulator transition \cite{evers2008anderson}, quantum chaos \cite{bohigas1991random, bohigas2005chaotic}, and 2D gravity \cite{di19952d}. In Mathematics, RMT even makes an unexpected appearance in number theory -- the zeros of the Riemann zeta function seem to possess the same spacing statistics as the eigenvalues of the Guassian unitary random matrix ensemble \cite{keating2000random, odlyzko1987distribution} -- as well as finding use in the study of random tilings and non-intersecting paths \cite{johansson2002non}.

\subsection{What will you learn here?}
The aim of these notes is not only to derive the canonical results of RMT. In Part 1 of these notes, each result is also discussed in the context of an application, which is introduced without the assumption of prior knowledge. For example, we discuss how Wigner's semicircle and surmise may be applied to nuclear spectra, how Girko's elliptic law may be used to determine the stability of complex dynamical systems, and how the Mar\v{c}enko-Pastur law relates to de-noising and statistical inference. Some select exercises are provided at the end of each section to illustrate some generalisations of the results or to make a connection with another part of the course. By reading Part 1, the reader should expect to become well-versed in RMT, the associated calculations, and the tremendous array of applications. 

While Part 1 proceeds via comparatively straightforward analytical means (namely, the cavity method), Part 2 introduces some of the more advanced techniques associated with RMT and the adjacent area of disordered systems physics. It is meant as a reference guide of random matrix methodology, and to provide extra information for those whose curiosity is piqued by the exercises and asides. Herein, we provide the \textit{simplest} possible examples of these more advanced calculations by rederiving some of the results already seen in Part 1 (with some additions), and we discuss under what circumstances each analytical method is necessary/advantageous. The objective is to lessen the intimidation level of these somewhat more sophisticated methods, which appear frequently in the random matrix and disordered systems literature, and to inspire the reader -- it is amazing what can be accomplished analytically if one is given the right tools!

Of course, many excellent texts exist with which these notes have some overlap. Works that have been a particular influence on the author are the book by Livan, Novaes, and Vivo \cite{livan2018introduction} (which provides a lively pedagogical introduction to many central results in RMT), that by Potters and Bouchaud \cite{potters2020first} (which is similarly pedagogical and introduces applications, particularly those of finance and inference, in detail), the Oxford Handbook of RMT \cite{akemann2011oxford} (which is a thorough encyclopedic guide to RMT results and applications by leading scientists/mathematicians in the field), the classic book by Mehta \cite{mehta2004random} (which introduces some fundamental results with great clarity and elucidates the classic physics applications), and the book by Tao \cite{tao2023topics} (which provides a rigorous mathematical perspective from one of the great minds on the subject).

\begin{figure}[H]
	\centering 
	\includegraphics[scale = 0.6]{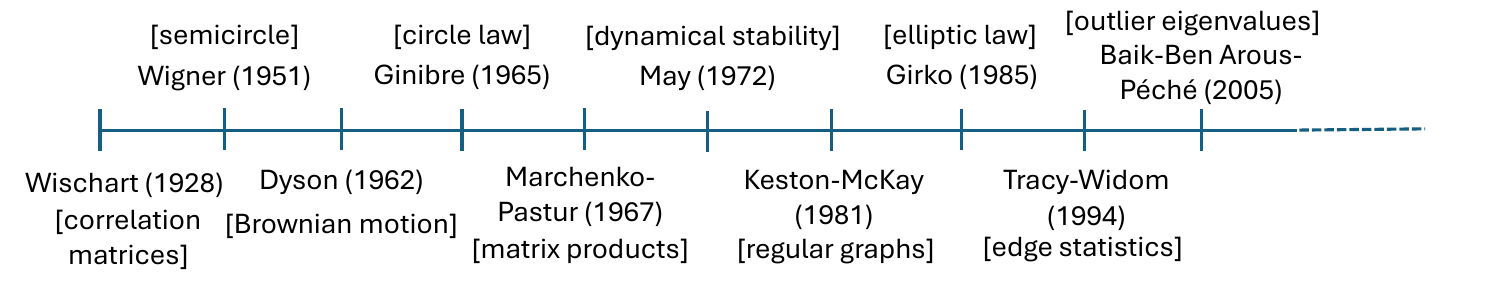} 
	\captionsetup{justification=raggedright,singlelinecheck=false}
	\caption{A (vastly incomplete) timeline of random matrix theory. }\label{fig:timeline}
\end{figure}

\section*{Acknowledgements}
The author would like to thank the Leverhulme Trust for support through the Leverhulme Early Career Fellowship scheme. I would also like to thank all of the attendants of the lecture course at the University of Bath, UK, (both in person and online). In particular, I would like to thank Jong Il Park and Pietro Valigi, for comments, discussions, and corrections. The course would not have been possible without the Taught Course Centre organisers both at the University of Bath and the University of Oxford. 

\newpage

\part{Central results and applications}

\section{Wigner's semicircle and his surmise}\label{section:semicircle}
\begin{quotation}The miracle of the appropriateness of the language of mathematics for the formulation of the laws of physics is a wonderful gift which we neither understand nor deserve. We should be grateful for it and hope that it will remain valid in future research and that it will extend, for better or for worse, to our pleasure, even though perhaps also to our bafflement, to wide branches of learning. -- Eugene Paul Wigner
\end{quotation}

\subsection{Symmetric random matrix ensembles}
As was discussed in the introduction, these notes will largely be concerned with the task of computing the spectral properties (i.e. the eigenvectors and eigenvalues) of large random matrices, and we will see along the way why this is a truly important and impactful mathematical pursuit. So that we can begin to understand the tools that are brought to bear on random matrices, we consider what is perhaps the simplest RMT problem, and one of the earliest on the timeline of results (see Fig. \ref{fig:timeline}). 

We start our journey with Wigner in the 1950s \cite{wigner1955characteristic,  wigner1957conference, wigner1958distribution, wigner1967random, wigner1993characteristic}, who was motivated to understand the energy level spacings in large nuclei. Thanks to his efforts, Wigner discovered the semicircle law and the rule for eigenvalue spacings (his `surmise) that now bear his name. For now, we consider only the mathematical task of deriving Wigner's results, removed from the physical context. However, we will later return to discuss how random matrices of the type that we now introduce apply to nuclear physics.

Let us consider a square matrix $\underline{\underline{J}}$ with random entries. Throughout these notes, we will indicate a matrix by using a double underline and, unless stated otherwise, the matrix dimension will be denoted $N$, which we take to be large. In the present instance, we will suppose that $\underline{\underline{J}}$ is symmetric, that all its entries are real, and that the matrix elements are all independent random variables (up to the constraint $J_{ij} = J_{ji}$). For example, we might draw the entries $J_{ij}$ from a Gaussian distribution. That is, let us suppose that the joint distribution of the matrix entries in this case is
\begin{align}
	P(\underline{\underline{J}}) = \frac{1}{\mathcal{Z}}\exp\left[ -\frac{N}{4} \mathrm{Tr} (\underline{\underline{J}}^2) \right], \label{goedef}
\end{align}
where $\mathcal{Z}$ is a normalisation constant. We thus see that the entries $J_{ij}$ are independent random variables (up to the symmetry constraint) with mean zero and variance $1/N$. The diagonal elements $J_{ii}$ are also i.i.d. centred Gaussian random variables, but with variance $2/N$. Matrices with elements drawn from this distribution are said to belong to the Gaussian Orthogonal Ensemble (GOE), which is a canonical random matrix ensemble that will appear many times in these notes. We note however that the results of this section will apply more generally than just to the GOE. 

\begin{figure}[H]
	\centering 
	\includegraphics[scale = 0.48]{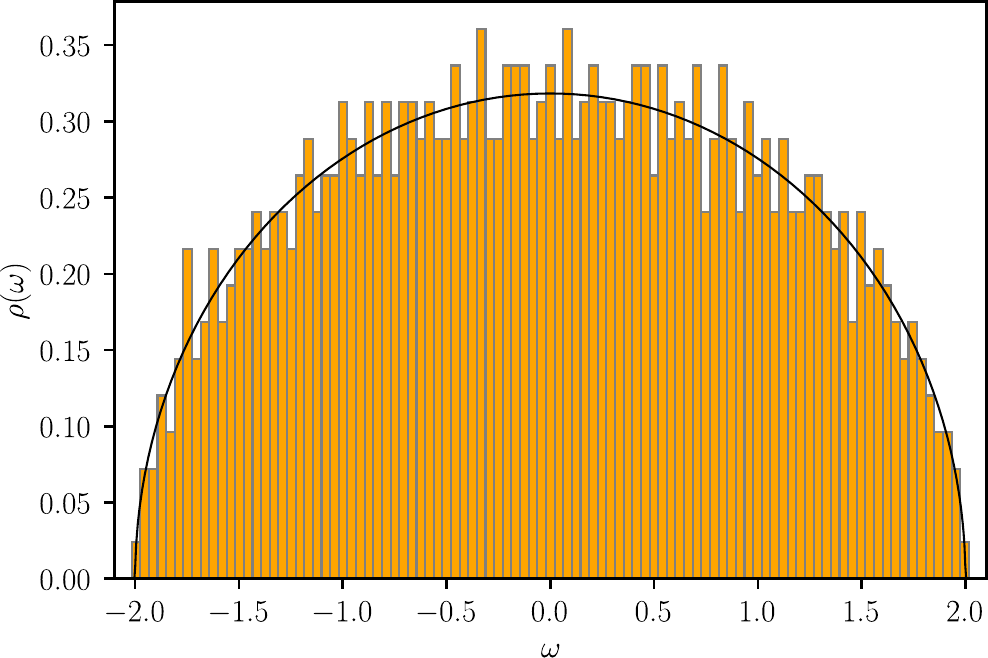} 
	\includegraphics[scale = 0.48]{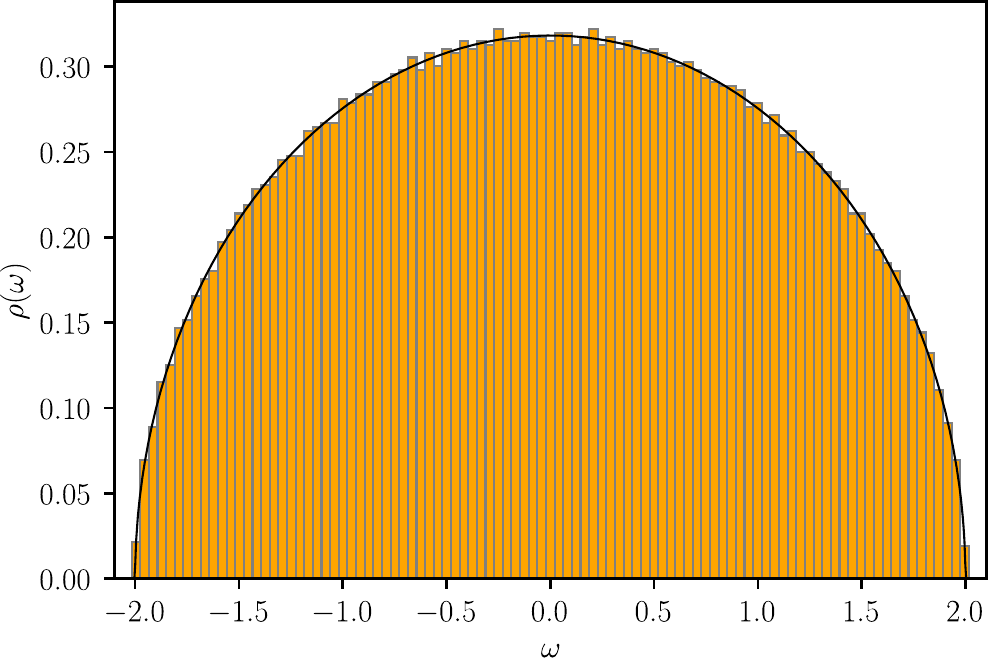} 
	\captionsetup{justification=raggedright,singlelinecheck=false}
	\caption{Histograms of the eigenvalues for a single symmetric Gaussian random matrix with statistics as in Eq.~(\ref{goedef}). The solid line is the Wigner semicircle law in Eq.~(\ref{semicircle}). (Left): $N= 10^3$, (Right): $N = 10^4$. The bin width of the histogram is the same in both panels. }\label{fig:semicircle}
\end{figure}

What we wish to understand, primarily, are the statistics of the eigenvalues of the matrix $\underline{\underline{J}}$. Since all the eigenvalues must be real (due to the Hermiticity of $\underline{\underline{J}}$), one might imagine that making a histogram of the $N$ eigenvalues on the real line would be useful place to start. Such histograms are shown in Fig. \ref{fig:semicircle}. What is particularly striking is the following. As we increase $N$, we see in Fig. \ref{fig:semicircle} that this histogram becomes smooth, and tends towards a limiting form. Our present task will be to understand why this is the case, and to derive this limiting expression.

To this end, we must formalise the quantity that we are trying to find. We thus introduce the concept of the `eigenvalue density'. Denoting the eigenvalues of the matrix $\underline{\underline{J}}$ as $\lambda_1<\lambda_2< \cdots < \lambda_N$, we define the empirical eigenvalue density at a point on the real line as
\begin{align}
	\rho(\omega) = \frac{1}{N}\sum_{\nu} \delta(\omega-\lambda_\nu ) ,\label{eigenvaluedensity}
\end{align}
where $\delta(\cdot)$ is the Dirac delta function. This is defined in such a way that we have the normalisation 
\begin{align}
	\int_{-\infty}^\infty d\omega \rho(\omega) = 1.
\end{align}
We note that in defining $\rho(\omega)$ as in Eq.~(\ref{eigenvaluedensity}) we consider a kind of `extreme' version of the histogram. In some sense, we have taken the bin width to be infinitesimal. As we will show below, a careful use of the limit $N \to \infty$ and a `regularisation' [i.e. a smoothing of the Dirac comb of Eq.~(\ref{eigenvaluedensity})] will yield the limiting form of the eigenvalue density that we desire. We further note that a crucial ingredient for the convergence with $N$ of the histograms in Fig. \ref{fig:semicircle} is the scaling of the variance of the random matrix entries with $1/N$. That is, we require $\mathrm{Var}(J_{ij}) = \frac{1}{N}$.

Now that we have introduced the problem, let us understand how we might proceed analytically. 

\subsection{The resolvent matrix}
We now seek to obtain an analytical expression for the quantity $\rho(\omega)$ in the limit $N \to \infty$. An exceptionally useful quantity, which will appear time and time again in these notes, is the so-called resolvent matrix. This is defined as
\begin{align}
	\underline{\underline{G}}(z) = \left( z \underline{\underline{\id}}_N -\underline{\underline{J}}  \right)^{-1},\label{resdef}
\end{align}
where $\underline{\underline{\id}}_N$ is the $N\times N$ identity matrix and where $z \in \mathbb{C}$. In particular, if we take the normalised trace of this quantity, we obtain what is variously referred to as the (1-point) Green's function, \textbf{the} resolvent, or the Stieltjes transform, which we denote $G(z)$ (i.e. without the double underline). We will show below that this is indeed the Stieltjes transform of the distribution $\rho(\omega)$ in a precise sense. 

To be specific, we define
\begin{align}
	G(z) \equiv N^{-1} \mathrm{Tr} \underline{\underline{G}}(z) = N^{-1} \sum_i G_{ii}(z), \label{restracedef}
\end{align}
The true utility of this object becomes clear once we write it in terms of the eigenvalues of $\underline{\underline{J}}$. Let $\psi_i^{(\nu)}$ be the $i$\textsuperscript{th} component of the eigenvector of $\underline{\underline{J}}$ corresponding to the eigenvalue $\lambda_\nu$ (normalised such that $\sum_{i} \vert \psi_{i}^{(\nu)}\vert^2 = 1$). We further define the matrix with these eigenvectors as columns  $\underline{\underline{P}}$, so that $P_{i\nu} =\psi_{i}^{(\nu)}$. Using the orthogonality of the eigenvectors of a real symmetric matrix $\underline{\underline{P}}^T \underline{\underline{P}} = \underline{\underline{\id}}_N$, we may write using simple matrix algebra
\begin{align}
	\underline{\underline{P}}^T \underline{\underline{G}}\, \underline{\underline{P}}\, \underline{\underline{P}}^T \left( z \underline{\underline{\id}}_N -\underline{\underline{J}}  \right)\, \underline{\underline{P}} = \underline{\underline{\id}}_N. 
\end{align}
By definition, the matrix $\underline{\underline{P}}$ diagonalises $\underline{\underline{J}}$, and we thus have
\begin{align}
	G_{ij}(z) = \sum_{\nu} \frac{\psi_i^{(\nu)} \psi_j^{(\nu)}}{z-\lambda_\nu}. \label{resvectors}
\end{align}
We finally see using the definition in Eq.~(\ref{eigenvaluedensity}) that
\begin{align}
	G(z) =\frac{1}{N} \sum_\nu \frac{1}{z-\lambda_\nu}  =  \int d\mu \frac{\rho(\mu)}{z-\mu} ,\label{resolventasstieltjes}
\end{align}
which is precisely the definition of the Stieltjes transform of $\rho(\omega)$ \cite{widder1938stieltjes}. The Stieltjes transform is a common tool used in the computation of probability measures generally. 

To complete the picture, we now show how we can perform the inverse Stieltjes transform, and thus extract the eigenvalue density from $G(z)$. Suppose we take the complex variable $z$ to be \textit{almost real} such that $z= \omega - i\epsilon$. Let us examine the imaginary part of $G(z)$ under the limit $\epsilon\to 0$. One has
\begin{align}
	&\lim_{\epsilon \to 0}\mathrm{Im}G(\omega-i\epsilon) = \lim_{\epsilon \to 0} \int d\mu \rho(\mu)\frac{\epsilon}{(\omega -\mu)^2 + \epsilon^2}
	= \int d\mu \rho(\mu) \pi \delta(\omega - \mu) = \pi \rho(\omega),
\end{align}
i.e., one has the crucial relation for the inverse Stieltjes transform

\begin{empheq}[box={\fboxsep=6pt\fbox}]{align}
	\rho(\omega) = \frac{1}{\pi} \lim_{\epsilon\to 0} \mathrm{Im} G(\omega - i\epsilon) . \label{densefromres} 
\end{empheq}
Alternatively, one may also use 
\begin{align}
	\rho(\omega) = \frac{1}{2\pi} \lim_{\epsilon\to 0} \left[ G(\omega - i\epsilon) - G(\omega + i\epsilon)\right] .
\end{align}
We thus see that by computing the resolvent $G(z)$, we may extract the eigenvalue density in which we are interested. 

But what is the point of introducing this object? The real utility of the transform $G(z)$ is 2-fold. First, we see that for finite $\epsilon$, the quantity $\pi^{-1}\mathrm{Im}G(\omega-i\epsilon) = \int d\mu \epsilon \rho(\mu)/[(\omega-\mu)^2 + \epsilon^2]$ is a smoothed version of the eigenvalue density defined in Eq.~(\ref{eigenvaluedensity}). In some sense, a finite value of $\epsilon$ corresponds to the finite bin width used in the histogram shown in Fig. \ref{fig:semicircle}. It is this smoothed eigenvalue density in which we are interested\footnote{Formally speaking, this means that we first take the limit $N \to \infty$ and then $\epsilon \to 0$ in our calculations. We must also take care to use a bin width that is much greater than $1/N$ (the average eigenvalue spacing) in numerical experiments.}. Furthermore, the smoothness of $G(z)$ away from the real line gives $G(z)$ desirable analytic properties. Second, the resolvent involves the matrix entries $J_{ij}$ in a mathematically advantageous way. That is, it is a much more simple object to deal with than, say, the characteristic polynomial $\mathrm{det}(\lambda \underline{\underline{\id}}_N - \underline{\underline{J}}) = 0$.

In summary, the trace of the resolvent matrix provides us with an extremely useful stepping stone for obtaining the eigenvalue density. If we compute the resolvent $G(z)$ for an arbitrary complex $z$ in the limit $N\to \infty$, we may subsequently take $z$ to the real line, and obtain the limiting $\rho(\omega)$ as the imaginary part of this object. Most computations in random matrix theory involve the resolvent matrix in some way. 

\begin{figure}[H]
	\centering 
	\includegraphics[scale = 0.7]{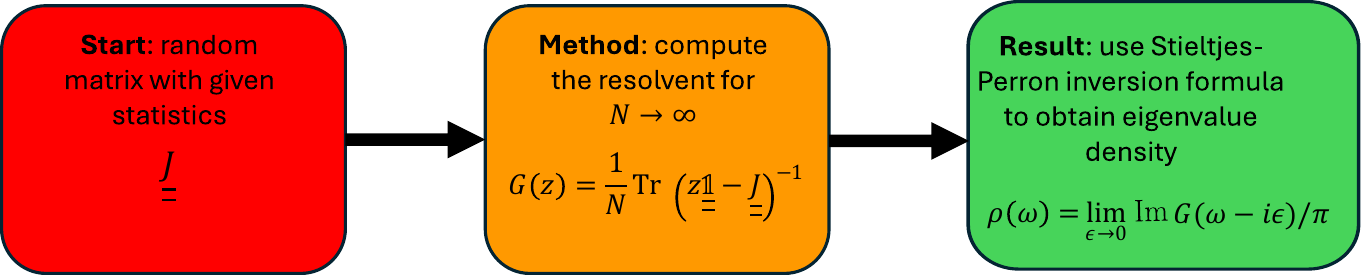}
	\captionsetup{labelformat=empty}
\end{figure}

\subsection{The cavity (Block Inversion/Schur complement) approach}\label{section:cavity}
Given the relation between the eigenvalue density and the resolvent in Eq.~(\ref{densefromres}), all that remains is for us to compute $G(z)$. As we shall see as we progress through these notes, there exist a plethora of methods for computing the resolvent, each with its own advantages and disadvantages, depending on the situation.

Perhaps the most elementary method is the one that we will discuss now, which we refer to as the cavity method \cite{kuhn2008spectra, rogers2008cavity}. What makes the cavity method simple in comparison to others is that it requires no consideration of combinatorics (as in Wigner's original approach in Section \ref{section:wignerapproach}), no auxiliary integral constructions (such as the replica, supersymmetric or path-integral methods -- also see Part \ref{part:additionaltools}), and no additional dynamics (as in the Dyson Brownian motion approach in Section \ref{section:dysonbrownian}). For this reason, we will rely on it as our primary analytical approach throughout Part 1 of these notes. With that being said, the aforementioned alternative methods are indispensable when considering more complicated calculations, and they can provide greater insight or an alternative interpretation of the results.

At its heart, our strategy exploits the standard formula for the inversion of a block matrix.

\begin{tcolorbox}[ colback=blue!10!white,colframe=blue!90!black,title=Lemma: Block matrix inverse]
	Suppose we have a block matrix
	\begin{align}
		\underline{\underline{M}} = \begin{bmatrix}
			\underline{\underline{S}} & \underline{\underline{T}} \\
			\underline{\underline{U}} & \underline{\underline{V}}
		\end{bmatrix},
	\end{align}
	where $\underline{\underline{M}}$ is an $N\times N$ matrix, $\underline{\underline{S}}$ is $L\times L$, $\underline{\underline{T}}$ is $L \times (N-L)$, $\underline{\underline{U}}$ is $(N-L) \times L$, and $\underline{\underline{V}}$ is $(N-L)\times (N-L)$. Assuming that $\underline{\underline{S}}$ and $\underline{\underline{V}}$ are both invertible, it is straightforward to verify by block matrix multiplication that the inverse of $\underline{\underline{M}}$ is given by
	\begin{align}
		\underline{\underline{M}}^{-1} =\begin{bmatrix}
			\left(\underline{\underline{S}}-\underline{\underline{T}}\underline{\underline{V}}^{-1}\underline{\underline{U}}\right)^{-1} & -\left(\underline{\underline{S}}-\underline{\underline{T}}\underline{\underline{V}}^{-1}\underline{\underline{U}}\right)^{-1} \underline{\underline{T}}\underline{\underline{V}}^{-1} \\
			-\left(\underline{\underline{V}}-\underline{\underline{U}}\underline{\underline{S}}^{-1}\underline{\underline{T}}\right)^{-1}\underline{\underline{U}}\underline{\underline{S}}^{-1} & \left(\underline{\underline{V}}-\underline{\underline{U}}\underline{\underline{S}}^{-1}\underline{\underline{T}}\right)^{-1}
		\end{bmatrix}. \label{blockinverse}
	\end{align}
\end{tcolorbox} 

Let us now consider the specific case where $\underline{\underline{M}} = \underline{\underline{G}}^{-1}= z\underline{\underline{\id}}- \underline{\underline{J}}$. We split the matrix $\underline{\underline{M}}$ into blocks in an extreme way such that $L = 1$. We then have that $\underline{\underline{S}}$ is the $1 \times 1$ matrix with element $z - J_{11}$, $\underline{\underline{T}}$ is the $1 \times (N-1)$ row vector $(J_{1,2}, J_{1,3}, \cdots, J_{1,N})$, $\underline{\underline{U}}$ is the $(N-1) \times 1$ column vector $(J_{1,2}, J_{1,3}, \cdots, J_{1,N})^T$, and $\underline{\underline{V}}$ is the $(N-1)\times (N-1)$ matrix formed by deleting the first row and column of $z\underline{\underline{\id}}- \underline{\underline{J}}$. One therefore finds by inspecting the top-left block of Eq.~(\ref{blockinverse})
\begin{align}
	G_{11} = \frac{1}{z - J_{11} - \sum_{j,k=2}^N J_{1j} G^{(1)}_{jk} J_{k1}}, 
\end{align}
where $\underline{\underline{G}}^{(1)}$ is the resolvent of the $(N-1)\times (N-1)$ matrix $\underline{\underline{J}}^{(1)}$, which is formed from $\underline{\underline{J}}$ by deleting its first row and column. It is sometimes referred to as the \textit{cavity} resolvent matrix. Since there is nothing special about the index $1$, we may also write
\begin{align}
	G_{ii} = \frac{1}{z - J_{ii} - \sum_{j,k\neq i}^N J_{ij} G^{(i)}_{jk} J_{ki}}. \label{cavitydiagonal}
\end{align}
Although the diagonal elements of the resolvent matrix are what we require for the computation of the eigenvalue density, we also must consider the off-diagonal elements, which appear in the sum in the denominator of Eq.~(\ref{cavitydiagonal}). Considering the top-right block of Eq.~(\ref{blockinverse}), one obtains in a similar fashion
\begin{align}
	G_{ij} = - \frac{\sum_{k\neq i} J_{ik} G^{(i)}_{kj}}{z - J_{ii} - \sum_{j,k\neq i} J_{ij} G^{(i)}_{jk} J_{ki}} = - G_{ii} \sum_{k\neq i} J_{ik} G^{(i)}_{kj} .\label{cavityoff}
\end{align}
The strategy for finding $G_{ii}$ now goes as follows. We examine the sum $\sum_{j,k\neq i}^N J_{ij} G^{(i)}_{jk} J_{ki}$, which appears in the denominators of both Eq.~(\ref{cavitydiagonal}) and (\ref{cavityoff}), and we split the sum into two parts: one involving only the diagonal elements $G^{(i)}_{jj}$ and one involving only the off-diagonals. First, one shows that the sum involving the diagonal elements $G^{(i)}_{jj}$ concentrates on its average (i.e. the random fluctuations between realisations of the matrix elements $J_{ij}$ is negligible), and that this average has a non-zero value. Secondly, one shows that one can neglect the off-diagonal elements of the resolvent $G^{(i)}_{kj}$ in the sum, since their contribution concentrates on a value of zero in the limit $N\to \infty$. Ultimately, one finds that $\sum_{j,k\neq i}^N J_{ij} G^{(i)}_{jk} J_{ki} \to\sum_{j\neq i}^N \langle J^2_{ij}\rangle G^{(i)}_{jj} \to G(z)$ for all values of $i$ when $N \to \infty$, which will enable us to solve for $G(z)$. We will thus be able to extract the eigenvalue density $\rho(\omega)$.

So as not to get sidetracked by technicalities, we will temporarily alleviate ourselves of this second task, and we will simply assume for now that we may neglect the elements $G_{jk}^{(i)}$ for $j \neq k$ in the sum in the denominator of Eq.~(\ref{cavitydiagonal}). We discuss why this is justified in more detail in Section \ref{section:offdiag}. As a brief argument, one can see at from Eq.~(\ref{cavityoff}) that if $G_{ii}(z) \approx \langle G_{ii} \rangle$ (which we demonstrate presently), then $\langle G_{ij} \rangle = 0$. With some more careful consideration, one can also see that the fluctuations of $G_{ij}$ are also vanishing for large $N$. One therefore expects that the off-diagonal cavity resolvent elements $G_{jk}^{(i)}$ ought to behave similarly, and are thus negligible. 


\subsection{Concentration of the large sum and universality}\label{section:cavityconcentration}
Using that the off-diagonal elements of the resolvent are not relevant in the thermodynamic limit $N \to \infty$, we have from Eq.~(\ref{cavitydiagonal})
\begin{align}
	G_{ii} \approx \frac{1}{z -J_{ii} - \sum_{j\neq i} J_{ij}^2 G_{jj}^{(i)}} . \label{cavity}
\end{align}
Let us then consider the sum in Eq.~(\ref{cavity}). Since we are dealing with a large sum of i.i.d. random variates, we will exploit the central limit theorem to show that we can ignore its statistical fluctuations, and we will derive its limiting value as $N \to \infty$.

Since we are using the central limit theorem, it becomes clear that the limiting result that we obtain will apply far more generally than just to the Gaussian random matrix ensemble introduced in Eq.~(\ref{goedef}). In fact, the only important information about the random matrix entries that we need is that each $J_{ij} = J_{ji}$ is drawn independently from a distribution (not necessarily Gaussian) with the following mean and variance
\begin{align}
	\langle J_{ij} \rangle = 0, \hspace{2cm} \left\langle J_{ij}^2\right\rangle = \frac{1}{N}, \label{statisticswigner}
\end{align}
where here the angular brackets indicate an average over realisations of the matrix entries. Crucially, we also assume that all higher order moments $\langle (J_{ij})^r\rangle$ with $r \geq 3$ decay more quickly with $N$ than $1/N$, and that the diagonal entries satisfy $\langle J_{ii}\rangle = 0$ and $\langle J_{ii}^2\rangle = a/N$. 

Under these circumstances, one obtains the same limiting eigenvalue density as in Fig. \ref{fig:semicircle}, regardless of the underlying distribution of the elements $\{J_{ij}\}$. This is demonstrated numerically later in Fig. \ref{fig:semicircleuniversal}. The phenomenon of the same result applying to many random matrix ensembles is referred to as \textit{universality}.

Let us now evaluate the large sum in Eq.~(\ref{cavity}). Since by definition $J_{ij}$ and $G^{(i)}_{jj}$ are statistically independent, the mean of this sum is given by (for large $N$)
\begin{align}
	\left\langle \sum_{j\neq i} J_{ij}^2 G_{jj}^{(i)} \right\rangle =  \sum_{j\neq i} \left\langle J_{ij}^2 \right\rangle  \left\langle G_{jj}^{(i)} \right\rangle = \frac{(N-1)}{N}  \left\langle G_{jj}^{(i)} \right\rangle \approx  \left\langle G_{jj}^{(i)} \right\rangle .
\end{align}
where we use the statistics given in Eq.~(\ref{statisticswigner}). Let us now consider the variance of the real part of the sum. Defining $\mathrm{Re}[G_{jj}^{(i)}] \equiv R_{jj}^{(i)}$, one finds
\begin{align}
	&\sum_{j,k\neq i} \left[\left\langle  J_{ij}^2 R_{jj}^{(i)}  J_{ik}^2 R_{kk}^{(i)} \right\rangle - \left\langle  J_{ij}^2 R_{jj}^{(i)} \right\rangle \left\langle  J_{ik}^2 R_{kk}^{(i)} \right\rangle \right]\approx \sum_{j\neq i} \left[ \left\langle J_{ij}^4\right\rangle \left\langle \left(R_{jj}^{(i)}\right)^2   \right\rangle - \left\langle J_{ij}^2\right\rangle^2  \left\langle R_{jj}^{(i)}   \right\rangle^2\right] \nonumber \\
	&= O\left(N \left\langle J_{ij}^4\right\rangle\right) . \label{vardiagcav}
\end{align}
In the second line, we have assumed that the correlators $\langle R_{jj}^{(i)}R_{kk}^{(i)} \rangle -\langle R_{jj}^{(i)} \rangle \langle R_{kk}^{(i)} \rangle$ (for $j\neq k$) vanish for $N \to \infty$.\footnote{One can convince oneself that this is reasonable by considering the power series [c.f. Eq.~(\ref{resdef})] $R_{kk}^{(i)}(z) = \frac{1}{z} + \frac{1}{z^2} J_{kk} + \frac{1}{z^3} \sum_{l\neq i} J_{kl} J_{lk} +  \frac{1}{z^4} \sum_{l,m\neq i} J_{km} J_{ml}J_{lk} + \cdots$. We see that the surviving terms in $\langle R_{jj}^{(i)}R_{kk}^{(i)} \rangle -\langle R_{jj}^{(i)} \rangle \langle R_{kk}^{(i)} \rangle$ involve quantities such as $\langle J_{kl}^4\rangle - \langle J_{kl}^2\rangle^2$, which indeed vanish when $N \to \infty$.} A similar computation to the above can be performed for the imaginary part of the sum $\sum_{j\neq i} J_{ij}^2 G_{jj}^{(i)}$. We therefore see that in the limit $N\to \infty$, the large sum tends towards its average value as long as $N\left\langle J_{ij}^4\right\rangle \to0$ as $N\to \infty$. 

We may simplify matters further by noticing that in the present case the elements $J_{ii} \sim 1/N$, are negligible in Eq.~(\ref{cavity}). Thus, having reasoned that the sum $ \sum_{j\neq i} J_{ij}^2 G_{jj}^{(i)}$ concentrates on its average, we see from Eq.~(\ref{cavity}) that all $G_{ii} = G(z)$ are equal, irrespective of the index $i$. If we also make the approximation that $G^{(i)} = N^{-1}\sum_j G_{jj}^{(i)} \approx G(z)$, since $G^{(i)}$ is just the resolvent of a random matrix with dimension $N-1$, one arrives at the following expression for $G(z)$ (valid for large $N$)
\begin{align}
	G(z) \approx \frac{1}{z - G(z)}. \label{reseqsemi}
\end{align}
We have thus succeeded in deriving an expression for the resolvent in the limit $N \to \infty$, which we can solve.

\newpage

\subsection{The Wigner semicircle law}\label{section:semicircleresult}
All that remains for us is to solve Eq.~(\ref{reseqsemi}) for $G(z)$ and use the inversion formula in Eq.~(\ref{densefromres}) to obtain the eigenvalue density. Solving for the resolvent, we obtain the expression $G(z) = \frac{1}{2}\left[ z \pm \sqrt{z^2-4} \right]$. 

Choosing the correct sign here is a bit tricky. In principle, we could stop here, and simply use our knowledge that the eigenvalue density must be positive to extract $\rho(\omega)$ from $G(z)$ using Eq.~(\ref{densefromres}). However, for the sake of completeness, and for future calculations where the correct form of $G(z)$ is actually important, we provide the correct form in full.

\begin{figure}[h]
	\centering 
	\includegraphics[scale = 0.47]{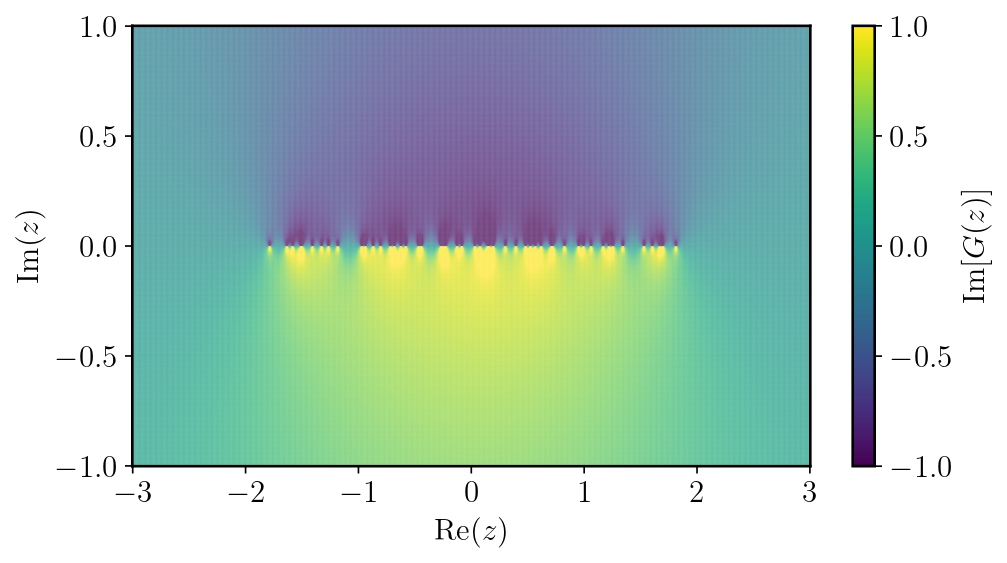} 
	\includegraphics[scale = 0.47]{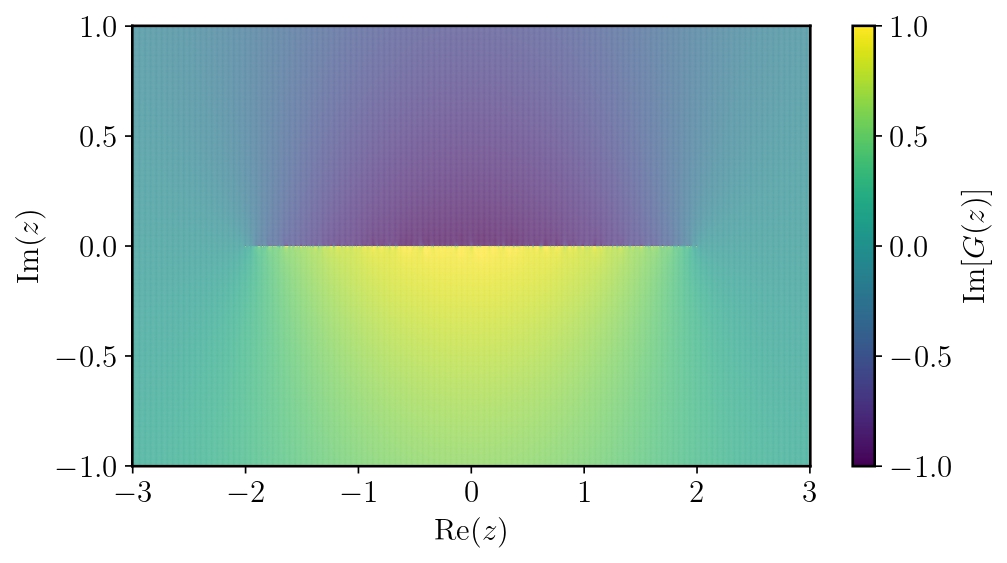} 
	\includegraphics[scale = 0.47]{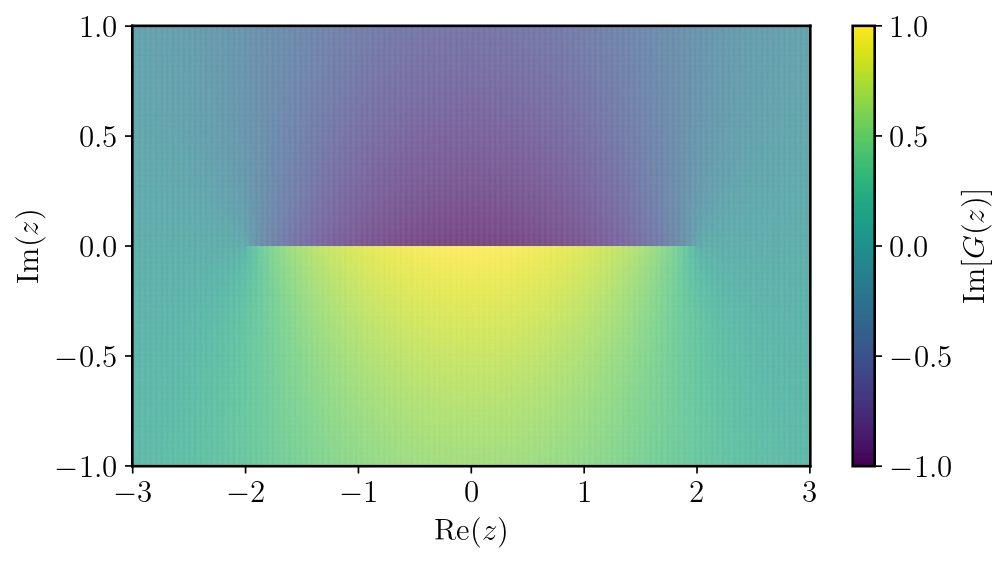} 
	\captionsetup{justification=raggedright,singlelinecheck=false}
	\caption{Comparison of $\mathrm{Im}[G(z)]$ for a single realisation of the random matrix for $N = 50$ (top-left) and $N = 1000$ (top-right) to the theory in Eq.~(\ref{ressemicircle}) (bottom). As $N$ is increased, the poles corresponding to the eigenvalues coalesce to form a smooth branch cut on the real axis between $-2$ and $2$. }\label{fig:resolvents}
\end{figure}

As is conventional, the multivalued function $f(z) = z^{1/2}$ is given a branch cut on the negative real axis of the $z$ plane. Following this convention, our function $G(z)$ would get a branch cut on the real axis for $\vert z\vert \leq2$. This region of non-analyticity corresponds to the locations of our eigenvalues. However, we would also obtain a branch cut on the whole imaginary axis if we were to blindly select one fixed sign in front of the radical, say, $G(z) = \frac{1}{2}\left[ z + (z^2-4)^{1/2} \right]$. The correct solution for $G(z)$, which has only one branch cut on the real axis and obeys the boundary condition $G(z) \to 1/z$ as $\vert z \vert\to \infty$ [as must be the case from our definition Eq.~(\ref{resolventasstieltjes})], is
\begin{align}
	G(z) = \frac{z - \mathrm{sign}[\mathrm{Re}(z)]\sqrt{z^2 - 4}}{2}. \label{ressemicircle}
\end{align}
We compare Eq.~(\ref{ressemicircle}) to the quantity $G(z)$ as obtained numerically from single instances of a random matrix in Fig. \ref{fig:resolvents}, thus verifying that we have indeed obtained the correct analytic structure of $G(z)$. We note that for $0<\mathrm{Re}(z)<2$ and $\mathrm{Im}(z)<0$, we have $\mathrm{Im}\sqrt{z^2 - 4}<0$, and for $-2<\mathrm{Re}z<0$ and $\mathrm{Im}(z)<0$, we have $\mathrm{Im}\sqrt{z^2 - 4}>0$, which leads to a positive imaginary part of $G(z)$ below the branch cut, and therefore a positive eigenvalue density $\rho(\omega)$, as required.

\begin{figure}[h]
	\centering 
	\includegraphics[scale = 0.48]{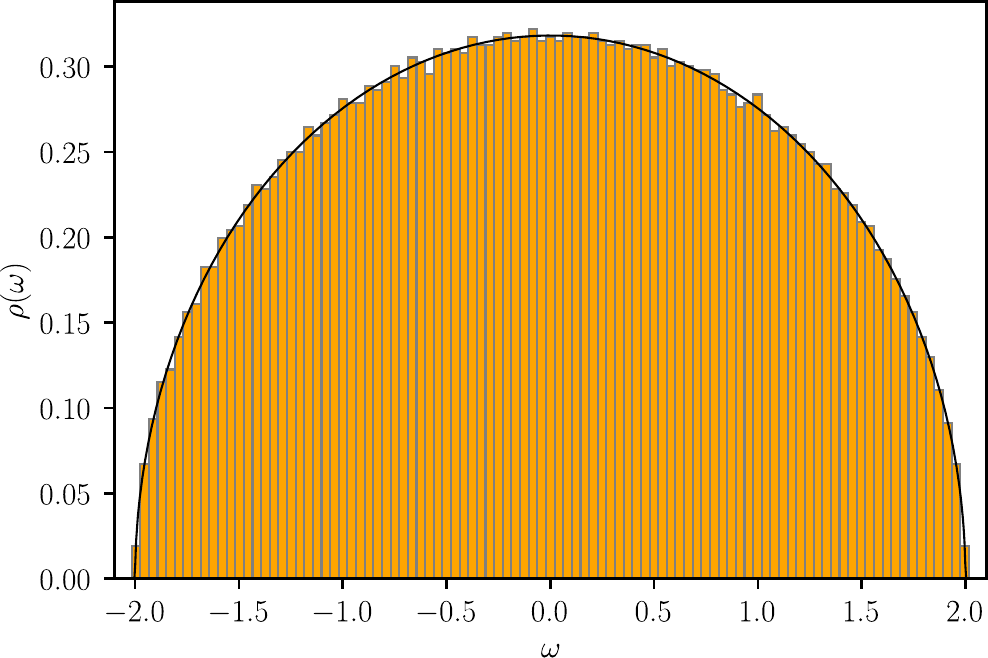} 
	\includegraphics[scale = 0.48]{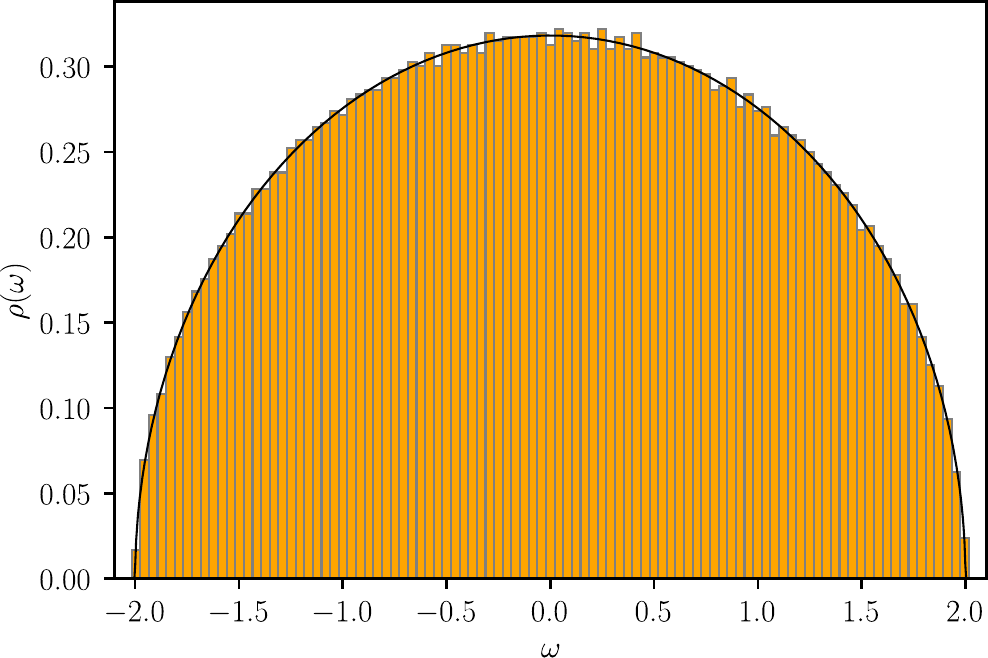} 
	\captionsetup{justification=raggedright,singlelinecheck=false}
	\caption{Universality of the semicircle law. As long as the elements $J_{ij}$ have the statistics in Eq.~(\ref{statisticswigner}) and the higher moments decay sufficiently quickly with $N$, the eigenvalue density obeys the Wigner semicircle law. We have $N = 10^4$ in both panels. (Left) $J_{ij} = \pm 1/\sqrt{N}$ with equal probability. (Right) $J_{ij}$ are uniform random variables distributed on the interval $[-\sqrt{3/N},\sqrt{3/N} ]$.}\label{fig:semicircleuniversal}
\end{figure}

Finally, therefore, using the inverse Stieltjes transform in Eq.~(\ref{densefromres}), we arrive at the celebrated Wigner semicircle law 
\begin{empheq}[box={\fboxsep=6pt\fbox}]{align}
	\rho(\omega) = \frac{1}{2\pi} \sqrt{4 -\omega^2} .\label{semicircle}
\end{empheq}
We have thus accomplished what set out to do. By using the cavity method, combined with the limit $N \to \infty$ and the Stieltjes inversion formula in Eq.~(\ref{densefromres}), we have shown how the resolvent $G(z)$ concentrates on its average, and thus that the eigenvalue density similarly approaches a limiting value. In finding a self-consistent expression for the average $\langle G(z) \rangle$ via the cavity approach, we were able to find this limiting value. We note that in doing this, we did not have to specify any particular distribution of $J_{ij}$ in our calculation. This `universality' of the semicircle law, which is a kind of matrix analogy to the law of large numbers or the central limit theorem, is verified in Fig. \ref{fig:semicircleuniversal}.

As a small side note, we observe that we have argued explicitly here that the eigenvalue density of an \textit{individual} matrix tends to the Wigner semicircle. When using many other approaches (such as the replica, supersymmetric or path integral methods discussed in Part 2 of these notes), one must calculate explicitly how the eigenvalue density fluctuates between realisations of the random matrix in order to demonstrate this fact. The necessary computation of two-point Green's functions to do this is highly non-trivial. The cavity method discussed above has the distinct advantage that the convergence of the diagonal resolvent elements to their mean for $N \to \infty$ is readily apparent.

\newpage

\subsection{The off-diagonal elements of the resolvent matrix are negligible}\label{section:offdiag}
To tie up the loose ends of the above calculation, let us now consider the contribution of the off-diagonal elements $G_{jk}^{(i)}$ to the sum in the denominator of Eq.~(\ref{cavitydiagonal}), and show that it is indeed negligible. The mean of this sum is, rather simply,
\begin{align}
	\left\langle \sum_{\substack{j,k\neq i\\ j\neq k}} J_{ij} G^{(i)}_{jk} J_{ki} \right\rangle = \sum_{\substack{j,k\neq i\\ j\neq k}} \langle J_{ij} \rangle \langle G^{(i)}_{jk} \rangle \langle J_{ki} \rangle = 0. \label{offdiagsum}
\end{align}
To understand the fluctuations between realisations of the matrix entries of this sum, we must be more careful. The variance of the real part is given by
\begin{align}
	&\left\langle \sum_{\substack{j,k\neq i\\ j\neq k}} \sum_{\substack{l,m\neq i\\ l\neq m}} J_{ij} R^{(i)}_{jk} J_{ki} J_{il} R^{(i)}_{lm} J_{mi} \right\rangle =  \frac{2}{N^2}\sum_{\substack{j,k\neq i\\ j\neq k}}  \left\langle\left(  R^{(i)}_{jk} \right)^2   \right\rangle \nonumber \\
	&= 2 \frac{N(N-1)}{N^2} \left\langle\left(  R^{(i)}_{jk} \right)^2   \right\rangle \approx   2 \left\langle\left(  R^{(i)}_{jk} \right)^2   \right\rangle.
\end{align}
All that remains is for us to find the value of $\left\langle\left(  R^{(i)}_{jk} \right)^2\right\rangle$ for $N \to \infty$, and similarly for the imaginary part. This is accomplished by inspecting Eq.~(\ref{cavityoff}), and it turns out that $\left\langle\left(  R^{(i)}_{jk} \right)^2\right\rangle\to 0$. In demonstrating this, we thus also demonstrate that the off-diagonal elements of the resolvent can be ignored in the computation of $G_{ii}$ in Eq.~(\ref{cavitydiagonal}).

We can show that $\left\langle\left(  R^{(i)}_{jk} \right)^2\right\rangle\to 0$ is a consistent solution as follows. If indeed it is the case that $\left\langle\left(  R^{(i)}_{jk} \right)^2\right\rangle \to 0$, then the sum $\sum_{\substack{j,k\neq i\\ j\neq k}} J_{ij} G^{(i)}_{jk} J_{ki}$ concentrates on its mean of zero. One then has that $\sum_{j,k\neq i} J_{ij} G^{(i)}_{jk} J_{ki} \to  \sum_{j\neq i} J_{ij}^2 G^{(i)}_{jj} \to \langle G^{(i)}_{jj} \rangle$. This means that we can write $G_{ij} \approx -\langle G_{ii} \rangle \sum_{k\neq i} J_{ik} G_{kj}^{(i)}$ for large $N$, which manifestly has zero mean. Computing the variance of the real part of $\sum_{k\neq i} J_{ik} G_{kj}^{(i)}$, we see that
\begin{align}
	 \mathrm{Var}\left(\sum_{k\neq i} J_{ik} R_{kj}^{(i)} \right) &= \sum_{k\neq i}  \left\langle J_{ik}^2 \right\rangle  \left\langle \left(R_{kj}^{(i)} \right)^2 \right\rangle +  \sum_{l \neq k\neq i} \left\langle J_{ik}\right\rangle \left\langle J_{il} \right\rangle  \left\langle \left(R_{kj}^{(i)} \right) \right\rangle \left\langle \left(R_{lj}^{(i)} \right) \right\rangle \nonumber \\
	&\approx  \left\langle \left(R_{kj}^{(i)} \right)^2 \right\rangle + \frac{1}{N}  \left\langle \left(R_{jj}^{(i)} \right)^2 \right\rangle  . 
\end{align}
Therefore, if $\left\langle \left(R_{kj}^{(i)} \right)^2 \right\rangle \approx   \left\langle \left(R_{ij} \right)^2 \right\rangle $ for large $N$ (which we would expect), then we have $\left\langle \left(R_{ij}\right)^2 \right\rangle \sim 1/N$, and we may draw the same conclusion about the imaginary part. That is, in summary, the off-diagonal elements have an average of zero, and their fluctuations about this value vanish as $N\to \infty$. We can therefore safely ignore them in our computations. 

\subsection{Statistics of eigenvalue spacings beyond the mean -- Wigner's surmise}\label{section:surmise}
Now that we have successfully performed our first random matrix theory calculation, we return to Wigner and his interest in nuclear physics. In finding the eigenvalue density, given by Wigner's semicircle law in Eq.~(\ref{semicircle}), we have essentially deduced the \textit{average} spacing of eigenvalues at a point on the real line, which is given by
\begin{align}
	\overline{\Delta} = \frac{1}{N\rho(\omega)}. 
\end{align}
For a finite but large value of $N$, however, the eigenvalues of an individual realisation of the matrix $\underline{\underline{J}}$ will take random values on the real line. If we look at the distances between each pair of neighbouring eigenvalues, these distances will fluctuate about the value $\overline{\Delta}$. 

In many applications, particularly those involving quantum mechanics, where the eigenvalues represent the energy levels of a complicated Hamiltonian, the statistics of the differences between eigenvalues is of central importance. It is these differences that determine the wavelength of light, for example, that is emitted by nuclei when they transition from one energy eigenstate to another. In fact, it was precisely this application that motivated Wigner to make his first forays into RMT. Specifically, to compare with experiment, we desire statistical information (beyond the mean) about the nearest-neighbour spacings. Let us consider how we might access this information. 

Wigner first presented his beautifully simple approximation (known as his `surmise') for the nearest-neighbour spacing statistics at a conference on nuclear physics in 1956 \cite{wigner1957conference}. Wigner introduces his approach to the eigenvalue spacings with the following:
\refquote{Perhaps 1 am now too courageous when I try
	to guess the distribution of the distances between
	successive levels. [...] Theoretically the situation is quite simple if one attacks it in a simple-minded fashion.}

Let us then proceed `simple-mindedly', as Wigner humbly put it. We take the case $N = 2$, and we return to the specific example of the GOE random matrix ensemble, defined in Eq.~(\ref{goedef}). We aim to compute the distribution of the spacing between the two eigenvalues of this simple matrix ensemble. Remarkably, we shall see that this special case can approximate the eigenvalue spacing distribution more generally. 

\begin{figure}[H]
	\centering 
	\includegraphics[scale = 0.6]{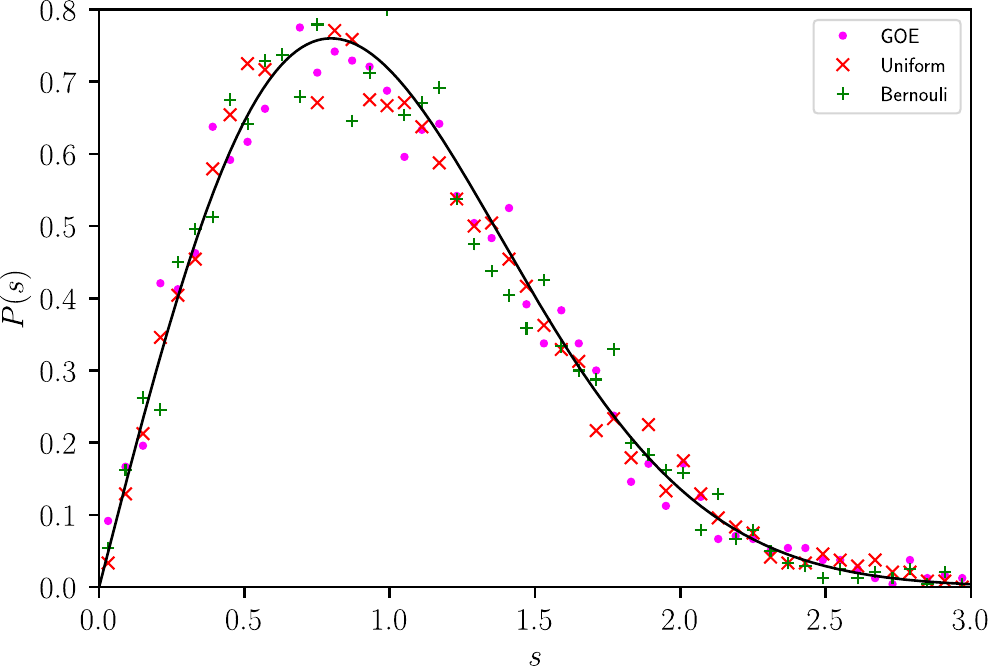} 
	\captionsetup{justification=raggedright,singlelinecheck=false}
	\caption{Comparison of numerical diagonalisation results to the Wigner surmise in Eq.~(\ref{wignersurmise}) for several random matrix ensembles, including the GOE and those described in the caption of Fig. \ref{fig:semicircleuniversal}. The histograms were made by unfolding eigenvalues of a single matrix of size $N = 4000$.}\label{fig:wignersurmise}
\end{figure}

The two eigenvalues in the special $2 \times 2$ case are given simply by
\begin{align}
	\lambda_\pm = \frac{1}{2}\left[ J_{11} + J_{22} \pm \sqrt{( J_{11} + J_{22})^2- 4 (J_{11}J_{22} - J_{12}^2)}\right],
\end{align}
so that the eigenvalue spacing is 
\begin{align}
	\Delta =  \sqrt{(J_{11} - J_{22})^2 +4 J_{12}^2}.
\end{align}
In the case of the GOE, $X = J_{11} - J_{22}$ is a Gaussian random variable with variance $4/N = 2$ and $Y = 2J_{12}$ is also a Gaussian random variable with variance $4/N = 2$. We can therefore compute the spacing distribution via
\begin{align}
	P(\Delta) &= \int dX dY \frac{1}{4 \pi } e^{-X^2/4} e^{-Y^2/4} \delta\left(\Delta - \sqrt{X^2 + Y^2} \right)  \nonumber \\
	&= \int^\Delta_{-\Delta} dX \frac{1}{2 \pi } e^{-X^2/4} e^{-(\Delta^2 -X^2)/4} \frac{\Delta}{\sqrt{\Delta^2 - X^2}} \nonumber \\
	&= \frac{\Delta}{2} e^{-\Delta^2/4}.
\end{align}
Of course, this result is exact only for the GOE and only in the case $N=2$. What we would like is to obtain an approximate result, based on this one, that is valid for larger $N$. We `surmise' that for arbitrary $N$ the spacing distribution is of the same form $P(\Delta) = a \Delta e^{-b\Delta^2}$. However, we now consider a slightly different variable -- the so-called `unfolded' eigenvalue spacing, defined as
\begin{align}
	s_i = N \langle\rho\left(\bar \lambda_i\right)\rangle\left(\lambda_{i+1} - \lambda_i\right), 
\end{align}
where $\langle \rho\left(\cdot\right)\rangle$ in the case of the GOE is given by the semicircle law in Eq.~(\ref{semicircle}), and we define the midpoint $\bar \lambda_i = (\lambda_{i+1}+\lambda_i)/2$. The quantity $s_i$ is the eigenvalue spacing rescaled such that the average spacing is unity. Now, choosing $a$ and $b$ such that the ansatz $P(s) = a s e^{-bs^2}$ is normalised and has unit mean, we obtain the Wigner surmise for the unfolded eigenvalue spacings
\begin{empheq}[box={\fboxsep=6pt\fbox}]{align}
	P(s) = \frac{\pi s}{2} e^{-\pi s^2/4}.\label{wignersurmise}
\end{empheq}
We can test this guess against numerical experiments (see Fig. \ref{fig:wignersurmise}). Astonishingly, we see that there is good agreement, not only for larger $N$ in the GOE case, but also for other non-Gaussian random matrix ensembles. The Wigner surmise has also been shown to be in very close agreement with the exact analytical result for the nearest neighbour spacings for large $N$, which has no simple closed form \cite{gaudin1961loi}.

We thus see how the eigenvalue spacings fluctuate about their mean in a highly universal fashion. The above derivation of Wigner's surmise for the level spacing is straightforward, but its accuracy may seem like something of a miracle at this stage. Later, we will discuss how this universal success can be better understood through Dyson's Brownian motion construction [see Section \ref{section:surmiserevisited}]. 

There are other related quantities -- the eigenvalue density correlations $\rho_c(\omega,\mu)\equiv\langle \rho(\omega) \rho(\mu)\rangle - \langle \rho(\omega) \rangle \langle \rho(\mu) \rangle$, for example -- which also encapsulate the fluctuations of the eigenvalues, but can be derived in a much more thorough manner \cite{efetov1983supersymmetry}. The results for the density fluctuations are also extraordinarily universal, as has been shown analytically in the literature \cite{erdHos2011universality}. Results derived for the GOE, for example, also hold for a very wide class of large symmetric real matrices that possesses the statistics in Eq.~(\ref{statisticswigner}), even those for which the higher moments are non-negligible \cite{mirlin1991universality}, and also those for which the elements are non-iid \cite{erdHos2011universality}. 

\subsection{Comparison with nuclear spectral data and other applications}
Quantum physics teaches us that, in studying the very small, we should promote observable quantities to operators. The eigenvalues of these operators are the possible values that we can observe when we take a measurement. For example, one defines a so-called Hamiltonian operator $\hat H$, whose eigenvalues tell us the possible energies of the system in question. The form of the Hamiltonian is system-dependent, and it captures the essential physics. If the system is subject to constraining potential, it is often the case that there are only a discrete set of allowed energy states -- the energy levels become \textit{quantised}. The nucleus of an atom is such a quantum system. The differences between the discrete energy levels of the nucleus can be detected by the gamma radiation that is emitted when transitions between levels occur, or via neutron scattering experiments \cite{weidenmuller2009random}.

The approach made famous by Wigner, Mehta \cite{mehta2004random}, Dyson \cite{dyson1962statistical, dyson1962threefold} and others was to suppose that the complexities of the nuclear interactions in large nuclei can be captured by a random Hamiltonian matrix, as long as the random matrix in question obeys the correct symmetries. Dyson in particular argued that the GOE ought to be suitable for `most purposes' \cite{dyson1962statistical3}, as we discuss in more detail in Section \ref{section:dysonbrownian}. 

In comparing RMT results to nuclear data, we unfortunately do not expect to see agreement between the overall density $\rho(\omega)$ and experiment \footnote{This is not to say that the Wigner semicircle doesn't have its uses in many other contexts, of course.}. The reason for this is that the Wigner semicircle is not sufficiently robust to changes of the random matrix model (see Section \ref{section:dysonbrownian} for examples of this), and thus the semicircle itself is not seen in data (Weidenm\"uller and Mitchell argue this point in Section II C2 of Ref. \cite{weidenmuller2009random}).  However, the eigenvalue spacing statistics do display a far greater robustness to changes in the random matrix ensemble, making them a much better candidate for comparison to experiment. 

While it is already fairly amazing that Wigner's surmise does such a good job of describing the eigenvalue separations of such a broad class of random matrices, what is truly remarkable is its ability also to capture the statistics of quantum energy level spacings, particularly in nuclear spectra. Decades after Wigner's initial proposal of the surmise, sufficiently accurate and plentiful nuclear data had finally been accumulated to make precise tests of RMT predictions. Figures 1.3 and 1.4 in Mehta's book \cite{mehta2004random} demonstrate the agreement with the GOE Wigner surmise. Haq, Pandey and Bohigas \cite{haq1982fluctuation} also found extraordinary agreement with GOE predictions by carefully unfolding the spectra of many heavy nuclei and amalgamating the results.

Agreement with the Wigner surmise has also been found in other contexts. Particularly, in quantum chaotic systems (the quantum billiard \cite{bohigas1984characterization} for example), the GOE-inspired Wigner surmise Eq.~(\ref{wignersurmise}) has been shown to agree with the level statistics. In general, eigenvalue separation statistics that obey RMT predictions are widely accepted as a signature of quantum chaos. Perhaps most bizarre of all, however, is the fact that the imaginary parts of the non-trivial zeros of the Riemann zeta function appear to obey RMT separation statistics \cite{keating2000random}. This was famously first discovered when Montgomery \cite{montgomery1973pair} showed his results to Dyson, who immediately recognised the expression for $\rho_c(\omega,\mu)$ for the GUE (see exercise below). The study by Odlyzko \cite{odlyzko1987distribution} compares the zeros with the GUE-inspired Wigner surmise with astounding agreement (see Fig. 4 in this reference in particular). 

\subsection{Exercises}
\begin{itemize}
	\item First of all, making a histogram of eigenvalues, and obtaining the semicircle law numerically, is a rite of passage for every random matrix theory practitioner. I would enthusiastically recommend the reader to reproduce Fig. \ref{fig:semicircle}.
\end{itemize}

As will become apparent in these notes, there exists a multitude of ways to derive the semi-circle law (as shown in particular in Part \ref{part:additionaltools}). Here, we briefly consider an alternative approach to the one discussed above, which is similar to the used in Ref. \cite{bray1979evidence}. We begin with the definition of the resolvent matrix in Eq.~(\ref{resdef}). 
\begin{itemize}
	\item Show that one can write \begin{align}
		\underline{\underline{G}} = \frac{1}{z} \underline{\underline{\id}}+ \frac{1}{z}\underline{\underline{G}}\, \underline{\underline{J}}.
	\end{align}
	Thus, by reinserting the expression for $\underline{\underline{G}}$ into the right-hand side, show that the normalised trace of the resolvent can be written as an infinite series (known as a Dyson series)
	\begin{align}
		G(z) =& \frac{1}{z} + \frac{1}{z^2N} \sum_i J_{ii} + \frac{1}{z^3 N} \sum_{i,j} J_{ij}J_{ji} \nonumber \\
		& + \frac{1}{z^4N} \sum_{i,j,k} J_{ij}J_{jk} J_{ki} + \frac{1}{z^5N} \sum_{i,j,k, l} J_{ij}J_{jk} J_{kl} J_{li} + \cdots .  \label{resseries}
	\end{align}
	\item Using the expression for the resolvent in Eq.~(\ref{ressemicircle}), which is valid in the limit $N\to \infty$, show that 
	\begin{align}
		G(z) = \frac{1}{z} + \frac{1}{z^3} + \frac{2}{z^5}  +\cdots. \label{expandedres}
	\end{align}
	The question is, could we have obtained this series directly from Eq.~(\ref{resseries})?
	\item Neglecting fluctuations about the mean, we assume that $G(z) = \langle G(z)\rangle$ in the limit $N\to \infty$. Argue that the terms with odd multiples of $\underline{\underline{J}}$ in Eq.~(\ref{resseries}) do indeed vanish [after taking the ensemble average], if $J_{ij} = J_{ji}$ obeys the statistics in Eq.~(\ref{statisticswigner}). 
	\item Show that 
	\begin{align}
		N^{-1}  \sum_{ijkl}\left\langle J_{ij} J_{jk}J_{kl}J_{li} \right\rangle = N^{-1}\sum_{(i,j, k)} \left[\left\langle J_{ij}^2 \right\rangle  \left\langle J_{ik}^2 \right\rangle + \left\langle J_{ij}^2 \right\rangle  \left\langle J_{jk}^2 \right\rangle \right]  + N^{-1}\sum_{(i,j)}\left\langle J_{ij}^4 \right\rangle,
	\end{align}
	where $(i,j,k)$ indicates a combination of the indices where none take the same value. Hence, show that Eq.~(\ref{expandedres}) and Eq.~(\ref{resseries}) agree up to terms proportional to $1/z^5$.
\end{itemize}
In Section \ref{section:wignerapproach}, we continue this approach with the assistance of Feynman diagrams, which help to keep track of the combinatorics, and we re-derive the semicircle law. This approach is closely related to Wigner's original derivation. \\

Let us now consider an alternative ensemble to the GOE, the so-called Gaussian Unitary Ensemble (GUE). In this case, the random matrix elements are independent complex numbers up to the Hermiticity constraint $J_{ij} = J_{ji}^\star$ drawn such that $J_{ij} = U_{ij} + i S_{ij}$ where $U_{ij}$ and $S_{ij}$ are independent Gaussian random variables with variance $1/(2N)$. The diagonal elements are given by $J_{ii} = \sqrt{2}U_{ii}$, so that $\langle J_{ii}^2 \rangle = 1/N$. The joint probability of the matrix elements can therefore be written
\begin{align}
	P(\underline{\underline{J}}) \propto \exp\left[ -\frac{N}{2} \mathrm{Tr} \left(\underline{\underline{J}}\,\underline{\underline{J}}^\dagger \right)\right],
\end{align}
where $\underline{\underline{\dagger}}$ is the Hermitian conjugate. 
\begin{itemize}
	\item Show that $\langle J_{ij}^2 \rangle = 0$ and $\langle J_{ij} J_{ji}\rangle = 1/N$.
	\item Beginning with Eq.~(\ref{cavitydiagonal}), argue that the eigenvalue density still satisfies the semi-circle law Eq.~(\ref{semicircle}).
	\item Show that the Wigner surmise in this case is instead
	\begin{align}
		P(s) = \frac{32 s^2}{\pi^2} e^{-4s^2/\pi}. 
	\end{align}
\end{itemize}
As it turns out, the imaginary parts of the zeros of the Riemann zeta function possess these spacing statistics \cite{odlyzko1987distribution}.

\newpage

\section{The elliptic law and May's model ecosystem}\label{section:ellipse}
\begin{quotation}Not only in research, but also in the everyday world of politics and economics, we would all be better off if more people realised that simple non-linear systems do not necessarily possess simple dynamical properties. -- Robert M. May\end{quotation}

\subsection{Non-Hermitian matrices}
In the previous section, we considered the eigenvalues of symmetric real random matrices. Because of the symmetry constraint, all the eigenvalues were real, and we were able to compute the density of eigenvalues on the real line using the resolvent method. However, in many applications, this symmetry constraint is not obeyed. For example, random matrices can be used to represent the interactions in a complex ecosystem. In this context, the constraint of symmetry has very little relevance. After all, why should it be that, say, a population of rabbits is affected by foxes in the same way that the foxes are affected by the rabbits?

When the matrix symmetry constraint is violated, the eigenvalues can invade the complex plane. This means that the methodology that we developed to understand the behaviour of the eigenvalues on the real axis must be modified. We must therefore adapt the resolvent/cavity method to handle a 2D eigenvalue density. In this way, we obtain analogous limiting results to the Wigner semicircle law (namely, the Girko elliptic law) for the eigenvalue density in the complex plane.

In this section, we focus on the application of complex ecosystems, but the results that we discuss here also apply to models of neural networks such as the firing rate model, for which the eigenvalues of a random matrix govern the transition from a quiescent to a chaotic state \cite{ rajan2006eigenvalue, sompolinsky1988chaos, aljadeff2015transition, molgedey1992suppressing}. Non-Hermitian random matrices are also important in quantum systems with dissipation or decay \cite{haake1992statistics, feinberg1997non}.

\subsection{May's model}
In the 1970s, Robert May \cite{may1972will} introduced his toy model of a complex ecosystem. His work `Will a large complex system be stable?' was intended as something of a polemic. The prevailing point of view at the time was that ecosystems are so intricate \textit{because} this helps them to be stable. The na\"ive thinking was that if the ecosystem interaction network were sufficiently complex, a perturbation due to the removal of a single species, for example, would be more easily `absorbed' by a large network, since the other components would have more options to adjust, and thus accommodate the perturbation. May's reasoning turns this argument on its head. He provides a counter example, using random matrix theory, where more interaction complexity tilts the system towards instability, not away from it. May provokes us to be more careful when seeking to understand stability in complex systems, prompting us towards `elucidating the devious strategies of nature which make for stability in enduring natural systems' \cite{may2001stability}.

Following May \cite{may1972will}, we imagine that we have an ecosystem with many species, whose abundances we label $x_i(t)$. The abundances undergo some (unspecified) coupled dynamics. Suppose now that there is some fixed point to the dynamics. We ask: under what conditions will this fixed point be stable? May's null model was to posit that the entries of the \textit{Jacobian matrix} of the system linearised about this hypothetical fixed point could each be treated as random variables. That is, if we denote the deviation from the hypothetical fixed point as $\delta x_i(t) = x_i(t) - x_i^\star$, then we have
\begin{align}
	\frac{d(\delta x_i)}{dt} = \sum_j \left(J_{ij} - \delta_{ij}\right) \delta x_j, \label{maysystem}
\end{align}
where $\delta_{ij}$ is the Kronecker delta. We have assumed that the diagonal elements of our hypothetical Jacobian matrix are such that the community would be stable without species interactions, and are hence negative (and we set $J_{ii} = 0$). 

May's great realisation was that we can take a statistical view of the interactions, and gain insight as to what statistics lead to stability or instability. We therefore assume that the off-diagonal elements $J_{ij}$ are drawn from a distribution with the following statistics
\begin{align}
	\langle J_{ij} \rangle = \frac{\mu}{N},\,\,\, \langle J_{ij}^2 \rangle - \langle J_{ij} \rangle ^2 = \frac{\sigma^2}{N}, \,\,\, \langle J_{ij} J_{ji} \rangle- \langle J_{ij} \rangle \langle J_{ji} \rangle  = \frac{\Gamma\sigma^2}{N} .\label{ellipsestats}
\end{align}
This could be achieved, for example, by drawing the matrix elements $J_{ij}$ and $J_{ji}$ from a joint Gaussian distribution, but our results will apply more generally than this. In May's original work, he fixed $\mu = \Gamma = 0$, and used $\sigma^2 \propto N$, but we wish to keep the consideration more general here, and use a finite-size scaling that is friendly to analysis. We note that the finite-size scaling can be changed after the result is obtained (see Section \ref{section:stabilityandscaling}).

\begin{figure}[H]
	\centering 
	\includegraphics[scale = 0.47]{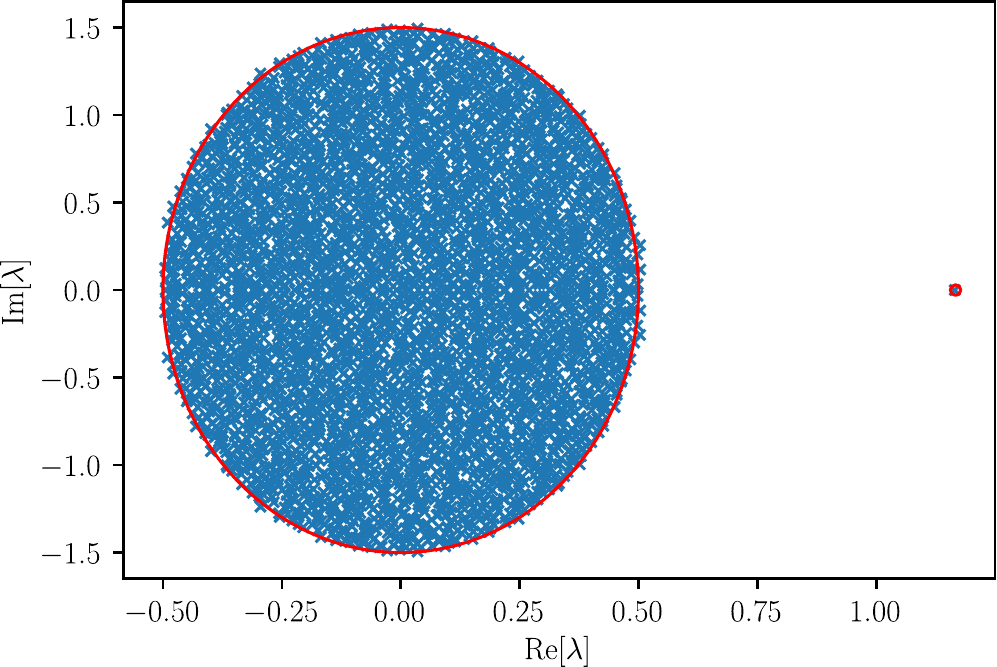} 
	\captionsetup{justification=raggedright,singlelinecheck=false}
	\caption{Example eigenvalue spectra of the matrix $\underline{\underline{J}}$ with statistics as defined in Eq.~(\ref{ellipsestats}) for $\mu = 1.5$, $\sigma = 1$, $\Gamma = -0.5$, $N = 4000$. The eigenvalues of a single realisation of the random matrix are shown as blue crosses. The elliptic law for the spectral boundary [see Eq.~(\ref{ellipticlawfull})] is depicted by the solid red line. The prediction for the outlier eigenvalue [see Eq.~(\ref{outlierellipse})] is shown as the hollow red circular marker.  }\label{fig:ellipseandoutlier}
\end{figure}

Since the eigenvalues of the matrix $\underline{\underline{J}} - \underline{\underline{\id}}$ in Eq.~(\ref{maysystem}) are simply the eigenvalues of $\underline{\underline{J}}$ shifted by $-1$, we conclude that our hypothetical fixed point is unstable to perturbations when any of the real parts of the eigenvalues of $\underline{\underline{J}}$ are greater than 1. Our goal now is thus to understand the nature of the eigenvalue spectrum of $\underline{\underline{J}}$, and under what circumstances the stability of our model ecosystem is possible. An example of the eigenvalue spectrum of $\underline{\underline{J}}$ are shown in Fig. \ref{fig:ellipseandoutlier}. In the following discussion, we will derive the elliptical boundary of the region to which the majority of the eigenvalues are contained, as well as the location of the single outlier, and thus obtain criteria for (in)stability in terms of the ecosystem interaction statistics.

\subsection{2D eigenvalue density and the non-Hermitian resolvent}

To find the eigenvalue density in the Hermitian case, we exploited the fact that all of the eigenvalues were real. However, in the present case where $\underline{\underline{J}}$ is no longer Hermitian, we must generalise our concept of the eigenvalue density, and modify our analytical approach accordingly.

Let us denote a general point in the complex plane as $z = \omega_x + i \omega_y$ and write the eigenvalues of the matrix $\underline{\underline{J}}$ as $\lambda_\nu = \lambda_\nu^{(x)} + i \lambda_\nu^{(y)}$. The 2D eigenvalue density in the complex plane then can be defined as
\begin{align}
	\rho^{(2)}(z) = \frac{1}{N} \sum_\nu \delta^{(2)}(z-\lambda_\nu) = \frac{1}{N} \sum_\nu \delta(\omega_x-\lambda^{(x)}_\nu) \delta(\omega_y-\lambda^{(y)}_\nu) . \label{2Ddens}
\end{align}
We construct this in such a way that we have the normalisation 
\begin{align}
	\int d^2z \rho^{(2)}(z) = \frac{1}{N} \sum_\nu\int_{-\infty}^\infty d\omega_x \int_{-\infty}^\infty d\omega_y \, \delta(\omega_x-\lambda^{(x)}_\nu) \delta(\omega_y-\lambda^{(y)}_\nu) = 1. \label{normalisation2d}
\end{align}
Now that we are considering a 2D eigenvalue density, this means that in the thermodynamic limit, the region of non-analyticity of $G(z)$ [the resolvent defined in Eq.~(\ref{restracedef})] will no longer be confined to the real axis, and thus the formula in Eq.~(\ref{densefromres}) (the inverse Stieltjes transform) no longer applies. The question is therefore whether we can relate the 2D eigenvalue density to something that we can compute easily, i.e. a non-Hermitian version of the Stieltjes transform. We employ a `trick' to this end. First, we introduce a regulariser $\eta$ so that the 2D Dirac delta can be understood as a limiting distribution \cite{nowak2018probing}
\begin{align}
	\delta^{(2)}(z) = \lim_{\eta\to 0} \frac{1}{\pi}\frac{\eta^2}{(\left \vert z\right\vert^2 + \eta^2)^2} .
\end{align}
One can check that, indeed, we have $\int d\omega_x \int d\omega_y \frac{1}{\pi}\frac{\eta^2}{( \omega_x^2 +\omega_y^2 + \eta^2)^2}  = 1$. This identity allows us to see that 
\begin{align}
	\delta^{(2)}(z) = \frac{1}{\pi} \lim_{\eta \to 0}\frac{\partial}{\partial z^\star} \frac{z^\star}{\vert z\vert^2 +\eta^2}, \label{2Ddelta}
\end{align}
where $\star$ indicates the complex conjugate. With this in mind, we can make an analogue of the inverse Stieltjes transform for the non-hermitian case. Inspired by the form of Eq.~(\ref{2Ddelta}), we consider the object 
\begin{align}
	C(z, z^\star) = \frac{1}{N} \mathrm{Tr} \left\{ (z^\star \underline{\underline{\id}} -\underline{\underline{J}}^T ) \left[(z \underline{\underline{\id}} -\underline{\underline{J}} )(z^\star \underline{\underline{\id}} -\underline{\underline{J}}^T ) + \eta^2 \underline{\underline{\id}}  \right]^{-1}\right\} . \label{cdef}
\end{align}
Let us introduce the matrix that has the right eigenvectors of $\underline{\underline{J}} $ as columns, denoted $\underline{\underline{P}}$. Defining $\underline{\underline{Q}} = \underline{\underline{P}}\underline{\underline{P}}^\dagger$ and the diagonal matrix $[\underline{\underline{\Lambda}}(z)]_{\mu \nu} = (z-\lambda_\nu)\delta_{\mu,\nu}$, one can show that
\begin{align}
	C(z, z^\star) &= \frac{1}{N} \mathrm{Tr} \left\{ [\underline{\underline{\Lambda}}(z)]^\star  \left[ \underline{\underline{\Lambda}}(z) [\underline{\underline{\Lambda}}(z)]^\star + \eta^2 \underline{\underline{Q}}[\underline{\underline{\Lambda}}^{-1}(z)]^\star \underline{\underline{Q}}^{-1} [\underline{\underline{\Lambda}}(z)]^\star\right]^{-1}\right\}\nonumber \\
	&\approx \frac{1}{N} \sum_{\nu} \frac{z^\star- \lambda_\nu^\star}{\left\vert z-\lambda_\nu\right\vert^2 + \eta^2}. \label{cintermsofeig}
\end{align}
where the approximation in the second line applies when $\eta\ll 1/N$ is satisfied\footnote{The approximation here works because, when $z$ is away from any eigenvalues, the term proportional to $\eta$ is negligible, and clearly the two expressions agree. When $\vert z-\lambda_\nu\vert < \eta$ on the other hand, the term proportional to $\eta$ dominates over $\underline{\underline{\Lambda}}(z) [\underline{\underline{\Lambda}}(z)]^\star$ due to the factor of $\underline{\underline{\Lambda}}^{-1}(z)$ (when considering the $\nu$\textsuperscript{th} entry). Inverting the term proportional to $\eta^2$ and carrying out the trace, one obtains simply $(z^\star - \lambda_\nu)/(N\eta^2)$, again in agreement with the second line. Alternatively, one may use that $C(z,z^\star) = -\partial_z\Phi(z,z^\star)$ where $\Phi(z,z^\star) = -\ln\det[\eta^2 \underline{\underline{\id}} +  (z^\star\underline{\underline{\id}} -\underline{\underline{J}}^T) (z\underline{\underline{\id}} -	\underline{\underline{J}})] = - \sum_\nu \ln[\eta^2 + s^2_\nu(z)]$, where $s_\nu(z)$ are the singular values of $z\underline{\underline{\id}} -	\underline{\underline{J}}$. Since $s_\nu(z)=0$ iff $z= \lambda_\nu$, one may approximate $s^2_\nu(z) \approx a\vert z-\lambda_\nu\vert^2$ in the vicinity of the eigenvalue, and we arrive at the same limiting result.}. Combining Eq.~(\ref{cintermsofeig}) with the 2D Dirac delta identity in Eq.~(\ref{2Ddelta}) and the definition of the 2D eigenvalue density in Eq.~(\ref{2Ddens}), we therefore have our desired analogue of the inverse Stieltjes transform in Eq.~(\ref{densefromres}) for the non-Hermitian case
\begin{align}
	\rho^{(2)}(z) = \frac{1}{\pi} \lim_{\eta \to 0}\frac{\partial C}{\partial z^\star} . \label{densefromresnonherm}
\end{align}
We note that in the limit $\eta \to 0$, and if $z$ is not equal to an eigenvalue, we have that $\lim_{\eta \to 0} 	C(z, z^\star) = \frac{1}{N} \mathrm{Tr} \left\{ \left[(z \underline{\underline{\id}} -\underline{\underline{J}} ) \right]^{-1}\right\} $, i.e. we see that there is an important relationship between $C$ and the resolvent that we used for the Hermitian case. While the region of non-analyticity of the resolvent was confined to the real line in the Hermitian case (where the eigenvalues were confined), we see from Eq.~(\ref{densefromresnonherm}) that the non-analytic region in the $z$-plane for $C(z,z^\star)$ now occupies a finite area of the complex plane, corresponding to the locations of the eigenvalues. 

\begin{figure}[h]
	\centering 
	\includegraphics[scale = 0.45]{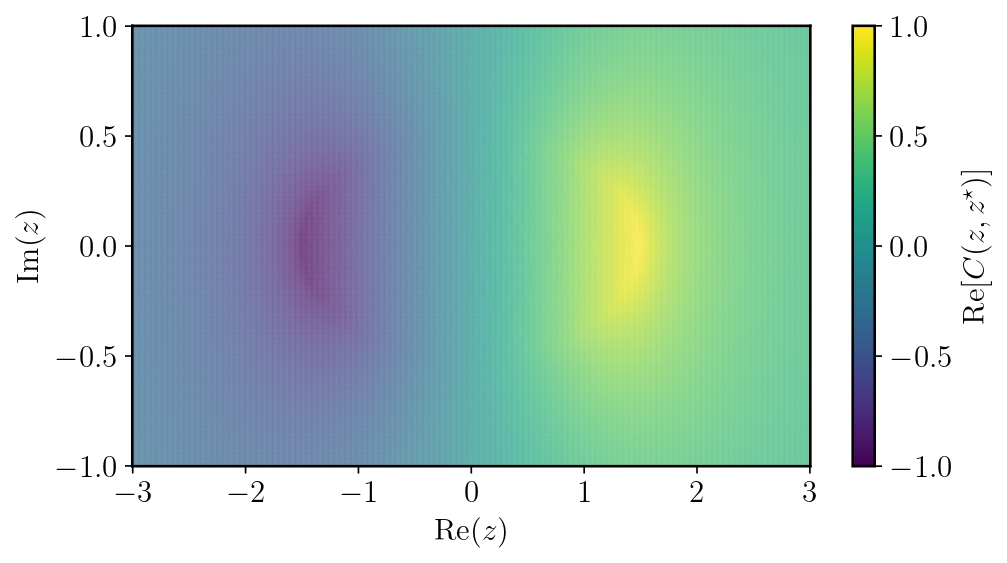} 
	\includegraphics[scale = 0.45]{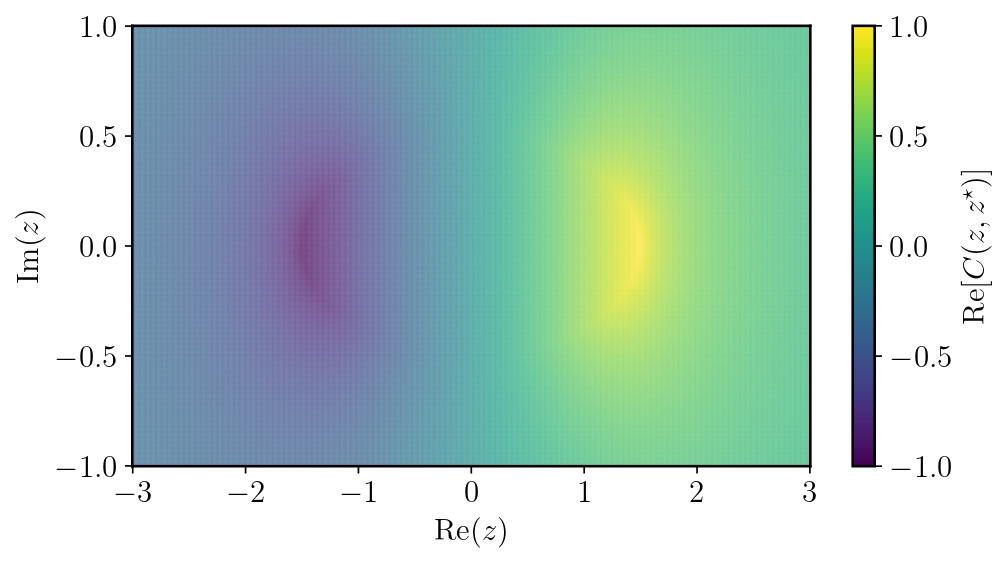} 
	\includegraphics[scale = 0.45]{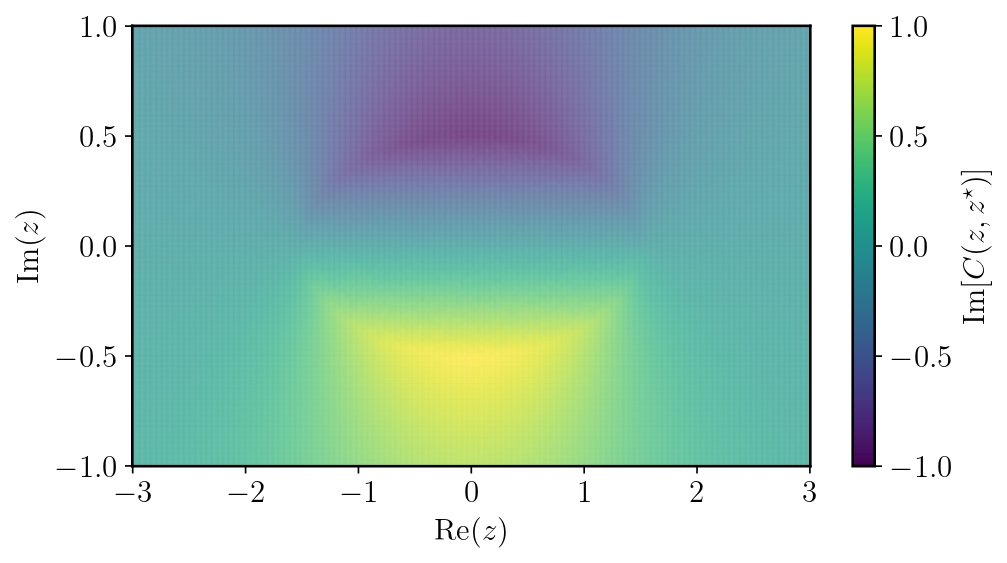} 
	\includegraphics[scale = 0.45]{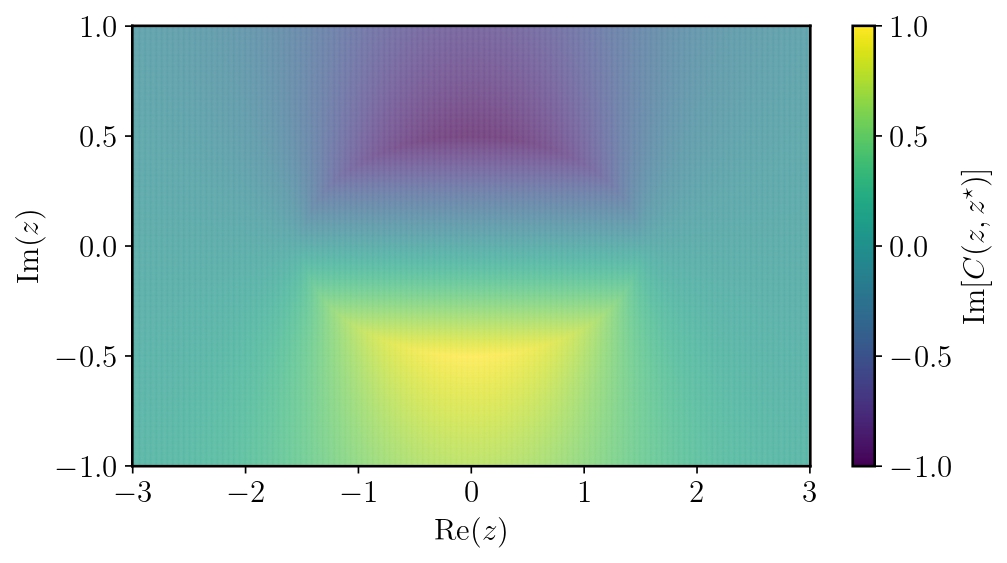} 
	\captionsetup{justification=raggedright,singlelinecheck=false}
	\caption{Comparison of $C(z,z^\star)$ for a single realisation with $N = 4000$ and $\eta = 0.01$ (left column) with theory [see Eqs.~(\ref{csolanalytic}) and (\ref{csolnonanalytic})] (right column).   }\label{fig:resolventsnonherm}
\end{figure}

Our task is now analogous to that of Section \ref{section:semicircle}. We wish to find the limit $N \to \infty$ of the object $C(z,z^\star)$. Once we have done so, we may use the formula in Eq.~(\ref{densefromresnonherm}), which is analogous to the inverse Stieltjes transform of the Hermitian case, and thus recover the limiting form of the eigenvalue density.

\subsection{Girko's trick}
The object $C(z,z^\star)$ in Eq.~(\ref{cdef}) is a somewhat more complicated object than the resolvent that we studied in the Hermitian case [see Eq.~(\ref{resdef})]. It would make our lives a lot easier if we could deal with objects that were more analogous to those that we constructed for the Hermitian case, so that we could follow an analysis along similar lines. The `trick', originally due to Girko \cite{girko1985circular,girko1986elliptic} and refined by Janik et al \cite{janik1997non} and Feinberg et al \cite{feinberg1997non}, is to consider a block matrix (sometimes referred to as the `Hermitised resolvent') from which the object $C(z,z^\star)$ can later be extracted. A similar strategy was also used by Sommers et al \cite{sommers1988spectrum} in conjunction with the replica method (see also Section \ref{section:replicas}).

Let us consider
\begin{align}
	\underline{\underline{\mathcal{H}}} \equiv \begin{bmatrix}
		i\eta \underline{\underline{\id}}_N & z\underline{\underline{\id}}_N -	\underline{\underline{J}} \\
		z^\star\underline{\underline{\id}}_N -\underline{\underline{J}}^T & i\eta \underline{\underline{\id}}_N
	\end{bmatrix}^{-1} \equiv  \begin{bmatrix}
		\underline{\underline{A}} & \underline{\underline{B}} \\
		\underline{\underline{C}} & \underline{\underline{D}}
	\end{bmatrix}. \label{hermresdef}
\end{align}
This is now more reminiscent of the resolvent matrix that we used in the Hermitian case to obtain the 1D eigenvalue density, since it is linear in $\underline{\underline{J}}$. The blocks $\underline{\underline{A}}$, $\underline{\underline{B}}$, $\underline{\underline{C}}$ and $\underline{\underline{D}}$ can be obtained via the block matrix inversion formula in Eq.~(\ref{blockinverse})
\begin{align}
	\underline{\underline{A}} &= -i\eta\left[\eta^2 \underline{\underline{\id}}_N +(z\underline{\underline{\id}}_N -	\underline{\underline{J}})  (z^\star\underline{\underline{\id}}_N -\underline{\underline{J}}^T)\right]^{-1}, \nonumber \\
	\underline{\underline{B}} &=(z\underline{\underline{\id}}_N -\underline{\underline{J}})\left[\eta^2 \underline{\underline{\id}}_N +(z^\star\underline{\underline{\id}}_N -\underline{\underline{J}}^T)(z\underline{\underline{\id}}_N -	\underline{\underline{J}})  \right]^{-1} , \nonumber \\
	\underline{\underline{C}} &=  (z^\star\underline{\underline{\id}}_N -\underline{\underline{J}}^T)\left[\eta^2 \underline{\underline{\id}}_N +(z\underline{\underline{\id}}_N -	\underline{\underline{J}})  (z^\star\underline{\underline{\id}}_N -\underline{\underline{J}}^T)\right]^{-1} , \nonumber \\
	\underline{\underline{D}} &= -i\eta\left[\eta^2 \underline{\underline{\id}}_N +  (z^\star\underline{\underline{\id}}_N -\underline{\underline{J}}^T) (z\underline{\underline{\id}}_N -	\underline{\underline{J}})\right]^{-1}.
\end{align}
We note that we use a notation in keeping with Section \ref{section:semicircle}, such that we denote the normalised traces $C(z,z^\star) = N^{-1} \mathrm{Tr} \underline{\underline{C}}$, and so on. We thus see that we can obtain the 2D eigenvalue density by first computing $\underline{\underline{\mathcal{H}}}$, taking the trace of the bottom-left block, and then using the analogue of the inverse Stieltjes transform given in Eq.~(\ref{densefromresnonherm}). 

As a technical point, we note that while a deviation away from the real axis $z= \omega - i\epsilon$ provided the `smoothing' of the eigenvalue density that was required to perform computations in the Hermitian case, here the smoothing is provided by the regulariser $\eta$, since the eigenvalues are already in the complex plane. One can actually regard the block matrix construction above as extending the consideration from complex numbers to quaternions \cite{rogers2010universal}, in direct analogy to the extension we made from the real line to complex numbers in the Hermitian case.

\subsection{Finding the block-matrix inverse}\label{section:blockmatrixinverseellipse}

All that remains for us now is to analyse the object $\underline{\underline{\mathcal{H}}}$. We proceed along lines that are exactly analogous with the `cavity' approach in Section \ref{section:cavity}, only now we take into account the block structure of $\underline{\underline{\mathcal{H}}}$. We emphasise that it is by examining the simpler $\underline{\underline{\mathcal{H}}}$ (rather than $\underline{\underline{C}}$ directly) that we are able to follow the same procedure as for the Hermitian case. 

For ease of notation, we introduce the following $2\times2$ matrices (for which we forgo the double underlines and instead use a calligraphic font)
\begin{align}
	\mathcal{H}_{ij} = \begin{bmatrix}
		A_{ij} & B_{ij}\\
		C_{ij} & D_{ij}
	\end{bmatrix}, \hspace{2cm}\mathcal{J}_{ij} = \begin{bmatrix}
		0 & J_{ij}\\
		J_{ji} & 0
	\end{bmatrix}, \hspace{2cm}\mathcal{Z} = \begin{bmatrix}
		i\eta & z\\
		z^\star & i \eta
	\end{bmatrix}.
\end{align}
It is helpful now to shift our perspective of the matrix $\underline{\underline{\mathcal{H}}}$ (which is a block matrix comprised of a $2\times 2$ grid of blocks, each of which is an $N\times N$ matrix), and instead consider the following rearranged version, which is comprised of a grid of $N \times N$ blocks, each of which is a $2 \times 2$ matrix 
\begin{align}
	\underline{\underline{\tilde{\mathcal{H}}}} \equiv\begin{bmatrix}
		\mathcal{Z}-\mathcal{J}_{11} & -\mathcal{J}_{12}  & \cdots & -\mathcal{J}_{1N} \\
		-\mathcal{J}_{21} & \ddots & \cdots & -\mathcal{J}_{2N} & \\
		\vdots & \vdots & \ddots &\vdots \\
		-\mathcal{J}_{N1} & -\mathcal{J}_{N2} & \cdots &\mathcal{Z}-\mathcal{J}_{NN}
	\end{bmatrix}^{-1} = \begin{bmatrix}
		\mathcal{H}_{11} & \mathcal{H}_{12}  & \cdots & \mathcal{H}_{1N} \\
		\mathcal{H}_{21} & \ddots & \cdots & \cdots & \\
		\vdots & \vdots & \ddots &\vdots \\
		\mathcal{H}_{N1} & \cdots & \cdots &\mathcal{H}_{NN}
	\end{bmatrix} ,
\end{align}
where the equality can be seen by simple inspection and comparison to the definition of $\underline{\underline{\mathcal{H}}}$ in Eq.~(\ref{hermresdef})\footnote{Another way of stating this is as follows. The matrix $\underline{\underline{\tilde{H}}}^{-1}$ can be obtained by performing a set of row swap operations and an identical series of column swaps on $\underline{\underline{H}}^{-1}$. Let us denote the series of $M$ row-swap transformations as $\underline{\underline{R}} =  \underline{\underline{S}}_1 \cdots \underline{\underline{S}}_M$. We may thus write $\underline{\underline{\tilde{H}}}^{-1} = \underline{\underline{R}} \underline{\underline{H}}^{-1} \underline{\underline{R}}^T$. Since $\underline{\underline{R}}^{-1} =  \underline{\underline{R}}^T$, we also see that $\underline{\underline{\tilde{H}}} = \underline{\underline{R}} \underline{\underline{H}} \underline{\underline{R}}^T$. That is, if we perform a set of row/column swap operations, one performs the same set of operations to the inverse of the original matrix to get the inverse of the rearranged matrix. This is also the principle behind Gaussian elimination.  }. 

Now, we again follow the cavity procedure, and use the block inversion formula in Eq.~(\ref{blockinverse}). This time, we choose $L = 2$, and we select $\underline{\underline{S}}$ to be the $2\times 2$ matrix $ \mathcal{Z}-\mathcal{J}_{11}$. Thus, $\underline{\underline{T}}$ is the $2 \times 2(N-1)$ matrix $(\mathcal{J}_{1,2}, \mathcal{J}_{1,3}, \cdots, \mathcal{J}_{1,N})$, and $\underline{\underline{U}}$ is the $2(N-1) \times 2$ matrix $(\mathcal{J}_{1,2}, \mathcal{J}_{1,3}, \cdots, \mathcal{J}_{1,N})^T$. Finally, the remaining $2(N-1)\times 2(N-1)$ matrix $\underline{\underline{V}}$ is $\underline{\underline{\tilde{\mathcal{H}}}}^{(1)}$. The matrix $\underline{\underline{\tilde{\mathcal{H}}}}^{(1)}$ is the version of $\underline{\underline{\tilde{\mathcal{H}}}}$ that we would obtain if we were computing the Hermitised resolvent of the $(N-1)\times(N-1)$ matrix $\underline{\underline{J}}^{(1)}$ (the matrix $\underline{\underline{J}}$ with the first row and column removed). 

Performing the block matrix inversion, we therefore arrive at
\begin{align}
	\mathcal{H}_{11} = \left( \mathcal{Z}-\mathcal{J}_{11} - \sum_{j,k=2}^N \mathcal{J}_{1j} \mathcal{H}^{(1)}_{jk}\mathcal{J}_{k1}\right)^{-1} . \label{withoff}
\end{align}
This is an analogue of Eq.~(\ref{cavitydiagonal}) where the resolvent $G$ and the random matrix entries have been replaced by their appropriate $2\times 2$ counterparts. We note that to compute the eigenvalue density we only require the diagonal entries of the matrix $\underline{\underline{C}}$, which are the bottom left entries of the $2\times2$ matrices $\mathcal{H}_{ii}$. All that remains now is to understand how Eq.~(\ref{withoff}) behaves in the limit $N \to \infty$.

\subsection{Concentration of the large sums in the non-Hermitian cavity equations}
We now deduce the diagonal entries of the matrix $\underline{\underline{C}}$. We do this by following identical arguments to Section \ref{section:cavityconcentration}. For simplicity, we set $\mu = 0$ for now. We deal with the single additional outlier eigenvalue that can come about for $\mu \neq 0$ subsequently.

We begin with the observation that we may neglect the elements of $\mathcal{H}^{(1)}_{jk}$ with $j\neq k$ when $N$ is large in Eq.~(\ref{withoff}), analogously to the Hermitian case (see Section \ref{section:offdiag}). We again also realise that the index $1$ was an arbitrary choice, and we may hence write 
\begin{align}
	\mathcal{H}_{ii} = \left( \mathcal{Z}-\mathcal{J}_{ii} - \sum_{j\neq i} \mathcal{J}_{ij} \mathcal{H}^{(i)}_{jj}\mathcal{J}_{ji}\right)^{-1} . \label{cavitynonherm}
\end{align}
More explicitly, we may multiply out the block matrices, and we obtain 
\begin{align}
	\sum_{j\neq i} \mathcal{J}_{ij} \mathcal{H}^{(i)}_{jj}\mathcal{J}_{ji} = \begin{bmatrix}
		\sum_{j\neq i} J_{ij}^2 D_{jj}^{(i)} & \sum_{j\neq i} J_{ij} C_{jj}^{(i)} J_{ji}\\
		\sum_{j\neq i} J_{ij} B_{jj}^{(i)} J_{ji} & \sum_{j\neq i}  A_{jj}^{(i)} J_{ji}^2
	\end{bmatrix} .\label{largesums}
\end{align}
Just as we found in the Hermitian case in Section \ref{section:semicircle}, each of these sums concentrates on its ensemble average. This means that for large $N$, each $2 \times 2$ matrix $\mathcal{H}_{ii}$ takes the same value regardless of $i$. We thus find
\begin{align}
	\mathcal{H}_{ii} \approx \mathcal{H} = 
	\begin{bmatrix}
		A & B\\
		C & D
	\end{bmatrix},
\end{align}
where the reader is reminded that we denote $A = N^{-1} \mathrm{Tr} \underline{\underline{A}}$, and so on. Further, we also expect (again) that $A^{(i)} \approx A$, $B^{(i)} \approx B$, etc., for large $N$. This has the consequence that the large sums in Eq.~(\ref{largesums}) concentrate on the following values
\begin{align}
	\sum_{j\neq i} \mathcal{J}_{ij} \mathcal{H}^{(i)}_{jj}\mathcal{J}_{ji} \approx \begin{bmatrix}
		\sigma^2  D & \Gamma \sigma^2 C\\
		\Gamma \sigma^2 B & \sigma^2 A
	\end{bmatrix} . \label{concentrated}
\end{align}
Finally, if we neglect the diagonal elements $\mathcal{J}_{ii}$, we therefore arrive at a set of 4 simultaneous equations by substituting Eq.~(\ref{concentrated}) into Eq.~(\ref{cavitynonherm})
\begin{align}
	A = \frac{1}{\Delta}(i \eta -\sigma^2 A),& \hspace{1cm} B = \frac{1}{\Delta}(\Gamma \sigma^2 C - z), \nonumber \\
	C = \frac{1}{\Delta}(\Gamma \sigma^2 B - z^\star),& \hspace{1cm} D = \frac{1}{\Delta}(i \eta -\sigma^2 D) , \nonumber \\
	\Delta = (i\eta -\sigma^2 A)(i\eta -\sigma^2 D) &- (\Gamma \sigma^2 B - z^\star)(\Gamma \sigma^2 C - z) = \frac{1}{AD -BC}
	. \label{simultellipse}
\end{align}
Our task has been reduced to solving these simultaneous equations for $C(z,z^\star)$ and extracting the eigenvalue density using Eq.~(\ref{densefromresnonherm}). 

\subsection{The elliptic law}
From the first of the simultaneous equations in Eqs.~(\ref{simultellipse}), we find that there are two solutions for $A$. Bearing in mind that we will subsequently take $\eta \to 0$, we consider $\eta$ to be small. Rearranging, we see that $A(1 + \Delta/\sigma^2) = \eta/\Delta$. We thus see that there is a solution for which $A \sim \eta$ and thus $\lim_{\eta \to 0}A = 0$. Another solution is the one for which $(1 + \Delta/\sigma^2) \sim \eta$ and thus $\Delta = -\sigma^2$ in the limit $\eta \to 0$, with $A$ taking a non-zero value. 

It turns out that these two solutions of the equations are valid in two different regions of the complex plane. For the first solution, $C$ is analytic and the eigenvalue density is nil. The other applies in the region to which the eigenvalue are confined, where $C$ is non-analytic. By setting these two solutions for $C$ equal to each other, we can find the set of values of $z$ on the border of these two regions, and thus find the region to which the eigenvalues are confined.

Let us first take the case where $A \sim \eta$. In this case, we see that $\Delta = - (\Gamma \sigma^2 B - z^\star)(\Gamma \sigma^2 C - z) + O(\eta)$, and thus
\begin{align}
	C &= \frac{1}{z -\Gamma \sigma^2 C} + O(\eta) , \nonumber \\
	\Rightarrow C &= \frac{1}{2 \Gamma \sigma^2} \left[ z -\mathrm{sign}[\mathrm{Re}(z)] \sqrt{z^2 - 4 \Gamma \sigma^2}\right] + O(\eta) . \label{csolanalytic}
\end{align}
The choice of sign in front of the radical ensures that there is no discontinuity on the imaginary axis, and that the correct asymptotics for $\vert z\vert \to \infty$ are obeyed [see the discussion around Eq.~(\ref{ressemicircle})]. 

Let us now consider the solution of Eq.~(\ref{simultellipse}) for which $A$ is of the order $\eta^0$. In this case, $\Delta = - \sigma^2 + O(\eta)$. Therefore, we have 
\begin{align}
	C &= B^\star = \frac{1}{\sigma^2} \left[ z^\star - \Gamma\sigma^2 B\right] + O(\eta), \nonumber \\
	\Rightarrow C &= \frac{z^\star - \Gamma z}{\sigma^2 (1-\Gamma^2)} + O(\eta). \label{csolnonanalytic}
\end{align}
By Eq.~(\ref{densefromresnonherm}), one therefore has a non-zero eigenvalue density in the region of the complex plane where this is the solution
\begin{align}
	\rho^{(2)}(z) = \frac{1}{\pi\sigma^2  (1-\Gamma^2)} . 
\end{align}
We note however that as $\vert z \vert \to \infty$, Eq.~(\ref{csolnonanalytic}) does not give the correct asymptotics $C(z,z^\star) \to 1/z$. Therefore, this solution can only apply in a bounded region of the complex plane. 

\begin{figure}[h]
	\centering 
	\includegraphics[scale = 0.45]{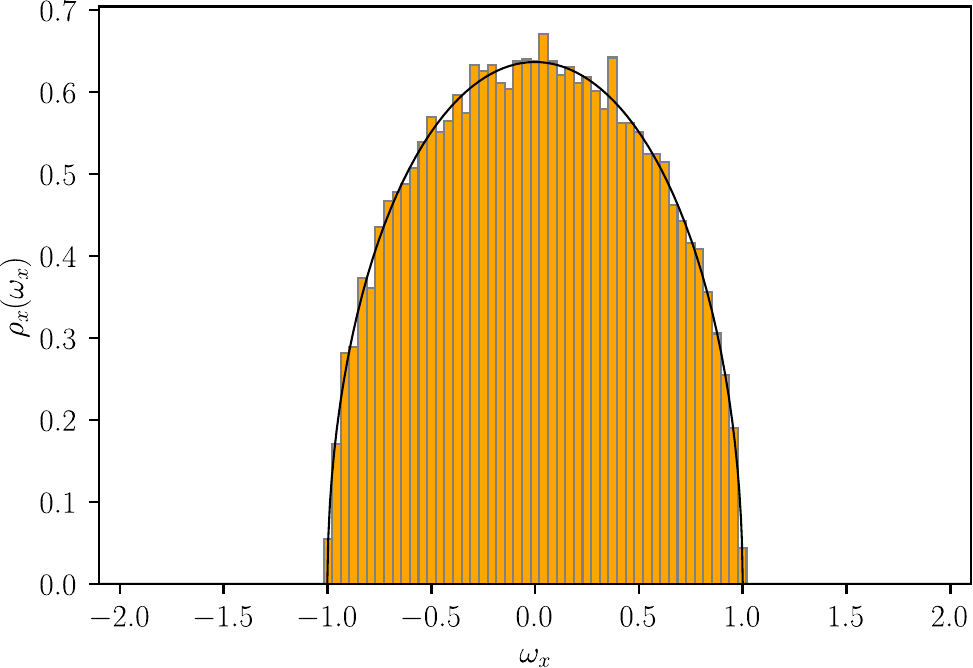}  
	\includegraphics[scale = 0.45]{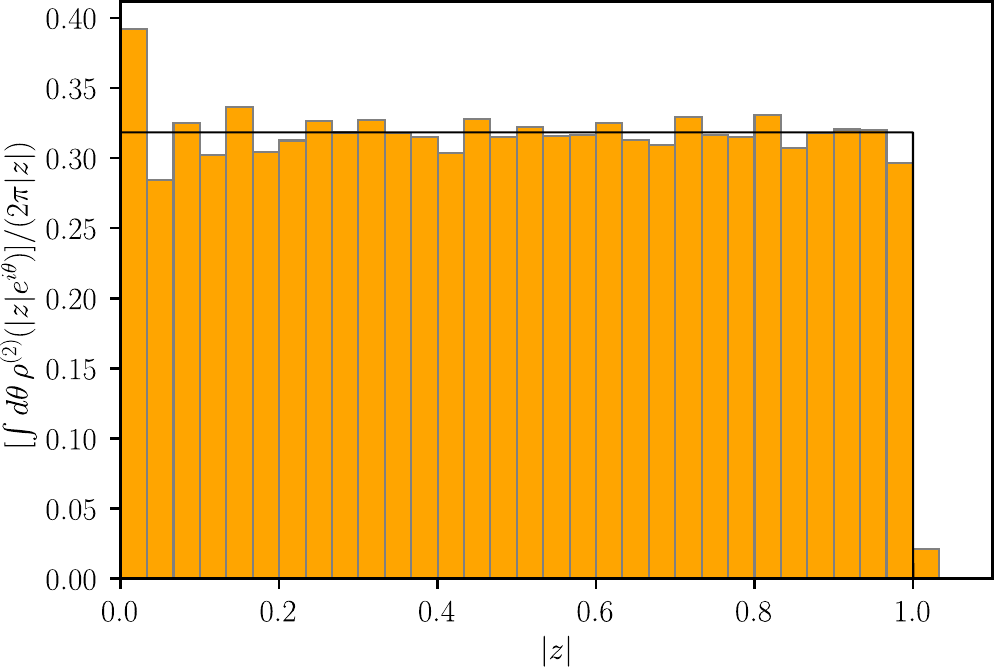}  
	\captionsetup{justification=raggedright,singlelinecheck=false}
	\caption{Tests of the uniform eigenvalue density given in Eq.~(\ref{ellipticlawfull}). (Left) Integrated eigenvalue density and comparison to the modified Wigner semicircle -- see exercises. (Right) Eigenvalue density as a function of distance from the origin, averaging over angle. Parameters: $N = 4000$, $\sigma = 1$, $\Gamma = 0$.}\label{fig:density}
\end{figure}

We now wish to find the boundary of the region of the complex plane where Eq.~(\ref{csolnonanalytic}) is the correct solution for $C(z,z^\star)$, and thus where we have a non-zero eigenvalue density. Since $C(z,z^\star)$ is expected to be a continuous function of $z$ (this is verified by the numerics in Fig. \ref{fig:resolventsnonherm}), the boundary is given by the set of values of $z$ that satisfy Eqs.~(\ref{csolanalytic}) and (\ref{csolnonanalytic}) simultaneously. Substituting the second of Eq.~(\ref{csolnonanalytic}) into the first of Eq.~(\ref{csolanalytic}), we find that the boundary of the region of non-analyticity of $C$ is given by
\begin{align}
	\left\vert z^\star - \Gamma z \right\vert^2 = (1-\Gamma^2)^2 \sigma^2.
\end{align}
Denoting $z = \omega_x + i \omega_y$, we thus finally obtain the elliptic law 
\begin{align}
	\left(\frac{\omega_x}{1+\Gamma} \right)^2 + \left(\frac{\omega_y}{1-\Gamma} \right)^2 = \sigma^2 .\label{ellipticlaw}
\end{align}
The eigenvalue density is therefore given by
\begin{empheq}[box={\fboxsep=6pt\fbox}]{align}
	\rho^{(2)}(z) = \begin{cases}
		\frac{1}{\pi\sigma^2 (1-\Gamma^2)} \hspace{1cm} \mathrm{if} \left(\frac{\omega_x}{1+\Gamma} \right)^2 + \left(\frac{\omega_y}{1-\Gamma} \right)^2 < \sigma^2 , \\
		0 \hspace{2cm} \mathrm{otherwise} .
	\end{cases}\label{ellipticlawfull}
\end{empheq}
One can check easily that this satisfies the normalisation condition given in Eq.~(\ref{normalisation2d}). We check the expressions for $C(z,z^\star)$ in Eqs.~(\ref{csolanalytic}) and (\ref{csolnonanalytic}) against results for a single realisation of the random matrix $\underline{\underline{J}}$ in Fig. \ref{fig:resolventsnonherm}, and we check the elliptic law in Eq.~(\ref{ellipticlaw}) against numerics in Figs. \ref{fig:ellipseandoutlier}, \ref{fig:density} and \ref{fig:instabilities}.

\subsection{Outlier eigenvalue}\label{section:outlier}
So far, we have identified the regions of analyticity and non-analyticity of $C(z,z^\star)$, and we have thus identified the elliptic region to which the majority of the eigenvalues are confined. However, there may exist another isolated eigenvalue [as we can see in Fig. \ref{fig:ellipseandoutlier}], which can protrude from the ellipse for certain values of $\mu$. 

We can develop some intuition for this as follows. Suppose we were to set $\sigma = 0$. In this case, the matrix $\underline{\underline{J}}$ would be a simple rank-1 matrix whose elements are all uniformly $\mu/N$. In this case, there is one eigenvalue at location $\lambda_1 = \mu$ with eigenvector $\underline{v}_1 = (1,1,\cdots,1)^T$, and $N-1$ degenerate eigenvalues at $\lambda_\nu = 0$. As we increase $\sigma$, the degenerate eigenvalues at the origin spread out and form the ellipse, while the outlier eigenvalue picks up a correction $O(\sigma^2)$, as we will show below. In general, low-rank perturbations give rise to a number of outlier eigenvalues less than or equal to the rank of the perturbation \cite{benaych2011eigenvalues, orourke2014low, baik2005phase}. We note in passing that random matrix statistics other than the mean can also influence the location and/or existence of the outlier eigenvalue \cite{baron2022eigenvalues, baron2023breakdown}. 

For $\mu = 0$, we have no outlier eigenvalue, since there is no rank-1 perturbation to the matrix. Supposing that there is an outlier eigenvalue that protrudes from the bulk, it must satisfy (by definition)
\begin{align}
	\mathrm{det}\left[ \underline{\underline{J}} - \lambda_\mathrm{outlier} \underline{\underline{\id}}_N\right] = 0.
\end{align}
Let us now separate the rank-1 perturbation proportional to $\mu/N$ from the random part of $\underline{\underline{J}}$, by writing $\underline{\underline{J}} = \underline{\underline{K}} + \mu N^{-1} \underline{u} \underline{u}^T$, where $\underline{u}$ is the length-$N$ vector whose entries are all equal to $1$. We then have
\begin{align}
	\mathrm{det}\left[ \underline{\underline{\id}}_N -\mu N^{-1} \underline{u} \, \underline{u}^T \left(\lambda_\mathrm{outlier} \underline{\underline{\id}}_N-\underline{\underline{K}}  \right)^{-1}\right] = 0,
\end{align}
where we have assumed the outlier coincides with none of the eigenvalues of $\underline{\underline{K}}$, which are confined to the ellipse. We now use the so-called Weinstein–Aronszajn (or Sylvester's) determinant identity
\begin{align}
	\mathrm{det}\left(\underline{\underline{\id}}_N + \underline{\underline{V}} \,\underline{\underline{U}} \right) =\mathrm{det}\left(\underline{\underline{\id}}_M + \underline{\underline{U}} \,\underline{\underline{V}} \right) ,\label{sylvester}
\end{align}
where $\underline{\underline{U}}$ is an $M \times N$ matrix and $\underline{\underline{V}}$ is an $N \times M$ matrix. Assigning $\underline{\underline{U}} = \underline{u}^T \left(\lambda_\mathrm{outlier} \underline{\underline{\id}}-\underline{\underline{K}}  \right)^{-1}$ and $\underline{\underline{V}} =\mu N^{-1} \underline{u} $, so that $M = 1$, we obtain
\begin{align}
	\frac{1}{N}\sum_{ij}\left[\left(\lambda_\mathrm{outlier} \underline{\underline{\id}}_N-\underline{\underline{K}}  \right)^{-1}\right]_{ij} = \frac{1}{\mu}.
\end{align}
Let us now turn our attention to the object $\left(\lambda_\mathrm{outlier} \underline{\underline{\id}}_N-\underline{\underline{K}}  \right)^{-1}$. As was explained in the discussion after Eq.~(\ref{densefromresnonherm}), we have that $\lim_{\eta \to 0} C(z,z^\star) = (z\underline{\underline{\id}}_N - \underline{\underline{J}})^{-1}$, as long as the value of $z$ that we choose is away from the eigenvalues. Therefore, $\left(\lambda_\mathrm{outlier} \underline{\underline{\id}}_N-\underline{\underline{K}}  \right)^{-1}$ is precisely equal to $\underline{\underline{C}}$ in the case $\mu = 0$.

\begin{figure}[h]
	\centering 
	\includegraphics[scale = 0.45]{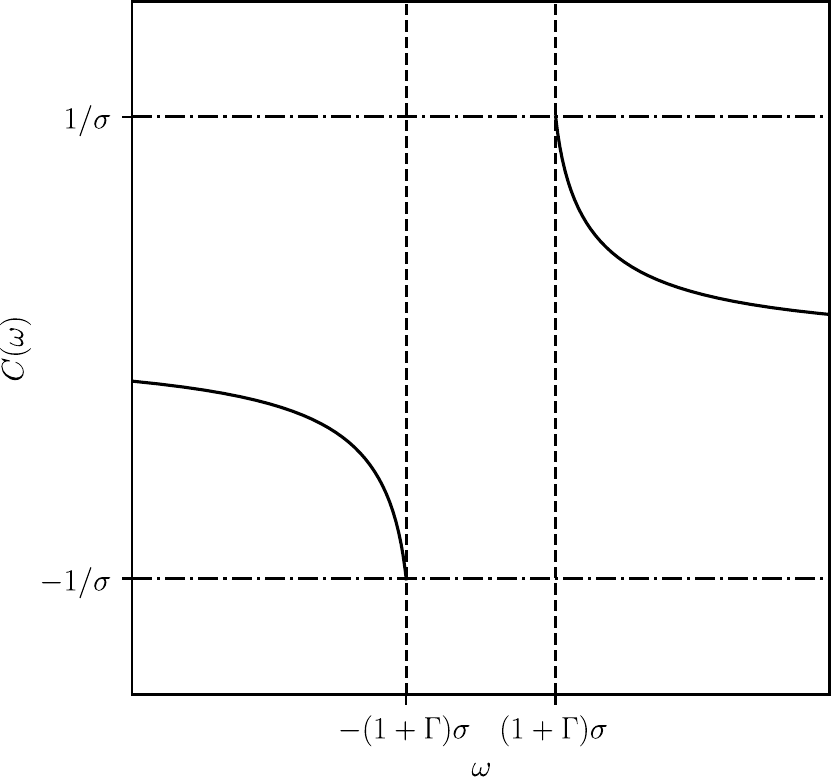}  
	\captionsetup{justification=raggedright,singlelinecheck=false}
	\caption{Sketch of $C(z,z^\star)$ on the real axis $z = \omega$ in the region where it is analytic and real. }\label{fig:resolventrealaxis}
\end{figure}

\begin{figure}[h]
	\centering 
	\includegraphics[scale = 0.47]{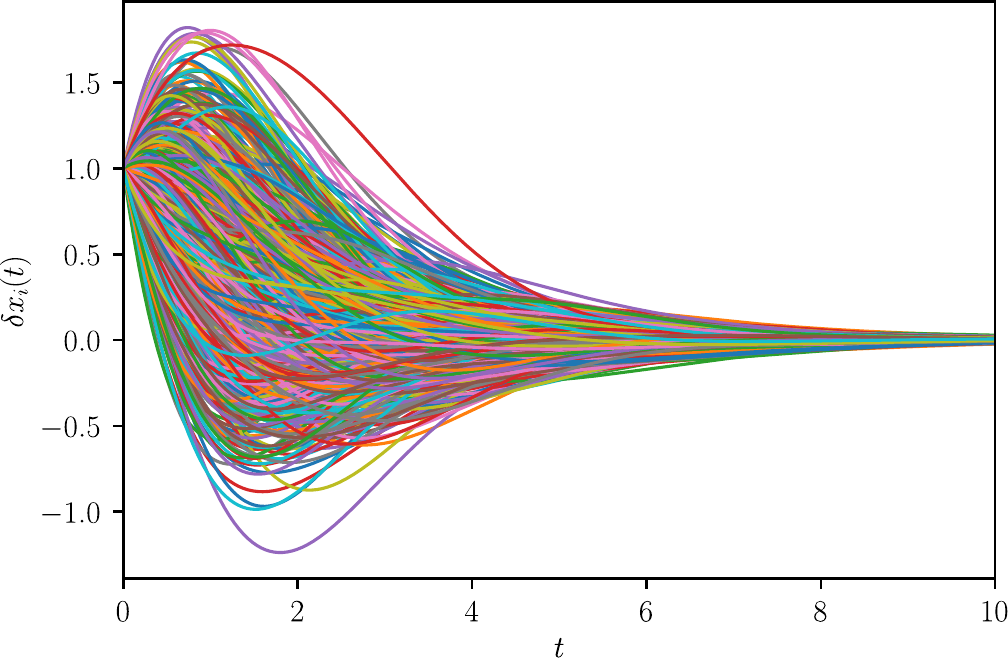}
	\includegraphics[scale = 0.47]{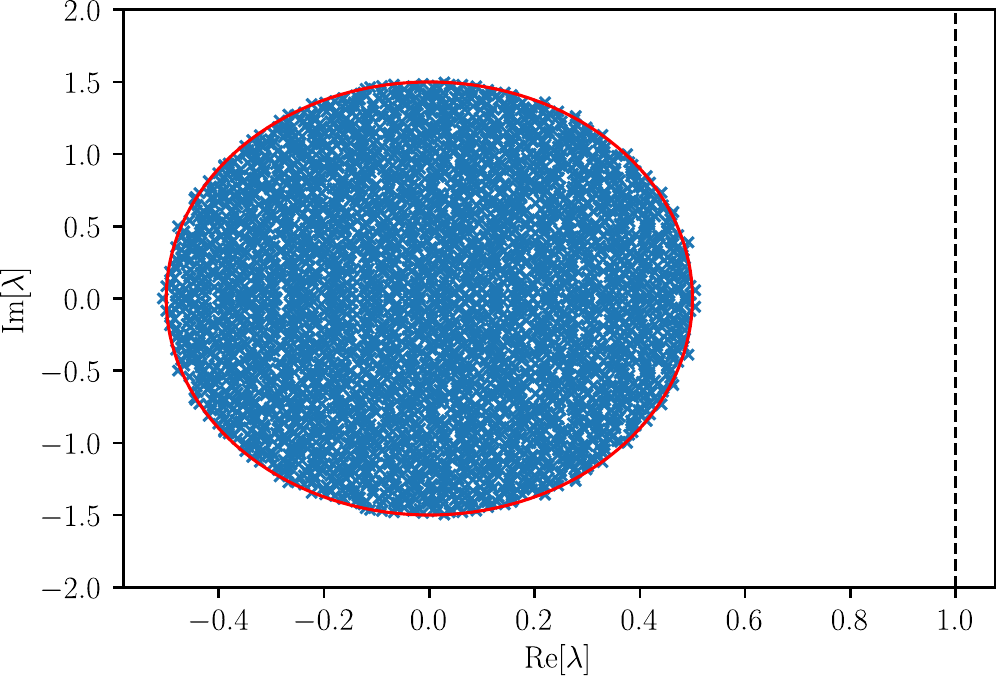}
	\includegraphics[scale = 0.47]{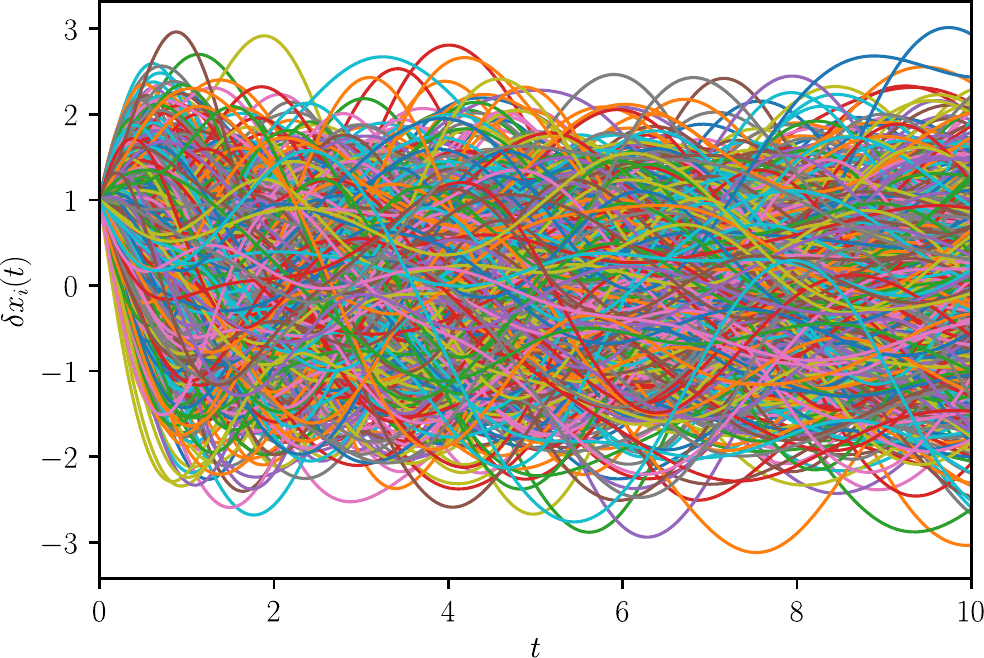} 
	\includegraphics[scale = 0.47]{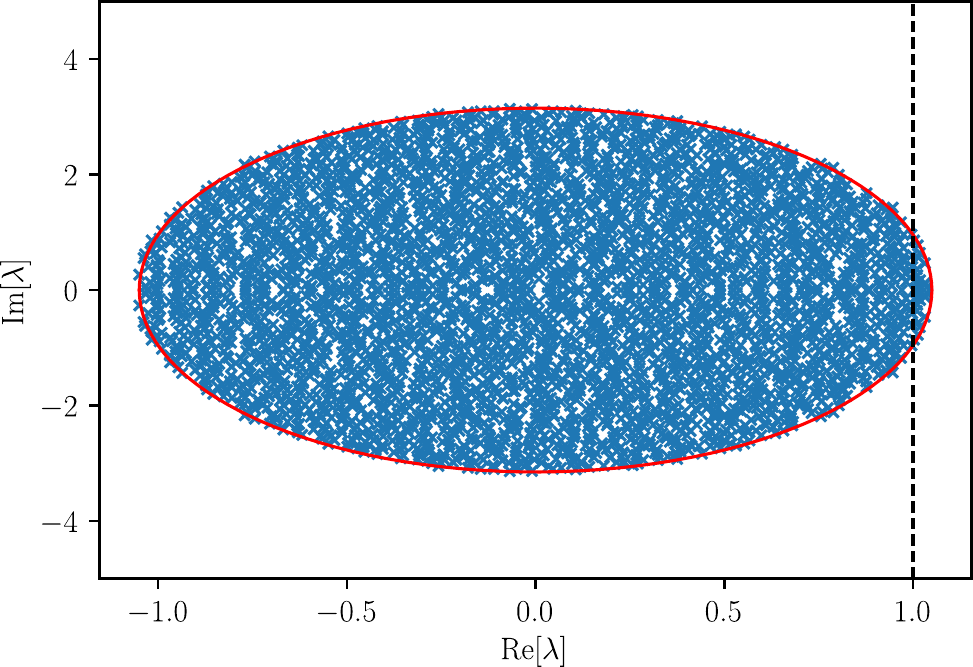}
	\includegraphics[scale = 0.47]{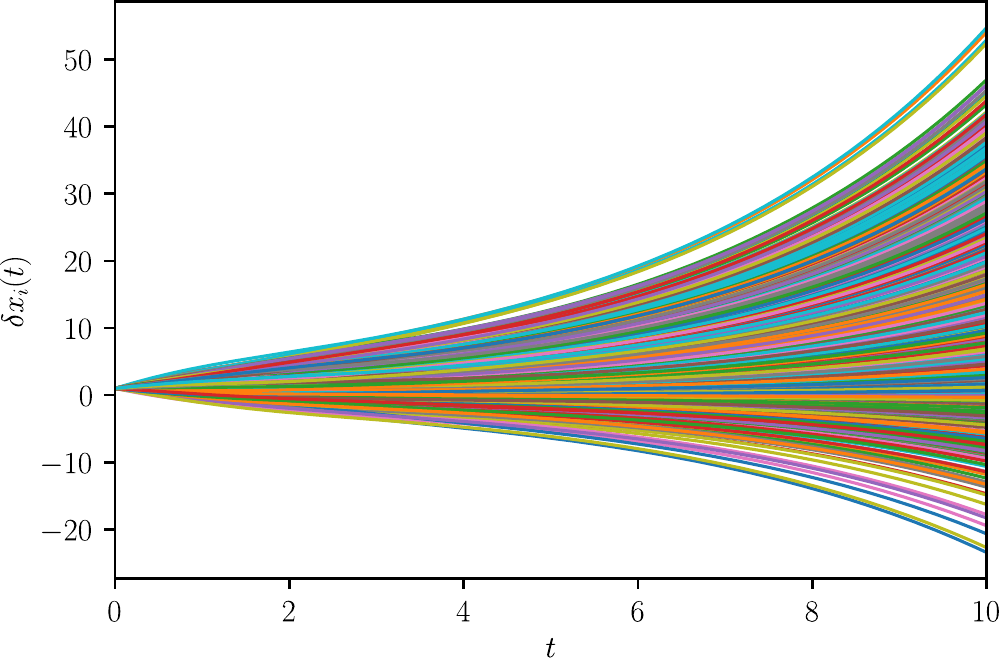} 
	\includegraphics[scale = 0.47]{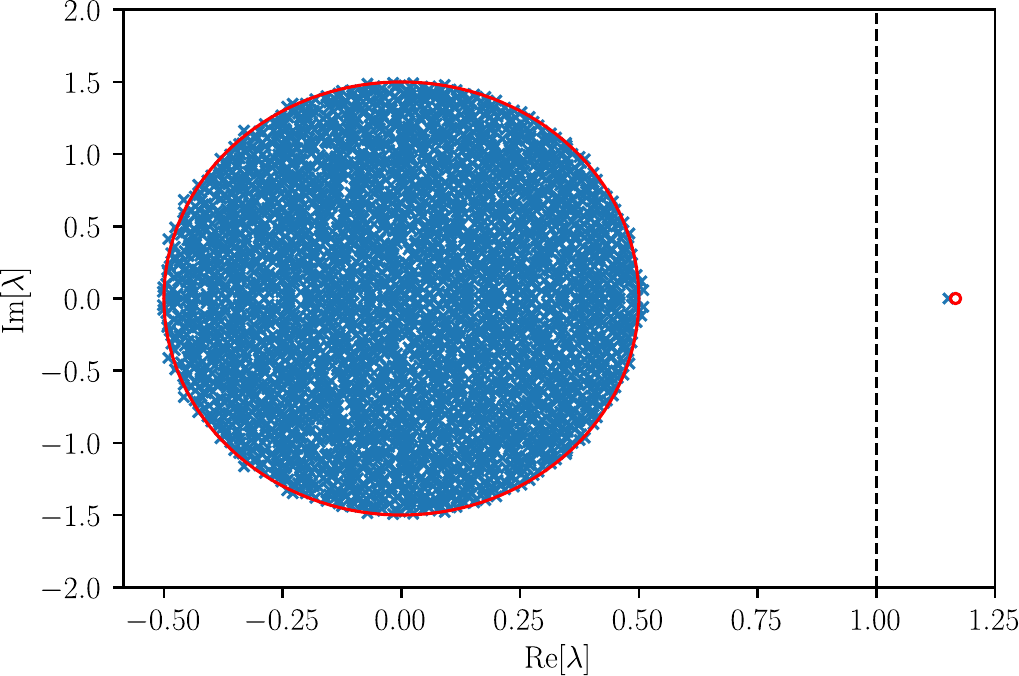}
	\captionsetup{justification=raggedright,singlelinecheck=false}
	\caption{The nature of the dynamics depending on the spectrum. (Top) Stable fixed point with parameters $\mu = 0.5$, $\sigma = 1$, $\Gamma = -0.5$. (Centre) Beyond the oscillatory transition with parameters $\mu = 0.5$, $\sigma = 2.1$, $\Gamma = -0.5$. (Bottom) Beyond the exponential growth transition with parameters $\mu = 1.5$, $\sigma = 1$, $\Gamma = -0.5$.}\label{fig:instabilities}
\end{figure}

Understanding that the off-diagonal elements of $\underline{\underline{C}}$ are negligible, we may use the result in Eq.~(\ref{csolanalytic}). That is,
\begin{align}
	C(\lambda_\mathrm{outlier}) = \frac{1}{\mu}, \label{cmusol}
\end{align}
and thus
\begin{align}
	\frac{1}{\mu} = \frac{1}{\lambda_\mathrm{outlier} - \frac{\Gamma \sigma^2}{\mu} },\hspace{1cm} \Rightarrow \hspace{1cm}
	\lambda_\mathrm{outlier} = \mu + \frac{\Gamma \sigma^2}{\mu}. \label{outlierellipse}
\end{align}
However, we note that this solution for the outlier eigenvalue cannot always be valid. We see this immediately by taking the limit $\mu \to 0$, whereupon the above expression for $\lambda_\mathrm{outlier}$ blows up, which cannot be. The reason for this is that the solution $C(z)$ that is valid outside the bulk ellipse [see Eq.~(\ref{csolanalytic})] is bounded for real $z$. The maximum value is taken as we approach the bulk region at $\omega_+ = (1+\Gamma)\sigma$ and the minimum is at $\omega_- = -(1+\Gamma)\sigma$. In these cases, we have $C(\omega_\pm) = \pm\frac{1}{\sigma}$ [see the sketch in Fig. \ref{fig:resolventrealaxis}]. We thus see that in order for Eq.~(\ref{cmusol}) to have a solution, and therefore for an outlier to exist outside the bulk region, we must have that
\begin{align}
	\vert\mu \vert\geq \sigma.
\end{align}
We see that the outlier eigenvalue touches the bulk region precisely when this bound is saturated. The expression in Eq.~(\ref{outlierellipse}) is tested in Figs. \ref{fig:ellipseandoutlier} and \ref{fig:instabilities}. 

We note that other approaches exist to obtain the outlier eigenvalue(s). Edwards and Jones computed the outlier as a $1/N$ correction to the spectrum explicitly \cite{edwards1976eigenvalue} using the replica method. Also using the replica method, one can perform an analysis to demonstrate a `Bose-Einstein-condensation-like' phenomenon, and one can deduce the outlier eigenvalue as the analogous ground state \cite{ikeda2023bose}. We also in Section \ref{section:marchenkopastur} show how the Woodbury identity may be used, and how this provides information on the corresponding eigenvector.

\subsection{Stability criteria and rescaling with $N$}\label{section:stabilityandscaling}
Now that we have computed the eigenvalue density, we return to May's model in Eq.~(\ref{maysystem}). Since the eigenvalues are all contained within an ellipse, with the possible exception of an outlier eigenvalue, we can predict the statistics of $\underline{\underline{J}}$ that will lead the dynamical system to be (un)stable in response to perturbations. 

There are two separate types of instability. Either the elliptical region can cross the line $\mathrm{Re}(z) = 1$, or the outlier eigenvalue can. As we show in Fig. \ref{fig:instabilities}, the character of these two instabilities is very different. When the bulk elliptical region crosses the critical line, the dynamics is oscillatory. In the case of the outlier however, a more straightforward exponential deviation from the fixed point occurs. As we discussed above in Section \ref{section:outlier}, and as we discuss in greater detail later in Section \ref{section:marchenkopastur}, this is because the eigenvector corresponding to the outlier eigenvalue has a significant overlap with the direction $(1,1,\cdots,1)$, whereas this is not the case for the other eigenvectors. We also demonstrate the qualitative difference between the two instabilities using dynamic mean-field theory in Section \ref{section:dmft}.

The criterion for stability are as follows. In order for the ellipse not to cross the line $\mathrm{Re}(z) = 1$, we must have that 
\begin{align}
	(1+\Gamma) \sigma <1.
\end{align}
On the other hand, for the outlier not to cross the line $\mathrm{Re}(z) = 1$, we require
\begin{align}
	\mu + \frac{\Gamma \sigma^2}{\mu} <1.
\end{align}

However, we would like to relate the stability of the system explicitly to the statistics of the Jacobian, and thus be more in line with May's original analysis \cite{may1972will}. To do this, we rescale with the system size. Let $\sigma^2 = N \sigma_0^2$ (so that $\sigma_0^2 = \mathrm{Var}(J_{ij})$ is the variance of the matrix elements) and $\mu = N \mu_0$ (so that $\mu_0 = \langle J_{ij}\rangle$ is the mean). The stability criteria now become
\begin{align}
	N(1 + \Gamma)^2\sigma_0^2 <1, \hspace{1cm} N \mu_0 + \frac{\Gamma \sigma_0^2}{\mu_0} <1.
\end{align}
From this we may draw a number of conclusions. First, if indeed there is a significant variance or mean interaction strength, we obtain an upper limit on the number of species $N$ that may interact together before instability sets in. This was May's main finding -- a more complex system, in this sense, is more unstable! Second, we see that the system will also be unstable if either the mean or variance of the interactions is much greater than $1/N$. Finally, a greater proportion of asymmetric interactions (i.e. negative $\Gamma$) has a stabilising effect. 

Remarkably, with modern improvements of experiments involving the tracking of microbial population abundances, the predictions of May's simple model have recently become testable in the lab. Indeed, by increasing the strength/complexity of interactions or the number of interacting species, Hu et al found a transition from stable to unstable dynamics in microbial communities \cite{hu2022emergent}.

\subsection{Generalisations of May's model}
In the above, we have supposed that all species in our model ecosystem interact with one another, and they do so with homogeneous statistics. Obviously, this is not in the slightest bit realistic. Real ecosystems have a species hierarchy (perhaps with trophic levels), and there is often a non-trivial interaction network structure. Let us discuss briefly how such considerations have been handled in the context of May's simple random matrix model.

\subsubsection{Network structure}
Let us suppose that each species interacts only with a subset of the full community. We can therefore think of the community as being associated with an interaction network, with each species being a node on such a network, and species interactions being represented by the network edges. We suppose then, given that two species interact, $(J_{ij}, J_{ji})$ are drawn with the following statistics
\begin{align}
	\langle J_{ij} \rangle = \mu_0,\,\,\,\,\, \langle J_{ij}^2 \rangle - \langle J_{ij} \rangle ^2 = \sigma^2_0, \,\,\,\,\, \langle J_{ij} J_{ji} \rangle- \langle J_{ij} \rangle \langle J_{ji} \rangle  = \Gamma\sigma^2_0 .
\end{align}

The number of other species with which each species interacts is referred to as its degree. For a given degree distribution $P_k$, the simplest network model for us to suppose is the configuration model  \cite{rodgers2005eigenvalue, baron2022networks}. In this case, one can show that in the simple case $\Gamma = 0$, and writing $z = \omega_x + i \omega_y$, we have the modified circular law \cite{baron2022networks}
\begin{align}
	\omega_x^2 + \omega_y^2 = \sigma^2_0 p (1 + s^2),\label{circularlawhetero}
\end{align}
where here $s^2$ is the so-called degree heterogeneity
\begin{align}
	s^2 = \sum_k P_k \frac{k^2 - p^2}{p^2},
\end{align}
and $p = \sum_k kP_k$ is the average degree of the network. We thus obtain a new criterion for stability
\begin{align}
	p \sigma_0^2 (1+s^2) <1. \label{networkmaybound}
\end{align}
We therefore see that, with this generalisation, it isn't the number of interacting species that matters for stability -- it is the average number of interactions per species. This was also noted by May \cite{may1972will}. Further, we see that as the heterogeneity of the interaction network increases, this tips the system towards instability. We note that as the system is made more sparsely interacting (i.e. as $p$ becomes small enough that $p/N \ll 1$), the above result breaks down and more careful calculations must be performed \cite{baron2023pathintegralapproachsparse, valigi2025eigenvalue}. We consider complex networks in more detail in Section \ref{section:networks}. 

\subsubsection{Hierarchical or trophic structure}
As one might imagine, a great many further generalisations of May's model have been made, so that one might understand the effect on stability of introducing various more realistic interaction structures. For example, one may include the effects of trophic structure by making the statistics of the matrix $\underline{\underline{J}}$ block structured \cite{baron2020dispersal}. A more continuous hierarchy of species can be reflected by using the cascade \cite{allesina2015predicting} or niche models \cite{poley2024eigenvalue}. One can also include the possibility of varying autoregulation by including random diagonal matrix elements \cite{barabas2017self}, and more generalised correlation structures \cite{baron2022eigenvalues} or delay \cite{pigani2022delay} and spatial effects \cite{gravel2016stability}.

\subsection{Applications in non-linear models}
\subsubsection{Generalised Lotka-Volterra}
One critique of May's approach is that it does not take into account the effects of non-linearities, nor is it at all obvious that one can conceive of a realistic ecosystem dynamics that would give rise to Eq.~(\ref{maysystem}) when linearised about its fixed point. 

Let us then consider a simple, but somewhat more realistic and complete, ecosystem model -- the generalised Lotka-Volterra equations (gLVE)
\begin{align}
	\dot x_i = x_i \left[1 + \sum_{j} (J_{ij} - \delta_{ij})x_j \right], \label{glves}
\end{align}
where we assume that $J_{ij}$ have the statistics in Eq.~(\ref{ellipsestats}), and $\delta_{ij}$ is the Kronecker delta. 

At the fixed point of the gLVE system, either $\bar x_i = 0$, or $\bar x_i = 1+ \sum_j J_{ij} \bar x_j$. The statistics of the fixed-point abundances $\bar x_i$ for $N \to \infty$ can be determined using dynamic mean-field theory \cite{bunin2016interaction, galla2018dynamically} (see also Section \ref{section:dmft}), and therefore the fraction of surviving species $\phi = N^{-1} \sum_i \Theta(\bar x_i)$ [where $\Theta(\cdot)$ is the Heaviside function] can also be determined. One can show (see exercises) that the stability of the fixed point is actually determined by the eigenvalues of the interaction matrix of the \textit{surviving} species $\underline{\underline{J}}'$, and that all of these eigenvalue must be less than 1 for stability \cite{stone2018feasibility, biroli2018marginally}. We therefore find that May's reasoning carries over to the gLVE system, and we arrive at a May-like stability criterion \cite{baron2023breakdown}
\begin{align}
	\sqrt{\phi} \sigma (1+\Gamma) <1.
\end{align}
\subsubsection{Firing-rate model of neural networks}
We briefly show here how our random matrix results also describe the transition from a `quiescent' state (fixed point) to a chaotic excited state in a model of neural networks \cite{rajan2006eigenvalue, sompolinsky1988chaos}. We suppose that we have a set of $N$ neurons, each with an associated activity $S_i(t)$. The neuron is activated by a local field $h_i(t)$ such that $S_i = \phi(h_i)$, where $\phi(\cdot)$ is a saturating function, say, $\phi(h) = \tanh(g h)$ for example, where $g$ is a positive constant. The fields then evolve according to 
\begin{align}
	\dot h_i = - h_i + \sum_j J_{ij} \phi(h_j). \label{firingrate}
\end{align}
This set of equations has a trivial fixed point $h_i^\star = 0$. The Jacobian about this fixed point is then $M_{ij} = -\delta_{ij} + g J_{ij}$. We see that if $J_{ij}$ has statistics as in Eq.~(\ref{ellipsestats}) with $\mu = 0$, then we once again obtain a May-like stability criterion
\begin{align}
	g\sigma (1+\Gamma) < 1.
\end{align}
\subsection{Exercises}
One might wonder, quite rightly, under what circumstances the results that we have derived here for asymmetric matrices reduce to what we found for the symmetric case. We show here that in the case $\Gamma = 1$, we recover the Wigner semicircle law.
\begin{itemize}
	\item By considering $\langle (J_{ij}- J_{ji})^2 \rangle$, show that we must have $J_{ij} = J_{ji}$ when $\Gamma = 1$.
	\item By integrating Eq.~(\ref{ellipticlawfull}) with respect to $\omega_y$, demonstrate that the density of the real parts of the eigenvalues is given by the generalised Wigner semicircle
	\begin{align}
		\rho_x(\omega_x) = \frac{2}{\pi \sigma^2 (1+\Gamma)^2} \sqrt{(1+\Gamma)^2 \sigma^2 - \omega_x^2}.
	\end{align}
	\item Hence, show that we recover the usual Wigner semicircle in the case of perfect symmetry. 
	\item Find the corresponding results for the imaginary parts of the eigenvalues [i.e. $\rho_y(\omega_y)$].
	\item I would thoroughly recommend verifying numerically Eq.~(\ref{ellipticlawfull}) and the marginal distributions that we have just derived.
\end{itemize}
There exist a range of other methods for finding the 2D eigenvalue density, just as there exist many ways to find the Wigner semicircle. For example, 
\begin{itemize}
	\item Show that the Dirac delta function in Eq.~(\ref{2Ddelta}) may be written
	\begin{align}
		\delta^{(2)}(z) = \frac{1}{\pi} \lim_{\eta\to 0}\frac{\partial^2}{\partial z \partial z^\star}\ln \left[\eta^2 + \vert z\vert^2\right] .
	\end{align}
	\item Proceeding with this in mind, we may seek to obtain the eigenvalue density in a similar form. Show that the resolvent in Eq.~(\ref{cintermsofeig}) may be written
	\begin{align}
		C(z,z^\star) = \frac{1}{N}\frac{\partial}{\partial z}\ln\left[\prod_{\nu}\left(\eta^2 + \vert z-\lambda_\nu\vert^2\right) \right],
	\end{align}
	and hence that
	\begin{align}
		\rho^{(2)}(z) &= -\frac{1}{\pi}  \lim_{\eta\to 0}\frac{\partial^2}{\partial z \partial z^\star} \Phi(z,z^\star) , \nonumber \\
		\Phi(z,z^\star) &= -\frac{1}{N}\ln\mathrm{det}\left[\eta^2 \underline{\underline{\id}} +  (z^\star\underline{\underline{\id}} -\underline{\underline{J}}^T) (z\underline{\underline{\id}} -	\underline{\underline{J}}) \right].\label{potential}
	\end{align}
	The quantity $\Phi(z,z^\star)$ is referred to as the eigenvalue potential \cite{sommers1988spectrum}. Hint: how is the determinant of a matrix related to its eigenvalues?
	\item Show that, if we define $E_x = 2 \mathrm{Re} C$ and $E_y = -2 \mathrm{Im}C$, we obtain Poisson's equation
	\begin{align}
		\nabla^2 \Phi \equiv \frac{\partial^2 \Phi}{\partial \omega_x^2} + \frac{\partial^2 \Phi}{\partial \omega_y^2} =- \nabla\cdot E \equiv - \frac{\partial E_x}{\partial \omega_x}  - \frac{\partial E_y}{\partial \omega_y}= - 4\pi \rho^{(2)}(\omega_x, \omega_y). 
	\end{align}
	We see here that there is a nice analogy with electrostatics, where the eigenvalues play the role of charges, the resolvent is related to the electric field, and the eigenvalue potential lives up to its name. Interestingly, the electrostatics analogies run quite deeply in random matrix theory, as we discuss in Section \ref{section:dysonbrownian}.
	
	The disorder-averaged eigenvalue potential can be computed using the replica method, as we show in Section \ref{section:replicanonherm}. This is one of many alternative ways of obtaining the elliptic law. The replica method is a arguably quicker and easier way of generating results (once you know how to use it), but it is considerably less rigorous.
	
\end{itemize}

Let us now go about showing how the stability of the generalised Lotka-Volterra equations in Eq.~(\ref{glves}) is determined by the so-called reduced interaction matrix \cite{baron2023breakdown, stone2018feasibility, biroli2018marginally}. 

\begin{itemize}
	\item Show that the Jacobian matrix of the generalised Lotka-Volterra equations linearised about the fixed point solution is
	\begin{align}
		M_{ij} = x_i^\star (J_{ij} -\delta_{ij}) + \delta_{ij} \left[1 + \sum_{k} (J_{ik} - \delta_{ik})x_k^\star \right]. 
	\end{align}
	
	\item Suppose that we relabel species in order of their abundance so that $x_1^\star \leq x_2^\star \leq \cdots \leq x_N^\star$. Show that the Jacobian matrix has a block structure, with the top-left block being a diagonal matrix with all negative diagonal entries, and the top-right block being entirely zeros.
	
	\item We assume that the leading eigenvalue of the Jacobian is real (which it happens to be in this case \cite{baron2023breakdown}). By considering the determinant of the block matrix, show that if a real eigenvalue of the Jacobian were to change sign due to a change in system parameters, then we require the change of sign of the leading (real) eigenvalue of the reduced Jacobian matrix 
	\begin{align}
		M_{ij}^\star = x_i^\star (J^\star_{ij} - \delta_{ij}) ,
	\end{align}
	where $\underline{\underline{J}}^\star$ is the interaction matrix with all extinct species removed. 
	
	\item Thus, argue that for a change in the sign of the leading eigenvalue of $\underline{\underline{M}}^\star$, one requires an eigenvalue of $\underline{\underline{J}}^\star$ to exceed 1 for the first time.
\end{itemize}

\newpage

\section{Disordered dynamical systems}\label{section:dmft}
\begin{quotation}The behaviour of large and complex aggregates of elementary particles, it turns out, is not to be understood in terms of a simple extrapolation of a few particles. Instead, at each level of complexity entirely new properties appear, and the understanding of the new behaviours requires research which I think is as fundamental in its nature as any other. -- Philip Anderson 
\end{quotation}

So far, we have discussed a number of different methods for deducing the properties of the eigenvalues of large random matrices, and we have shown in particular how these results can be applied to deduce the stability of dynamical systems with fixed random interactions. In this section, we will discuss how we can understand the stability of many-component dynamical systems with random interactions using a somewhat different approach. In so doing, we will shed some further light on the results that we have found already, and we will develop some tools that cater specifically to dynamical phenomena, which are particularly useful in non-linear systems.

Specifically, we will introduce the so-called dynamic mean-field theory (DMFT) of a disordered dynamical system (i.e. a system with fixed random interactions). Using the mean-field approximation to the dynamics (which becomes exact in the limit $N \to \infty$), we will be able to compute quantities such as the correlations and response functions, as well the power-spectrum of fluctuations (all to be defined below). From these quantities, we deduce the circumstances under which the dynamical system tends to a simple equilibrium solution, and investigate the nature of the dynamics when transitions away from this solution occur. We also discuss how these predictions correspond to the results of the previous sections, and how we can extend our analysis to more complicated non-linear systems such as the spherical $p$-spin model. While random matrices (or tensors) are still defining ingredients in more complicated models such as the spherical $p$-spin, the simple spectral analysis above that led us to the semicircle and elliptic law reaches the limits of its applicability. Dynamic mean-field theory thus becomes an indispensable tool.

\subsection{Stochastic differential equations}
In this section, as well as Sections \ref{section:dysonbrownian} and \ref{section:pathintegral}, we will consider dynamical systems of equations that are also subject to noise. For a detailed introduction to stochastic differential equations, the reader is directed to works such as the book by Gardiner \cite{gardiner2009stochastic}. We briefly summarise the main ideas here. 

Let us first consider the following single-component discretised dynamics
\begin{align}
	x(t + \delta t) = x(t) + \Delta t f(x(t), t) + \sqrt{\Delta t}\eta(t) ,
\end{align}
where here $f(x(t), t)$ is some arbitrary function of the variable $x(t)$ and also of time explicitly, and $\eta(t)$ (for each time $t$) are independent Gaussian random numbers zero mean with time-dependent variance $2 D(x(t),t)$, such that $\langle \eta(t) \eta(t') \rangle = 2 D(x(t),t) \delta_{t,t'}$ (where $\delta_{t,t'}$ is the Kronecker delta). By scaling the noise term with $\sqrt{\Delta t}$, we ensure that the distribution of $x(t)$ approaches a sensible limit as $\Delta t \to 0$. We note that we are using the It\^o convention for the noise term, as we do throughout these notes.

Taking the limit $\Delta t \to 0$, we may thus write
\begin{align}
	\dot x = f(x(t), t) + \xi(t), \label{sde1}
\end{align} 
where we now have $\langle \xi(t) \xi(t') \rangle = 2 D(x(t),t)\delta(t-t')$. The natural thing to do now is to compute the probability distribution of $x(t)$, $P(x,t)$. This is given by the Fokker-Planck equation
\begin{align}
	\frac{\partial P(x(t),t)}{\partial t} = \frac{\partial}{\partial x}\left[ \frac{\partial }{\partial x}\left( D(x,t)P(x,t)\right) - f(x,t) P(x,t) \right].
\end{align}
We can easily extend our considerations to systems of many variables. Analogously to Eq.~(\ref{sde1}), we can write the set of $N$ coupled SDEs
\begin{align}
	\dot x_i = f_i(\underline{x}, t) + \xi_i(t),
\end{align}
where the noise term again has zero mean, but now has the correlator $\langle \xi_i(t) \xi_j(t') \rangle = 2 D_{ij}(\underline{x},t) \delta(t-t')$. The corresponding Fokker-Planck equation in this case is given by
\begin{align}
	\frac{\partial P(\underline{x},t)}{\partial t} = \sum_{i= 1}^N\frac{\partial }{\partial x_i}\left[- f_i(\underline{x}, t) P(\underline{x},t) +\sum_{j=1}^N\frac{\partial}{\partial x_j}\left( D_{ij}(\underline{x},t) P(\underline{x},t)\right)\right] . \label{fpedef}
\end{align} 
One notes that in order for the Fokker-Planck equation to be valid, the noise term must be `white'. That is, $\xi_i(t)$ and $\xi_j(t')$ cannot be correlated when $t \neq t'$. In the course of this section, we shall see how `coloured' noise (i.e. time-correlated noise) arises naturally in DMFT as a consequence of there being many coupled interactions.

\subsection{Test case: A coupled linear system}\label{section:stability1}
In this section, we wish to understand the complex dynamic phenomena that are induced due to the presence of quenched (fixed) disordered interactions amongst many components. To this end, we first consider a linear dynamical system, similar to May's model, for the purposes of illustration
\begin{align}
	\dot x_i = - r x_i + \sum_{j} J_{ij} x_j + \xi_i(t) + h_i(t) .\label{linearsystem}
\end{align}
We will later discuss how the analysis can be extended easily to non-linear models. Here, $r>0$ is a constant, and $J_{ij}$ are random matrix elements with statistics as in Section \ref{section:ellipse}
\begin{align}
	\langle J_{ij} \rangle = \frac{\mu}{N},\,\,\,\,\, \langle J_{ij}^2 \rangle - \langle J_{ij} \rangle ^2 = \frac{\sigma^2}{N}, \,\,\,\,\, \langle J_{ij} J_{ji} \rangle- \langle J_{ij} \rangle \langle J_{ji} \rangle  = \frac{\Gamma\sigma^2}{N} .\label{ellipsestatsdyn}
\end{align}
In contrast to May's model, we add noise. The $\xi_i(t)$ again represent Gaussian noise variables with statistics
\begin{align}
	\langle \xi_i(t) \rangle_\xi = 0, \hspace{1cm} 	\langle \xi_i(t)\xi_j(t') \rangle_\xi = \sigma^2_T\delta(t-t') \delta_{ij}, 
\end{align}
where here $\sigma_T^2$ is just a constant. The $h_i(t)$ are arbitrary external fields that we include for the purpose of analysis, but we will later set these to zero. 

The trivial fixed-point solution of this system of equations in the absence of noise or external fields (i.e. for $\xi_i(t) = h_i(t) = 0$) is $x_i(t)= 0$. As we discussed in Section \ref{section:ellipse}, we can deduce the stability of such a deterministic system using the elliptic law and its associated outlier eigenvalue. We have two criteria for stability
\begin{align}
	(1+\Gamma)\sigma <r, \hspace{1cm}	\mu + \frac{\Gamma \sigma^2}{\mu} <r. \label{stabilitycriteria}
\end{align}
Although spectral analysis provides us with the point of instability, we would like to understand more about the properties of the dynamics on the approach to instability, and take into account noise (i.e. $\sigma^2_T>0$). The dynamic mean-field theory (DMFT) method accomplishes exactly this. We will discuss precisely what DMFT is momentarily.

We note that while we have chosen a simple linear system for the sake of clarity, DMFT can facilitate a stability analysis in non-linear systems for which a standard random matrix spectral analysis of the Jacobian matrix is difficult, if not impossible. As such, DMFT is extremely useful for understanding the generalised Lotka-Volterra system \cite{galla2018dynamically, bunin2017ecological, biroli2018marginally, baron2023breakdown,roy2019numerical, altieri2021properties}, for instance. Above all, DMFT is indispensable for detecting dynamic phenomena such as ageing \cite{altieri2020dynamical}, ergodicity breaking \cite{crisanti1993spherical} and chaos \cite{sompolinsky1988chaos,molgedey1992suppressing}. We illustrate this in the case of the spherical $p$-spin model, which exhibits an ergodicity-breaking dynamical transition.

\subsection{Dynamic mean-field theory (DMFT)}
\subsubsection{General overview}
The logic of dynamic mean-field theory is not dissimilar to that of traditional mean-field theory, to which one is usually introduced in the context of simple problems in statistical mechanics such as the Ising model \cite{sakthivadivel2022magnetisation}. 



One constructs a mean-field theory by considering a single component (e.g. a single Ising spin), and imagining how this component experiences its interaction with its `environment' (i.e. the rest of the system). Using the microscopic rules of the model in question, one determines the statistical behaviour of the single component \textit{given} some overall statistical properties of the environment (for example, the average magnetisation, the temperature and any external fields). The key realisation is then to observe that the environment consists of many such components, which are statistically identical to the test component. One thus must match the statistics of the environment and the individual component in a \textit{self-consistent} manner. The mean-field solution is the one that correctly matches the statistics of the individual component and the environment. Crucially, one treats all components as statistically independent (i.e. one neglects correlations) to do this -- this is the mean-field approximation. 

The dynamic mean-field theory of disordered systems, which uses this same philosophy, was first developed in the context of spin glasses by Sompolinsky and Zippelius \cite{sompolinsky1981dynamic, sompolinsky1982relaxational} with the ground work having been laid by de Dominicis \cite{dedominicis1978dynamics}. While these works used the  Martin-Siggia-Rose-Janssen-de Dominicis \cite{martin1973statistical,janssen1976lagrangean,dominicis1976techniques} (MSRJD) path-integral formalism (see also Section \ref{section:pathintegral}), we here use the dynamic cavity approach \cite{roy2019numerical}, which is analogous to the cavity computations that we have met in Sections \ref{section:semicircle} and \ref{section:ellipse}.

\begin{figure}[h]
	\centering 
	\includegraphics[scale = 0.45]{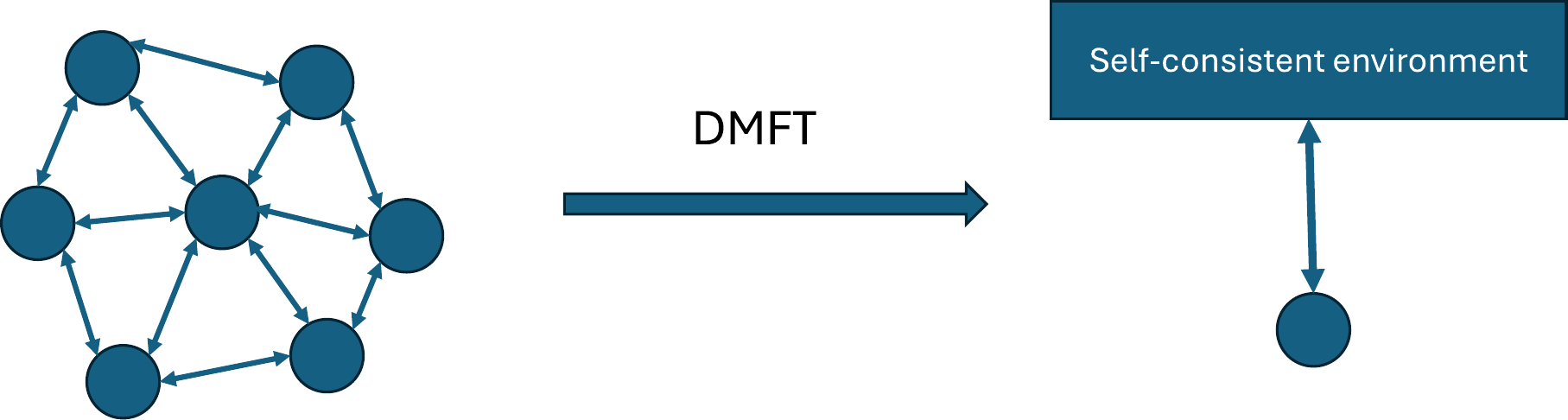}  
	\captionsetup{justification=raggedright,singlelinecheck=false}
	\caption{Schematic illustration of the idea of mean-field theory. }\label{fig:dmft}
\end{figure}

\subsubsection{Introducing the `test' component}
We begin with Eq.~(\ref{linearsystem}). Let us suppose that we introduce a single new component to the system $x_0$. This will be the focal single component of our consideration in the spirit of the above discussion. This new component has dynamics
\begin{align}
	\dot x_0 = -r x_0 + \sum_{j\neq 0} J_{0j}x_j + \xi_0 + h_0. \label{testcomp}
\end{align} 
Let us also consider the effect of the introduction of the new component on the old components. The sum in Eq.~(\ref{linearsystem}) receives an additional term $J_{i0}x_0(t)$ when the new component is introduced. Since this is a small effect for each component of the system, we can consider the linear response due to the introduction of the new component on the trajectories of the old components.

To do this, we must introduce the functional derivative, $\frac{\delta x^{(0)}_i(t)}{\delta h_i(t')}$. The object $\frac{\delta x^{(0)}_i(t)}{\delta h_i(t')}$ can be understood as follows. We imagine that we increment the external field by $\delta h_i(t')$ at only the specific time-point $t'$, and we measure the corresponding change of the dynamic variable $\delta x_i(t)$ at a later time. The object $\frac{\delta x^{(0)}_i(t)}{\delta h_i(t')}$ thus encapsulates the `response' of the system to perturbations. We emphasise that these will eventually be set to zero, and hence we will evaluate the objects $\frac{\delta x^{(0)}_i(t)}{\delta h_i(t')}$ at $h_i = 0$.

The additional term $J_{i0}x_0(t)$ in the sum in Eq.~(\ref{linearsystem}) can be thought of as an increment to $h_i(t)$ at every time point $t$. Using the definition of the functional derivative, we therefore have for each $i \neq 0$  
\begin{align}
	x_i(t) = x_i^{(0)}(t) + \int_0^t dt' \frac{\delta x^{(0)}_i(t)}{\delta h_i(t')} J_{i0} x_0(t') , \label{linearresponse}
\end{align}
where $x_i^{(0)}(t)$ is the trajectory of the component $i$ without the additional component $0$. The higher-order terms in the expansion in Eq.~(\ref{linearresponse}) would vanish for $N \to \infty$\footnote{In principle, we should also include the linear response due to the perturbations of the other elements here, i.e. $\frac{\delta x^{(0)}_i(t)}{\delta h_k(t')}$ for $i\neq k$. However, we ignore these `off-diagonal elements' in the same way that we did when we used the cavity method before (see Section \ref{section:offdiag}). One can make arguments along similar lines as to why this is valid. However, perhaps the simplest way to see that the components do in fact `decouple' is via the path integral approach of Section \ref{section:pathintegral}}. Inserting Eq.~(\ref{linearresponse}) into Eq.~(\ref{testcomp}), we have
\begin{align}
	\dot x_0 = -r x_0 + \sum_{j\neq 0} J_{0j}x_j^{(0)} + \int_0^t dt' \sum_{j\neq 0} J_{0j}\frac{\delta x^{(0)}_j(t)}{\delta h_j(t')} J_{j0} x_0(t') + \xi_0 + h_0. \label{testspecies}
\end{align}
We have thus succeeded in writing the dynamics of the component $0$ in terms of quantities from which it is statistically independent, since the trajectories $x_j^{(0)}(t)$ are by definition independent of the trajectory $x_0(t)$. This was possible because the effect of introducing the component $0$ was small for large $N$, owing to the scaling of $J_{0j}$ with $N$. We thus see how the rest of the components $i\neq0$, which constitute the `environment' of $x_0$, can be treated as statistically independent from component $0$ for large $N$. This is how we can begin to build the mean-field theory.

\subsubsection{Statistical behaviour of the interaction terms}

Let us now consider how the terms in Eq.~(\ref{testspecies}) vary between realisations of the new random matrix entries $\{J_{0j}, J_{j0}\}$ associated with the test component. The following analysis is very similar to that performed in Section \ref{section:cavity}, where we used the cavity approach to evaluate the resolvent, with the only difference being that instead of removing one component, we add anew one. The procedure here also bears a resemblance to the methodology of Thouless, Anderson and Palmer \cite{thouless1977solution} in their `solution of a solvable model of a spin glass'.

Let us first consider the sum $\sum_{j\neq 0} J_{0j}x_j^{(0)}(t)$ in Eq.~(\ref{testspecies}). Since this is a large sum of random quantities (noting that $J_{0j}$ and $x_j^{(0)}$ are independent), it obeys the central limit theorem. It thus suffices to compute $\langle \sum_{j\neq 0} J_{0j}x_j^{(0)}(t) \rangle_0$ and $\mathrm{Cov}_0\left(\sum_{j\neq 0} J_{0j}x_j^{(0)}(t), \sum_{k\neq 0} J_{0k}x_k^{(0)}(t) \right)$, where here we average only over realisations of the coupling constants $\{J_{0j}, J_{j0}\}$, imagining the others to be fixed, and highlight this with a subscript $0$.

Firstly, for the mean, we may write
\begin{align}
	\left \langle \sum_{j\neq 0} J_{0j}x_j^{(0)}(t) \right\rangle_0 = \sum_{j\neq 0}\left \langle  J_{0j}\right\rangle_0  x_j^{(0)}(t) = \frac{\mu}{N} \sum_{j} x_j^{(0)} \equiv \mu M^{(0)}(t).  
\end{align}
We can also determine the covariance of the sum $\sum_{j\neq 0} J_{0j}x_j^{(0)}(t)$ at different time points
\begin{align}
	\mathrm{Cov}_0\left(\sum_{j\neq 0} J_{0j}x_j^{(0)}(t), \sum_{k\neq 0} J_{0k}x_k^{(0)}(t') \right) 
	=&\sum_{j\neq 0} \left \langle J_{0j}^2 \right\rangle_0 x_j^{(0)}(t)x_j^{(0)}(t')+ \sum_{j\neq k\neq 0} \left\langle J_{0j} \right\rangle_0 x_j^{(0)}(t) \left\langle J_{0k} \right\rangle_0 x_k^{(0)}(t) \nonumber \\
	&-\mu^2  M^{(0)}(t) M^{(0)}(t') \nonumber \\
	=&   \sigma^2 C^{(0)}(t,t')  ,
\end{align}
where $C^{(0)}(t,t') \equiv \frac{1}{N} \sum_j  x_j^{(0)}(t)x_j^{(0)}(t')$. We therefore see that the sum $\sum_{j\neq 0} J_{0j}x_j^{(0)}(t)$ can be thought of as Gaussian random variable that fluctuates between realisations of the interaction coefficients $\{J_{0j}, J_{j0}\}$. It is also correlated in time (i.e. it is a \textit{coloured} noise). Notably, its mean, variance and correlations are dependent on the statistics of the rest of the components in the ensemble. We are beginning to see how our desired mean-field theory is emerging. 

Let us now turn our attention to the quantity $\sum_{j\neq 0} J_{0j}\frac{\delta x^{(0)}_j(t)}{\delta h_j(t')} J_{j0} x_0(t')$ in Eq.~(\ref{testspecies}). Let us compute its mean (averaging only over the interactions  $\{J_{0j}, J_{j0}\}$)
\begin{align}
	\left\langle \sum_{j\neq 0} J_{0j}\frac{\delta x^{(0)}_j(t)}{\delta h_j(t')} J_{j0} x_0(t') \right\rangle_0 &=  \sum_{j\neq 0}\left\langle J_{0j} J_{j0}\right\rangle_0 \frac{\delta x^{(0)}_j(t)}{\delta h_j(t')} x_0(t') \nonumber \\
	&\approx \frac{\Gamma\sigma^2}{N} \sum_{j\neq 0} \frac{\delta x^{(0)}_j(t)}{\delta h_j(t')} x_0(t')  \equiv \Gamma \sigma^2 R^{(0)}(t,t') x_0(t') . 
\end{align}
where we ignore a term proportional to $\frac{\mu^2}{N^2}$, which is subleading in $N$. Computing the covariance of the sum $\sum_{j\neq 0} J_{0j}\frac{\delta x^{(0)}_j(t)}{\delta \xi_j(t')} J_{j0} x_0(t')$, we obtain
\begin{align}
	&\mathrm{Cov}\left( \sum_{j\neq 0} J_{0j}\frac{\delta x^{(0)}_j(t)}{\delta h_j(t')} J_{j0} x_0(t') ,  \sum_{k\neq 0} J_{0k}\frac{\delta x^{(0)}_k(t)}{\delta h_k(t'')} J_{k0} x_0(t'')\right) \nonumber \\
	&\approx \sum_{j\neq 0} \left[\left\langle (J_{0j}J_{j0})^2 \right\rangle - \left\langle (J_{0j}J_{j0}) \right\rangle^2\right]  \frac{\delta x^{(0)}_j(t)}{\delta h_j(t')} x_0(t') \frac{\delta x^{(0)}_j(t)}{\delta h_j(t'')} x_0(t'') .
\end{align}
So we see that if $\left\langle (J_{0j}J_{j0})^2 \right\rangle - \left\langle (J_{0j}J_{j0}) \right\rangle^2 \sim N^{-(1+\epsilon)}$ for $\epsilon >0$, then the above covariance is of the order $N^{-\epsilon}$ and thus vanishes in the thermodynamic limit $N\to\infty$. Therefore, as long as the fourth moments of $J_{0j}$ decay sufficiently quickly with $N$, the term $\sum_{j\neq 0} J_{0j}\frac{\delta x^{(0)}_j(t)}{\delta h_j(t')} J_{j0} x_0(t')$ can be treated as a deterministic (self-averaging) quantity. 

Let us now take a brief moment to understand the significance of each the terms in Eq.~(\ref{testspecies}), whose statistics we have computed. The term $\sum_{j\neq 0} J_{0j} x_j^{(0)}$ encapsulates the straightforward influence of environment on the test component -- the na\"ive mean-field theory. We saw that it has a non-zero average (proportional to $N\langle J_{0j} \rangle = \mu$), but it can vary depending on the particular realisation of the $N$ interaction coefficients $J_{0j}$. Ultimately, this means that because each component in the system $i$ has a different set of interaction coefficients $\{J_{ij}\}$, each component will take a different trajectory, under the influence of a heterogeneous coupling with the wider system.

The term involving $\sum_{j\neq 0} J_{0j}\frac{\delta x^{(0)}_j(t)}{\delta h_j(t')} J_{j0} x_0(t')$ is perhaps more subtle. This term represents how the trajectory of the test component perturbs the environment, and how this perturbation feeds back to the test component itself in a non-negligible fashion, modifying the na\"ive mean-field. We saw that this term does not fluctuate between realisations of the $2N$ coefficients $\{J_{0j},J_{j0}\}$. This `reaction' term is sometimes referred to as an `Onsager' term, after Lars Onsager's work on Polar liquids \cite{onsager1936electric}. 

\subsubsection{Result: effective process and interpretation}
From the above reasoning, we can therefore rewrite the process in Eq.~(\ref{testspecies}) as 
\begin{align}
	\dot x_0 = -rx_0 + \mu M^{(0)} + \Gamma \sigma^2 \int_0^t dt' R^{(0)}(t,t') x_0(t') + \eta_0(t) + \xi_0(t) + h_0(t) ,
\end{align}
where we have defined an additional Gaussian noise variable $\eta_0(t)$, which has zero mean and autocorrelation
\begin{align}
	\left \langle \eta_0(t) \eta_0(t')\right\rangle_0 = \sigma^2 C^{(0)}(t,t') \equiv \frac{\sigma^2}{N} \sum_{j\neq  0} x_j^{(0)}(t) x_j^{(0)}(t'), \label{orderp01}
\end{align}
and where we recall
\begin{align}
	M^{(0)}(t) \equiv \frac{1}{N} \sum_{j\neq 0} x_j^{(0)}(t), \hspace{2cm}
	R^{(0)}(t,t') \equiv \frac{1}{N} \sum_{j\neq 0} \frac{\delta x_j^{(0)}(t)}{\delta h_j(t')} .\label{orderp02}
\end{align}
As we have demonstrated, such a process has identical statistics (for $N \to \infty$) to Eq.~(\ref{testcomp}). Let's review what we have found so far. We began with the dynamical system in Eq.~(\ref{linearsystem}), and introduced an additional `test' component $x_0$. We then treated the rest of the system as the `environment' of the new component, and we computed the behaviour of the test component given the statistics of the environment [$M^{(0)}(t)$, $C^{(0)}(t,t')$ and $R^{(0)}(t,t')$].

The crucial realisation to make now is that the test component $x_0$ is in fact not special in any real sense. It is representative of the behaviour of any component in the system. Further, the quantities $C^{(j)}(t,t')$, $M^{(j)}(t)$ and $G^{(j)}(t,t')$ differ only by an amount $O(1/N)$ across components. This is the key to determining the behaviour of the full system self-consistently. 

We therefore write for an \textit{arbitrary} component in the system the so-called \textbf{effective process}
\begin{empheq}[box={\fboxsep=6pt\fbox}]{align}
	\dot x = -r x + \mu M(t) + \Gamma\sigma^2 \int_0^t dt' R(t,t') x(t') + \eta(t) + \xi(t) + h(t).\label{effectiveprocess}
\end{empheq}
This is to be interpreted as follows. The interaction of the each component with the rest of the system has now been replaced by an effective Gaussian noise term $\eta(t)$, an average interaction term $\mu M(t)$, and the Onsager reaction term $\Gamma\sigma^2 \int dt' R(t,t') x(t')$. These terms are to be thought of collectively as the `self-consistent environment' in Fig. \ref{fig:dmft}. The noise is a time-correlated (i.e. coloured) zero-mean Gaussian noise with correlator
\begin{align}
	\left\langle \eta(t) \eta(t') \right\rangle_\eta = \sigma^2 C(t,'t) ,
\end{align}
where $\left\langle \cdot \right\rangle_\eta$ indicates an average over realisations of the Gaussian random variables $\{\eta(t)\}$. Averaging over realisations of this noise is equivalent (in the $N \to \infty$ limit) to averaging over the different sets of interaction coefficients $J_{ij}$ (i.e. the different environmental effect) that each component $i$ possesses.

Bearing in mind that each component in the original system experiences a separate white noise term $\xi_i(t)$, the so-called `order parameters' $M(t)$, $R(t,t')$ and $C(t,t')$ are obtained by averaging over realisations of the noises $\eta(t)$ and $\xi(t)$, so that [c.f. Eqs.~(\ref{orderp01}) and (\ref{orderp02})]
\begin{align}
	M(t) = \left\langle x(t) \right\rangle_{\eta,\xi}, \hspace{1cm} R(t,t') = \left\langle \frac{\delta x(t)}{\delta h(t')} \right\rangle_{\eta,\xi}, \hspace{1cm}  C(t,t') = \left\langle x(t)x(t') \right\rangle_{\eta,\xi} .
\end{align}
In words, one may refer to $M(t)$ as a `magnetisation', $R(t,t')$ as a `response function' and $C(t,t')$ as a `correlation function'. If we can determine $M(t)$, $R(t,t')$ and $C(t,t')$ such that when we average over many realisations of the noise in Eq.~(\ref{effectiveprocess}), then we will have succeeded in producing a representative \textit{single-component} process (i.e. a mean-field theory), which reproduces the statistical behaviour of the original coupled system in Eq.~(\ref{linearsystem}). So, once we have $M(t)$, $R(t,t')$ and $C(t,t')$, we will in some sense have solved the problem, and we will have a full characterisation of the dynamical properties of the system. 

\subsection{Computing the dynamic order parameters}
\subsubsection{ODE form for order parameters}
Let us now compute the quantities $M(t)$, $R(t,t')$ and $C(t,t')$ from Eq.~(\ref{effectiveprocess}) taking the external field to be $h(t) = 0$. Beginning with $M(t)$, we can take the average over the Gaussian noises in Eq.~(\ref{effectiveprocess}) to obtain
\begin{align}
	\dot M(t) = - rM(t) + \mu M(t) +\Gamma \sigma^2 \int_0^t dt' R(t, t') M(t').\label{mode}
\end{align}
We may also functionally differentiate Eq.~(\ref{effectiveprocess}) and then take the average to obtain
\begin{align}
	\partial_t R(t,T) = -r R(t,T) + \Gamma \sigma^2 \int_0^t dt' R(t, t') R(t',T) + \delta(t-T) .\label{gode}
\end{align}
Finally, we consider the correlation function. One must treat this object slightly more carefully. One has
\begin{align}
	\partial_t C(t,T) =& - r C(t,T) + \mu M(t) M(T) + \Gamma \sigma^2 \int_0^t dt' R(t,t') C(t', T) \nonumber \\
	&+ \left\langle \eta(t) x(T) \right\rangle_{\eta, \xi}  + \left\langle \xi(t) x(T) \right\rangle_{\eta, \xi}. 
\end{align}
We therefore must compute the objects $\left\langle \eta(t) x(T) \right\rangle_{\eta, \xi}$ and $\left\langle \xi(t) x(T) \right\rangle_{\eta, \xi}$. This can be accomplished by recognising that $R(t,t')$ can be related to $C(t,t')$ via these correlators [see Eqs.~(\ref{xetaav}) and (\ref{xxiav})]. It is helpful for this purpose to recognise that $\partial x(t)/\partial h(t')\vert_{h=0} = \partial x(t)/\partial \eta(t')\vert_{h=0}$.

\begin{tcolorbox}[colback=blue!10!white,colframe=blue!90!black,title=Lemma: Correlators  $\left\langle \eta(t) x(T) \right\rangle_{\eta, \xi}$ and $\left\langle \xi(t) x(T) \right\rangle_{\eta, \xi}$]
	Writing the average over the noise explicitly and integrating by parts, one has
	\begin{align}
		R(T,t') &= \int D\eta(t)\, \frac{1}{\mathcal{N}}e^{-\frac{1}{2\sigma^2} \int dt dt' \eta(t) C^{-1}(t,t') \eta(t') } \frac{\delta x(T)}{\delta \eta(t')} \nonumber \\
		&= \frac{1}{\sigma^2}\int dt  \left\langle x(T) \eta(t) \right\rangle C^{-1}(t, t') ,
	\end{align} 
	where we define $C^{-1}(t,t')$ such that $\int dT C^{-1}(t,T) C(T,t') = \delta(t-t')$, the measure $D\eta(t)$ indicates an integral over all possible realisations of the noise $\eta(t)$, and $\mathcal{N}$ is a normalisation constant. The second equality is obtained by integrating by parts. Technically, the integration over trajectories should be interpreted in the sense that we first discretise the dynamics, perform the computation, and subsequently take the continuous time limit. Finally, using the definition of $C^{-1}(t,t')$, we obtain
	\begin{align}
		\left\langle x(T) \eta(t) \right\rangle = \sigma^2 \int_0^{T} dt' R(T,t') C(t', t). \label{xetaav}
	\end{align}
	Using similar reasoning, we also have 
	\begin{align}
		\langle  x(T) \xi(t)\rangle = \sigma^2_T\int_0^T dt' R(T,t') \delta(t'- t) = \sigma^2_T R(T,t). \label{xxiav}
	\end{align}
\end{tcolorbox} 

The correlation function therefore obeys
\begin{align}
	\partial_t C(t,T) =& - r C(t,T) + \mu M(t) M(T) + \Gamma \sigma^2 \int_0^t dt' R(t,t') C(t', T) \nonumber \\
	&+ \sigma^2 \int_0^T dt' R(T,t') C(t', t) + \sigma^2_T R(T,t) . \label{code}
\end{align}
These self-consistent equations for the dynamic order parameters are sometimes referred to as the Cugliandolo-Kurchan equations \cite{cugliandolo1993analytical}, which were first developed for the spherical $p$-spin model (see Section \ref{section:pspin}).

\subsubsection{Explicit formulae for the order parameters}
The present linear model under consideration affords us the opportunity to obtain closed-form solutions for the order parameters. Assuming a time-translation-invariant (TTI) solution for the response function $R(t,T) = R(t-T)$, we may take the Laplace transform of Eq.~(\ref{gode}) to obtain
\begin{align}
	\hat R(u) = \frac{1}{2\Gamma \sigma^2}\left[r+u - \mathrm{sign}[\mathrm{Re}(r+u)] \sqrt{(r+u)^2-4\Gamma \sigma^2} \right],\label{rhateq}
\end{align}
where $\hat R(u) = \int dt e^{-ut} R(t)$. One notes that this coincides with the analytic formula for the resolvent in Eq.~(\ref{csolanalytic}). This is no accident. The response functions of a linear dynamical system are exactly related to the resolvent of the random matrix of the couplings, and can therefore be used to compute the eigenvalue density \cite{baron2022eigenvalues, baron2023pathintegralapproachsparse} (see Section \ref{section:pathintegral}).

In the case where $\Gamma = 0$, one can obtain an explicit formula for the response function. In the more general case above, one must settle for $R(t)$ obtained via the inverse Laplace transform. One has simply that $\hat R(u) = \frac{1}{r + u}$ for $\Gamma = 0$, and therefore
\begin{align}
	R(t) = e^{-rt} \Theta(t),
\end{align}
where $\Theta(\cdot)$ is the Heaviside theta function. One notes that the response function $R(t)$ must be zero for negative $t$ due to causality. 

\begin{figure}[H]
	\centering 
	\includegraphics[scale = 0.48]{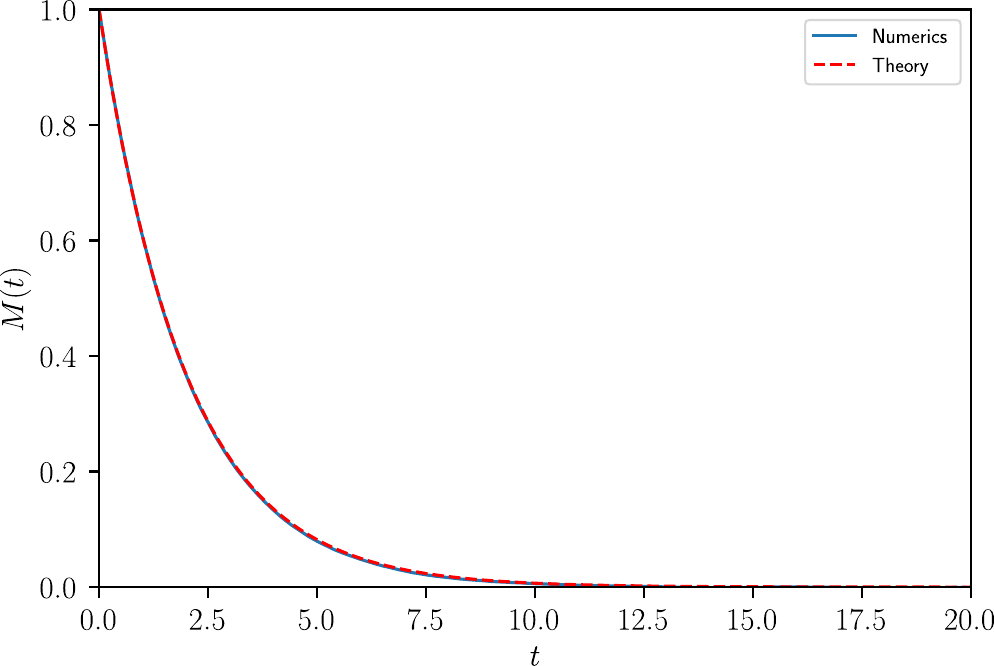} 
	\includegraphics[scale = 0.48]{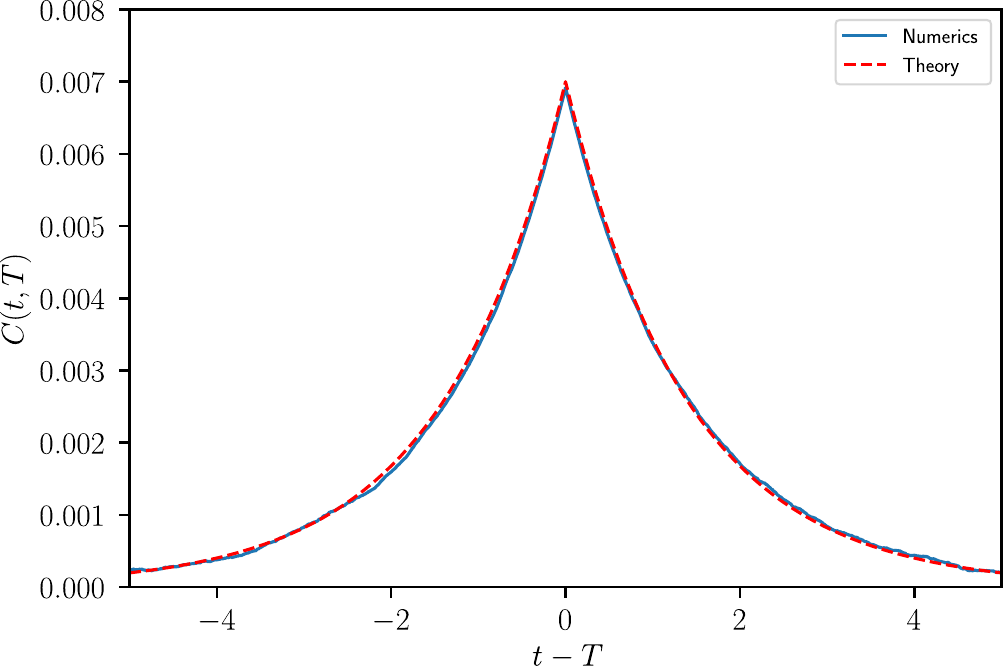}
	\captionsetup{justification=raggedright,singlelinecheck=false}
	\caption{Relaxational dynamics of the system in Eq.~(\ref{linearsystem}) compared with the DMFT predictions in Eqs.~(\ref{mgamma0}) and (\ref{cgamma0}). The system parameters are $\Gamma = 0$, $\sigma =0.7$, $r = 1$, $\sigma_T = 0.1$, $\mu = 0.5$ and $N = 4000$. Numerical results were averaged over 10 trials. }\label{fig:orderparameters}
\end{figure}

Similarly, we can obtain $M(t)$ via the Laplace transform
\begin{align}
	\hat M(u) = \frac{M(0)}{u + r -\mu - \Gamma \sigma^2 \hat R(u)} . 
\end{align}
In the case $\Gamma = 0$, we also obtain a simple solution 
\begin{align}
	M(t) = M(0) e^{-(r-\mu)t}. \label{mgamma0}
\end{align}
Finally, for the correlation function we have in the long-time limit [i.e. when $M(t) = 0$]
\begin{align}
	\tilde C(\omega) = \frac{\sigma^2_T\tilde R(-\omega)}{i\omega +r-\sigma^2\Gamma \tilde R(\omega) - \sigma^2\tilde R(-\omega)} , \label{ctransform}
\end{align}
where $\tilde C(\omega) = \int_{-\infty}^\infty dt e^{i\omega t} C(\omega)$ is the Fourier transform of the correlation function. In the case $\Gamma = 0$, we have
\begin{align}
	C(t-T) = \sqrt{\frac{\sigma^4_T}{4(r^2-\sigma^2)}}\exp\left[- \sqrt{r^2 -\sigma^2} \left\vert t-T\right\vert \right] . \label{cgamma0}
\end{align}
The expressions in Eqs.~(\ref{mgamma0}) and (\ref{cgamma0}) are tested against numerics in Fig.~\ref{fig:orderparameters}. 

\subsection{Power-spectrum of fluctuations}
One other common quantity of interest is the so-called power spectrum of fluctuations \cite{opper1992phase} $\left \langle \left \vert \tilde x(\omega) \right\vert^2 \right \rangle$, where again $\tilde x(\omega)$ is the Fourier transform. This quantity is familiar to us from the study of forced oscillations of a damped oscillator in classical mechanics \cite{young1996university}, for example. It is particularly useful to detect to the presence of (quasi-)oscillatory behaviour \cite{galla2009intrinsic,brett2013stochastic}, which is indicated by a peak of the power spectrum at a non-zero dominant frequency of oscillation, and can also be used to detect the onset of dynamical instability \cite{opper1992phase}.

\begin{figure}[H]
	\centering 
	\includegraphics[scale = 0.6]{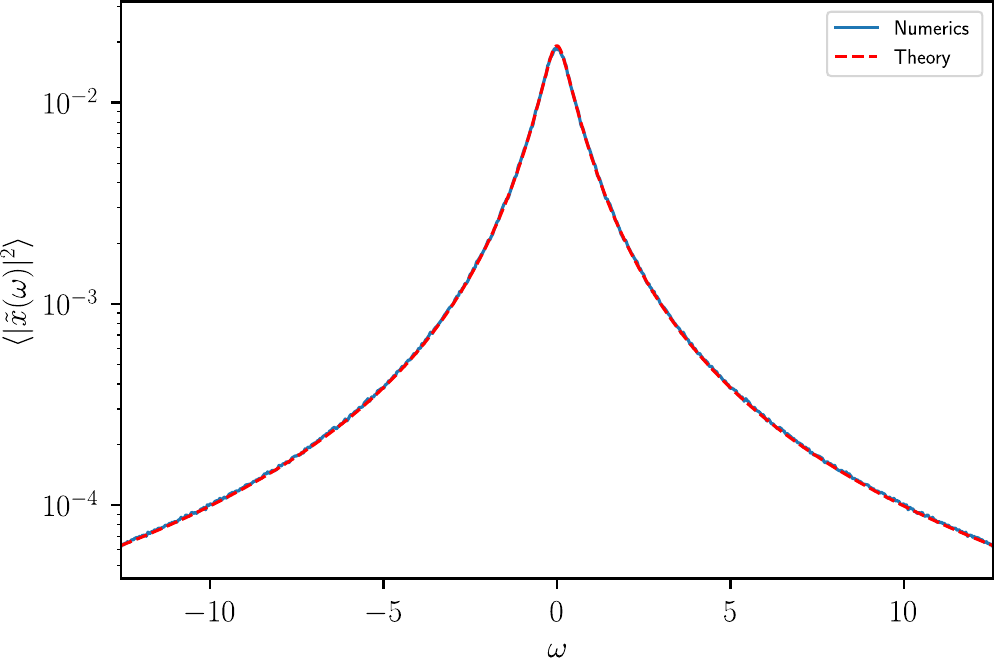} 
	\captionsetup{justification=raggedright,singlelinecheck=false}
	\caption{Power spectrum of fluctuations for the system in Eq.~(\ref{linearsystem}) compared with the DMFT prediction in Eqs.~(\ref{powerspectrum}). The system parameters are $\Gamma = 0.9$, $\sigma =0.4$, $r = 1$, $\sigma_T = 0.1$, $\mu = 0$ and $N = 1000$. Numerical results were averaged over 10 trials. }\label{fig:powerspectrum}
\end{figure}

From Eq.~(\ref{effectiveprocess}), one obtains in the long-time limit (in which $M(t) \to 0$)
\begin{align}
	\left \langle \left \vert \tilde x(\omega) \right\vert^2 \right \rangle = \frac{\sigma^2_T}{\left \vert i \omega + r - \Gamma \sigma^2 \tilde R(\omega) \right\vert^2 - \sigma^2} = \frac{\sigma^2_T}{\left \vert \tilde R(\omega) \right\vert^{-2} - \sigma^2}, \label{powerspectrum}
\end{align}
where in the last equality we have used Eq.~(\ref{gode}). In the case $\Gamma = 0$, we once again obtain a great simplification, and we have
\begin{align}
	\left \langle \left \vert \tilde x(\omega) \right\vert^2 \right \rangle = \frac{\sigma^2_T}{\omega^2 + r^2 - \sigma^2}. \label{powerspectrumgamma0}
\end{align}
We test Eq.~(\ref{powerspectrum}) against numerics in Fig. \ref{fig:powerspectrum}. 

\subsection{Dynamical transitions and comparison with predictions from the eigenvalue spectrum}\label{section:dynamicalinstabilities}
We showed in Section \ref{section:stability1} that the trivial fixed point of the deterministic system, which is recovered for $\sigma_T \to 0$, becomes unstable under two circumstances [see Eq.~(\ref{stabilitycriteria})], depending on whether it is the bulk ellipse or the outlier eigenvalue that gains a positive real part. Let us now understand how these instabilities manifest in the dynamics of the stochastic system in Eq.~(\ref{linearsystem}). 

In the case of the outlier eigenvalue crossing the imaginary axis, one finds that the average $M(t)$ diverges exponentially from its initial value. One can see this as follows. We make an exponential ansatz $M(t) = M(0) e^{\lambda t}$ in Eq.~(\ref{mode}), and we obtain
\begin{align}
	\lambda + r - \mu - \Gamma \sigma^2 \hat R(\lambda) = 0.
\end{align}
Using that $\Gamma \sigma^2 [\hat R(\lambda)]^2 + 1 - (r + \lambda)\hat R(\lambda) = 0$, we obtain $\lambda =-r+ \mu + \Gamma \sigma^2/\mu$ [i.e. exactly the outlier eigenvalue from Eq.~(\ref{outlierellipse}) with $r=1$], and thus we see that $M(t)$ blows up exponentially beyond the point where $\lambda = 0$. 

Let us now consider instead the point at which the ellipse crosses the imaginary axis. In this case, one finds that the power spectrum of fluctuations and the 2-point correlator $C(t,t')$ diverge. We can see this by inspecting Eq.~(\ref{ctransform}) and Eq.~(\ref{powerspectrum}). In both cases, one sees that the denominator is zero when
\begin{align}
	\sigma^2\left\vert \tilde R(\omega) \right\vert^2 = 1.
\end{align}
Considering the mode $\omega = 0$ in particular (which corresponds to the time average), and using that $\Gamma \sigma^2 [ \tilde R(0)]^2 - r  \tilde R(0) + 1 = 0$, one indeed finds that the power spectrum and correlator diverge when $(1+\Gamma)\sigma = r$. We note that both of these transitions occur in a manner that is independent from $\sigma_T$, which is why we recover the results from the deterministic analysis.

We note that the different nature of the dynamical instabilities when the outlier and the bulk cross the imaginary axis correspond to the differing nature of the eigenvectors. It can be shown (see Section \ref{section:marchenkopastur}) that the eigenvector of the outlier eigenvalue has a non-vanishing overlap with the vector $(1,1,\cdots,1)^T$. On the other hand, the eigenvectors of eigenvalues belonging to the bulk region have a vanishing overlap with such a vector. 

In the case of both of the dynamical transitions discussed above, the system no longer attains a steady state solution, and time-translational invariance (TTI) is violated. We now discuss the spherical $p$-spin model, for which the TTI violation is even more non-trivial and interesting.

\subsection{Spherical $p$-spin model}\label{section:pspin}
Although we certainly gained more insight into the dynamics of the linear system Eq.~(\ref{linearsystem}) by using DMFT, we saw that many system properties (particularly the locations of the dynamical transitions in parameter space) could simply be understood using the spectrum of $\underline{\underline{J}}$. Where DMFT really comes into its own is in the study of non-linear disordered systems, particularly those that exhibit exotic temporal phenomena. 

Here, we take a paradigmatic example of a model of glassy behaviour: the spherical $p$-spin model (SPSM). The SPSM was first concocted as a tractable toy model of a spin glass \cite{gross1984simplest, binder1986spin, mezard1987spin}. It drew a good deal of attention and excitement due to the fact that its DMFT equations \cite{kirkpatrick1987p} resembled those derived from the Mode Coupling Theory approach \cite{castellani2005spin, reichman2005mode} to the structural glass transition, and it displayed the ergodicity-breaking properties that are also characteristic of the (spin) glass transition \cite{crisanti1993spherical}. Here, `ergodicity breaking' means that the system fails to decorrelate over time when the temperature is below a certain threshold, meaning that the system is no longer able to explore the full phase space. Speaking roughly, this means that the spins get stuck in the vicinity of some locally preferred arrangement (of which there can be many). 

Although spin glass theory might be considered a somewhat niche subject in the physics corpus, the consequences of this line of inquiry have been very far-reaching indeed. Mathematical techniques such as DMFT and replica theory, as well as concepts such as complex energy landscapes, jamming and ageing, have had huge consequences for many areas of research outside of spin glasses physics. Indeed, the fruits of labours in spin glass theory and beyond led to the Nobel prize for Giorgio Parisi in 2021. We discuss the replica technique and its relation to spin glasses in Section \ref{section:replicas}.

\subsubsection{Model}
The relaxational dynamics of the SPSM are given as follows
\begin{align}
	\dot s_i = -r(t) s_i - \frac{\partial H}{\partial s_i } + \xi_i(t) , \label{pspin}
\end{align}
where the Hamiltonian is given by 
\begin{align}
	H = -\sum_{i}h_i s_i-\sum_{i_1<\cdots<i_p} J_{i_1 \cdots i_p} s_{i_1}\cdots s_{i_p} , \label{pspinhamiltonian}
\end{align}
for $p\geq 2$, and the Gaussian white noise has statistics
\begin{align}
	\langle \xi_i(t) \rangle = 0, \hspace{1cm} \langle \xi_i(t) \xi_i(t') \rangle = 2T\delta(t-t') .
\end{align}
The random coefficients are symmetric under an exchange of indices and are drawn from a distribution with statistics
\begin{align}
	\left\langle J_{i_1 \cdots i_p} \right\rangle =0, \hspace{1cm} \left\langle J_{i_1 \cdots i_p}^2 \right\rangle = \frac{p! J^2}{2 N^{p-1}} . \label{pspinstats}
\end{align}
The time-varying parameter $r(t)$ is a Lagrange multiplier chosen to enforce the spherical constraint $\sum_i s_i^2(t) = N$. We note that the stationary distribution of the spins $\{s_i\}$ (assuming that it is reachable) is given by the Boltzmann distribution 
\begin{align}
	P(\{s_i\}) \propto \delta\left(\sum_i s_i - N\right) e^{-\frac{H}{T}}.
\end{align}
\subsubsection{DMFT effective process and dynamic order parameter}
We may perform a completely analogous DMFT calculation as we did for the linear model in Eq.~(\ref{linearsystem}). As a result, we again obtain an effective process. That is, in the $N\to \infty$ limit, the statistics of the set of spins $\{s_i\}$ is replicated by the following single-spin process
\begin{align}
	\dot s = -r(t) s + \frac{p(p-1)J^2}{2} \int_{0}^t dt' R(t,t') C^{p-2}(t,t') s(t') + \eta(t)+ \xi(t) + h(t),
\end{align}
where the additional coloured noise has statistics
\begin{align}
	\langle \eta(t) \rangle_\eta = 0, \hspace{1cm} \langle \eta(t) \eta(t') \rangle_\eta = \frac{p J^2}{2} C^{p-1}(t,t'),
\end{align}
and the response and correlation functions are, as usual, defined self-consistently as
\begin{align}
	R(t,t') = \left\langle \frac{\delta s(t)}{\delta h(t')} \right\rangle_{\eta, \xi}, \hspace{1cm} C(t,t') = \left\langle s(t) s(t')\right\rangle_{\eta, \xi}.
\end{align}
We may also obtain closed-form equations for the correlation and response functions, analogous to Eqs.~(\ref{gode}) and (\ref{code}), as were obtained by Cugliandolo and Kurchan \cite{cugliandolo1993analytical}
\begin{align}
	\partial_t C(t,t') =& - r(t) C(t,t') + \frac{p(p-1)J^2}{2}\int_{0}^t ds C^{p-2}(t,s) R(t,s) C(s,t') \nonumber \\
	&\hspace{1cm} + \frac{pJ^2}{2}\int_{0}^{t'} ds C^{p-1}(t,s)R(t',s) + 2 T R(t',t), \nonumber \\
	\partial_t R(t,t') =& - r(t) R(t,t') + \frac{p(p-1)J^2}{2}\int_{0}^t ds C^{p-2}(t,s) R(t,s) R(s,t') +  \delta(t'-t). \label{randceqs}
\end{align}
All that remains is to close the set of equations is to find the Lagrange multiplier $r(t)$. This is obtained by enforcing the spherical constraint $\sum_i s_i^2(t) = N$, which connotes $C(t,t) = 1$. This is accomplished using It\^{o}'s lemma, which allows us to handle functions of stochastic processes [see Eq.~(\ref{ito})]. It\^{o}'s lemma tells us that
\begin{align}
	\frac{ds(t)^2}{dt} = 2 s(t) \frac{ds}{dt} + 2 T. 
\end{align}
Setting $\langle s^2(t) \rangle = 1$ and $\langle \frac{ds(t)^2}{dt}\rangle = 0$, we therefore have 
\begin{align}
	-T = \left\langle s(t) \frac{ds}{dt} \right\rangle &= - r(t) +  \frac{p(p-1)J^2}{2} \int_{0}^t dt' R(t,t') C^{p-1}(t,t')  + \langle \eta(t) s(t) \rangle \nonumber \\
	&=  - r(t) + \frac{p^2J^2}{2} \int_{0}^t dt' R(t,t') C^{p-1}(t,t'), \label{itoed}
\end{align}
where for the \^Ito noise convention we use $\langle s(t) \xi(t) \rangle = 0$. Hence,
\begin{align}
	r(t) = T + \frac{p^2J^2}{2} \int_{0}^t dt' R(t,t') C^{p-1}(t,t') . \label{lagrange}
\end{align}
The Eqs.~(\ref{randceqs}) and (\ref{lagrange}) are a closed set of equations that can, in principle, be integrated to yield the response and correlation functions.

\subsubsection{Ergodicity-breaking dynamical transition}

\begin{figure}[H]
	\centering 
	\includegraphics[scale = 0.6]{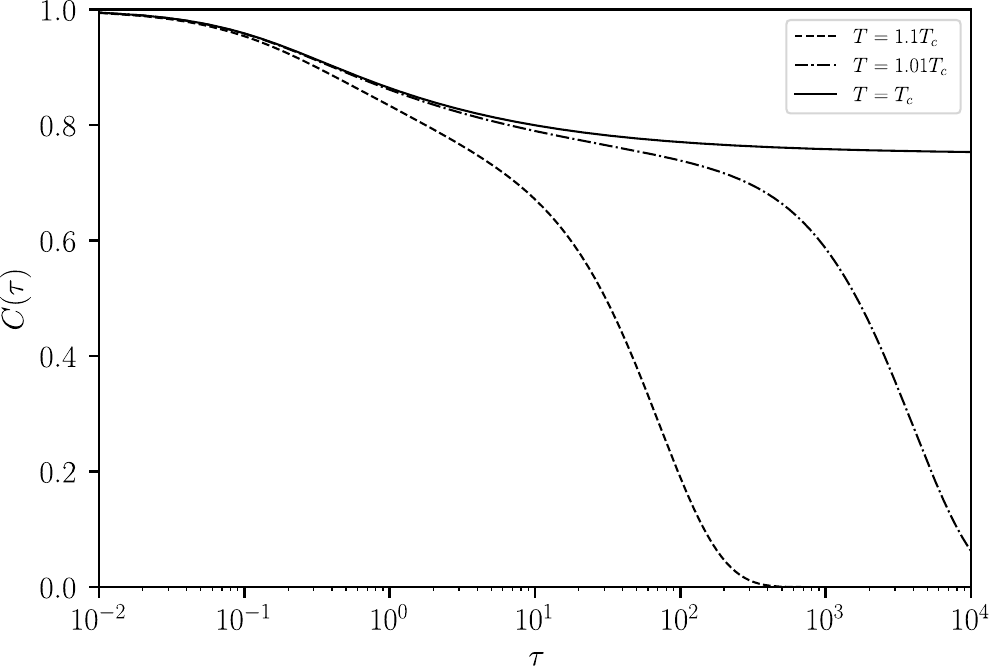} 
	\captionsetup{justification=raggedright,singlelinecheck=false}
	\caption{Solutions to Eq.~(\ref{cergodic}) for various $T$ where $T_c$ is given by Eq.~(\ref{tc}). Here, we use $p = 5$ and $J = 1$. }\label{fig:ergodicitybreaking}
\end{figure}

Now, we demonstrate that there exists a dynamical transition in these equations. First, we assume time translational invariance (TTI), as we did in the linear case, and then we show where our assumptions break down. That is, we assume
\begin{align}
	C(t,t') = C(t-t'), \hspace{1cm} R(t,t') = R(t-t'),
\end{align}
and we also assume that $\lim_{\tau \to \infty}C(\tau) = 0$. Crucially, we also use the fluctuation dissipation theorem (FDT), which applies for systems whose stationary distributions have a Boltzmann form, which our system in Eq.~(\ref{pspin}) does. For this to be the case, we require the symmetry of the object $J_{i_1, \cdots, i_p}$ under the exchange of indices. We note that the FDT is only valid when the system is in equilibrium (see Section \ref{section:fdt} for a derivation using the MSRJD formalism). The FDT is simply
\begin{align}
	R(\tau) = -\frac{1}{T} \frac{dC(\tau)}{d\tau}.\label{fdt}
\end{align}
The utility of the FDT is clear -- it allows us to replace occurances of $R(t,t')$ in the first of Eq.~(\ref{randceqs}) and thus obtain a closed equation for the correlation function. We can then integrate the first integral term in the first of Eqs.~(\ref{randceqs}) by parts  to obtain
\begin{align}
	\frac{p(p-1)J^2}{2}\int_{0}^t ds C^{p-2}(t,s) R(t,s) C(s,t')
	&= \frac{pJ^2}{2T}\int_{0}^t ds  \frac{d}{ds}\left[ C^{p-1}(t-s)\right] C(s-t') \nonumber \\
	&= \frac{pJ^2}{2T} C(t-t')  - \frac{pJ^2}{2T}\int_{0}^t ds   C^{p-1}(t-s) \frac{d C(s-t')}{ds}. \label{integral}
\end{align}
where we use the long time limit (with $t-t'$ remaining finite) so that $C(t-t')$ is finite, but $C(t) \to 0$. We may also compute the Lagrange multiplier in Eq.~(\ref{lagrange})
\begin{align}
	r(t) = T + \frac{p^2J^2}{2T} \int_{0}^t dt' \frac{d C(t-t')}{dt'} C^{p-1}(t-t') = T + \frac{pJ^2}{2T} . \label{lagrangeeval}
\end{align}
Inserting Eqs.~(\ref{integral}) and (\ref{lagrangeeval}) into the first of Eqs.~(\ref{randceqs}), we obtain (writing $\tau = t-t'$ and $u = s-t'$)
\begin{align}
	\dot C(\tau) = -T C(\tau) - \frac{pJ^2}{2T}\int_{0}^\tau du C^{p-1}(\tau -u) \dot C(u) .\label{cergodic}
\end{align}
We can easily integrate this equation numerically to find $C(t)$. We notice that as we decrease the temperature, the curve develops a plateau (see Fig. \ref{fig:ergodicitybreaking}). Eventually, for sufficiently low $T$, we obtain a non-zero asymptotic value of $C(t)$. This indicates that the system fails to decorrelate even at large times -- i.e. ergodicity is broken and our assumptions of equilibrium and $C(\tau \to \infty) \to 0$ fail. Let us now try to find this critical temperature. 

We search for the value of $T$ at which a non-zero constant solution for $C(t\to\infty) \equiv \bar C$ is possible. Taking the Laplace transform of Eq.~(\ref{cergodic}), we obtain
\begin{align}
	\mathcal{L}_t[C(t)](s)=\hat C(s) = \frac{1}{s + T \gamma(s)}, \hspace{1cm} \gamma^{-1}(s) = 1 + \frac{pJ^2}{2 T} \mathcal{L}_t[C^{p-1}(t)](s).  
\end{align}
If there is indeed a non-zero solution, then we have $\mathcal{L}_t[C^{p-1}(t)](s) = \bar C^{p-1}/s$ and $\hat C(s) = \bar C/s$. We thus obtain using the final value theorem for Laplace transforms
\begin{align}
	\frac{2T^2 \bar C}{p J^2} = \bar C^{p-1} (1-\bar C) .
\end{align}
It is obvious that $ \bar C = 0$ is a solution. Let us then understand when another solution such that $0<\bar C<1$ emerges. We note that the function $f(\bar C) =  \bar C^{p-2} (1-\bar C)$ has a maximum at $\bar C = (p-2)/(p-1)$, which is the only maximum between 0 and 1. We thus obtain a non-zero solution for $\bar C$ when $T$ is lowered to such an extent that $f[(p-2)/(p-1)] = \frac{2T^2 }{p J^2}$. 
We therefore find that the critical temperature is \cite{kirkpatrick1987p, castellani2005spin}
\begin{align}
	T_\mathrm{dyn} = \sqrt{\frac{p J^2}{2}\frac{(p-2)^{p-2}}{(p-1)^{p-1}} } . \label{tc}
\end{align} 
One notes that for $p\to 2$, we obtain $T_\mathrm{dyn} = J$. This corresponds to our previous result in Section \ref{section:dynamicalinstabilities} $(1+\Gamma)\sigma < r$, if we make the substitutions $r = T + \frac{p J^2}{2 T}$, $\Gamma = 1$ and $\sigma = J$. We note here however that the nature of the dynamics beyond the instability is different to the linear system in Eq.~(\ref{linearsystem}) due to the dynamic spherical constraint, which prevents $C \to \infty$ in the SPSM. Instead, at the transition, the SPSM aligns with the eigenvector of the largest eigenvalue. We discuss the $p$-spin model, for both $p =2$ and $p>2$, in greater detail in Section \ref{section:rsb} in the context of the replica method.

\subsection{Exercises}
Another example of a non-linear system for which DMFT is extremely useful is the generalised Lotka-Volterra equations, which is a simple model of a complex ecological community
\begin{align}
	\dot x_i = x_i \left[ 1- x_i + \sum_j J_{ij} x_j \right] , 
\end{align}
and where $J_{ij}$ have statistics as in Eq.~(\ref{ellipsestatsdyn}). 
\begin{itemize}
	\item Show that the effective process [analogous to Eq.~(\ref{effectiveprocess}) for the linear system] is \begin{align}
		\dot x = x\left[1 - x + \mu M + \Gamma \sigma^2 \int dt' R(t,t') x(t') + \eta \right],
	\end{align} 
	where $\eta$ is a centred and coloured Gaussian noise with correlator $\langle \eta(t) \eta(t') \rangle = \sigma^2 \langle x(t) x(t') \rangle$. 
	\item In the case of the Lotka-Volterra equations, the fixed point is non-trivial. By assuming $x(t\to\infty) = x^\star$, $\eta(t\to\infty) = \eta^\star$ and $\int_0^\infty d\tau R(\tau) = \chi$, show that either $x^\star = 0$ or 
	\begin{align}
		x^\star = \frac{1+\mu M + \eta^\star}{1 - \Gamma \sigma^2 \chi}. 
	\end{align}
	\item Thus, given that the species abundances $x_i$ must be non-negative at all times, show that the distribution of species abundances at the fixed point is given by
	\begin{align}
		P(x^\star) = (1-\phi) \delta(x^\star) +\frac{ \exp\left[ - \left(x^\star - \frac{1+\mu M}{1-\Gamma\sigma^2 \chi} \right)^2 \frac{(1-\Gamma \sigma^2 \chi)^2}{2 \sigma^2 q}\right]}{\sqrt{2\pi \sigma^2 q/(1-\Gamma\sigma^2\chi)^2}}.
	\end{align}
	where $q = \langle (x^\star)^2 \rangle$ and $\phi = \int_{0^+}^\infty dx^\star P(x^\star)$.
	\item In the case $\Gamma = 0$ and $\mu = 0$, thus find an explicit integral equation for the order parameter $q= \int dx^\star (x^\star)^2 P(x^\star)$. 
\end{itemize}
As we mentioned briefly in the discussion around Eq.~(\ref{rhateq}), it is possible to derive some of the results of earlier sections using the dynamic approach. 
\begin{itemize}
	\item By considering the functional derivative of Eq.~(\ref{linearsystem}), show that for $R_{ij} \equiv \frac{\delta x_i(t)}{\delta h_j(0)}$, we have
	\begin{align}
		\hat R_{ij}(u) \equiv \mathcal{L}_t[R_{ij}(t)](u) = \left[\left((u+r)\underline{\underline{\id}} - \underline{\underline{J}}\right)^{-1}\right]_{ij}.
	\end{align}
	\item Thus, given the expression in Eq.~(\ref{rhateq}) that was derived using DMFT, rederive the Wigner semicircle law. 
\end{itemize}
Consider again the firing rate model [c.f. Eq.~(\ref{firingrate})] in the case of additional external noise 
\begin{align}
	\dot h_i = - h_i + \sum_{j}J_{ij} \phi(h_j) + \xi_i, \hspace{1cm} \langle \xi_i (t) \xi_j(t')\rangle = 2T\delta_{ij}\delta(t-t').
\end{align}
\begin{itemize}
	\item Derive the DMFT equations for this process. Show, assuming small $T$, that the perturbations about the fixed point $h_i^\star = 0$ follow
	\begin{align}
		\delta \dot h &\approx - \delta h + g \Gamma \sigma^2 \int_0^t dt' R(t,t') \delta h(t') + \delta \eta(t) + \xi(t) , \nonumber \\
		\langle \delta \eta(t) \delta \eta(t') \rangle &= \sigma^2 g^2\langle  \delta h(t) \delta h(t') \rangle ,
	\end{align}
	where  $R(t,t') \approx g \langle \delta h(t) / \delta \eta(t')\rangle$.
	\item By functionally differentiating this effective process, show that
	\begin{align}
		\partial_t R(t,t') \approx - R(t,t') + g \Gamma \sigma^2 \int_0^t dt'' R(t,t'')R(t'', t') + g \delta (t,t') .
	\end{align}
	\item Demonstrate also that the power spectrum of fluctuations is given by 
	\begin{align}
		\langle \vert\delta \tilde h(\omega)\vert^2\rangle \approx \frac{2 T}{\left \vert i \omega + 1 - g\Gamma \sigma^2 \tilde R(\omega)\right\vert^2 - g^2\sigma^2}. 
	\end{align}
	\item Thus, by considering the case $\omega \to 0$, show that we recover the stability criterion $g(1+\Gamma)\sigma<1$ as long as $T$ is small.
\end{itemize}
DMFT is an extremely efficient way of deriving the temporal behaviour of disordered dynamical systems. Let us briefly explore one other possible route of deriving the power spectrum of fluctuations of the linear model in Eq.~(\ref{linearsystem}). 
\begin{itemize}
	\item By taking the Fourier transform of Eq.~(\ref{linearsystem}) directly, show that
	\begin{align}
		N^{-1}\sum_i \langle \vert \tilde x_i(\omega) \vert ^2 \rangle_\xi = \frac{\sigma^2_T}{N} \sum_{ij}\left[(i\omega + r)\underline{\underline{\id}} - \underline{\underline{J}} \right]^{-1}_{ij}  \left[(-i\omega + r)\underline{\underline{\id}} - \underline{\underline{J}} \right]^{-1}_{ij}. 
	\end{align}
	\item Using the expansion $G_{ij} =\left[z\underline{\underline{\id}} - \underline{\underline{J}} \right]^{-1}_{ij} = \frac{1}{z}\delta_{ij} +\frac{1}{z^2} J_{ij} + \frac{1}{z^3} \sum_k J_{ik}J_{kj}+\frac{1}{z^4} \sum_{kl} J_{ik}J_{kl}J_{lj} + \cdots $, show that in the case $\Gamma = 0$ one must have (averaging over $J_{ij}$) in the limit $N \to \infty$
	\begin{align}
		N^{-1}\sum_i \langle \vert \tilde x_i(\omega) \vert ^2 \rangle_{\xi,J} = \frac{\sigma_T^2}{(\omega^2+r^2)} + \frac{\sigma_T^2 \sigma^2}{(\omega^2+r^2)^2} + \frac{\sigma_T^2 \sigma^4}{(\omega^2+r^2)^3}  \cdots.
	\end{align}
	\item Hence, recover Eq.~(\ref{powerspectrumgamma0}). One notes that in the case $\Gamma \neq 0$, resumming the series becomes much more involved! We thus see how elegant DMFT is in comparison.
\end{itemize}

\newpage

\section{Mar\v{c}enko-Pastur law: sample covariance matrices and the BBP transition}\label{section:marchenkopastur}

\begin{quotation}If you torture the data long enough, it will confess.--Ronald H. Coase \end{quotation}

\subsection{Principal component analysis (PCA)}
At the core of statistical data analysis is the principle of dimensional reduction. That is, in order to understand the `patterns' present in a large data set, we seek to describe the data in terms of only a few key quantities. However, if we are agnostic of any underlying model, the key variables that explain the statistical behaviour of the data may not be obvious. How can we best find these key variables in a systematic and unbiased fashion?

Principal component analysis seeks to resolve precisely this issue \cite{greenacre2022principal}. More precisely, suppose that we have $N$ measurable quantities (which we index $i$), and we take $n$ independent measurements of each of these quantities (labelling the measurements with index $\alpha$). We label the $\alpha$-th measurement of the $i$-th quantity $X_{i}^{\alpha}$. We assume that the data has been standardised such that the sample mean of each component is nil. We wish to find the so-called \textit{principal components} for this data. These are the vector `directions' in $N$-dimensional space along which the data varies the most. That is, we wish to find a set of basis vectors $\underline{w}^{(i)}$ [where $i = 1, \cdots, N$ and $(\underline{w}^{(i)})^T\underline{w}^{(i)} = 1$] such that the variance of the data is greatest in the direction $\underline{w}^{(1)}$, then second greatest in the direction $\underline{w}^{(2)}$, and so on. 

Let us consider \textit{the} principal component $\underline{w}^{(1)}$. It is defined by
\begin{align}
	\underline{w}^{(1)} =\mathrm{arg} \max_{\vert\vert \underline{w}\vert\vert = 1}\left\{\sum_{i,\alpha} \left(X_{i}^{\alpha} w_i\right)^2 \right\} = \mathrm{arg} \max_{\vert\vert \underline{w}\vert\vert = 1}\left\{\underline{w}^T\underline{\underline{X}}\, \underline{\underline{X}}^T \underline{w} \right\} ,
\end{align}
where $\underline{\underline{X}}$ is the $N\times n$ matrix containing all the data. Now, if $\underline{u}^{(\nu)}$ is the eigenvector of $\underline{\underline{X}}\, \underline{\underline{X}}^T$ with eigenvalue $\lambda_\nu$, then we may perform a change of basis and write $w_i = \sum_\nu w_\nu u^{(\nu)}_i$. Thus, $\underline{w}^T\underline{\underline{X}} \,\underline{\underline{X}}^T \underline{w} = \sum_\nu \lambda_\nu w_\nu^2$ if the eigenvectors are normalised to unity. Therefore, the direction that maximises the quadratic form $\underline{w}^T\underline{\underline{X}} \,\underline{\underline{X}}^T \underline{w}$, and therefore provides the principal component $\underline{w}^{(1)}$, is simply the normalised eigenvector corresponding to the largest eigenvalue of $\underline{\underline{X}} \underline{\underline{X}}^T$. 

To find the $k$-th principal component, one proceeds by subtracting the first $k-1$ principal components of the data via
\begin{align}
	\underline{\underline{\hat{X}}}^{(k)} = \underline{\underline{X}} - \sum_{r= 1}^{k-1} \underline{w}^{(r)}\left(\underline{w}^{(r)}\right)^T\underline{\underline{X}} \, .
\end{align}
One then simply performs the same procedure again, finding the direction along which the variance of the remaining data is maximised
\begin{align}
	\underline{w}^{(k)} =  \mathrm{arg} \max_{\vert\vert \underline{w}\vert\vert = 1}\left\{\underline{w}^T \underline{\underline{\hat{X}}}^{(k)}\left(\underline{\underline{\hat{X}}}^{(k)}\right)^T \underline{w} \right\}.
\end{align}
By a similar eigenvector decomposition, one ultimately finds that the $k$-th principal component corresponds exactly to the eigenvector of the $k$-th largest eigenvalue of $\underline{\underline{X}} \underline{\underline{X}}^T$ \cite{greenacre2022principal}. 

\begin{figure}[H]
	\centering 
	\includegraphics[scale = 0.48]{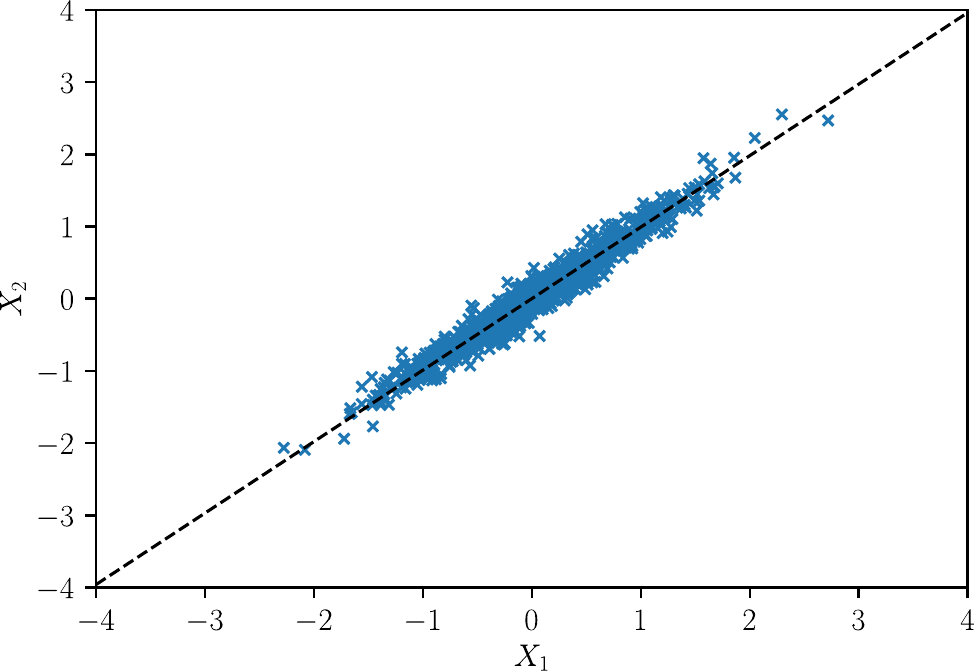}	
	\includegraphics[scale = 0.48]{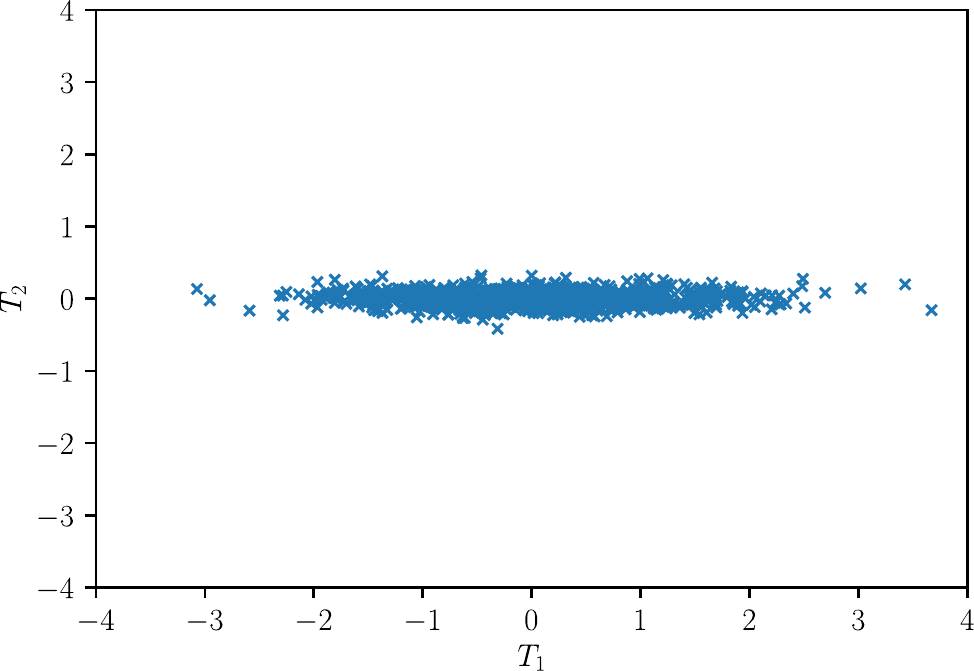} 
	\captionsetup{justification=raggedright,singlelinecheck=false}
	\caption{(Left): Some quasi-1-dimensional data. The location of a point in the $(X_1,X_2)$ plane is mostly determined by the distance along a straight line. The principal component direction is indicated by the dashed line. (Right) The same data subjected to a PCA.  }\label{fig:PCAexample}
\end{figure}

Letting $\underline{\underline{W}}$ be the matrix with the normalised eigenvectors of $\underline{\underline{X}} \underline{\underline{X}}^T$ as its columns, the principal component decomposition of the data $\underline{\underline{X}}$ may be written
\begin{align}
	\underline{\underline{T}} = \underline{\underline{W}}^T \underline{\underline{X}} .
\end{align}
An example of principal component decomposition in the case $N = 2$ is given in Fig. \ref{fig:PCAexample}. We see that the principal component aligns with the data, which is quasi 1-dimensional. By specifying the value of $T_1$ of a data point, we more-or-less specify its location in the plane. Including the information $T_2$ only provides a small correction. One may therefore dispense of the $T_2$ data, if one so chooses, and therefore perform a dimensional reduction of the data. This principle naturally extends to higher dimensions.

An entertaining example of PCA in action is the so-called `big 5' personality traits in psychology \cite{goldberg1993structure}. Suppose a researcher asks a fixed set of questions to many individuals. These questions ask the person to assign a number to how much they agree with a statement about their personality (`I enjoy parties', `I find it hard to confront others', `I am usually quite happy', etc.). Alternatively, one can ask people to describe themselves in words, and monitor the occurrence of the various descriptors that they use. Cross-culturally, and over a variety of studies, it has been found that 5 principal components are adequate to describe the data. These 5 factors have been interpreted as personality traits: openness to experience, conscientiousness, extraversion, agreeableness, and neuroticism. Of course, a certain amount of scepticism has been levelled at this simple categorisation, but it is nevertheless remarkable that the principal components appear to be fairly robust \cite{franic2014big}.

We note that while PCA assumes an underlying linear structure to the data, a range of more sophisticated methods for dimensional reduction exist that can take into possible nonlinearities \cite{van2009dimensionality}. 

\subsection{PCA and RMT}
We saw in Fig. \ref{fig:PCAexample} that when there is a clear principal direction, the PCA deftly identifies it. However, if the data were more noisy, and deviated from the principal direction to a greater extent, our analysis would fail to provide any meaningful information. Further, if we did not have sufficiently many data points in comparison to the dimension of the data space $N$, the PCA would also fail. An obvious question is thus the following: under what circumstances does the PCA detect the \textit{true} underlying principal components of the data, and separate them from the noise?

It turns out that random matrix theory is very useful for answering this question \cite{paul2014random}. For the sake of understanding the theoretical limits of PCA, we posit an idealised case. In this scenario, there is a single true principal component in $N$-dimensional space $\underline{v}$ along which the data vary. However, the data is also subject to random noise. In other words
\begin{align}
	X_i^{\alpha} = z^{\alpha} v_i + x_i^{\alpha}, \label{decomposition}
\end{align}
where $z^{\alpha}$ and $x_i^{\alpha}$ are each supposed to be independent random variables with statistics
\begin{align}
	\left\langle  \left(z^{\alpha}\right)^2 \right\rangle = \beta, \hspace{1cm} \left\langle  \left(x_i^{\alpha}\right)^2 \right\rangle = \sigma^2 .
\end{align}
If we had infinite samples $n \to \infty$ with $N$ finite, then the sample covariance matrix $n^{-1}\underline{\underline{X}}\underline{\underline{X}}^T$ would tend to the so-called \textit{population covariance matrix}
\begin{align}
	\underline{\underline{\Sigma}} = \sigma^2\underline{\underline{\id}} + \beta \underline{v} \underline{v}^T . \label{spikedcovariance}
\end{align}
The question is, under what circumstances does the eigenvector of the rightmost eigenvalue of the sample covariance matrix $n^{-1}\underline{\underline{X}}\underline{\underline{X}}^T$ align with the true principal component $\underline{v}$? In particular, if the number of samples $n$ is of the same order of magnitude as the data dimension $N$ (when $N$ is large), or if $\sigma^2$ is comparable to $\beta$, do we still obtain the correct principal component?

We proceed by first computing the eigenvalue spectrum of the matrix $n^{-1}\underline{\underline{X}}\underline{\underline{X}}^T$ in the simpler case $\beta = 0$\footnote{The ensemble of random matrices of this form is sometimes referred to as the Wishart ensemble. Wishart first studied these matrices all the way back in 1928.}.  We then include the effect of the principal component direction $\beta >0$ as a modification. In the latter case, we compute the rightmost eigenvalue and investigate under what circumstances the eigenvector aligns appreciably with $\underline{v}$. We note that the analysis here can also be extended to the case where there is a generic number of (but finitely many) principal component directions \cite{baik2005phase}. 

\subsection{Block-structured resolvent trick for matrix products and the Mar\v{c}enko-Pastur law}\label{section:mplaw}
The trick of considering a block-structured resolvent matrix, which resembled the classic resolvent matrix in Eq.~(\ref{resdef}), was exploited to deal with the case of non-Hermitian random matrices in Section \ref{section:ellipse}. We show here how a similar set-up can be used to extract the eigenvalue density of symmetric matrix products like the sample covariance matrix $n^{-1}\underline{\underline{X}}\,\underline{\underline{X}}^T$. For simplicity, we first take the noise-dominated case $\beta = 0$, and we thus consider $n^{-1}\underline{\underline{x}}\,\underline{\underline{x}}^T$.

Consider the object (using $z = \omega - i\epsilon$)
\begin{align}
	\underline{\underline{\mathcal{H}}} = \begin{bmatrix}
		z\underline{\underline{\id}}_N & n^{-1/2}\underline{\underline{x}} \\
		n^{-1/2}\underline{\underline{x}}^T & \underline{\underline{\id}}_n
	\end{bmatrix}^{-1} \equiv \begin{bmatrix}
		\underline{\underline{A}} & \underline{\underline{B}} \\
		\underline{\underline{C}} & \underline{\underline{D}}
	\end{bmatrix} .\label{blockdef}
\end{align}
Using the block matrix inversion formula in Eq.~(\ref{blockinverse}), we obtain 
\begin{align}
	\underline{\underline{A}} &= \left[	z \underline{\underline{\id}}_N - n^{-1}\underline{\underline{x}} \,\underline{\underline{x}}^T \right]^{-1}, \nonumber \\
	\underline{\underline{B}} &= -\left[	z \underline{\underline{\id}}_N - n^{-1}\underline{\underline{x}}\, \underline{\underline{x}}^T \right]^{-1} \frac{1}{\sqrt{n}} \underline{\underline{x}}, \nonumber \\
	\underline{\underline{C}} &= -\left[	z \underline{\underline{\id}}_n - n^{-1}\underline{\underline{x}}^T \underline{\underline{x}} \right]^{-1}\frac{1}{\sqrt{n}}\underline{\underline{x}}^T = -\frac{1}{\sqrt{n}}\underline{\underline{x}}^T\left[	z \underline{\underline{\id}}_N - n^{-1}\underline{\underline{x}} \,\underline{\underline{x}}^T \right]^{-1}, \nonumber \\
	\underline{\underline{D}} &= z\left[z \underline{\underline{\id}}_n - n^{-1}\underline{\underline{x}}^T \underline{\underline{x}} \right]^{-1}.
\end{align}
We thus see that the top left block of the matrix $\underline{\underline{\mathcal{H}}}$ yields the resolvent in which we are interested in this case, which is $\underline{\underline{G}} = \underline{\underline{A}} = \left[	z \underline{\underline{\id}} - n^{-1}\underline{\underline{x}}\, \underline{\underline{x}}^T \right]^{-1}$. We note that since $\underline{\underline{x}}$ is of dimension $N \times n$, these blocks are of different sizes. Specifically, $\underline{\underline{A}}$ is $N \times N$, $\underline{\underline{B}}$ is $N \times n$, $\underline{\underline{C}}$ is $n \times N$, and $\underline{\underline{D}}$ is $n \times n$.

We now proceed along very similar lines to Section \ref{section:blockmatrixinverseellipse}. However, since the blocks $\underline{\underline{A}}$, $\underline{\underline{B}}$, $\underline{\underline{C}}$ and $\underline{\underline{D}}$ are different sizes, we have to be somewhat more careful about how we go about taking the matrix inverse in Eq.~(\ref{blockdef}). That is, we can no longer rearrange the matrix $\underline{\underline{\mathcal{H}}}$ into $2\times2$ blocks. However, some simplification does occur, owing to the structure of the matrix $\underline{\underline{\mathcal{H}}}$ and the independence of the rows of  $\underline{\underline{x}}$, which allows us to proceed. 

Let us now consider the element $A_{11}$. Consulting the block matrix inverse formula in Eq.~(\ref{blockinverse}), we assign
\begin{align}
	\underline{\underline{S}}&=  z, \nonumber \\
	\underline{\underline{T}} &= n^{-1/2}(0,0, \cdots, 0; 0, x_{1}^{2},  x_{1}^{3}, \cdots,  x_{1}^{n}), \nonumber \\
	\underline{\underline{U}} &= n^{-1/2}(0,0, \cdots, 0; 0,  x_{1}^{2}, x_{1}^{3}, \cdots, x_{1}^{n})^T, \nonumber \\
	\underline{\underline{V}}^{-1} &= \underline{\underline{\mathcal{H}}}^{(1)},
\end{align}
where $\underline{\underline{\mathcal{H}}}^{(1)}$ is the version of $\underline{\underline{\mathcal{H}}}$ that we obtain by deleting the first row of $\underline{\underline{x}}$ (i.e. $x_1^{\alpha}$ for all $\alpha = 1, \cdots, n$). We thus obtain 
\begin{align}
	A_{11} = \left( z - n^{-1}\sum_{\alpha,\beta = 1}^n x_{1}^{\alpha} D_{\alpha \beta}^{(1)}x_{1}^{\beta}\right)^{-1}. \label{cavitya11}
\end{align}
One notes that this expression is independent of $B_{11}$, $C_{11}$ and $D_{11}$, which greatly simplifies the analysis. Similarly, we can consider $D_{nn}$, and we obtain
\begin{align}
	D_{nn} = \left(1 - n^{-1}\sum_{j,k = 1}^{N} x_{j}^{n} A_{jk}^{(n)} x_{k}^{n}\right)^{-1}. 
\end{align}
Following the usual argument, we neglect the elements of $\mathcal{H}_{jk}^{(i)}$ with $j \neq k$, and we reason that the sums $ n^{-1}\sum_{\alpha=1}^n (x_{1}^{\alpha})^2 D_{\alpha \alpha}^{(1)}$ and $n^{-1}\sum_{j=1}^N (x_{j}^{n})^2 A_{jj}^{(n)} $ concentrate on their averages. Using the usual notation $A = N^{-1} \mathrm{Tr} \underline{\underline{A}}$ and $D = n^{-1} \mathrm{Tr} \underline{\underline{D}}$, and using the fact that $A_{ii} = A$ and $D_{ii}=D$, we have 
\begin{align}
	A &= \left( z - \sigma^2 D\right)^{-1}, \nonumber \\
	D &= \left(1 - \gamma \sigma^2  A\right)^{-1} , \label{aandd}
\end{align}
where $\gamma = N/n$, which we take to be of order $O(N^0)$. Happily, we see that there is no dependence on the matrices $\underline{\underline{B}}$ and $\underline{\underline{C}}$ here. We can thus solve these equations easily for $A(\omega) = G(\omega)$ (the trace of the resolvent matrix in which we are interested) and extract the eigenvalue density using the usual inverse Stieltjes transform in Eq.~(\ref{densefromres}). For later use, let us briefly observe that (again using the block inversion formula)
\begin{align}
	B_{1\beta} = -A_{11} \sum_\alpha x_{1}^\alpha D^{(1)}_{\alpha \beta}/\sqrt{n}. \label{neglectoff}
\end{align}
Thus, by arguments similar to those by which we neglected the off-diagonal elements in Section \ref{section:offdiag}, we may safely approximate $\underline{\underline{B}} = \underline{\underline{C}}^T = \underline{\underline{0}}$.

In solving for the resolvent, we must again be careful about our choice of branch, just as we were in the case of the Wigner semicircle law. The meaningful solution is the one for which $G(\omega) \to 1/\omega$ as $\vert\omega \vert \to \infty$ that has a single branch cut on a bounded range of the real axis. Here, this is
\begin{align}
	G(z) &= \frac{(\gamma-1)\sigma^2 + z - \mathrm{sign}[\mathrm{Re}(z- [\lambda_+ + \lambda_-]/2)]\sqrt{(z-\lambda_+)(z-\lambda_-)}}{2\sigma^2 \gamma z} , \nonumber \\
	\lambda_\pm &= (1 \pm \sqrt{\gamma})^2 \sigma^2.
\end{align}
Using Eq.~(\ref{densefromres}), we therefore obtain the celebrated Mar\v{c}enko-Pastur law \cite{marvcenko1967distribution}
\begin{empheq}[box={\fboxsep=6pt\fbox}]{align}
	\rho(\omega) &= \begin{cases}
		(1-\gamma^{-1})\delta\left( \omega\right) +\frac{\sqrt{(\lambda_+-\omega)(\omega - \lambda_-)}}{2 \gamma \sigma^2 \pi \omega} \hspace{1cm} \mathrm{if} \; \gamma >1 \\
		\frac{\sqrt{(\lambda_+-\omega)(\omega - \lambda_-)}}{2 \gamma \sigma^2 \pi \omega} \hspace{2.9cm} \mathrm{otherwise}
	\end{cases} . \label{marchenkopastur}
\end{empheq}
One notes the atom contribution, which appears for $\gamma>1$. This contribution emerges due to the degenerate eigenvalues at $\lambda = 0$ that occur due to rank deficiency. That is, we know that $\underline{\underline{x}}\underline{\underline{x}}^T$ and $\underline{\underline{x}}^T\underline{\underline{x}}$ have the same non-zero eigenvalues. However, since $\underline{\underline{x}}^T$ has dimension $n\times N$, when $N>n$ we see that $\underline{\underline{x}}^T$ must have a non-empty kernel, and therefore that $\underline{\underline{x}}\underline{\underline{x}}^T$ must have precisely $N-n$ zero eigenvalues, which corresponds to the delta mass at zero.

\subsection{Outlier eigenvalue due to population covariance spike}
Let us now consider the case where not all of the variates are independent, and there is a single `principal component' direction. That is, let us now consider the case where the population covariance matrix is of the form in Eq.~(\ref{spikedcovariance}), but now with $\beta>0$. This is often referred to as a \textit{spiked} covariance matrix, the spike being the rank-1 perturbation. 

As we described before in Eq.~(\ref{decomposition}), another way to think of is to treat the random variates $X_i^\alpha$ as the sum of two independent random contributions: $x_i^{(\alpha)}$, which are centred Gaussian random variables with variance $\sigma^2$, and $z^{(\alpha)}v_i$, where $z^{(\alpha)}$ are independent centred Gaussian random variables with variance $\beta$. 

The spike direction $\underline{v}$ is an arbitrary vector, which we take to be of unit magnitude such that $\underline{v}^T\underline{v} = 1$. One can indeed check that the ensemble average $\langle \underline{ X}^{(\alpha)} (\underline{ X}^{(\alpha)})^T \rangle$, using $X_i^{(\alpha)}$ in Eq.~(\ref{decomposition}), yields the population covariance matrix in Eq.~(\ref{spikedcovariance}). So each realisation of the data $\underline{ X}^{(\alpha)}$ can be thought of as a random position along a single 1-D axis in the direction $\underline{v}$, plus some completely random $N$-dimensional noise. The data in Fig.~\ref{fig:PCAexample} is an example with $N = 2$.

Inspecting Eq.~(\ref{spikedcovariance}), we see that the matrix $\underline{\underline{\Sigma}}$ has $N-1$ degenerate eigenvalues at $\omega = \sigma^2$ and an additional eigenvalue, corresponding to the spike, at $\sigma^2 + \beta$. We therefore expect the sample covariance matrix to have a Mar\v{c}enko-Pastur contribution, which would shrink to give the degenerate $N-1$ eigenvalues at $\sigma^2$ as $\gamma \to 0$, and a single additional outlier eigenvalue in the vicinity of $\sigma^2 + \beta$. As long as $\beta$ is sufficiently large (or $\gamma$ sufficiently small), the outlier should protrude from the Mar\v{c}enko-Pastur bulk.

Since the principle component contribution is a rank-1 perturbation to $\underline{\underline{x}}$, we expect to be able to use the approach the we previously used to calculate the outlier in Section \ref{section:outlier}. However, there are some additional complications here. First, the rank-1 perturbation has been made to $\underline{\underline{x}}$ and not the matrix $n^{-1} \underline{\underline{x}}\underline{\underline{x}}^T$, which is the matrix on which we have just performed the spectral analysis to obtain the Mar\v{c}enko-Pastur law. Secondly, the rank-1 perturbation is random in this case, rather than deterministic. 

\begin{tcolorbox}[colback=blue!10!white,colframe=blue!90!black,title=Lemma: Block matrix determinant]
	Consider again the generic matrix 
	\begin{align}
		\underline{\underline{M}} = \begin{bmatrix}
			\underline{\underline{S}} & \underline{\underline{T}} \\
			\underline{\underline{U}} & \underline{\underline{V}}
		\end{bmatrix}.
	\end{align}
	If block $\underline{\underline{S}} $ is invertible, then the determinant is given by
	\begin{align}
		\mathrm{det}\underline{\underline{M}} = \mathrm{det} \underline{\underline{S}} \,\mathrm{det}\left(\underline{\underline{V}} - \underline{\underline{U}}  \underline{\underline{S}}^{-1} \underline{\underline{T}} \right).
	\end{align}
	Alternatively, if block $\underline{\underline{V}}$ is invertible, it is given by 
	\begin{align}
		\mathrm{det}\underline{\underline{M}} = \mathrm{det} \underline{\underline{V}} \,\mathrm{det}\left(\underline{\underline{S}} - \underline{\underline{T}}  \underline{\underline{V}}^{-1} \underline{\underline{U}} \right).
	\end{align}
\end{tcolorbox} 

To overcome these difficulties, we exploit the same block-matrix trick as before. Supposing that a single new outlier eigenvalue is produced as a result of the rank-1 perturbation, we may write for the location of the outlier
\begin{align}
	\mathrm{det}\begin{bmatrix}
		\lambda_\mathrm{outlier} \underline{\underline{\id}} & n^{-1/2}\underline{\underline{ X}} \\
		n^{-1/2}\underline{\underline{ X}}^T & \underline{\underline{\id}}
	\end{bmatrix} = \mathrm{det}\left[ \lambda_\mathrm{outlier} \underline{\underline{\id}} - n^{-1} \underline{\underline{ X}}\, \underline{\underline{ X}}^T \right] = 0 .
\end{align}
We may therefore write
\begin{align}
	\mathrm{det}\begin{bmatrix}
		\lambda_\mathrm{outlier} \underline{\underline{\id}} & n^{-1/2}\underline{\underline{ x}} + n^{-1/2}  \underline{v} \, \underline{z}^T \\
		n^{-1/2}\underline{\underline{ x}}^T  + n^{-1/2}  \underline{z}\,\underline{v}^T  & \underline{\underline{\id}}
	\end{bmatrix}
	=  \mathrm{det} \begin{bmatrix}
		\underline{\underline{\id}} & n^{-1/2} \underline{v} \, \underline{z}^T \underline{\underline{D}} \\
		n^{-1/2}  \underline{z}\,\underline{v}^T \underline{\underline{A}}& \underline{\underline{\id}}
	\end{bmatrix} \mathrm{det} \underline{\underline{ \mathcal{H}}}^{-1},
\end{align}
where the matrices $ \underline{\underline{ \mathcal{H}}}$, $ \underline{\underline{ A}}$ and $ \underline{\underline{ D}}$ are as defined in Eq.~(\ref{blockdef}) with $z= \lambda_\mathrm{outlier}$. We also use that $\underline{\underline{B}} = \underline{\underline{C}} = \underline{\underline{0}}$ for $N \to \infty$ [see the discussion around Eq.~(\ref{neglectoff})], and we assume that the outlier eigenvalue is outside the region of non-analyticity of $A(z)$ and $D(z)$. 

Using Sylvester's determinant identity in Eq.~(\ref{sylvester}), we therefore obtain
\begin{align}
	\mathrm{det} \begin{bmatrix}
		\underline{\underline{\id}} & n^{-1/2}  \underline{v} \, \underline{z}^T \underline{\underline{D}} \\
		n^{-1/2}  \underline{z}\,\underline{v}^T \underline{\underline{A}}& \underline{\underline{\id}}
	\end{bmatrix} &= \mathrm{det} \left[\underline{\underline{\id}} - n^{-1}  \underline{v} \, \underline{z}^T \underline{\underline{D}}\, \underline{z}\,\underline{v}^T \underline{\underline{A}} \right] \nonumber \\
	&= 1 - n^{-1}  \, \underline{z}^T \underline{\underline{D}}\, \underline{z}\,\underline{v}^T \underline{\underline{A}}\,\underline{v} = 0.
\end{align}
As we discussed in Section \ref{section:mplaw}, the matrices $\underline{\underline{A}}$ and $\underline{\underline{D}}$ can be approximated in the limit $N \to \infty$ by diagonal matrices, where the diagonal elements are all uniform. Using the fact that $\underline{v}$ is a unit vector, and the fact that $\sum_{\alpha} (z^{(\alpha)})^2/n \approx \beta + O(n^{-1/2})$ by the central limit theorem, we have finally
\begin{align}
	A(\lambda_\mathrm{outlier}) D(\lambda_\mathrm{outlier}) = \frac{1}{\beta}. \label{outliereqad}
\end{align}
We need merely now to use the expressions for $A(z)$ and $D(z)$ that we obtained previously in Eq.~(\ref{aandd}). We obtain 
\begin{align}
	(1 + \sigma^2 A D)	(1 + \gamma\sigma^2 A D) - z A D = 0.
\end{align}
Imposing analyticity, we obtain the solution
\begin{align}
	A(z) D(z) = \frac{z - \sigma^2 (1+\gamma) - \mathrm{sign}[\mathrm{Re}(z- \frac{\lambda_+ + \lambda_-}{2})]\sqrt{[z-\sigma^2 (1+\gamma)]^2- 4 \gamma \sigma^2}}{2 \gamma \sigma^4}.
\end{align}
This function has a maximum value of $A(\lambda_+) D(\lambda_+) = \frac{1}{\sqrt{\gamma}\sigma^2}$ and a minimum of $A(\lambda_-) D(\lambda_-) = -\frac{1}{\sqrt{\gamma}\sigma^2}$. We therefore finally obtain the outlier eigenvalue
\begin{align}
	\lambda_\mathrm{outlier} = \frac{1}{AD} \left( 1 + \sigma^2 AD\right) \left( 1 + \gamma\sigma^2 AD\right) = \left(\sigma^2 + \beta\right) \left(1 + \frac{\gamma \sigma^2}{\beta} \right) , \label{bbpoutlier}
\end{align}
which is valid provided that $\gamma< \beta^2/\sigma^4$. This outlier eigenvalue is often called the BBP eigenvalue after Baik, Ben Arous and P\'ech\'e \cite{baik2005phase}. We see that as $\gamma \to 0$ the eigenvalue in Eq.~(\ref{bbpoutlier}) tends towards that of the spiked population covariance matrix, as we expect. Notably, however, we only observe an outlier eigenvalue at all when the number of samples is sufficiently high, so that $\gamma< \beta^2/\sigma^4$. We now discuss what this means for the PCA.

\begin{figure}[H]
	\centering 
	\includegraphics[scale = 0.6]{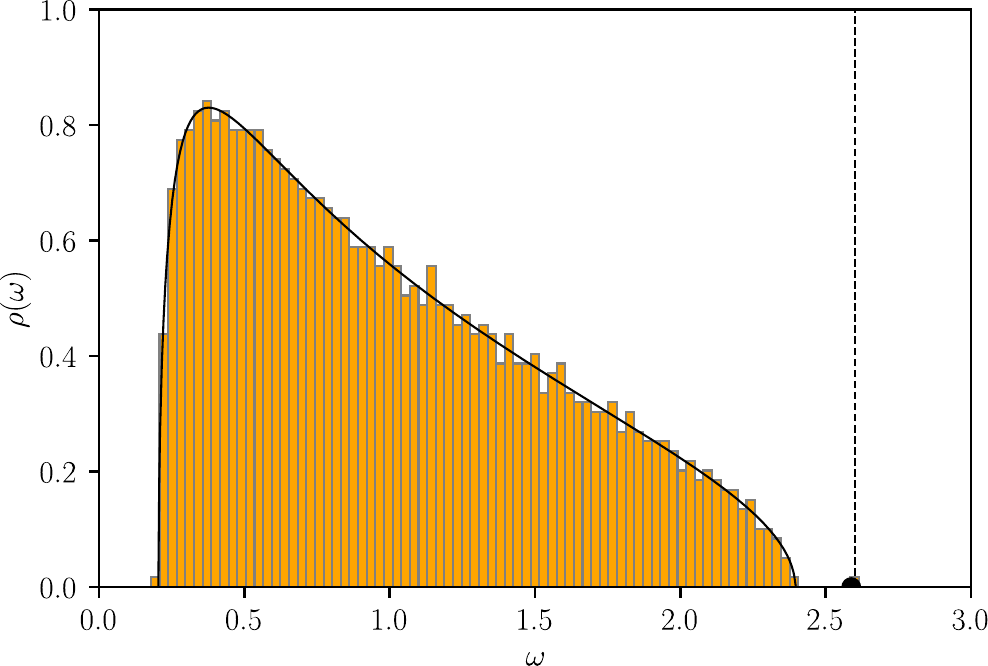} 
	\captionsetup{justification=raggedright,singlelinecheck=false}
	\caption{Verification of the Mar\v{c}enko-Pastur law and associated outlier in Eqs.~(\ref{marchenkopastur}) and (\ref{bbpoutlier}). Numerical results are for a single realisation -- there is no average over different realisations of the random matrix. The numerical outlier eigenvalue is represented by a black dot, while the theory is represented by the vertical dashed line. Parameters: $N = 2000$, $\beta = 1$, $\sigma = 1$, $\gamma = 0.3$. }\label{fig:mphistogram}
\end{figure}

\subsection{Principal component overlap with population spike}
We now seek to compute the extent to which the eigenvector corresponding to the principal eigenvalue (i.e. the principal component obtained from PCA) aligns with the true principal component direction $\underline{v}$. This quantifies the degree to which the PCA procedure reflects the ground truth. 

To do this, we exploit Eq.~(\ref{resvectors}), which relates the diagonal elements of the resolvent to the square magnitude of the eigenvectors:$G_{ij}(z) = \sum_{\nu} \frac{w_i^{(\nu)} w_j^{(\nu)}}{z-\lambda_\nu}$ in the present notation. Denoting the empirical principal component again by $\underline{w}^{(1)}$ (i.e. the eigenvector corresponding to the rightmost eigenvalue), we see that the overlap that we desire is given by 
\begin{align}
	\vert \underline{v}^T \underline{w}^{(1)}  \vert^2 = \mathrm{Res}\left[ \underline{v}^T \underline{\underline{G}}( \lambda_\mathrm{outlier}) \underline{v} \right], \label{residue}
\end{align}
where $\mathrm{Res}$ indicates the residue. That is, expanding about $\omega = \lambda_\mathrm{outlier}$, we have
\begin{align}
	\underline{v}^T \underline{\underline{G}}(\omega) \underline{v} = \frac{\vert  \underline{v}^T \underline{w}^{(1)}  \vert^2}{\omega - \lambda_\mathrm{outlier}} + \cdots . \label{overlapresolvent}
\end{align}
Our task is thus to find the resolvent matrix in the presence of a spike and compute the above product. We may use a very similar construction to the previous section. That is, we write 
\begin{align}
	\underline{\underline{\tilde{\mathcal{H}}}}^{-1}  \equiv \underline{\underline{\mathcal{H}}}^{-1} +  \begin{bmatrix}
		\underline{\underline{0}} & n^{-1/2}\underline{v} \, \underline{z}^T \\
		n^{-1/2}  \underline{z}\,\underline{v}^T & \underline{\underline{0}}
	\end{bmatrix} ,
\end{align}
where $\underline{\underline{\mathcal{H}}}$, obeying Eq.~(\ref{blockdef}), is the block resolvent without the population spike (i.e. $\beta = 0$).
\begin{tcolorbox}[colback=blue!10!white,colframe=blue!90!black,title=Lemma: Woodbury's identity]
	Woodbury's identity allows us to compute matrix inverses of the form
	\begin{align}
		\left(\underline{\underline{S}} + \underline{\underline{U}}\underline{\underline{T}}\underline{\underline{V}} \right)^{-1} = \underline{\underline{S}}^{-1} - \underline{\underline{S}}^{-1} \underline{\underline{U}} \left(\underline{\underline{T}}^{-1} + \underline{\underline{V}} \underline{\underline{S}}^{-1} \underline{\underline{U}}\right)^{-1} \underline{\underline{V}} \underline{\underline{S}}^{-1},
	\end{align}
	where $\underline{\underline{S}}$ is an $r\times r$ matrix, $\underline{\underline{T}}$ is a $k \times k$ matrix, $\underline{\underline{U}}$ is an $r\times k$ matrix, and $\underline{\underline{V}}$ is a $k \times r$ matrix. This identity may be verified easily by multiplying both sides by $\underline{\underline{S}} + \underline{\underline{U}}\underline{\underline{T}}\underline{\underline{V}}$. This is particularly useful when $k$ is small.
\end{tcolorbox} 

In order to find $\underline{\underline{\tilde{\mathcal{H}}}}$, we exploit Woodbury's matrix identity. That is, we write
\begin{align}
	\begin{bmatrix}
		\underline{\underline{0}} & n^{-1/2} \underline{v} \, \underline{z}^T \\
		n^{-1/2}  \underline{z}\,\underline{v}^T & \underline{\underline{0}}
	\end{bmatrix} = \sqrt{\frac{1}{n}}\begin{bmatrix}
		\underline{v} &  \, \underline{0}  \\
		\underline{0} & \underline{z}
	\end{bmatrix} \begin{bmatrix}
		0 &  \, 1  \\
		1 & 0
	\end{bmatrix} \begin{bmatrix}
		\underline{v}^T &  \, \underline{0}^T  \\
		\underline{0}^T & \underline{z}^T
	\end{bmatrix},
\end{align} 
and using Woodbury's result above with $r = N+n$ and $k = 2$, we may invert $\underline{\underline{\mathcal{H}}}^{-1}$ to find
\begin{align}
	\underline{\underline{\tilde{\mathcal{H}}}} = \underline{\underline{\mathcal{H}}} - \sqrt{\frac{1}{n}} \underline{\underline{\mathcal{H}}} \begin{bmatrix}
		\underline{v} &  \, \underline{0}  \\
		\underline{0} & \underline{z}
	\end{bmatrix} \underline{\underline{\mathcal{M}}}^{-1} \begin{bmatrix}
		\underline{v}^T &  \, \underline{0}^T  \\
		\underline{0}^T & \underline{z}^T
	\end{bmatrix}  \underline{\underline{\mathcal{H}}} ,
\end{align}
where we define
\begin{align}
	\underline{\underline{\mathcal{M}}} = \begin{bmatrix}
		0 &  \, 1  \\
		1 & 0
	\end{bmatrix} +\sqrt{\frac{1}{n}} \begin{bmatrix}
		\underline{v}^T &  \, \underline{0}^T  \\
		\underline{0}^T & \underline{z}^T
	\end{bmatrix}\underline{\underline{\mathcal{H}}}  \begin{bmatrix}
		\underline{v} &  \, \underline{0}  \\
		\underline{0} & \underline{z}
	\end{bmatrix} .
\end{align}
Multiplying out the matrices, and again using that in thermodynamic limit $B=C = 0$, one arrives at
\begin{align}
	\tilde{\underline{\underline{A}}} = \underline{\underline{A}} + \frac{\beta D}{(1-\beta A D)} \underline{\underline{A}} \,\underline{v}\,	\underline{v}^T \underline{\underline{A}}, \label{modifieda}
\end{align}
which is the block in which we are interested. When the rightmost eigenvalue is given by the outlier, which satisfies Eq.~(\ref{outliereqad}), we see that the second term in Eq.~(\ref{modifieda}) has a pole at $\omega = \lambda_\mathrm{outlier}$. By differentiating Eq.~(\ref{bbpoutlier}), we find in the vicinity of the pole
\begin{align}
	\underline{v}^T\tilde{\underline{\underline{A}}}\,\underline{v} = A + \frac{\beta D A^2}{(1-\beta AD)} 
	&\approx \frac{1}{\omega - \lambda_\mathrm{outlier}} \frac{\beta D A^2}{ \beta A D} \left[ \omega -\sigma^2 (1+\gamma)  - 2 \gamma \sigma^4 A D \right] \bigg \vert_{\omega = \lambda_\mathrm{outlier}} \nonumber \\
	&= 	\frac{1}{\omega - \lambda_\mathrm{outlier}}\frac{1-\gamma \sigma^4/\beta^2}{1+\gamma \sigma^2/\beta}.
\end{align}
Using Eq.~(\ref{residue}), we then obtain $\vert  \underline{v}^T \underline{w}^{(1)}  \vert^2 = \frac{1-\gamma \sigma^4/\beta^2}{1+\gamma \sigma^2/\beta}$. However, one notes that this expression becomes negative for $\gamma>\beta^2/\sigma^4$, which cannot be the case for the quantity $\vert \underline{v}^T \underline{w}^{(1)} \vert^2$. This coincides with the region of validity for the outlier eigenvalue expression that we found before. In the case where there is no outlier eigenvalue, the rightmost edge of the Mar\v{c}enko-Pastur bulk is the rightmost eigenvalue. Since there is no simple pole of $\underline{v}^T\tilde{\underline{\underline{A}}}\,\underline{v}$ at this location (in the thermodynamic limit), we must have that the overlap with the population principle component $\underline{v}$ in this case is nil. 

\subsection{The BBP transition}
Summarising the findings, the rightmost eigenvalue and the corresponding eigenvector overlap with the true principal component direction $\underline{v}$ are given by
\begin{empheq}[box={\fboxsep=6pt\fbox}]{align}
	\lambda_\mathrm{max} &= \begin{cases}
		\left(\sigma^2 + \beta\right)\left(1 + \frac{\gamma \sigma^2}{\beta} \right) \hspace{1cm} \mathrm{for} \hspace{1cm} \gamma < \beta^2/\sigma^4  ,\\
		(1+\sqrt{\gamma})^2\sigma^2 \hspace{2.3cm} \mathrm{for} \hspace{1cm} \gamma \geq \beta^2/\sigma^4 .
	\end{cases} \nonumber \\
	\vert \underline{v}^T \underline{w}^{(1)}  \vert^2 &= \begin{cases}
		\frac{1-\gamma \sigma^4/\beta^2}{1+\gamma \sigma^2/\beta} \hspace{1cm} \mathrm{for} \hspace{1cm} \gamma < \beta^2/\sigma^4 ,\\
		0 \hspace{2.3cm} \mathrm{for} \hspace{1cm} \gamma \geq \beta^2/\sigma^4 .
	\end{cases} \label{overlap}
\end{empheq}
We see that there is a critical point $\gamma_c = \beta^2/\sigma^4 $ at which the overlap $\vert  \underline{v}^T \underline{w}^{(1)} \vert^2$ changes from zero to non-zero continuously. This is therefore a second-order phase transition, known as the BBP transition \cite{baik2005phase}. 

In the context of PCA, the value $n_c = N/\gamma_c = N\sigma^4/\beta^2$ is the critical number of samples above which PCA yields some alignment with the true population principal component direction. We also see that the alignment increases as the number of samples is increased. Alternatively, if the signal strength $\beta$ is increased or the noise strength $\sigma^2$ is reduced, the alignment also improves. Above all, we see that below a threshold number of samples, there is absolutely no hope whatsoever of obtaining any information of the true population behaviour from PCA.

\begin{figure}[H]
	\centering 
	\includegraphics[scale = 0.48]{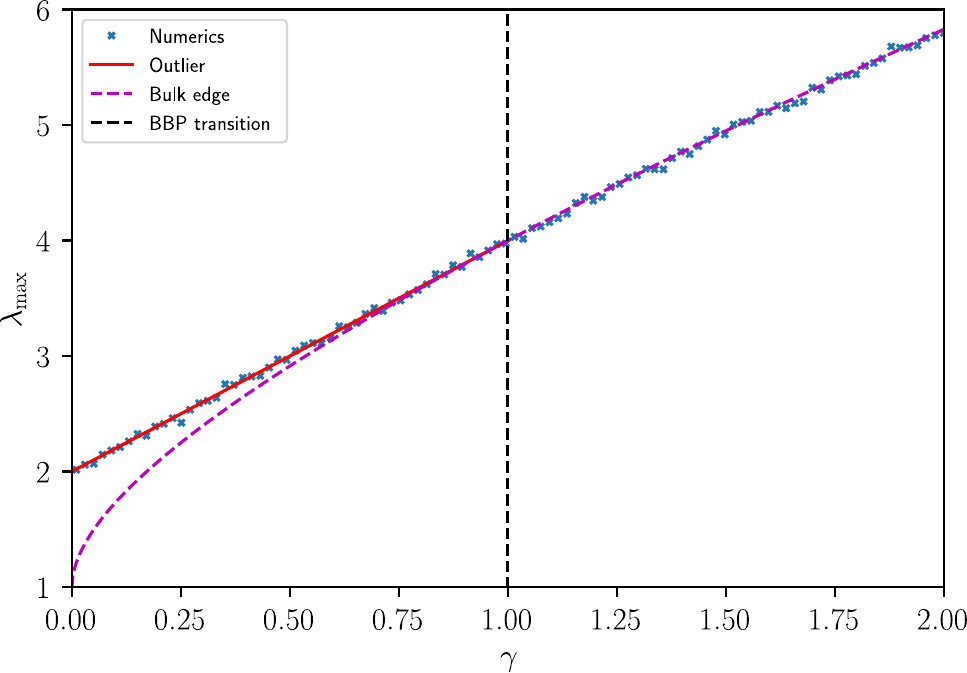} 
	\includegraphics[scale = 0.48]{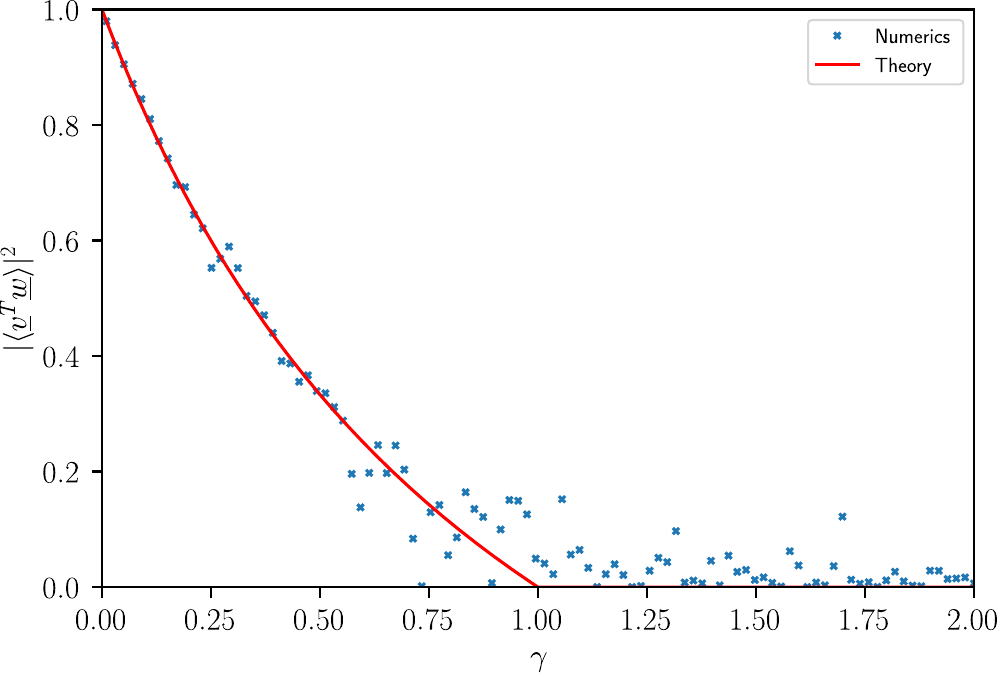} 
	\captionsetup{justification=raggedright,singlelinecheck=false}
	\caption{Comparison of numerical diagonalisation results to Eq.~(\ref{overlap}). Numerical results are for a single realisation -- there is no average over different realisations of the random matrix. Parameters: $N = 2000$, $\beta = 1$, $\sigma = 1$. }\label{fig:bbp}
\end{figure}

\subsection{Mar\v{c}enko-Pastur relation for arbitrary covariance matrices}
For the sake of illustrating the BBP transition, we took a simple example for the structure of the hypothetical correlation matrix: signal plus homogeneous noise. However, in reality, true correlation matrices in applications are unlikely to be so simple. We would therefore like to have a more general relation between the population covariance matrix $\underline{\underline{\Sigma}}$ and the spectrum of the sample covariance matrix. Such a relation actually dates back to the original work by Mar\v{c}enko and Pastur \cite{marvcenko1967distribution}. So, the question is: supposing that we know the eigenvalue spectrum of the matrix $\underline{\underline{\Sigma}}$, can we deduce what the sample covariance matrix spectrum will be?

Since we wish not to make any assumptions on the precise structure of $\underline{\underline{\Sigma}}$, we now wish to compute the trace of $\underline{\underline{A}}$, and relate this generically to $\underline{\underline{\Sigma}}$. For this task, the cavity method becomes impractical. This is because the cavity method relies on the matrices $\underline{\underline{A}}$ and $\underline{\underline{D}}$ being diagonal. Since we no longer assume $\underline{\underline{\Sigma}}$ itself to be diagonal (or close to being so), we must be more careful. 

Two natural ways to proceed are by using free probability theory (see Ref. \cite{bun2017cleaning} or Section \ref{section:freeprob} for this approach) or by using a combinatorial approach. We refer the interested reader to Section \ref{section:mpgeneral} for a derivation along combinatorial lines using Feynman diagrams. For now, we simply state the final result, which is
\begin{align}
	A(z) = \int d\nu \frac{\rho_\Sigma(\nu)}{z - \nu\left( 1-\gamma +z\gamma A(z)\right)}. \label{generalmplaw}
\end{align}
This is an implicit equation that can, in principle, be solved for $A(z)$, which yields the sample covariance matrix eigenvalue density via $\rho(\omega) = \pi^{-1}\lim_{\epsilon\to 0}\mathrm{Im} A(\omega-i\epsilon)$ as usual. One can check that one recovers the classic Mar\v{c}enko-Pastur law in Eq.~(\ref{marchenkopastur}) and the outlier in Eq.~(\ref{bbpoutlier}) by using $\rho_\Sigma(\nu) = (1-N^{-1})\delta(\nu-\sigma^2) + N^{-1}\delta(\nu-\sigma^2 -\beta)$.

What is rather exciting about Eq.~(\ref{generalmplaw}) is that it is tantalisingly close to a formula by which we can `de-noise' the eigenvalue spectrum for arbitrary covariance matrices. That is, it suggests the inverse problem -- if we are given an eigenvalue spectrum from a sample covariance matrix, what is the population covariance matrix? This question turns out to be very pressing, with one particularly prominent application being in finance \cite{bun2017cleaning, laloux2000random, potters2020first,potters2005financial, laloux1999noise}, as we now discuss.

\subsection{Mar\v{c}enko-Pastur in finance: optimal portfolios and sample covariance matrix `cleaning'}
Given the fundamental nature of the problem, it is not difficult to see that PCA, or a general statistical understanding of sample covariance matrices, would have a very broad range of applications. 

Let us now turn the PCA question on its head somewhat. We asked before under what circumstances a `signal' from a spiked covariance matrix would be detectable from PCA. That is, we wished to understand how the structure of the population covariance matrix led to a definitive principle component in the sample covariance matrix spectrum. Now, let us suppose that we have measured a sample covariance matrix and we have the empirical spectrum. What then is our best estimate of the true covariance matrix? Indeed, it has been shown that a Mar\v{c}enko-Pastur bulk, with additional outliers indicating market trends, is a good fit for a range of representative financial data sets \cite{laloux2000random, laloux1999noise}, suggesting that random matrix theory may well have a lot to say about this problem.

We show now how this question is imperative in the context of finance, where more accurate knowledge of the covariance matrix minimises financial risk \cite{bun2017cleaning}. 

\subsubsection{Markowitz' optimal portfolio theory}
To illustrate one example where an accurate knowledge of the population covariance matrix is of great importance, let us consider the following simple thought experiment due to Markowitz in the 1950s \cite{markowitz1952modern, markowitz2008portfolio, markowitz2010portfolio}, which despite its simplicity, remains a cornerstone of financial theory. Suppose we wish to construct a portfolio of financial assets out of a possible set labelled $i = 1, \cdots, N$, and suppose we have perfect information about these assets. In particular, we imagine that we know the expected returns of the assets over a fixed time period $R_i$, as well as the covariances of these returns $\Sigma_{ij}$. We wish to construct an optimal portfolio such that we minimise the total `risk' (i.e. the total variance of our total return over the time period) for a fixed given average return. That is, supposing that we choose to assign a fraction of our total investment $p_i$ to asset $i$, we must minimise 
\begin{align}
	\sigma_p^2 = \sum_{ij} p_i \Sigma_{ij} p_j,
\end{align}
subject to the constraints $R_p = \sum_i p_i R_i$ and $\sum_i p_i =1$. This is a simple constrained optimisation problem, for which we can use the method of Lagrange multipliers. Defining the Lagrange function $f(\{p_i\} ; \lambda_1, \lambda_2) = \sum_{ij} p_i \Sigma_{ij} p_j + \lambda_1 (R_p - \sum_{i} p_i R_i)+ \lambda_2 (1-\sum_i p_i)$, we obtain from $\partial f/\partial p_i = 0$
\begin{align}
	p_i^\star = \frac{1}{2} \sum_j \Sigma^{-1}_{ij} (\lambda_1 R_j + \lambda_2) . \label{optimalportfolio}
\end{align}
Enforcing the constraints, we obtain the Lagrange multipliers 
\begin{align}
	\frac{\lambda_1}{2} = \frac{R_p S_{11}- S_{1R}  }{S_{11}S_{RR} - S_{1R}^2}, \hspace{1cm} \frac{\lambda_2}{2} = \frac{S_{RR} - R_p S_{1R}}{S_{11}S_{RR} - S_{1R}^2}, \nonumber \\
	S_{RR} = \sum_{ij} R_i \Sigma_{ij}^{-1} R_j, \hspace{1cm} S_{1R} = \sum_{ij} \Sigma_{ij}^{-1} R_j, \hspace{1cm}  S_{11} = \sum_{ij} \Sigma_{ij}^{-1}.
\end{align}
Additionally, we can thus compute the minimised risk as a function of $R_P$
\begin{align}
	\sigma_\star^2(R_p) = \sum_{ij} p_i^\star \Sigma_{ij} p_j^\star = \frac{R_P^2 S_{11} - 2 R_P S_{1R} + S_{RR}}{S_{11}S_{RR} - S_{1R}^2} .
\end{align}
The hyperbola $\sigma_\star(R_p)$ is known as the `efficient frontier'. This tells us the minimum amount of risk that we must endure if we wish to obtain a given expected return. 

We see from Eq.~(\ref{optimalportfolio}) that the optimal portfolio is one that places the a high weight on those assets with the highest expected returns, with the proviso that they have low risk. We thus see that an accurate estimate of $\underline{\underline{\Sigma}}$ is crucial. If we underestimate the risk of an asset due to a finite-time measurement window, for example, one could be led to overinvest in what are actually risky stocks. We would thus like to understand how to obtain an optimal estimate of the population covariance matrix for a given sample covariance matrix. 

\subsubsection{Optimal rotationally invariant estimator of the population covariance matrix}
Now that we understand the importance of having an estimate of the population covariance matrix that is as accurate as possible, we turn our attention to finding such an optimal estimator. The na\"ive estimator of the population covariance matrix would of course be the sample covariance matrix. However, we have seen that the relationship between spectra of the population and sample covariance matrices is predictable (albeit non-trivial). This gives us some hope that an improvement on the `noisy' sample covariance matrix might be possible.

If we have no special information on any preferred directions of the eigenvectors (which may come from some prior knowledge of the market, for example), then we should seek to obtain an estimator that is not biased towards any particular direction (i.e. it is rotationally invariant). Under such assumptions, it can be shown that our best guess at the eigen-basis of the population covariance matrix is simply to assume that it is the same as that of the sample covariance matrix \cite{bun2017cleaning, bun2016rotational}. This is intuitive, since if we have no special knowledge of any preferred directions, we simply have to work with what we have.

Therefore, we reduce our problem of finding the best estimator of the population covariance matrix to a simpler problem: that of modifying the eigenvalues of the sample covariance matrix to minimise the `distance' to the population matrix. More precisely, our rotationally invariant estimator (RIE) of the population covariance matrix is given by 
\begin{align}
	\Sigma_{ij}^\mathrm{RIE} = \sum_{\nu} \xi_\nu w_i^{(\nu)} w_j^{(\nu)}, \label{RIE}
\end{align}
where $\underline{w}^{(\nu)}$ is the eigenvector of $\underline{\underline{X}}\underline{\underline{X}}^T$ with eigenvalue $\lambda_\nu$. We emphasise that $\xi_\nu \neq \lambda_\nu$ in general. We wish to find the set $\{\xi_\nu\}$ such that the Frobenius norm $\mathrm{Tr}[(\underline{\underline{\Sigma}} -\underline{\underline{\Sigma}}^\mathrm{RIE} )^2]$ is minimised. Performing the simple calculus, one thus obtains the rescaled eigenvalues, subjected to so-called \textit{optimal shrinkage}
\begin{align}
	\xi_\nu = (\underline{w}^{(\nu)})^T\underline{\underline{\Sigma}}\,\underline{w}^{(\nu)} .
\end{align}
What is remarkable is that we are able to obtain $\xi_\nu$, which are currently dependent on the \textit{population} covariance matrix, entirely in terms of quantities that are available to us from the \textit{sample} covariance matrix. 
\begin{figure}[t]
	\centering 
	\includegraphics[scale = 0.48]{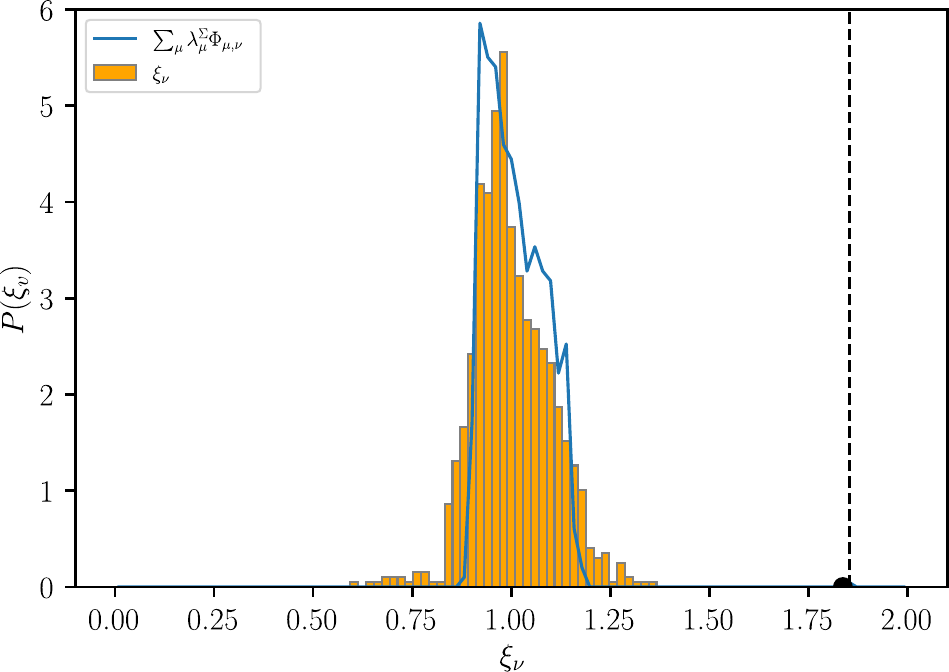} 
	\includegraphics[scale = 0.48]{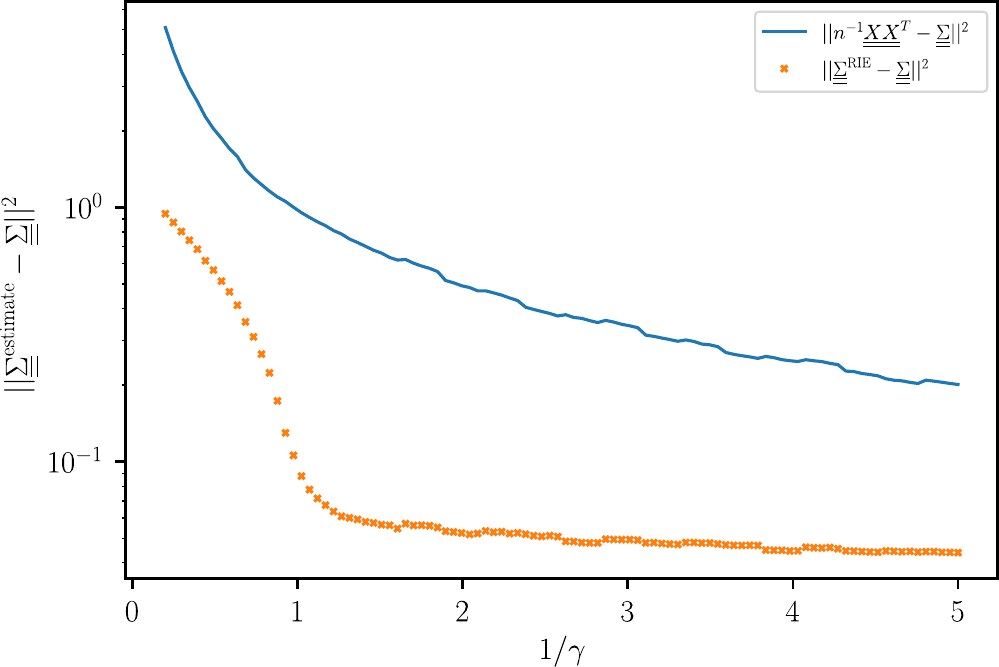} 
	\captionsetup{justification=raggedright,singlelinecheck=false}
	\caption{(Left) Comparison of the rescaled eigenvalues $\xi_\nu$ found using the formula in Eq.~(\ref{ledoitpeche}) versus the desired $\sum_\mu \lambda_\mu^\Sigma \Phi_{\nu,\mu}$. Here, $\gamma = 0.5$. (Right) The Frobenius norm distance of the RIE and the sample covariance matrix from the population covariance matrix. The RIE is more accurate by orders of magnitude. In both plots, a single random matrix is used, $N = 1000$, the regulariser $\epsilon = 0.01$ was used, and the population covariance matrix consisted of a sum of a diagonal matrix, with elements drawn from a uniform distribution on the interval $[0.6,1.4]$, and a rank-1 perturbation $1.5\underline{u}\underline{u}^T$, where $\underline{u}$ is a random unit vector. }\label{fig:oracle}
\end{figure}
Let us then compute the rescaled eigenvalues. We begin by defining the squared overlap $\Phi_{\nu,\mu} = \left[ (\underline{w}^{(\nu)})^T \underline{v}^{(\mu)}\right]^2$, where $\underline{v}^{(\mu)}$ is the eigenvector of $\underline{\underline{\Sigma}}$ corresponding to the eigenvalue $\lambda^\Sigma_\mu$. As we saw before in Eq.~(\ref{overlapresolvent}), the squared overlap can be written in terms of the resolvent. We therefore have 
\begin{align}
	\xi_\nu = \sum_\mu \lambda^\Sigma_\mu \Phi_{\nu,\mu} = \lim_{\epsilon\to 0}\frac{\mathrm{Im}\sum_\mu \lambda^\Sigma_\mu \, (\underline{v}^{(\mu)})^T \underline{\underline{G}}(\lambda_\nu-i\epsilon) \underline{v}^{(\mu)} }{\mathrm{Im} \mathrm{Tr}\underline{\underline{G}}(\lambda_\nu-i\epsilon) } .
\end{align}
Using the general Mar\v{c}enko-Pastur formula in Eq.~(\ref{generalmplaw}) [see also Eq.~(\ref{helpfullp}) in Section \ref{section:mpgeneral}], we have (writing $z_\nu = \lambda_\nu - i\epsilon$)
\begin{align}
	&\frac{1}{N}\sum_\mu\lambda^\Sigma_\mu \, (\underline{v}^{(\mu)})^T \underline{\underline{G}}(z_\nu) \underline{v}^{(\mu)} = \frac{1}{N}\sum_\mu \frac{\lambda^\Sigma_\mu}{z_\nu - \lambda^\Sigma_\mu [1-\gamma + \gamma z_\nu A(z_\nu)]} \nonumber \\
	&= \frac{z_\nu A(z_\nu)}{1-\gamma + \gamma z_\nu A(z_\nu)} -\frac{1}{1-\gamma + \gamma z_\nu A(z_\nu)} \nonumber \\
	&= \frac{1}{\gamma} \left[  1 -\frac{1}{1-\gamma + \gamma z_\nu A(z_\nu)}\right] .
\end{align}
Taking the imaginary parts, we finally arrive at the Ledoit-P\'ech\'e-Wolf formula for optimal shrinkage \cite{ledoit2015spectrum,ledoit2012nonlinear,ledoit2011eigenvectors}
\begin{empheq}[box={\fboxsep=6pt\fbox}]{align}
	\xi_\nu \approx \frac{\lambda_\nu}{\left\vert 1 - \gamma + \gamma z_\nu  A_E(z_\nu) \right\vert^2} , \label{ledoitpeche}
\end{empheq}
where we choose a small regulariser $\epsilon \sim N^{-1}$ to smooth the empirical resolvent
\begin{align}
	A_E(z) \equiv \frac{1}{N} \sum_\nu \frac{1}{z - \lambda_\nu}.
\end{align}
We thus have obtained, miraculously, a best estimator for the population covariance matrix that involves only quantities that are measurable from the sample covariance matrix. We test this estimator in Fig. \ref{fig:oracle}.

\subsection{Exercises}
In our analysis above, we distinguished carefully between the properties of the population and those of the sample. In particular, the principal component direction corresponded to a rank-1 perturbation to the population covariance matrix $\underline{\underline{\Sigma}}$ and not the sample covariance matrix $\underline{\underline{X}}\underline{\underline{X}}^T$. Na\"ively, one might expect a rank-1 perturbation to the sample covariance to give the same results. We show here that this is not the case. 

Let us suppose that we can approximate the effect of the additional principal component direction as a deterministic rank-1 perturbation such that $n^{-1}\underline{\underline{x}} \underline{\underline{x}}^T \to n^{-1}\underline{\underline{x}}\underline{\underline{x}}^T + \beta \underline{v}\underline{v}^T$. 
\begin{itemize}
	\item Use Woodbury's identity with $r = N+n$ and $k = 1$ to show that 
	\begin{align}
		\underline{\underline{\tilde{A}}} = \left[	( \omega -i \epsilon) \underline{\underline{\id}} - n^{-1}\underline{\underline{x}} \underline{\underline{x}}^T -  \beta \underline{v}\underline{v}^T\right]^{-1} = \underline{\underline{A}} + \frac{\beta}{1 - \beta A} \underline{\underline{A}} \, \underline{v} \underline{v}^T \underline{\underline{A}}. 
	\end{align}
	\item By identifying the location of the pole of $\underline{v}^T \underline{\underline{\tilde{A}}}\,\underline{v}$ from this equation, and using Eq.~(\ref{aandd}), show that the predicted outlier eigenvalue is 
	\begin{align}
		\lambda_\mathrm{outlier} = \beta + \frac{\sigma^2}{1 - \frac{\gamma \sigma^2}{\beta}} .
	\end{align}
	\item From Eq.~(\ref{aandd}), show by implicit differentiation that
	\begin{align}
		A'(\omega) = - A^2 \left[ 1 - \frac{\gamma\sigma^4}{(1-\gamma \sigma^2 A)^2} A'(\omega)\right] .
	\end{align}
	Hence, show that the predicted eigenvector overlap in this case is 
	\begin{align}
		\vert \underline{v}^T \underline{w} \vert^2 = 1 - \frac{\gamma \sigma^4/\beta^2}{(1-\gamma \sigma^2/\beta)^2} .
	\end{align}
\end{itemize}
Clearly these results differ from those obtained by being more careful with the introduction of the population covariance spike. 

We may also perform a similar analysis for the Wigner semicircle law and the outlier that emerges when we perturb $\underline{\underline{J}}$ [with statistics as defined in Eq.~(\ref{statisticswigner})] by a rank-1 perturbation $\frac{\mu}{N} \underline{u} \underline{u}^T$, where in this case $\underline{u}^T = (1,1,\cdots, 1)$. This corresponds to giving the matrix elements a non-zero mean $\langle J_{ij} \rangle = \frac{\mu}{N}$.
\begin{itemize}
	\item Again using Woodbury's identity with $r = N$ and $k = 1$, show that 
	\begin{align}
		\underline{\underline{\bar{G}}} = \left[	( \omega -i \epsilon) \underline{\underline{\id}} - \underline{\underline{J}}  - \frac{\mu}{N} \underline{u}\underline{u}^T\right]^{-1} = \underline{\underline{G}} + \frac{\mu}{N}\frac{1}{1 - \mu G} \underline{\underline{G}} \, \underline{u} \underline{u}^T \underline{\underline{G}}. 
	\end{align}
	\item By using the expression $G = \frac{1}{\omega - G}$, show that $\bar{G} = N^{-1} \mathrm{Tr}\underline{\underline{\bar{G}}}$ has a pole at
	\begin{align}
		\lambda_\mathrm{outlier} = \mu + \frac{1}{\mu}.
	\end{align}
	\item Using that $N^{-1} \sum_{ijk} G_{ij}G_{ki}= -\partial_\omega G(\omega)$ [which is an identity that is generally true of the resolvent of symmetric matrices], show that close to the pole we may write
	\begin{align}
		\bar{G} \approx \frac{1}{N}\frac{1}{\omega - i\epsilon- \lambda_\mathrm{outlier}} .
	\end{align}
	Hence, noting the $G(\omega)$ is real for $\epsilon \to 0$ for $\vert \omega\vert>2$ and complex otherwise, show that the Wigner semicircle is modified only by a contribution $O(1/N)$, and that the eigenvalue density is given by 
	\begin{align}
		\rho(\omega) = \frac{1}{2 \pi} \sqrt{4 -\omega^2} + \frac{1}{N} \delta\left( \omega - \mu - \mu^{-1}\right) .
	\end{align}
	\item Using that $\partial_\omega G(\omega) = -G^2/(1-G^2)$, show that the overlap between the eigenvector corresponding to the outlier $\underline{w}$ and the uniform vector $N^{-1/2}\underline{u}^T=N^{-1/2} (1, 1, \cdots, 1)$ is
	\begin{align}
		\frac{1}{N}\vert  \underline{w}^T \underline{u}  \vert^2 = 1-\frac{1}{\mu^2} ,
	\end{align}
	and show that the overlap with any eigenvector in the bulk is nil. This result helps us see that dynamical instabilities governed by eigenvalues in the bulk have different characteristics to instabilities governed by outlier eigenvalues. This is as we found using the dynamic approach in Section \ref{section:dynamicalinstabilities}. 
	
\end{itemize}

\newpage

\section{Dyson's Brownian motion approach and the $\beta$-ensembles}\label{section:dysonbrownian}

\begin{quotation}I found it difficult ...... to keep my mind from wandering into the magical world of random matrices. -- Freeman J. Dyson \end{quotation}

We have seen how one can compute the \textit{average} density of eigenvalues in the limit of large $N$ in a number of circumstances, and we have briefly touched on the Wigner surmise in Section \ref{section:semicircle}, which approximates the distribution of distances between neighbouring eigenvalues. However, we will show now that for certain classes of random matrix, one can understand much more completely the statistics of the eigenvalues, even without taking the limit $N\to \infty$. In fact, we can obtain the \textit{full} joint distribution of the eigenvalues. This enables us to study objects of interest, such as the eigenvalue correlations or large spectral fluctuations, which would be very difficult to analyse otherwise.

A particularly elegant way of achieving this feat is the beautiful method due to Dyson \cite{mehta2004random, dyson1962brownian}, whose series of papers \cite{dyson1962statistical, dyson1962statistical2, dyson1962statistical3, dyson1963statistical4, mehta1963statistical5} on RMT very much revolutionised the field, making clear the impact of RMT particularly in physics \cite{dyson1962threefold}. Dyson's strategy is as follows. We imagine that the random matrix entries each follow a stochastic process in time. We choose the stochastic process so that after a long time the matrix entries each have a desired distribution. We then imagine that we freeze the process. We have thus produced the random matrix that we desire. 

This may seem like a protracted way to produce a random matrix. However, this method has one distinct advantage. One can consider the concomitant time evolution of the eigenvalues, and thus also extract their joint statistics at stationarity. 

Before we demonstrate this, we examine Dyson's reasons for being interested in random matrices, and we introduce the so-called $\beta$-ensembles that will be at the centre of much of our discussion here.

\subsection{Dyson's threefold way}
Dyson's papers on RMT contributed to physics in a number of ways, but perhaps his most salient observation was that there are 3 matrix symmetry classes that are most relevant for traditional quantum mechanics\footnote{This has been extended to certain non-Hermitian systems \cite{kawabata2019symmetry} in recent years to give rise to a 38-fold symmetry classification!}. 

The identification of these symmetry classes is related to Frobenius' theorem of real division algebras \cite{palais1968classification}. Frobenius finds that every finite-dimensional associative division algebra over the real numbers is isomorphic to one of the following: the real numbers, the complex numbers and the quaternions. Elements of these sets can each be represented by 1 real dimension, 2 real dimensions and 4 real dimensions respectively. 

Similarly, Dyson (following Weyl and Wigner) argued that there are 3 important symmetry classes of the Hamiltonian of a quantum system \cite{dyson1962threefold}. These are systems that: obey time reversal symmetry, break time reversal symmetry, have half-integer spin and do not break the time-reversal symmetry, respectively. Knowing Wigner's work on random matrices and heavy nuclei, Dyson was aware that a statistical representation of the interactions in complex quantum systems was extremely useful. His insight was that, owing to the symmetry classifications of the Hamiltonian and Frobenius' theorem, the random matrix representations of quantum systems ought also to fall into 3 broad classifications. These are: real symmetric matrices, complex hermitian matrices, and quaternion self-adjoint matrices. Matrix ensembles that satisfy these properties are known as $\beta$ ensembles, and are associated with values of $\beta = 1$, $2$ and $4$, respectively. The significance of the number $\beta$ will become more clear momentarily. The aforementioned random matrix ensembles are invariant to changes of basis that involve orthogonal, unitary and symplectic rotation matrices respectively.

The point of this classification was that physical observables, such as the energy level separations, might be expected to share the statistics of these random matrix ensembles, depending on the symmetry constraints of the system. As we will show below, Dyson both provides a means of deriving the level statistics, and of demonstrating that such statistics ought to be common to all random matrices sharing the symmetry class (universality).

\subsection{Rotationally invariant ensembles}
To give a concrete example, the Gaussian $\beta$-ensembles are used ubiquitously in the RMT literature as convenient test cases. For each value of $\beta$, the joint distribution of the matrix entries is given by 
\begin{align}
	P(\underline{\underline{J}}) = \frac{1}{Z} e^{-\frac{\beta N}{4 \sigma^2} \mathrm{Tr} \underline{\underline{J}}^2} , \label{gbetae}
\end{align}
but we have that $\underline{\underline{J}}$ is a symmetric real matrix for $\beta = 1$, $\underline{\underline{J}}$ is complex hermitian matrix for $\beta = 2$, and $\underline{\underline{J}}$ is self-adjoint quaternionic for $\beta = 4$. More precisely, we may define the auxiliary variables $X_{ij}$ such that
\begin{align}
	X_{ij} = \begin{cases}
		\mathcal{N}(0, 1) \hspace{0.5cm} \mathrm{for} \hspace{0.5cm} \beta = 1\\
		\mathcal{N}(0, 1/2) + i\mathcal{N}(0, 1/2) \hspace{0.5cm} \mathrm{for} \hspace{0.5cm} \beta = 2\\
		\mathcal{N}(0, 1/4) + i\mathcal{N}(0, 1/4) + j\mathcal{N}(0, 1/4) + k\mathcal{N}(0, 1/4) \hspace{0.5cm} \mathrm{for} \hspace{0.5cm} \beta = 4
	\end{cases} \label{xmat}
\end{align}
where $i$, $j$, and $k$ are the quaternionic basis elements obeying $i^2 = j^2 = k^2 = ijk = -1$. The matrix $\underline{\underline{J}}$ is then constructed via $J_{ij} = \frac{1}{2}(X_{ij} + X_{ji}^\star)$. This means that the entries have statistics
\begin{align}
	\langle J_{ii}^2 \rangle = \frac{2\sigma^2}{\beta N}, \hspace{1cm} \langle \vert J_{ij} \vert^2 \rangle = \frac{\sigma^2}{ N} , \label{gbetaestats}
\end{align}
where we note that $J_{ii}$ is real. One can indeed show that the above-defined Gaussian ensembles obey the appropriate symmetries. In the $\beta = 1$ case, the distribution in Eq.~(\ref{gbetae}) is invariant under the rotation $\underline{\underline{J}} \to \underline{\underline{O}}\, \underline{\underline{J}}\, \underline{\underline{O}}^T$, where $\underline{\underline{O}}^{-1} = \underline{\underline{O}}^T$. Similarly, for $\beta = 2$ we have invariance under a unitary transformation $\underline{\underline{J}} \to \underline{\underline{U}}\, \underline{\underline{J}}\, \underline{\underline{U}}^T$, where $\underline{\underline{U}}^{-1} = \underline{\underline{U}}^\dagger$. One notes that the $\beta = 1$ case would not be invariant under such a transformation, since the matrix $\underline{\underline{J}}$ would cease to be real valued. Finally, in the $\beta = 4$ case, each of the the quaternionic entries can be written in terms of complex $2 \times 2$ matrices. The compact symplectic rotation matrices belonging to the group $\mathrm{Sp}(2N; \mathbb{R})$ then preserve the probability distribution in Eq.~(\ref{gbetae}). For this reason, the Gaussian $\beta$ ensembles are usually referred to as the Gaussian Orthogonal Ensemble (GOE), the Gaussian Unitary Ensemble (GUE) and the Gaussian Symplectic Ensemble (GSE) for $\beta = 1$, $\beta = 2$ and $\beta = 4$ respectively. 

One may conceive of more general rotationally invariant ensembles belonging to the aforementioned symmetry classes. These ensembles have joint distributions of their entries that take the following form
\begin{align}
	P(\underline{\underline{J}}) = \frac{1}{\mathcal{Z}}\exp\left[-\frac{\beta N}{\sigma^2}\mathrm{Tr}V\left(\underline{\underline{J}}\right) \right],\label{rotationallyinvarientensemble}
\end{align}
where the so-called potential $V(\cdot)$ is a polynomial with real coefficients, and $\mathcal{Z}$ is a normalisation constant. One can check rather simply that indeed rotational invariance is preserved here. Notably, the Gaussian ensembles are recovered by using $V(x) = x^2/4$. 

\subsection{Stochastic process to produce rotationally invariant ensembles}
Now, following Dyson \cite{dyson1962brownian}, we wish to construct a stochastic process for the random matrix entries $J_{ij}$ that has the joint distribution in Eq.~(\ref{rotationallyinvarientensemble}) as its stationary distribution. We therefore imagine that each matrix entry $J_{ij} = J_{ji}^\star$ follows a stochastic process
\begin{align}
	J_{ij}(t+\Delta) - J_{ij}(t)	&=- \Delta \left[ V'\left(\underline{\underline{J}}\right) \right]_{ij}+ \sqrt{\Delta} B_{ij}(t), \label{evolutionofJ}
\end{align}
where $\Delta$ is a small time increment, $V'(x) = \partial_xV(x)$ and $B_{ij}(t)$ are Gaussian noise terms that are drawn from the $\beta$-ensemble in Eq.~(\ref{gbetae}). That is, at each point in time, we draw an independent realisation of the matrix $\underline{\underline{X}}(t)$, according to Eq.~(\ref{xmat}), and we construct $\underline{\underline{B}}(t) = \frac{1}{2}\left[\underline{\underline{X}}(t) + \underline{\underline{X}}^\dagger(t) \right] $. 

We emphasise that the time $t$ is completely fictitious, and without physical meaning. The temporal evolution is merely a clever tool that we introduce to exploit the tools of statistical physics. We now show that the stationary distribution of the above process satisfies Eq.~(\ref{rotationallyinvarientensemble}).

Since $J_{ij}(t)$ is comprised of different components if $\beta = 2$ or $4$, it is convenient to write $J_{ij}^\alpha(t)$, where $\alpha =1, \cdots, \beta$. Taking the limit $\Delta \to 0$, the corresponding Fokker-Planck equation \cite{risken1989fokker} [see also Eq.~(\ref{fpedef})] for the joint distribution of $J^\alpha_{ij}(t)$ is then
\begin{align}
	\partial_t P(\underline{\underline{J}}, t) &=\sum_{ij\alpha} \frac{\partial}{\partial J^\alpha_{ij}} \left[P(\underline{\underline{J}},t) \left[ V'\left(\underline{\underline{J}}\right) \right]_{ij}^\alpha +\frac{\sigma^2}{2N\beta} \frac{\partial P}{\partial J^\alpha_{ij}}   \right].
\end{align}
where $\left[ V'\left(\underline{\underline{J}}\right) \right]_{ij}^\alpha$ denotes the $\alpha$-th component of $\left[ V'\left(\underline{\underline{J}}\right) \right]_{ij}$, where again $\alpha = 1, \cdots , \beta$. We therefore arrive at the stationary distribution by setting $\partial_t P(J, t) = 0$. If we imagine that the stationary distribution $P_\mathrm{st}(\underline{\underline{J}})$ satisfies a zero-flux boundary condition at $J^\alpha_{ij} \to \pm \infty$, then we obtain for each $(\alpha,i,j)$
\begin{align}
	P_\mathrm{st}(\underline{\underline{J}}) \left[ V'\left(\underline{\underline{J}}\right) \right]_{ij}^\alpha + \frac{\sigma^2}{2N\beta} \frac{\partial P_\mathrm{st}}{\partial J^\alpha_{ij}} = 0 . \label{statfpe}
\end{align}
One can show that $\left[ V'\left(\underline{\underline{J}}\right) \right]_{ij}^\alpha = \frac{1}{2}\frac{\partial  \mathrm{Tr}V}{\partial J^\alpha_{ij}}$\footnote{This is done by using the chain rule and the fact that $\frac{\partial J_{lm}}{\partial J_{lm}^{\alpha}}= 1, i, j$ or $k$ (where here $i$, $j$, and $k$ are the quaternionic basis elements) depending on $\alpha$, and similarly for $\frac{\partial J_{ml}}{\partial J_{lm}^{\alpha}}$.}, and so we can solve Eq.~(\ref{statfpe}) to obtain Eq.~(\ref{rotationallyinvarientensemble}) as we desired.

\subsection{Stochastic evolution of the eigenvalues}
Now, let us consider the temporal evolution of the eigenvalues of the matrix $\underline{\underline{J}}(t)$. Suppose we know the eigenvalues at some time $t$, $\{\lambda_i(t)\}$. From Eq.~(\ref{evolutionofJ}), we see that the matrix only changes by a small amount to the next time step. We therefore expect the eigenvalues to change only by a small amount also. As a result, we may use standard results from quantum mechanical perturbation theory \cite{griffiths2018introduction}.

\begin{tcolorbox}[colback=blue!10!white,colframe=blue!90!black,title=Lemma: QM-style perturbation theory]
	Suppose we know the eigenvalues $\{\lambda_\mu\}$ and eigenvectors $\underline{v}^{(\mu)}$ of an `unperturbed Hamiltonian' (which is just a matrix, for us) $\underline{\underline{H_0}}$. We perturb the Hamiltonian with an `external potential' $\alpha \underline{\underline{U}}$, where $\alpha$ is a small `perturbation parameter'. We then seek to find the eigenvalues $\{\tilde \lambda_\mu\}$ of the modified Hamiltonian $\underline{\underline{H}} =\underline{\underline{H_0}} +\alpha \underline{\underline{U}}$. By systematically matching orders of $\alpha$, one finds 
	\begin{align}
		\tilde\lambda_\mu  = \lambda_\mu +  \alpha \sum_{ij} (v^{(\mu)}_i)^\star U_{ij} \, v_j^{(\mu)} + \alpha^2\sum_{\nu \neq \mu} \frac{\left\vert \sum_{ij} (v_i^{(\nu)})^\star U_{ij} v_j^{(\mu)} \right\vert^2}{\lambda_\mu - \lambda_\nu} + O(\alpha^3). \label{perturbedeigenvalues}
	\end{align}
	We identify $\underline{\underline{J}}(t+\Delta)=\underline{\underline{H}}$, $\underline{\underline{J}}(t)=\underline{\underline{H_0}}$ and $\alpha \underline{\underline{U}} = - \Delta V'(\underline{\underline{J}}) + \sqrt{\Delta} \underline{\underline{B}}(t)$.
\end{tcolorbox} 
We must understand the statistical behaviour of each of the terms appearing in Eq.~(\ref{perturbedeigenvalues}). We exploit the rotational invariance of the Gaussian $\beta$-ensemble from which we draw $\underline{\underline{B}}(t)$. Defining $B^{\mu\nu}(t) = (\underline{v}^{(\mu)})^\dagger \underline{\underline{B}}(t) \underline{v}^{(\nu)}$, we see that the transformed variables $B^{\mu\nu}(t)$ are obtained by rotating $\underline{\underline{B}}(t)$ into the eigenbasis of $\underline{\underline{J}}$. Due to the rotational invariance of the Gaussian $\beta$-ensembles, the statistics of $B^{\mu\nu}(t)$ and those of $B_{ij}(t)$ are identical, and we therefore have [c.f. Eq.~(\ref{gbetaestats})]
\begin{align}
	\langle B^{\mu\mu}(t)^2 \rangle &= \left\langle \left[\sum_{ij}(v^{(\mu)}_i)^\star B_{ij}(t) v_j^{(\mu)} \right]^2\right\rangle = \frac{\sigma^2}{ N},\nonumber \\
	\langle \vert B^{\mu\nu}(t)\vert^2 \rangle &=\left\langle \left\vert \sum_{ij} (v_i^{(\nu)})^\star B_{ij} v_j^{(\mu)} \right\vert^2 \right\rangle = \frac{2 \sigma^2}{\beta N}.
\end{align}
We therefore see that the first term in Eq.~(\ref{perturbedeigenvalues}) is a real Gaussian random variable with mean
\begin{align}
	\left\langle \alpha \sum_{ij} (v^{(\mu)}_i)^\star U_{ij} \, v_j^{(\mu)} \right\rangle &= - \Delta  V'(\lambda_\mu) , 
\end{align}
and variance
\begin{align}
	\mathrm{Var} \left[\alpha \sum_{ij} (v^{(\mu)}_i)^\star \, U_{ij} \, v_j^{(\mu)} \right] = \frac{2\Delta \sigma^2}{\beta N}.
\end{align}
The remaining term concentrates for $\Delta \to 0$, and we find
\begin{align}
	\left\langle\alpha^2\left\vert \sum_{ij} (v_i^{(\nu)})^\star U_{ij} v_j^{(\mu)} \right\vert^2\right\rangle 
	&=  \frac{\Delta \sigma^2}{N}, \nonumber \\
	\mathrm{Var}\left[\alpha^2\left\vert \sum_{ij} (v_i^{(\nu)})^\star U_{ij} v_j^{(\mu)} \right\vert^2\right] &\to 0.
\end{align}
Since $\underline{\underline{B}}(t)$ and $\underline{\underline{B}}(t')$ are independent random matrices, all correlators of similar quantities at different times are zero. Therefore, taking the continuous time limit, one finally arrives at
\begin{align}
	\frac{d \lambda_\mu}{dt} = -  V'(\lambda_\mu) + \frac{\sigma^2}{N} \sum_{\nu\neq \mu} \frac{1}{\lambda_\mu-\lambda_\nu} + \sqrt{\frac{2\sigma^2}{\beta N}}\xi_\mu, \label{dbmeq}
\end{align}
where the Gaussian white noise term here is taken in the It\^o sense, and is uncorrelated such that
\begin{align}
	\langle \xi_\mu(t) \xi_\nu(t')\rangle = \delta(t-t') \delta_{\mu \nu} .
\end{align}

\begin{figure}[h]
	\centering 
	\includegraphics[scale = 0.54]{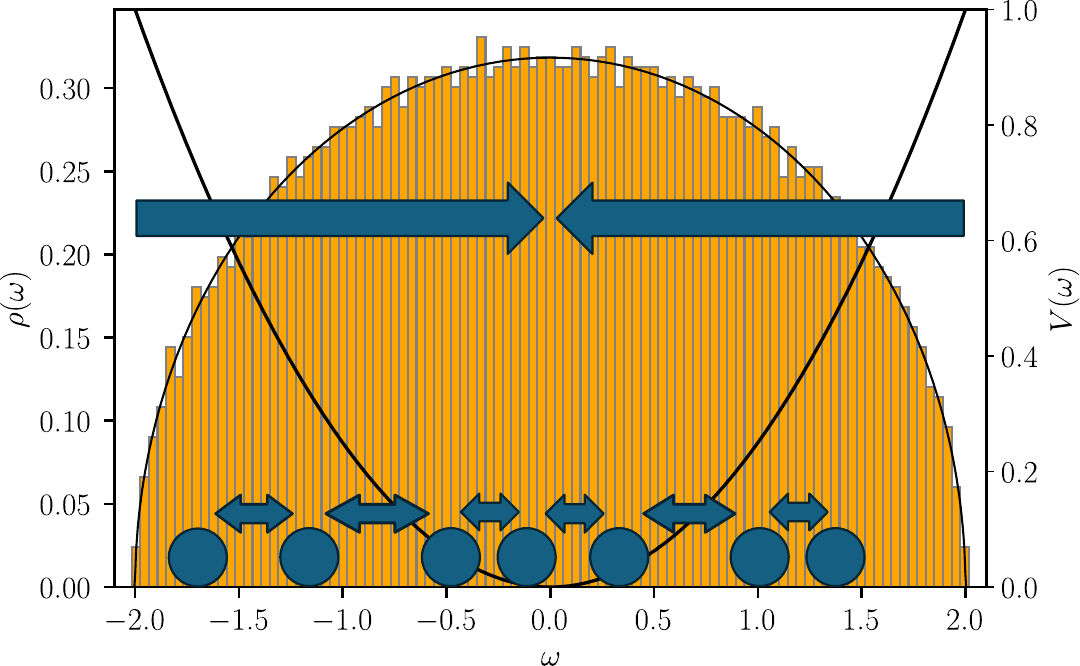} 
	\captionsetup{justification=raggedright,singlelinecheck=false}
	\caption{Eigenvalues that form the Wigner semicircle can be thought of as `charged particles' moving in a quadratic potential, subject to logarithmic inter-particle repulsion.}\label{fig:loggas}
\end{figure}

\subsection{Joint distribution of eigenvalues: Coulomb gas interpretation}
To obtain the stationary distribution of the eigenvalues, we may once again use the corresponding Fokker-Planck equation. Precisely, we have
\begin{align}
	\frac{\partial P(\{\lambda_\mu\}, t)}{\partial t} = \sum_\mu\frac{\partial }{\partial \lambda_\mu}\left[ \left(V'- \frac{\sigma^2}{N} \sum_{\nu\neq \mu} \frac{1}{\lambda_\mu-\lambda_\nu}\right) P + \frac{\sigma^2}{\beta N} \frac{\partial P}{\partial \lambda_\mu} \right]. \label{jointfpe}
\end{align}
Again assuming that the stationary distribution has a zero-flux boundary condition at $\lambda \to \pm\infty$, we finally obtain the joint distribution of the eigenvalues
\begin{empheq}[box={\fboxsep=6pt\fbox}]{align}
	P(\{\lambda\}) \propto \exp\left[ -\frac{\beta N}{\sigma^2}\sum_{\mu}V(\lambda_\mu) + \beta \sum_{\mu<\nu} \ln \left\vert \lambda_\nu -\lambda_\mu \right\vert)\right] .\label{jointdistribution}
\end{empheq}
The form of this solution invites a comparison with the Boltzmann distribution from Statistical Mechanics. If the joint energy of a set of particles in a state $S$ is $E_S$, and the system is in contact with a thermal bath of temperature $T$, the Boltzmann distribution tells us the probability of observing the particles in state $S$
\begin{align}
	P_S \propto e^{-\frac{E_S}{k_B T}} .
\end{align}
Rather nicely, we therefore see by comparing with Eq.~(\ref{jointdistribution}) that the eigenvalues can be thought of as a set of $N$ like-charges (e.g. electrons). These charges are subject to logarithmic Coulomb repulsion\footnote{One notes that the Coulomb repulsion potential is the Green's function of the spatial Laplacian, and is hence logarithmic in 2D space.} between themselves, are confined to the real axis, and move in an external confining potential $N V/\sigma^2$ at temperature $1/\beta$. This is often referred to as the \textit{Coulomb gas} interpretation. We recall that there was another analogy with 2D electrostatics analogy that was explored in the exercises of Section \ref{section:ellipse}. 

With this interpretation in mind, and also examining the fictitious motion of the eigenvalues in Eq.~(\ref{dbmeq}), we see that eigenvalues are inclined not to occupy the same spaces -- this is referred to as \textit{eigenvalue repulsion}. We also note that the form in Eq.~(\ref{jointdistribution}) is close to the Wigner surmise and indeed coincides for $N = 2$. We will explore this connection in greater detail below.

We remark that while the Brownian motion model that we have considered here gives rise only to values of $\beta = 1,2,4$, it is actually possible to conceive of random matrix ensembles whose joint distribution of eigenvalues takes the form of Eq.~(\ref{jointdistribution}), but with general real $\beta>0$. Specifically, one considers ensembles of symmetric tri-diagonal random matrices with i.i.d. normally-distributed random variables on the diagonal and chi-distributed elements on the off-diagonal. This construction was first found by Dumitriu and Edelman in 2002 \cite{dumitriu2002matrix}.

\subsection{Mean eigenvalue density}\label{section:densitydbm}
From the joint distribution in Eq.~(\ref{jointdistribution}) we can find the marginal distribution of a single eigenvalue, which is equivalent to the average eigenvalue density, and hence we can recover the Wigner semicircle law, and obtain more generalised expressions for the eigenvalue density for non-quadratic $V(\cdot)$. We use It\^o's lemma \cite{gardiner2009stochastic} to accomplish this, which allows us to understand the time evolution of \textit{functions} of stochastic variables.

\begin{figure}[h]
	\centering 
	\includegraphics[scale = 0.45]{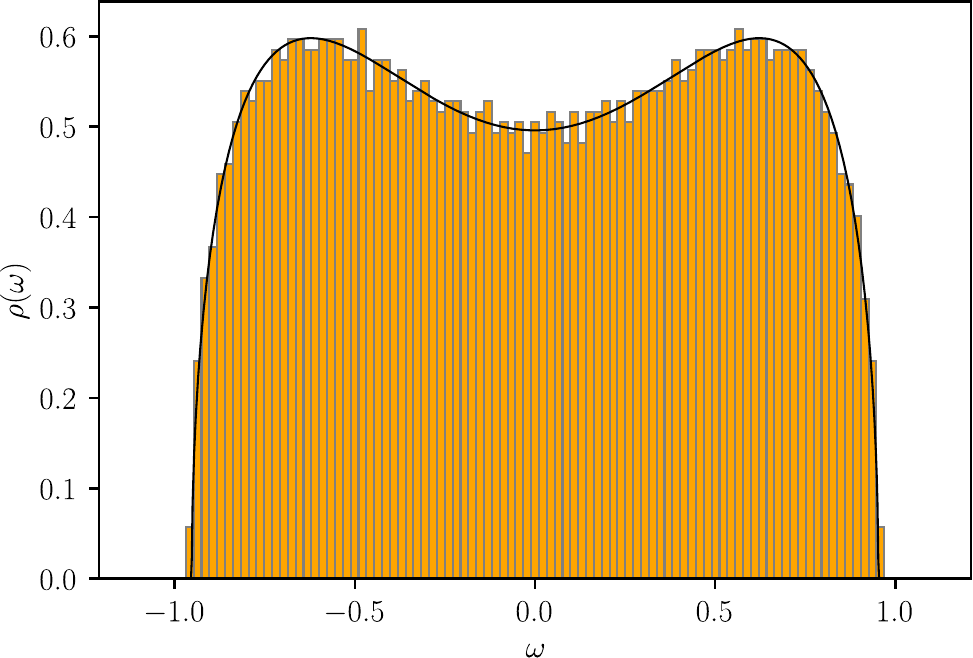} 
	\includegraphics[scale = 0.45]{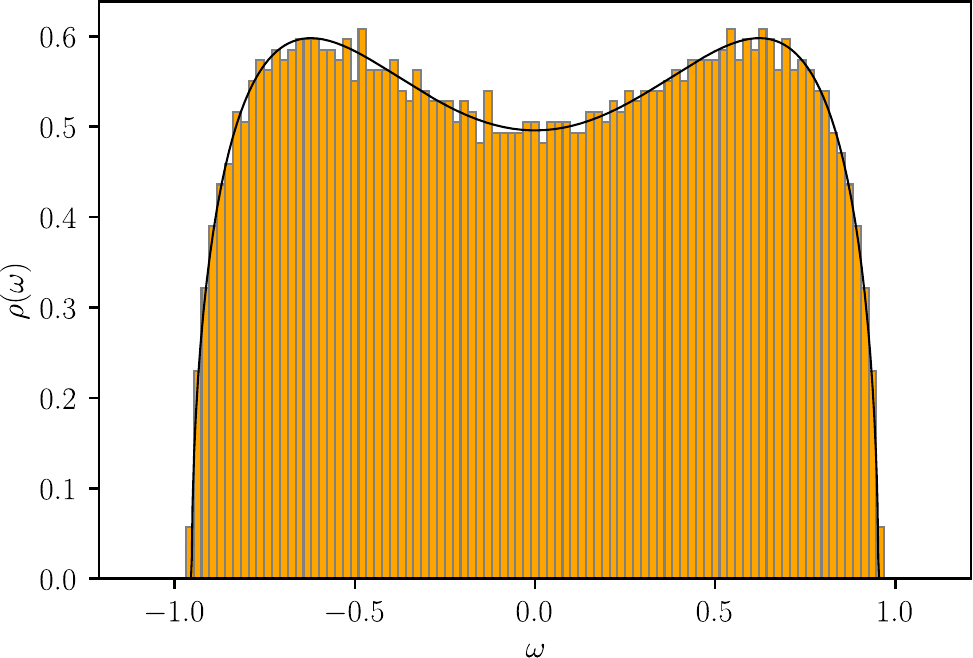} 
	\captionsetup{justification=raggedright,singlelinecheck=false}
	\caption{Eigenvalue density for the potential $V(z) = \frac{1}{2}\left[ \frac{x^2}{2} + a\frac{x^4}{4}\right]$, with $\sigma = 1$ and $a = 5$. We see clear deviations from the semicircle law. The numerical results from which the histogram was constructed were obtained by numerically integrating Eq.~(\ref{evolutionofJ}) with $N = 4000$. The black line is the theory derived in the exercises [see Eq.~(\ref{phi4density})]. (Left) $\beta = 1$, (Right) $\beta=2$. We see that the choice of symmetry class has no bearing on the overall eigenvalue density. }\label{fig:phi4}
\end{figure}

\begin{tcolorbox}[colback=blue!10!white,colframe=blue!90!black,title=Lemma: It\^o's lemma]
	Supposing $x(t)$ obeys an It\^o stochastic process $\dot x = a(x) +\sqrt{b(x)} \xi(t)$ with $\langle \xi(t) \xi(t')\rangle = \delta(t-t')$. The time evolution of $f(x(t))$ is then given by 
	\begin{align}
		\frac{df}{dt} = \dot x  \frac{\partial f}{\partial x} + \frac{b(x)}{2} \frac{\partial^2 f}{\partial x^2}.\label{ito}
	\end{align}
\end{tcolorbox} 
Using It\^o's lemma and Eq.~(\ref{dbmeq}), we may thus show that the resolvent $G(z,t) = \frac{1}{N} \sum_\nu \frac{1}{z-\lambda_\nu(t)}$ obeys \cite{allez2012invariant}
\begin{align}
	\frac{\partial G}{\partial t} = \frac{1}{N} \sum_\nu \frac{1}{(z-\lambda_\nu(t))^2} \frac{d \lambda_\nu}{dt} + \frac{2\sigma^2}{\beta N^2} \sum_\nu \frac{1}{(z-\lambda_\nu(t))^3},
\end{align}
Next, we make the observation that 
\begin{align}
\frac{1}{N^2}\sum_{\nu\neq \mu} \frac{1}{\lambda_\mu-\lambda_\nu} \frac{1}{(z-\lambda_\mu)^2} &= \frac{1}{N^2}\sum_{\nu<\mu} \frac{1}{(z-\lambda_\nu)(z-\lambda_\mu)}\left[\frac{1}{z-\lambda_\mu}+\frac{1}{z-\lambda_\nu}\right] \nonumber \\
&= -\frac{1}{2}\frac{\partial G^2}{\partial z}.
\end{align}
Hence, in the case $V(x) = x^2/4$, i.e. for the Gaussian $\beta$-ensembles, we find
\begin{align}
	\frac{\partial \langle G\rangle}{\partial t} = \frac{1}{2}\frac{\partial z \langle G\rangle}{\partial z} - \frac{\sigma^2}{2} \frac{\partial\langle G^2\rangle}{\partial z} + \frac{\sigma^2}{N\beta} \frac{\partial^2 \langle G\rangle}{\partial z^2}.
\end{align}
Crucially, we see that the term involving $\beta$ is subleading in $N$. We thus neglect this term, and we see that we ought to expect the eigenvalue density to be independent of $\beta$. That is, the GOE, GUE, and GSE each have the same eigenvalue density. This is also true more generally for other potentials $V(\cdot)$, since the noise term is the only one in Eq.~(\ref{dbmeq}) that is dependent on $\beta$, and this does not change when we alter $V(\cdot)$. This is verified in Fig. \ref{fig:phi4} for a quartic potential. 

Finally, using that $\langle G^2 \rangle \approx \langle G\rangle^2$ for large $N$, the boundary condition $\langle G\rangle \to 1/z$ for $z\to\infty$, and by imposing stationarity, we obtain the usual semicircle law for $N,t \to \infty$. We discuss how a more general potential may be handled in the exercises.

\subsection{2-point eigenvalue density correlations}
One of the advantages of the Dyson Brownian motion approach is the fact that information beyond the average eigenvalue density is available to us. For example, we may wish to obtain information on the eigenvalue correlations. We may integrate Eq.~(\ref{jointfpe}) to obtain the two-point function
$P_2(\lambda_1, \lambda_2) = \int \prod_{\mu=3}^N d\lambda_\mu P(\{\lambda_\mu\}) =\langle \rho(\lambda_1) \rho(\lambda_2)\rangle$, and we find
\begin{align}
	\frac{\partial P_2}{\partial t} = \sum_{\mu=1}^2\frac{\partial }{\partial \lambda_\mu}\bigg[& \left(V'- \frac{\sigma^2}{N} \sum_{\nu\neq \mu} \frac{1}{\lambda_\mu-\lambda_\nu}\right) P_2 - \frac{\sigma^2}{N} \int d\lambda_3 \frac{P_3(\lambda_1, \lambda_2,\lambda_3)}{\lambda_\mu-\lambda_3}  + \frac{\sigma^2}{\beta N} \frac{\partial P_2}{\partial \lambda_\mu} \bigg]. \label{p2}
\end{align}
Unfortunately, we see that we do not obtain a closed equation. Instead, we find a dependence on the 3-point functions, which in turn will be dependent on the higher order correlations. However, for small separations $r= \lambda_1-\lambda_2 \ll N^{-1}$, some simplification can be had. In this case, the terms proportional to $P_3$ and $V'$ are subleading, and we obtain
\begin{align}
	\frac{d^2P_2(r)}{dr^2} = \beta \frac{d}{dr} \frac{P_2}{r} .
\end{align}
We therefore see that for small separations $r$, we have (to leading order in $r$)
\begin{align}
	P_2(r) \sim \vert r\vert^\beta .\label{corrbeta}
\end{align}
The fact that $P_2(r)$ shrinks to zero as $r$ does indicates the `eigenvalue repulsion' that we identified previously. To get the full expression for the eigenvalue correlations, it is necessary to use more sophisticated methods, such as the supersymmetry \cite{efetov1983supersymmetry} or polynomial methods \cite{brezin1993universality}. 

\begin{figure}[h]
	\centering 
	\includegraphics[scale = 0.6]{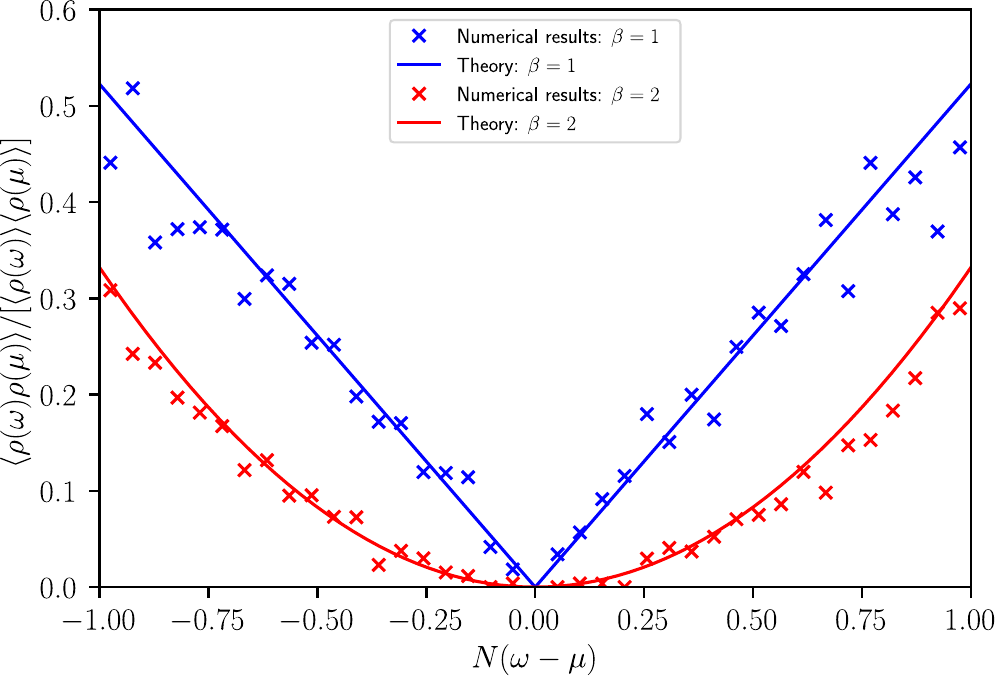} 
	\captionsetup{justification=raggedright,singlelinecheck=false}
	\caption{Eigenvalue density correlations in the cases $\beta = 1$ and $\beta = 2$. Here, we average the results of $10^6$ realisations of GOE and GUE matrices with $N = 100$. The theoretical results are $\langle \rho(\omega)\rho(\mu)\rangle/[\langle \rho(\omega)\rangle\langle\rho(\mu)\rangle] = C_\beta [N(\omega-\mu) \rho(\mu)]^\beta$ [see Eq.~(\ref{corrbeta})]. The constants of proportionality, and indeed the full sine-kernel expressions for the correlation functions, can be derived using the supersymmetric method \cite{efetov1983supersymmetry} or (skew-)orthogonal polynomial methods \cite{mehta2004random}. They are $C_1 =\pi^2/6$ and $C_2 = \pi^2/3$.}\label{fig:correlations}
\end{figure}

\subsection{DBM flow and universality}
We have already seen in our use of the cavity method that many of our results for the eigenvalue density have a degree of universality. That is, due to the central limit theorem, the sum that appears in the denominator of the cavity equations [see Eqs.~(\ref{cavity}), (\ref{cavitynonherm}) and (\ref{cavitya11}) for example] concentrates as long as the higher-order moments of the random matrix entries vanish sufficiently quickly with large $N$. Additionally, we demonstrated that quantities that describe the correlations or distance between eigenvalues also have a universal quality. This was explored in the context of the Wigner surmise in Section \ref{section:surmise}. The origin of this universality however was left as something of a mystery at the time. We aim here to demonstrate why the eigenvalue correlations ought to be thought of in some sense as even more universal than the eigenvalue density. We discuss why the Wigner surmise is successful for $N>2$ in the next section. 

Let us imagine that we have a matrix $\underline{\underline{H}}_0$, which corresponds to a particular symmetry class $\beta$. We imagine that we construct the following time-dependent matrix of the same symmetry class
\begin{align}
	\underline{\underline{H}}(t) = e^{-t/2} \underline{\underline{H}}_0 + \sqrt{1-e^{-t}} \underline{\underline{G}},
\end{align}
where here $\underline{\underline{G}}$ is a Gaussian matrix of the appropriate $\beta$-ensemble. The matrix $\underline{\underline{H}}(t)$ thus interpolates between $\underline{\underline{H}}_0$ and a Gaussian matrix as $t$ is increased. We can also construct the matrix $\underline{\underline{H}}(t)$ using the Dyson Brownian construction
\begin{align}
	\underline{\underline{H}}(t+\Delta) - \underline{\underline{H}}(t) = - \frac{\Delta}{2} \underline{\underline{H}}(t) + \sqrt{\Delta} \underline{\underline{B}},
\end{align}
where $\underline{\underline{H}}(0) = \underline{\underline{H}}_0$ and once again $\underline{\underline{B}}$ is drawn from the Gaussian $\beta$-ensemble given in Eq.~(\ref{gbetae}). The corresponding evolution of the eigenvalues is
\begin{align}
	\dot \lambda_\mu &= -\frac{1}{2}\lambda_\mu + \frac{\sigma^2}{N} \sum_{\nu\neq \mu} \frac{1}{\lambda_\mu-\lambda_\nu} + \sqrt{\frac{2\sigma^2}{\beta N}}\xi_\mu , \nonumber \\
	\langle \xi_\mu(t) \xi_\nu(t')\rangle &= \delta(t-t') \delta_{\mu \nu} .\label{gaussiandb}
\end{align}
What we  can show from this construction is that the \textit{local} statistics become Gaussian quickly, after a time $t \sim N^{-1}$, whereas the \textit{global} statistics hardly change in this same amount of time. That is, if wish to compute the two-point correlations as we did in Eq.~(\ref{p2}), the term involving $V'$ is irrelevant, as we saw. The time scale associated with equilibration of the variable $r$ is thus $\sim 1/N$. On the other hand, the time scale associated with a change in the wider spectrum (i.e. a change in the potential $V$) occurs on a time scale $\sim N^0$. For this reason, we see that a very broad range of random matrices possess the same microscopic statistics as the Gaussian ensembles. This kind of reasoning can be refined to demonstrate microscopic universality of the eigenvalue density correlations more formally and more generally \cite{erdHos2011universality,erdHos2012universality}. 

We note that other methods also offer a greater insight into universality. In particular, the supersymmetric approach has been used to great effect not only to derive expressions for the 2-point eigenvalue density correlations \cite{efetov1983supersymmetry}, but also to prove their universality beyond Gaussian and even beyond rotationally invariant ensembles \cite{mirlin1991universality}.

\subsection{The Wigner surmise revisited}\label{section:surmiserevisited}

We can also see, informally, from the Dyson Brownian motion of the eigenvalues why the Wigner surprise for the distribution of the nearest-neighbour eigenvalue spacings (see Section \ref{section:surmise}) might be expected to hold more generally than for the Gaussian $N = 2$ case from which it was derived. 

We wish to understand the relative Dyson motion of two neighbouring eigenvalues: $\lambda_1$ and $\lambda_2$. We imagine that all eigenvalues, including $\lambda_1$ and $\lambda_2$, only fluctuate a small amount about their equilibrium positions. Subtracting the equations for $\dot \lambda_1$ from $\dot \lambda_2$, given by Eq.~(\ref{dbmeq}), and defining $r = \lambda_2 - \lambda_1$, we may expand for small $r$ to find
\begin{align}
	\dot r \approx -\left[V''(\lambda_2) + \sum_{\nu\neq 1,2} \frac{1}{(\lambda_2 - \lambda_\nu)^2}\right] r + \frac{2\sigma^2}{N} \frac{1}{r} +  \sqrt{\frac{4 \sigma^2}{\beta N}} \xi ,
\end{align}
where $\xi(t)$ is a Gaussian white noise of unit amplitude. Making the somewhat coarse approximation that the quantity $V''(\lambda_1)+ \sum_{\nu\neq 1,2} \frac{1}{(\lambda_1 - \lambda_\nu)^2}$ can be considered effectively constant in comparison to the fluctuations of $r$, we may write
\begin{align}
	\dot r \approx -A r + \frac{2\sigma^2}{N} \frac{1}{r} +  \sqrt{\frac{4 \sigma^2}{\beta N}} \xi ,
\end{align}
where $A$ is a positive constant that is dependent on the average position of $\lambda _1$ and $\lambda_2$ in the spectrum. Thus, we may write the single-variable Fokker-Planck equation
\begin{align}
	\frac{\partial P(r,t)}{\partial t} = \frac{\partial }{\partial r}\left[\left(A r - \frac{2 \sigma^2}{N}\frac{1}{r} \right) P(r,t) + \frac{2 \sigma^2}{\beta N}\frac{\partial P(r,t)}{\partial r} \right]
\end{align}
Solving this for the stationary distribution, we obtain
\begin{align}
	P(r) = C_1 r^\beta e^{-C_2 r^2},
\end{align}
where the constants $C_1$ and $C_2$ may depend on the position in the spectrum $(\lambda_1 + \lambda_2)/2$. We have thus arrived at the general form of the Wigner surmise. Using the `unfolded' eigenvalue separations, as detailed in Section \ref{section:surmise}, we may then arrive at the usual Wigner surmise by imposing normalisation and unit mean. Noticing how this reasoning is independent of the particular form of the potential $V$, we can therefore see why the Wigner surmise applies for more general $N$, and also for more general ensembles than the $2\times 2$ Gaussian one that Wigner considered originally. 

For reference, the full expressions for the Wigner surmise in the cases $\beta =1,2,4$ are [recalling that $s_i = N \langle\rho\left(\bar \lambda_i\right)\rangle\left(\lambda_{i+1} - \lambda_i\right)$ with $\bar \lambda_i = (\lambda_{i+1}+\lambda_i)/2$]
\begin{empheq}[box={\fboxsep=6pt\fbox}]{align}
	P(s) = \begin{cases}
		\frac{\pi}{2} s e^{-\pi s^2/4} \hspace{1cm} \mathrm{for} \hspace{1cm} \beta = 1 \\
		\frac{32}{\pi^2} s^2 e^{-4 s^2/\pi} \hspace{1cm} \mathrm{for} \hspace{1cm} \beta = 2 \\
		\frac{2^{18}}{3^6 \pi^3} s^4 e^{-64 s^2/(9 \pi) } \hspace{1cm} \mathrm{for} \hspace{1cm} \beta = 4 
	\end{cases}.\label{surmises}
\end{empheq}
We saw in Section \ref{section:semicircle} that the $\beta = 1$ surmise has a remarkable ability to describe the level spacing statistics in nuclear data. Rather beautifully, there have been experimental efforts to realise microwave systems that possess the appropriate symmetry properties that ought to yield $\beta = 2,4$ statistics (see Ref. \cite{stoffregen1995microwave} and \cite{rehemanjiang2016microwave} respectively), even showing the crossover from $\beta = 1$ to $\beta = 2$ as time-reversal symmetry is broken \cite{so1995wave}. The agreement with appropriate surmise in all cases is very convincing.  

\subsection{Large deviations}
Given the joint distribution of eigenvalues in Eq.~(\ref{jointdistribution}), yet more interesting questions can be asked. For example, one may wonder what the probability is of a finite fraction $c\neq 1/2$ of the eigenvalues being positive. This can be tackled by imposing a hard constraint that precisely $cN$ of the eigenvalues be positive, which is rather like imposing a reflecting boundary condition on the Coulomb gas of eigenvalues at the origin. The problem may thus be solved using a saddle-point approximation for large $N$ \cite{majumdar2009index,cavagna2000index}. 

Using similar large deviation techniques, one may also study the top eigenvalue of the random matrix \cite{majumdar2014top}. While the typical top-eigenvalue fluctuations about the edge of the Wigner semicircle (for example) obey Tracy-Widom statistics \cite{tracy1994level} and are of the order $N^{-2/3}$, one may also study large fluctuations of the order $N^{0}$. In both instances, this is accomplished by placing a hard-wall boundary at some value $\omega$ and asking what difference this makes to the joint probability distribution of the Coulomb gas.

\subsection{Exercises}
Rather than the simple case of the quadratic potential, $V(x) = x^2/4$, which yields the Wigner semicircle, let us now instead consider the case $V(x) = \frac{1}{2}\left(\frac{x^2}{2} + a\frac{x^4}{4} \right)$. By following along the lines of the discussion in Section \ref{section:densitydbm}, we now compute the eigenvalue density in this quartic case.
\begin{itemize}
	\item By assuming that the eigenvalue spectrum is symmetric about the origin, show that for $N,t\to \infty$
	\begin{align}
		0 &= \frac{1}{2}\frac{\partial z \langle G\rangle}{\partial z} + \frac{a}{2}\frac{\partial z^3 \langle G\rangle}{\partial z} - a z  - \frac{\sigma^2}{2} \frac{\partial\langle G^2\rangle}{\partial z} + \frac{\sigma^2}{2N} \frac{\partial^2 \langle G\rangle}{\partial z^2} .
	\end{align}
	Solve this to show that
	\begin{align}
		\langle G \rangle = \frac{z + a z^3 \pm\sqrt{-4 \sigma^2 (c + a z^2) + (z + a z^3)^2}}{2 \sigma^2},
	\end{align}
	where in this case, $-c$ is the constant of integration, which cannot be determined by the boundary condition $\langle G \rangle \to 1/z$ at $z\to \pm\infty$. We note that members of this family of solutions could in principle have more than one branch cut.
	\item Since we expect the resolvent to have a single branch cut on the real axis, and be even in the variable $z$, we suppose that $c$ takes a value such that we may write the solution in the following form
	\begin{align}
		\langle G \rangle = \frac{z + a z^3 -\mathrm{sign}[\mathrm{Re}(z)](\alpha z^2 + \beta)\sqrt{z^2-L^2}}{2 \sigma^2}.
	\end{align}
	Hence, by requiring $\langle G \rangle\to 1/z$ for $z\to \pm\infty$, show that
	\begin{align}
		\alpha = a, \hspace{1cm} \beta = 1+ \frac{a L^2}{2}, \hspace{1cm} L^2 =  \frac{2(\sqrt{1+12 a \sigma^2}-1)}{3a} .
	\end{align} 
	One can check that the choice $c = \frac{-1 + 36 a \sigma^2 + (1 + 12 a \sigma^2)\sqrt{1 + 12 a \sigma^2}}{54 a \sigma^2}$ is consistent with this. Hence, show that one obtains the eigenvalue density
	\begin{align}
		\rho(\omega) = \frac{(a \omega^2 + a L^2/2 +1)}{2\sigma^2} \sqrt{L^2 -\omega^2}. \label{phi4density}
	\end{align}
	This is checked in Fig. \ref{fig:phi4}.
	\item Show that the eigenvalue density goes from having a single peak to two peaks at a critical value of $a = \frac{1}{4 \sigma^2}$.

\end{itemize}

\newpage

\section{Anderson localisation}\label{section:anderson}

\begin{quotation}
	Very few believed [localization] at the time, and even fewer saw its importance; among those who failed to fully understand it at first was certainly its author. It has yet to receive adequate mathematical treatment, and one has to resort to the indignity of numerical simulations to settle even the simplest questions about it.--Philip W. Anderson 
\end{quotation}

\subsection{A little context: conductors and insulators}
Electrical current can be understood as the movement of electrons through a medium. As we all know, some materials conduct electricity well, and others do not. A natural question is what distinguishes these two classes. As we have discovered over the decades, the reason for electrical conduction/insulation is, perhaps surprisingly, not one-size-fits-all. 

The canonical explanation for this duality is the one provided by electronic band theory \cite{ashcroft1976solid, kittel2018introduction}. In this scenario, electrons in conductors may be excited by small external electric fields to contribute to electric current, whereas an insulators, a band `gap' exists. Overcoming the band gap requires a much larger external electric field to allow the electron to travel as a delocalised quantum mechanical Bloch wave. Some materials (e.g. Cobalt(II) oxide), however, defy this interpretation. A supplementary mechanism, put forward by Mott \cite{mott1949basis}, suggested that electrical transport may also be inhibited by sufficiently strong electron-electron interactions.  

Anderson's mechanism \cite{anderson1958absence} for electron (de)localisation is of a more exotic kind. Anderson considers the situation in which impurities are added to a semi-conductor or insulator material. These impurities `donate' electrons, and the electrons may tunnel from impurity to impurity. At high impurity concentrations, an `impurity band' can be formed, and conduction occurs. Due to the random positioning of the impurities, the on-site energies of the electrons can be considered effectively random (or `disordered'). What Anderson showed was that there exists a critical concentration of impurities below which (or a critical amount of disorder above which) electrons remain localised about individual impurities and conductance does not occur. This is known as the Anderson transition \cite{evers2008anderson, Chulaevsky2014}. 

Anderson's model is one of the earliest examples of what would be considered by physicists to be a `disordered system'. Signifying his deep connection with the field, Anderson's name is also attached to another archetypal disordered system model, namely the Edwards-Anderson model of a spin glass \cite{edwards1975theory}. For his work, Anderson was awarded the Nobel prize jointly with Mott and Van Vleck (who was, incidentally, Anderson's graduate school supervisor) in 1977 \cite{lagendijk2009fifty}. We will show here how Anderson's model quickly reduces to a random matrix problem, and how our tools, with some some additional ingenuity, may be applied to provide us an understanding of the model's rich phenomenology. Indeed, the phenomenology of localisation is common to many other models aside from Anderson's original one, and is therefore of general interest to the RMT practitioner.

\subsection{Anderson tight-binding model on the random regular graph}\label{section:andersonmodel}
For the sake of completeness, we provide the quantum mechanical Hamiltonian of the Anderson model here. However, we emphasise that this set-up quickly yields a straightforward random matrix problem. The Hamiltonian of the tight-binding Anderson model describes a single electron that may hop between sites, which each have an associated local potential that may vary from site to site. It is similar to that of the classic Hubbard model \cite{arovas2022hubbard} of electron transport, but it has no electron repulsion and possesses an additional disordered on-site potential term. The Anderson Hamiltonian is given by
\begin{align}
	\mathcal{H} = - \sum_{\langle i,j\rangle} t (c_i^\dagger c_j + c_j^\dagger c_i) + \sum_{i=1}^N \epsilon_i c^\dagger_i c_i
\end{align}
where $c_i^\dagger$ and $c_i$ are the usual creation and annihilation operators \cite{griffiths2018introduction} of spinless fermions on site $i$ respectively, and $\sum_{\langle i,j\rangle}$ indicates a sum over all pairs of neighbouring sites. The two contributions to the Hamiltonian above (from left to right) should be interpreted as a hopping term, where $t$ is the hopping amplitude, and the remaining term encapsulates the on-site energies $\epsilon_i$, that vary from site to site. 

Since the Hamiltonian above is for a single, non-interacting fermion, diagonalising this Hamiltonian, and thus finding the energy eigenvalues/eigenstates, simply amounts to diagonalising a matrix with the following elements
\begin{align}
	H_{ij} = -tA_{ij}+ \epsilon_i \delta_{ij} , \label{hamiltoniananderson}
\end{align}
where here $A_{ij}$ is the adjacency matrix of the network of sites. That is, $A_{ij} = 1$ if the electron is able to hop between the two sites, and $A_{ij}= 0$ otherwise. As a generalisation, one could consider the case where $A_{ij}$ have non-trivial weights, but we only take the simple unweighted case here. Hence, we see that we arrive at a random matrix problem, which we have the tools to solve. 

An eigenvalue $E_\nu$ of the matrix $\underline{\underline{H}}$ represents the quantised energy level that an electron may occupy, and the eigenvector $\underline{v}^{(\nu)}$ represents the corresponding quantum state. Physically, the probability of observing the electron at site $i$ when it is in state $\nu$ is given by $ \vert v^{(\nu)}_i \vert^2$. Thus, if the eigenvector $\underline{v}^{(\nu)}$ is well-spread across many sites, the electron is said to be in a conducting or delocalised state, whereas if $\underline{v}^{(\nu)}$ has most of its weight on one site, the electron is said to be in an insulating or localised state.

The network of sites that we consider in this section, which is represented by the adjacency matrix $\underline{\underline{A}}$, is the random regular graph (RRG). This choice is again made for analytical tractability, and because it is a canonical case considered in the literature. A random regular graph is a network for which each node $i$ has exactly the same number of connections ($\sum_{j}A_{ij} = K+1$ for all $i$), with no self-connections ($A_{ii} = 0$), and each possible realisation of $\underline{\underline{A}}$ has the same probability of occurring. The RRG has the desirable property that it can be considered to be \textit{locally tree-like}. That is, for large $N$, the typical length of loops goes as $\ln N/\ln K$ \cite{wormald1999models}. This means that short cycles are rare, and so we can treat the graph as a tree for our purposes. This will turn out to be very useful in our analysis. We consider a number of other networks in Section \ref{section:networks}.

Let us take a moment to understand what we should expect to see when we perform the spectral analysis of the random matrix $\underline{\underline{H}}$. On one hand, if we were to set $t = 0$, then we would have a purely diagonal matrix. The eigenvalues would thus be the set $\{\epsilon_i\}$ and the eigenvectors would be $v_j^{(i)} = \delta_{ij}$. That is, the eigenvectors would be fully \textit{localised} on a single site, and the eigenvalues would each be independent random numbers, not experiencing any eigenvalue repulsion. On the other hand, if we were to set all the $\epsilon_i = 0$, then we would just have the adjacency matrix of a random graph. It is not obvious \textit{a priori} how the spectrum should be in this case (we will of course investigate below), but what we find is that the phenomenology is similar to the other symmetric random matrix ensembles that we have studied. That is, there is a single continuous finite support to the eigenvalue spectrum and the eigenvectors are \textit{extended} across all sites, roughly evenly. 

We will see that there is a non-trivial interplay between these two effects. In fact, in some circumstances, there is an upper limit to the degree of disorder, beyond which the extended states give way to a full set of localised eigenvectors, even when $t>0$. This is the transition to full localisation first identified by Anderson \cite{anderson1958absence}.

\subsection{Cavity equation closure for tree-like graphs}
We wish now to understand the eigenvalues and eigenvectors of $\underline{\underline{H}}$. As usual, we will proceed via the resolvent, which we now denote $\underline{\underline{G}}(E) = \left[ E\underline{\underline{\id}} - \underline{\underline{H}} \right]^{-1}$. We now use the argument $E$ since the eigenvalues of $\underline{\underline{H}}$ are the energy levels of the quantum Hamiltonian. 

We once again use the cavity approach [c.f. Eq.~(\ref{cavity})], and we thus obtain
\begin{align}
	G_{ii} = \frac{1}{E - \epsilon_i - t^2\sum_{j \neq i } A_{ij}^2  G_{jj}^{(i)}}= \frac{1}{E - \epsilon_i - t^2\sum_{j \in \partial_i }  G_{jj}^{(i)}}, \label{cavityanderson}
\end{align}
where we use $\partial_i$ to denote the set of nodes that are connected to node $i$. As a reminder, we refer to $G_{jj}^{(i)}$ as a `cavity resolvent' entry. Unlike the previous cases that we have studied, the sum over $j$ in the denominator of Eq.~(\ref{cavityanderson}) is not large, and therefore does not concentrate. Further, the diagonal elements $\epsilon_i$ are heterogeneous in a non-trivial way. We therefore cannot use our usual approach to obtain a closed equation for $G = N^{-1}\mathrm{Tr} \underline{\underline{G}}$. 

We require another way to close the equations for $G_{ii}$. To this end, we exploit the locally tree-like nature of the random regular graph. The comparative simplicity of disordered-systems problems on trees was noticed by Abou-Chacra, Anderson and Thouless in their seminal work \cite{abou1973selfconsistent}. The message-passing-related idea that we will explore below has been applied in the study of spin glasses \cite{mezard1987spin, mezard2001bethe}, random matrix theory \cite{kuhn2008spectra,rogers2008cavity, metz2019spectral}, and in belief-propagation algorithms in machine learning and other areas \cite{yedidia2003understanding}.

\begin{figure}[h]
	\centering 
	\includegraphics[scale = 0.3]{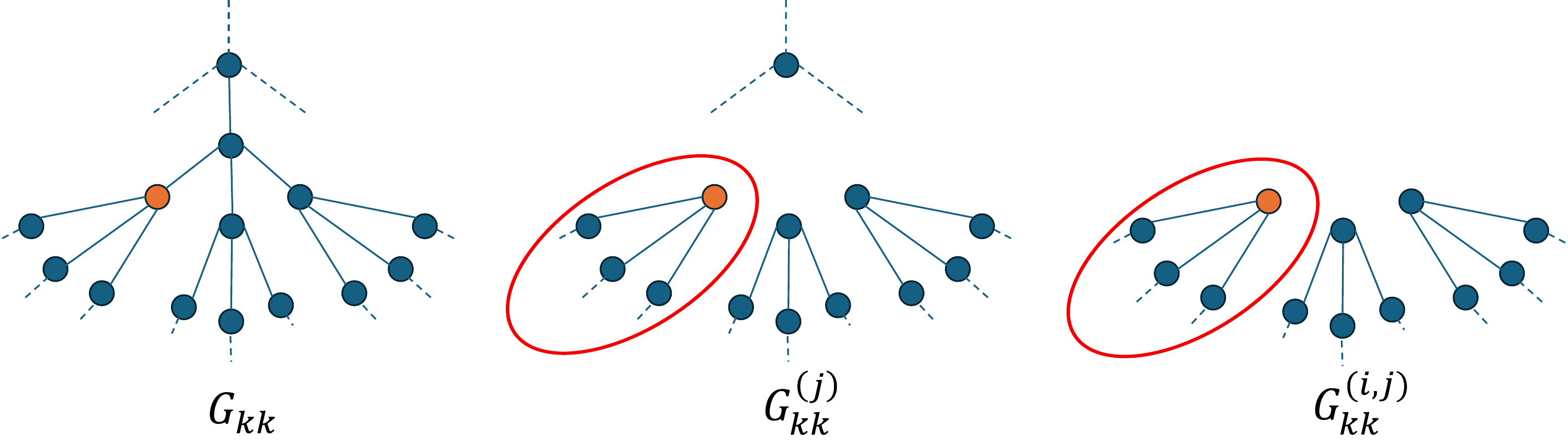} 
	\captionsetup{justification=raggedright,singlelinecheck=false}
	\caption{Illustration of the cavity factorisation on a tree-like graph. The highlighted node is labelled $k$. When we remove the node $j$ (of which $k$ is a neighbour), the graph factorises. This is only true because there are no loops. The red oval indicates the connected subgraph to which $k$ belongs. If we now remove another neighbour of $j$, this does not affect the subgraph to which $k$ belongs. Since these subgraphs correspond to separate diagonal blocks of the matrix $\underline{\underline{H}}^{(j)}$, we see that when we invert $E\underline{\underline{\id}}  - \underline{\underline{H}}^{(j)}$ to obtain $\underline{\underline{G}}^{(j)}$, it makes no difference to the value of $G_{kk}^{(j)}$ if we also remove the node $i$. Hence, $G_{kk}^{(i,j)} = G_{kk}^{(j)}$. }\label{fig:cavity}
\end{figure}

Let us iterate the block-matrix procedure once more to find an expression for $G_{jj}^{(i)}$. We obtain
\begin{align}
	G_{jj}^{(i)} = \frac{1}{E-\epsilon_j - t^2\sum_{k\in\partial_j^{/i}}  G_{kk}^{(i,j)}},
\end{align}
where $\partial_j^{/i}$ is the set of nodes neighbouring $j$, excluding $i$, and $\underline{\underline{G}}^{(i,j)}$ is the resolvent matrix corresponding to the matrix $\underline{\underline{H}}$ with the rows and columns $i$ and $j$ removed. 

The crucial observation to make is the following \cite{susca2021cavity, kuhn2008spectra, rogers2008cavity}: the cavity resolvent at site $k$ with $j$ removed is the same as the resolvent at $k$ with both $i$ and $j$ removed, i.e., $G_{kk}^{(j)} = G_{kk}^{(i,j)}$. The reason for this is that, with the node $i$ removed, it is possible to rearrange the matrix $\underline{\underline{H}}^{(i)}$ into a block form in which the off-diagonal blocks are all zero. Since the inverse of such a matrix just involves taking the inverse of each block individually, removing a row and column from one block does not affect the inverse of any other. See Fig. \ref{fig:cavity} for a graphical representation of this principle.

Hence, we may write the cavity resolvent entries in terms of the other cavity resolvent entries
\begin{align}
	G_{jj}^{(i)} = \frac{1}{E-\epsilon_j - t^2\sum_{k\in\partial_j^{/i}}  G_{kk}^{(j)}}, \label{cavityanderson2}
\end{align}
which now constitute a closed set of equations. These equations were first derived using the replica method, which also allows one to understand the factorisation of the graph \cite{kuhn2008spectra, rogers2008cavity}. They can be solved iteratively in principle \cite{rogers2008cavity}, or by using the population dynamics algorithm [see e.g. Refs. \cite{kuhn2008spectra, susca2021cavity} or Section \ref{section:popdyn}]. Once the cavity resolvents have been found, these then yield the true resolvent diagonal entries via Eq.~(\ref{cavityanderson}). Our strategy to solve the equations will be to consider the \textit{statistical} ensemble of the cavity resolvent elements. This will allow us, ultimately, to obtain approximate (but closed-form) expressions for the spectral observables in which we are interested.

\subsection{Kesten-McKay law in the case of zero on-site disorder}
Let us first take the simple example where the on-site disorder is zero, i.e. everywhere on the graph $\epsilon_i = 0$. In this case, we are able to obtain a closed-form solution for the eigenvalue density from the cavity equations, due to the symmetry between the sites. 

Ignoring edge effects or loops (which is valid for the large locally tree-like graph), we expect the cavity resolvent (and the true resolvent) elements to be the same on all sites, since all sites are identical, for all intents and purposes. Using this, we may write using Eq.~(\ref{cavityanderson2})
\begin{align}
	G_{jj}^{(i)} = G_\mathrm{cav} = \frac{1}{E - Kt^2 G_\mathrm{cav}},
\end{align}
which we may solve easily to obtain
\begin{align}
	G_\mathrm{cav} = \frac{E \pm \sqrt{4Kt^2 - E^2}}{2 Kt^2}.
\end{align}
Now, examining Eq.~(\ref{cavityanderson}), we have
\begin{align}
	G_{ii} \equiv G = \frac{1}{E - (K+1)t^2 G_\mathrm{cav}} = \frac{1}{2} \frac{(K-1) E \pm (1+K) \sqrt{E^2- 4 K t^2}}{(K+1)^2 t^2 - E^2} .
\end{align}
Hence, using the usual inverse Stieltjes transform in Eq.~(\ref{densefromres}), we obtain the Kesten-McKay law \cite{kesten1958symmetric,mckay1981expected}
\begin{empheq}[box={\fboxsep=6pt\fbox}]{align}
	\rho(E) = \frac{1}{2\pi} \frac{1+K}{(1+K)^2t^2 - E^2} \sqrt{4 K t^2 - E^2}.
\end{empheq}
If we take $t = 1/\sqrt{K}$, and we take the limit $K \to \infty$, one sees that we recover the Wigner semicircle of Eq.~(\ref{semicircle}).

\begin{figure}[h]
	\centering 
	\includegraphics[scale = 0.48]{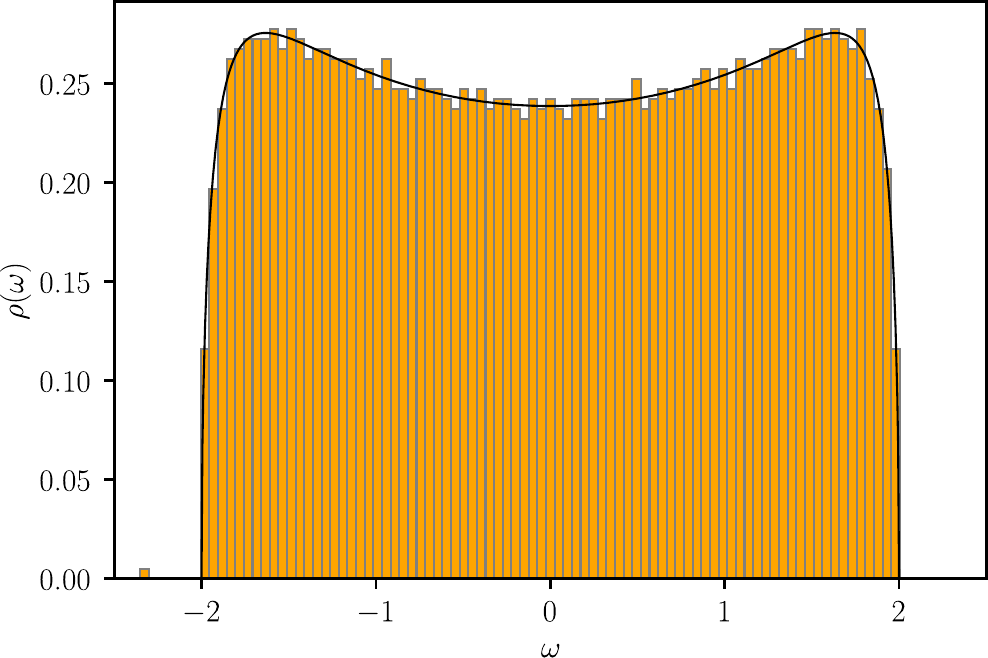} 
	\includegraphics[scale = 0.48]{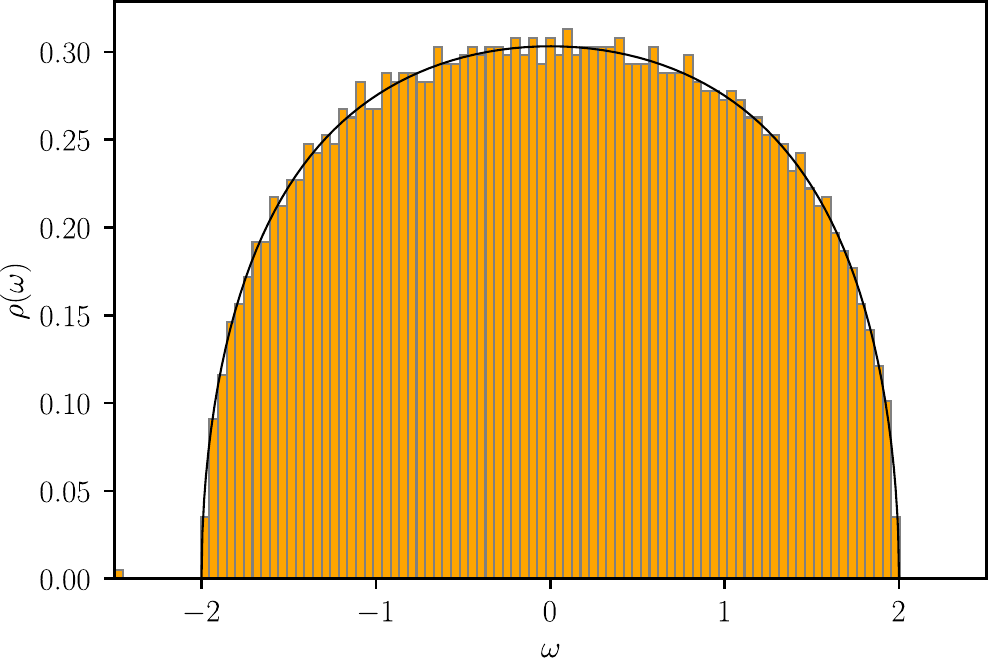} 
	\captionsetup{justification=raggedright,singlelinecheck=false}
	\caption{Kesten-McKay law for $K = 3$ (left) and $K=20$ (right) with $t = 1/\sqrt{K}$ and $N = 4000$. Results were averaged over one trial. We see that the Kesten-McKay law converges to the Wigner semicircle as the connectivity $K$ is increased. }\label{fig:kestenmckay}
\end{figure}

\subsection{Eigenvalue spectral density for non-zero disorder}\label{section:pertexpand}
Now, let us introduce the on-site disorder. For certain distributions of the disorder, one may obtain fully closed-form solutions for the spectral density. The canonical choice is to use a uniform distribution for the on-site disorder \cite{anderson1958absence,abou1973selfconsistent, biroli1999single, pietracaprina2016forward}. Indeed, one can close the equations in this case easily if one uses a large-$K$ approximation \cite{abou1973selfconsistent}. What is particularly nice about this example is that there exists a finite limit for the breadth of the uniform distribution, beyond which all eigenstates become `localised'. Before this point, there is a finite range of energies for which states are extended. We return to this example below in Section \ref{section:uniform}. 

For the sake of choosing a simple example that contains most of the interesting possible phenomenology, we mostly focus here on the Weibull distribution \cite{weibull1939statistical}
\begin{align}
	P(\epsilon) = \frac{a}{2 W} \left(\frac{\vert \epsilon\vert}{W} \right)^{a-1} \exp\left[- \left(\frac{\vert \epsilon\vert}{W} \right)^{a} \right]. \label{weibull}
\end{align} 
This is a nice example for us, because the tail of this distribution is sufficiently fat that we may hope to have a measurable number of localised states in numerics, even for small $W$. This allows us to make some simple approximations, which will allow us to understand the physics a bit more directly. We also discuss the canonical case of the uniform distribution below. 

For simplicity, let us consider the case where $W$ is small, so that the change in the spectrum in comparison to the case with zero disorder is small. We also consider the case where $t = 1/\sqrt{K}$ and $K$ is sufficiently large that we may use the semicircle law for the eigenvalue density in the $W\to 0$ limit. We now seek to obtain the lowest order correction to the semicircle law.

We may proceed by expanding the cavity equations Eqs.~(\ref{cavityanderson}) and (\ref{cavityanderson2}) perturbatively in $\epsilon_i$, which is typically small. One obtains
\begin{align}
	G_{ii} \approx \frac{1}{E -  G} + \frac{\epsilon_i}{(E -  G)^2} + \frac{\epsilon_i^2}{(E -  G)^3} ,
\end{align}
where we use the fact that $G \approx G_\mathrm{cav}$ for large $K$, as we found above. Averaging over sites $i$, and denoting $\langle \epsilon^2 \rangle = \alpha$ (which is proportional to $W^2$), we obtain
\begin{align}
	G (E-G) \approx 1  + \alpha G^2 + O(\alpha^2), \label{gapprox}
\end{align}
where we have used the fact that we can perform a replacement of $(E -  G)^{-1}$ with $G$ in the second term on the right-hand side, and we only contribute to the $O(\alpha^2)$ error. We have thus succeeded in finding an approximation for the resolvent that is accurate to first order in $\alpha$.

Finally, we may solve the above quadratic in $G$ to obtain (using a complex argument $z = E- i\eta$)
\begin{align}
	G(z) \approx \frac{z - \mathrm{sign}[\mathrm{Re}(z)] \sqrt{z^2 - 4 (1+\alpha)}}{2(1+\alpha)}
\end{align}
Using the inverse Stieltjes transform in Eq.~(\ref{densefromres}) as usual, we obtain the eigenvalue density
\begin{align}
	\rho(E) \approx \frac{2}{\pi E_c^2} \sqrt{E^2_c -E^2}, \hspace{1cm} E_c^2 \approx 4(1+\alpha), \label{bulkdensity}
\end{align}
which takes into account the slight deformation of the usual semicircle law due to non-zero on-site disorder. This is tested in Fig. \ref{fig:densitylog}. This calculation is similar in spirit to that carried out by Kim and Harris for sparse graphs \cite{kim1985density} where a $1/K$ expansion was obtained (see also Refs. \cite{rodgers1988density,baron2023pathintegralapproachsparse}). We could of course continue this treatment to higher orders in $\alpha$. In principle, we could also take into account the effects of finite $K$, which is explored in the exercises. 

We thus see that non-zero disorder acts to broaden the semicircle-like bulk of the eigenvalue spectrum. We note however that we have, in a sense, only taken into account here the effect of `typical' fluctuations in the diagonal disorder. We must now understand how atypical fluctuations (i.e. $\epsilon_i$ that come from the tails of the Weibull distribution) lead to a different contribution to the eigenvalue spectrum that is not captured by this perturbative approach.

\subsection{Point-like spectrum in the localised phase: single-defect approximation}\label{section:SDA}
Let us consider the case now where $W$ is still small, but a particular $\epsilon_i$ is atypically large. In this case, it becomes possible for the local Green's function $G_{ii}$ to have a pole, which then gives rise to a Dirac delta contribution in the eigenvalue spectrum. That is, in the case $t = 1/\sqrt{K}$ with large $K$, we have
\begin{align}
	G_{ii}(E) \approx \frac{1}{E - \epsilon_i - G}, \label{localres}
\end{align}
where we take $G(E)$ to satisfy Eq.~(\ref{gapprox}). That is, we imagine that the neighbouring sites of $i$ do not have similarly atypical values of $\epsilon_j$. This is reasonable, since it would be extremely unlikely to have two rare events in the vicinity of each other. We refer to this approximation as the \textit{single-defect approximation} \cite{biroli1999single, semerjian2002sparse, valigi2025eigenvalue}. It was first developed to handle defects in network topology, which we will also address in Section \ref{section:networks}. 

We therefore see that there could be a pole of $G_{ii}(E)$ if there is a value $E= E_i$ that satisfies
\begin{align}
	E_i = \epsilon_i + G(E_i) .
\end{align}
Using the expression for $G(E)$ in Eq.~(\ref{gapprox}), we may thus obtain
\begin{align}
	E_i \approx \epsilon_i + \frac{1}{\epsilon_i} + \frac{\alpha}{\epsilon_i^3} . \label{outlierloc}
\end{align}
However, as we have found previously for poles in the resolvent that come about due to other kinds of perturbation (see e.g. Section \ref{section:outlier}), since the function $G(E)$ only takes a bounded range of real values, the above solution for $E_i$ is only valid for a certain range of $\epsilon$. The maximum value of $G(E)$ occurs at $E = 2(1+\alpha/2)$ [i.e. on the edge of the continuous spectrum described by Eq.~(\ref{bulkdensity})], where it attains a value of $\approx 1-\alpha/2$. One thus obtains that $\epsilon_i \gtrapprox 1+3\alpha/2$, and similar statements can be made for the minimum value of $G(E)$ and negative outlier eigenvalues. 

\begin{figure}[H]
	\centering 
	\includegraphics[scale = 0.6]{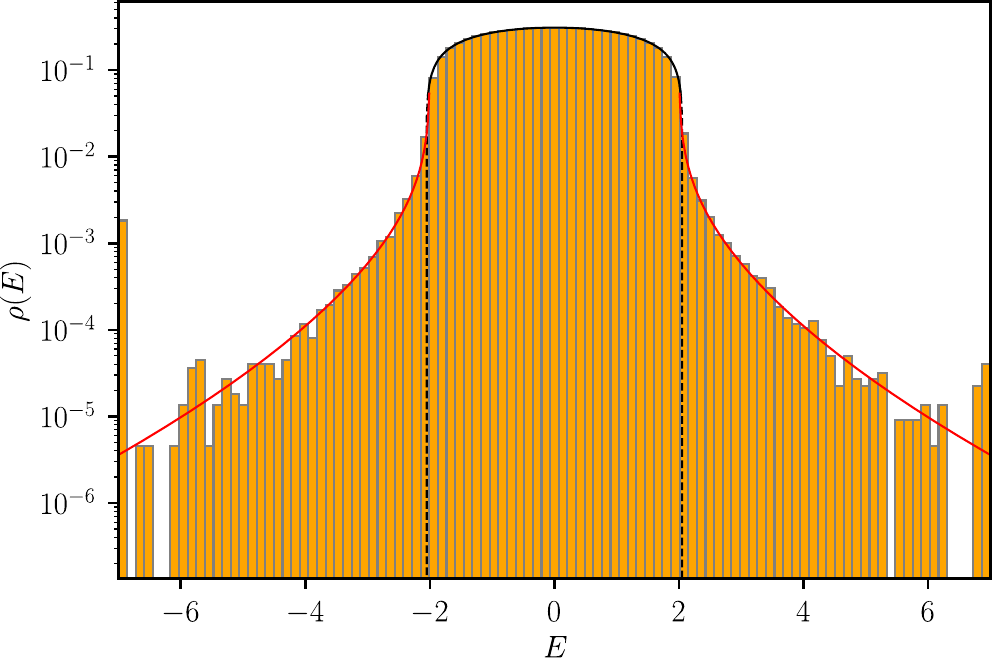} 
	\captionsetup{justification=raggedright,singlelinecheck=false}
	\caption{The bulk and tail regions of the eigenvalue spectrum. Here, the prediction in Eq.~(\ref{taildensity}) is indicated with a red line, while that of Eq.~(\ref{bulkdensity}) is indicated in black. The system parameters are $K = 30$, $N = 4000$, $t = 1/\sqrt{K}$, and the Weibull distribution in Eq.~(\ref{weibull}) was used with $W = 0.05$ and $a = 0.5$. Data was averaged over $100$ trials. }\label{fig:densitylog}
\end{figure}

This means that outside the range $\vert E\vert <E_c$, it is possible to have outlier eigenvalues at locations given by Eq.~(\ref{outlierloc}). Since the location of an outlier is uniquely determined by its corresponding value of $\epsilon$, we may write for the eigenvalue density in the region outside the semicircle-like bulk [ignoring the $O(\alpha)$ correction here]
\begin{align}
	\rho_\mathrm{tail}(E) &= \int_{-\infty}^\infty d\epsilon \, \delta(E- \epsilon - 1/\epsilon) P(\epsilon) = \frac{P(\epsilon(E))}{1- \epsilon(E)^{-2}}, \nonumber \\
	\epsilon(E) &= \frac{E - \mathrm{sign}(E)\sqrt{E^2-4}}{2}.\label{taildensity}
\end{align}
Spectral tails of outlier eigenvalues such as these (whose eigenvectors we are localised, as we shall soon see) are often referred to as Lifshitz tails \cite{lifshitz1965energy,kuhn2008spectra, khorunzhiy2006lifshitz, bapst2011lifshitz}. We compare the predictions of Eqs.~(\ref{bulkdensity}) and (\ref{taildensity}) to numerics in Fig. \ref{fig:densitylog}. 

\subsection{Exponential localisation and inverse participation ratio (IPR)}\label{section:localisation}
As we have also seen previously, outlier eigenvalues often have a signature of their atypicality in their associated eigenvectors (for example, the overlaps discussed in Section \ref{section:marchenkopastur}). In the present case, we expect the eigenvectors associated with eigenvalues outside of the semicircle-like bulk to be `localised' on the node that bears an atypically large on-site energy $\epsilon_i$, as we discussed above in Section \ref{section:andersonmodel}. Let us see how we may understand the structure of such eigenvectors in more detail. 

To this end, we define the so-called Inverse Participation Ratio (IPR), which indicates the degree to which eigenvectors are localised. For a particular eigenstate (eigenvector) $\nu$, it is defined as
\begin{align}
	q^{-1}_\nu = \sum_i \vert v_i^{(\nu)} \vert^4.
\end{align}
We note that since we normalise the eigenvectors according to $ \sum_i \vert v_i^{(\nu)} \vert^2 = 1$, those vectors that have components that are (roughly) evenly distributed across the $N$ sites have $q^{-1}_\nu \sim N^{-1}$, whereas those vectors that have their weight mostly localised over $\sim N^0$ sites have $q^{-1} \sim N^0$. A non-zero value of $q^{-1}$ for $N \to \infty$ is therefore a good indicator of localisation. 

\begin{figure}[h]
	\centering 
	\includegraphics[scale = 0.6]{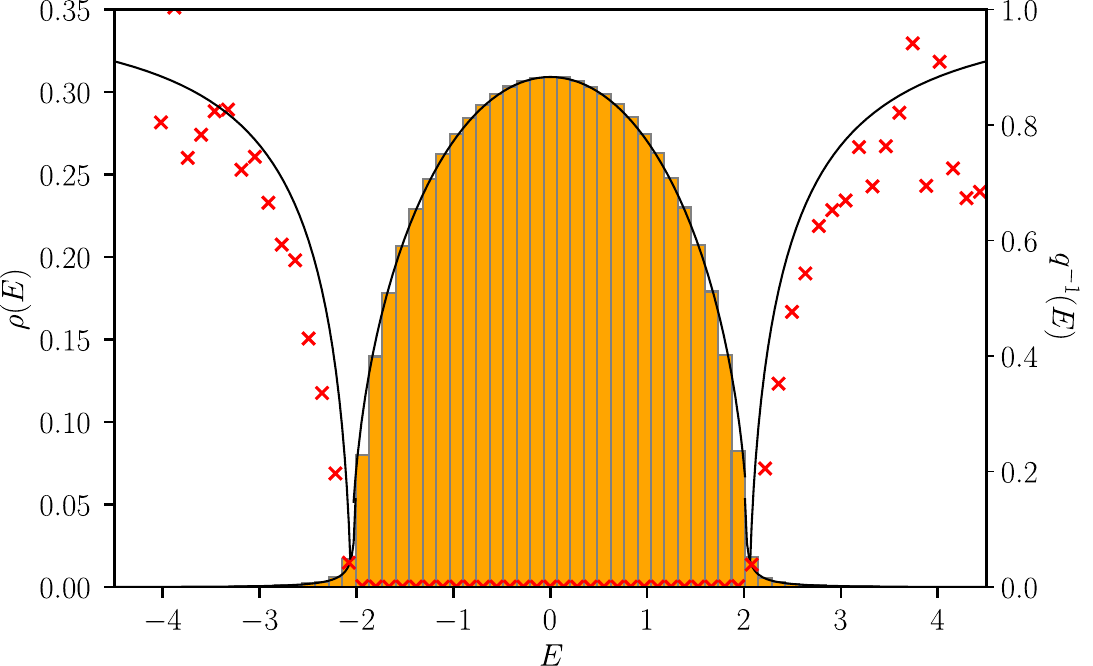} 
	\captionsetup{justification=raggedright,singlelinecheck=false}
	\caption{The eigenvalue density (numerical results represented by bars) and the IPR (red crosses for numerics). We see that in the deformed semicircular region, the IPR effectively vanishes, while in the tail regions it grows to non-zero values. The system parameters are $K = 30$, $N = 4000$, $t = 1/\sqrt{K}$, and the Weibull distribution in Eq.~(\ref{weibull}) was used with $W = 0.05$ and $a = 0.5$. Data was averaged over $100$ trials. }\label{fig:densityandIPR}
\end{figure}

To compute $q^{-1}$, we once again exploit the relation between the resolvent and the eigenvector entries Eq.~(\ref{resvectors}): $G_{ij}(z) = \sum_{\nu} \frac{v_i^{(\nu)} v_j^{(\nu)}}{z-E_\nu}$. This allows us to see that we may extract the squared eigenvector components via \begin{align}
	G_{ii}(E) \approx \vert v_i^{(\nu)}\vert^2/(E-E_\nu), \label{eigenvectorresidue}
\end{align}
as long as $E \approx E_\nu$. We therefore see that to obtain the eigenvector component magnitude corresponding to an anomalously high $\epsilon$, we need to find the residue of Eq.~(\ref{localres}) at the eigenvalue in question. 

Supposing that we have a large on-site energy at a particular site $0$, we wish to compute the corresponding localised eigenvector, which for now we label $\underline{v}^{(0)}$. By simple differentiation of Eq.~(\ref{localres}), we may compute the reside of the local resolvent at this site, and we thus have that
\begin{align}
	\vert v_i^{(0)}\vert^2 \approx \frac{1}{1- G'},
\end{align}
where $G' = dG(E)/dE$. Now, in order to evaluate $q^{-1}$, we require the components of this eigenvector on other sites. To find these, we exploit the following property of the cavity equations Eqs.~(\ref{cavityanderson}) and (\ref{cavityanderson2}).
\begin{tcolorbox}[colback=blue!10!white,colframe=blue!90!black,title=Lemma: Propagation of the resolvent along the tree]
	We may write the local resolvents on a neighbouring nodes in terms of their respective cavity resolvent entries 
	\begin{align}
		G_{jj} = \frac{G_{jj}^{(i)}}{G_{ii}^{(j)}} G_{ii}. 
	\end{align}
	One can see this by writing $(G_{ii})^{-1} = (G_{ii}^{(j)})^{-1} - t^2 G_{jj}^{(i)}$ and $(G_{jj})^{-1} = (G_{jj}^{(i)})^{-1} - t^2 G_{ii}^{(j)}$ and rearranging.
\end{tcolorbox} 
Keeping $K$ finite, we see that if $G_{ii}$ has a pole at $E = E_0$, then in the vicinity of this pole, $G_{ii}^{(j)} \approx (t^2 G_{jj}^{(i)})^{-1}$. Consequently, we have 
\begin{align}
	G_{jj}(E) \approx t^2 [G(E_0)]^2 G_{ii}(E)
\end{align}
for $E \approx E_0$ and where $G(E)$ is the expression for the resolvent without the inclusion of rare poles in Eq.~(\ref{gapprox}). We note that this deduction is only reasonable when there are no sites of similar on-site energy in the vicinity of one another.

We thus see that since $G_{ii}$ has a pole at energy $E_0$, then so do the neighbouring $G_{jj}$. Thus, by again using Eq.~(\ref{eigenvectorresidue}), we obtain
\begin{align}
	\vert v_j^{(0)}\vert^2 \approx t^2 [G(E_i)]^2 \vert v_i^{(0)}\vert^2 . \label{vectorrecursion}
\end{align}
Thus, the eigenvector $\underline{v}^{(0)}$ has a maximum at site $0$, and also has non-zero values on subsequent nodes on the tree-like graph, with the magnitude being modified by the same factor $ t^2 [G(E_0)]^2 $ on every step away from the node $0$. We thus see that we have an exponential localisation of the eigenvectors about the focal node $0$. 

We notice at this point that we have not had to use the on-site energy $\epsilon_0$ at any point in this argument -- the properties of the eigenvector can be written entirely in terms of the eigenvalue $E_0$ and the resolvent $G(E_0)$. Using the recursion in Eq.~(\ref{vectorrecursion}), we may thus evaluate the IPR for a general energy $E$ as
\begin{align}
	q^{-1}(E) &\approx \frac{1}{(1-G'(E))^{2}} \left[1+ K t^4 [G(E)]^4+ K^2 t^8 [G(E)]^8 +\cdots \right] \nonumber \\
	&=  \frac{1}{(1-G'(E))^{2}} \frac{1}{1 - K t^4 [G(E)]^4}, 
\end{align}
where we have used that there are a factor $K$ more nodes on each generation of the tree moving further away from the focal node, and we have evaluated the resulting geometric series. This expression is checked against numerics in Fig. \ref{fig:densityandIPR}. 

One observes that we only obtain a non-zero value of the eigenvector entries, and therefore $q^{-1}$, when there is a simple pole of the resolvent. Since, in our approximation scheme at least, the resolvent in semicircle-like bulk region does not have simple pole-like non-analyticities, the eigenvalues in the bulk region cannot possess localised eigenvectors. Only those eigenvalues that fall outside the continuous bulk region have localised eigenvectors. Hence, the edge of the bulk semicircle-like region corresponds to the so-called \textit{mobility edge}, dividing localised and extended states. This is seen in the numerics in Fig. \ref{fig:densityandIPR}.

\subsection{Level statistics: Poisson to Wigner-Dyson}
Another indicator of the transition from extended to localised states is the statistics of the eigenvalues. The reason for this is as follows. We have seen that large on-site energies give rise to eigenvalues at values $E_i$ that are determined entirely by $\epsilon_i$, within our approximation schemes. Since the value $\epsilon_i$ is an i.i.d. random variable, we expect the values $E_i$ also to be i.i.d., as long as $E_i$ is large. That is, the distances between large eigenvalues ought to obey Poisson statistics.

On the contrary, more typically small values of $\epsilon_i$ give rise to extended states within the continuous part of the eigenvalue spectrum, where we expect the usual Wigner-Dyson statistics to hold. That is, we expect the eigenvalue separations to be described by the Wigner surmise for energies $\vert E \vert \lessapprox 2$.

\begin{figure}[h]
	\centering 
	\includegraphics[scale = 0.6]{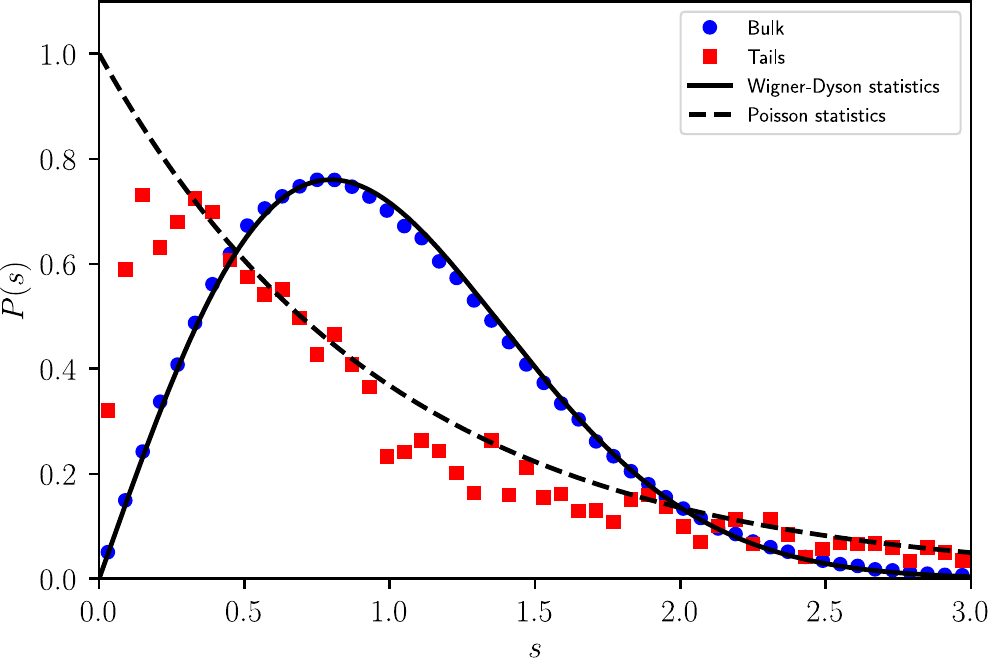} 
	\captionsetup{justification=raggedright,singlelinecheck=false}
	\caption{Crossover from Wigner-Dyson statistics (in the continuous bulk part of the spectrum) to Poisson statistics (in the Lifshitz tail). System parameters: $N = 10^4$, $K = 50$, $t=1/\sqrt{K}$, with on-site disorder drawn from the Weibull distribution in Eq.~(\ref{weibull}) with $a = 0.5$ and $W = 0.05$. The solid line is the $\beta =1$ Wigner surmise in Eq.~(\ref{surmises}), while the dashed line the curve $P(s) = e^{-s}$. The blue circles are results using eigenvalue with $\vert E\vert<1.9$ and the red squares are for $\vert E \vert >2.1$. }\label{fig:levelstats}
\end{figure}

Indeed, we see in Fig. \ref{fig:levelstats} that this is the case. In short, in the localised region of the spectrum (the tails), eigenvalues are independent events, and thus the probability of a particular eigenvalue separation is simply exponential. However, in the bulk region, eigenvalues experience repulsion (as usual), and we obtain GOE Wigner-Dyson statistics obeying the appropriate Wigner surmise.

\subsection{ Abou-Chacra--Anderson--Thouless criterion}\label{section:uniform}
We would be remiss if we did not discuss the content of Anderson's original work \cite{anderson1958absence}, which was later simplified in the subsequent work together with Abou-Chacra and Thouless \cite{abou1973selfconsistent} by using tree-like graphs. In these papers, a uniform distribution for $\epsilon$ is used on the interval $[-W/2,W/2]$, which is a common choice of disorder distribution due to tractability. What was particularly focussed on in these works was the value of the disorder above which no extended states could be found at all (or below which extended states emerged), since this would indicate the point of vanishing/emergence of the impurity conduction band. The fully localised phase occurs for large $W$, which we have not been able to access with the perturbative approach considered above.

To find the transition point, one assumes that we are in the fully localised phase, and thus that the eigenvectors (states) decay with distance from a localisation centre, as we have seen in our considerations above. In fact, the statement is made stronger than this: \textit{all} eigenvectors in the truly localised phase must have components at a distance $r$ away on the lattice that satisfy $[\ln \vert v_i^{(\nu)} \vert^2]/(2r) < -1/(2 \xi)$ for some $\xi$, which we call the localisation length \cite{pietracaprina2016forward}, in order to qualify as localised. This then gives us a stringent criterion for localisation. When we cannot find such an $\xi$, extended states emerge. 

The problem is approached using the so-called forward approximation \cite{pietracaprina2016forward} as follows. If we neglect the hopping term  proportional to $t$ in Eq.~(\ref{hamiltoniananderson}), then the eigenvector with eigenvalue $\epsilon_0$ would have component at $j$ equal to simply $v_j^{(0)} = \delta_{0j}$. Using perturbation theory and truncating at leading orders of $t$, one arrives at the following approximate expression for the magnitude of the eigenvector component at a site that is a distance $L$ away on the tree
\begin{align}
	v_L^{(0)} \approx \prod_{j=1}^L \frac{t}{\epsilon_0 - \epsilon_j}, \label{forwardapprox}
\end{align}
where here the sites $j=1, \cdots, L$ all fall on one particular continuous path along the tree. This can also be derived more carefully using a path expansion \cite{pietracaprina2016forward}. From this, we now wish to derive the value of $W$ at which we observe onset of extended states, according to the aforementioned definition. We choose to focus on eigenstates with $\epsilon_0 = 0$, since this is the point in the spectrum at which extended states will first occur. The procedure is this: (1) We find the distribution $P(x_L)$ of the quantity $x_L = \ln \vert v_L^{(0)}\vert^2$ for a given value of $L$. (2) At a given distance $L$ from the focal node $0$, there are $K^L$ nodes. We consider the set of all values of $x_L$ on this set of nodes, and we ask what is the expected value $x_L^\star$ of the \textit{largest} among them. The answer is estimated by $K^L P(x_L^\star) \sim 1$. (3) We then require that this largest possible value of $x_L^\star$ is less than $L$ if the state is to be localised (as discussed above). When this latter requirement cannot be satisfied, we must have the onset of extended states.

From this analysis, one obtains the following for the critical disorder in the case of tree-like graphs with uniformly distributed on-site disorder
\begin{align}
	W_c = 2 t e K \ln\left( \frac{W_c}{2t}\right) . \label{upperbound}
\end{align}
We enter into the details of this calculation in greater detail in the exercises below. Indeed, the existence of such an upper bound has been confirmed with a semi-analytical treatment of the cavity equations using population dynamics techniques \cite{biroli2010anderson}.

\subsection{Variations on the model and localisation in context}
It should be emphasised that the example that we focussed on mostly here, involving the Weibull on-site energy distribution and a well-connected RRG graph, is a somewhat contrived one. This was done for the sake of providing a simplified illustration of the phenomenology, which would fit in with the other topics of these notes. It departs from the models that are commonly considered in the literature in two important aspects. 

Firstly, we used a very carefully curated distribution for the on-site energies, which enabled us to use perturbative treatments and the single-defect approximation. As we have mentioned, the canonical distribution of the on-site energies is often taken to be a `box' distribution (uniform on the interval $[-W/2,W/2]$), and another important case is the Cauchy distribution, for which the distribution of the Green's functions is exactly solvable \cite{abou1973selfconsistent}. Importantly, our approximation schemes did not give us access to the large disorder regime, where extended states can be entirely eliminated. However, we were able to understand much of the phenomenology of both phases, including the level statistics, the structure of the spectrum, and exponential localisation.

Secondly, we considered the case of a tree-like graph, mostly in the high-connectivity limit. While this is certainly a very useful example for making analytical progress in a simple way, there has been a lot of success in understanding the model on sparse networks \cite{biroli2010anderson}, in power-law random banded matrices \cite{mirlin1996transition}, and on hyper-cubic lattices \cite{pietracaprina2016forward}, and field theoretically using the non-linear sigma model \cite{zirnbauer1986anderson}. Indeed, the critical behaviour has been well understood in many cases \cite{evers2008anderson}, with the eigenstates at criticality being shown to be multifractal \cite{mirlin2000multifractality}. In particular, it has been demonstrated that localisation always occurs in 1D chains for minimal disorder \cite{anderson1958absence,mott1961theory}, and 2D lattices present a kind of marginal case \cite{abrahams1979scaling}, beyond which \textit{bona fide} extended states are possible \cite{slevin2014critical}. 

Since the time of Anderson's original work, the phenomenon of localisation has also been successfully realised in experiments. This has not been without difficulty however, since the metal-insulator transition of the Anderson type is difficult to distinguish experimentally from the Mott variety, since the electron screening can be affected by the level of doping \cite{poduval2023anderson}. Only surprisingly recently has a disorder-induced metal-insulator transition of the Anderson variety been clearly identified in so-called `phase change materials' with special thermal and electronic properties \cite{siegrist2011disorder,ying2016anderson,bragaglia2016metal}. However, there are many other systems apart from the classic doped semiconductors that exhibit characteristic Anderson localisation phenomena \cite{lagendijk2009fifty}, in particular, disordered optical systems \cite{segev2013anderson} and ultra-cold atoms \cite{white2020observation}.

In recent times, one of the great extensions of Anderson's original simple non-interacting model has been the inclusion of interactions, whereupon it has been found that localisation of the many-body wave-functions (in Fock space \cite{roy2020fock,tarzia2020many}) may occur \cite{basko2006metal,altshuler1997quasiparticle}. The possibility of the inclusion of interactions was actually speculated upon by Anderson in his original work \cite{anderson1958absence}. There continues to be much ongoing progress and success towards demonstrating many-body localisation from the theoretical standpoint \cite{parameswaran2018many}, in simulations \cite{smith2016many} (the XXZ spin chain model is a paradigmatic example \cite{imbrie2016many,imbrie2016diagonalization}) and even in experiment \cite{choi2016exploring,bordia2016coupling,schreiber2015observation}. 

\subsection{Exercises}
\begin{figure}[h]
	\centering 
	\includegraphics[scale = 0.5]{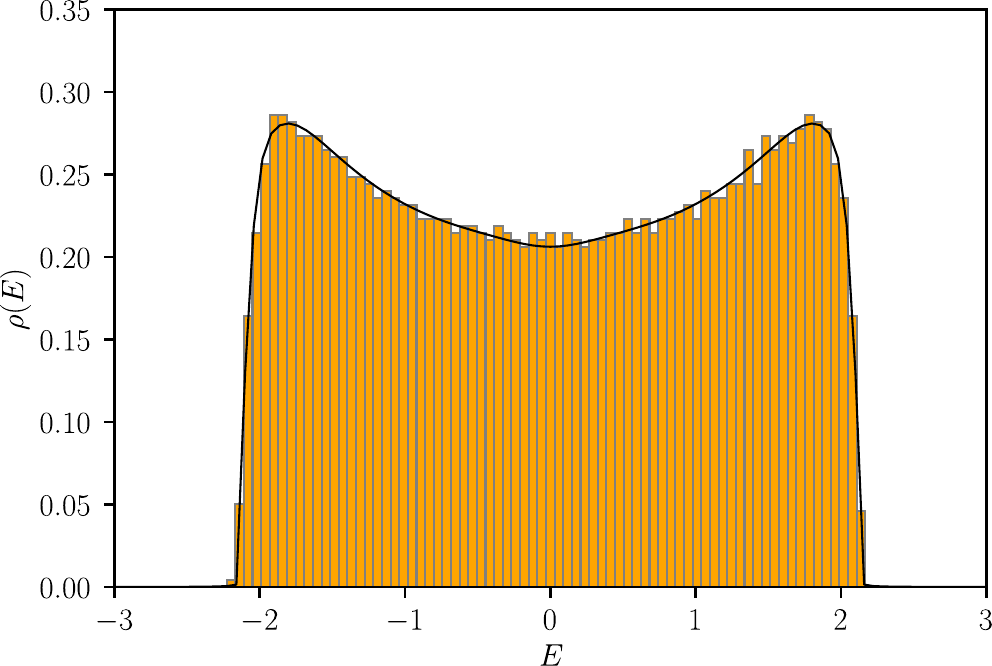} 
	\captionsetup{justification=raggedright,singlelinecheck=false}
	\caption{Eigenvalue density for $N = 4000$, $K = 2$, $t= 1/\sqrt{K}$, and $W = 1$ in the case of uniform disorder. }\label{fig:exercise}
\end{figure}

In the above, we derived a modified semicircle law, where the deformation was due to a small variance in the diagonal elements $\langle \epsilon_i^2 \rangle = \alpha$. We also derived the Kesten-McKay law for arbitrary $K$ and zero disorder. Let us now consider how a perturbative approach might be taken to account for the finite-$\alpha$ corrections to the Kesten-McKay law.
\begin{itemize}
	\item Beginning with Eq.~(\ref{cavityanderson2}), we now see that we cannot assume that the sum in the denominator concentrates. By writing $G_\mathrm{cav}(E) = \langle G_{jj}^{(i)} \rangle$ and $\delta G_{jj}^{(i)} =  G_{jj}^{(i)} - G_\mathrm{cav}$, show that to leading order in $v$ we have
	\begin{align}
		G_{jj}^{(i)} \approx \frac{1}{E - K t^2 G_\mathrm{cav}} + \frac{\epsilon_j +t^2 \sum_{k\in \partial_j^{/i}}\delta G_{kk}^{(j)}}{(E - K t^2 G_\mathrm{cav})^2} +\frac{\left(\epsilon_j + t^2 \sum_{k\in \partial_j^{/i}}\delta G_{kk}^{(j)}\right)^2}{(E - K t^2 G_\mathrm{cav})^3} .
	\end{align}
	\item Hence, show that 
	\begin{align}
		\langle [\delta G_{jj}^{(i)}]^2 \rangle \approx G_\mathrm{cav}^4\frac{\alpha}{1-t^4 K G_\mathrm{cav}^4} .
	\end{align}
	\item With this in mind, derive the following simultaneous expressions for the average cavity resolvent and the average true resolvent elements
	\begin{align}
		G_\mathrm{cav} &\approx \frac{1}{E - t^2 K G_\mathrm{cav}} + \alpha G_\mathrm{cav}^3 + \frac{\alpha K t^4 G_\mathrm{cav}^7 }{1- t^4 K G_\mathrm{cav}^4}, \nonumber\\
		G &\approx \frac{1}{E - t^2 (K+1) G_\mathrm{cav}} + \frac{\alpha}{\left(E - t^2 (K+1) G_\mathrm{cav} \right)^3}\left[ 1 + \frac{ (K+1) t^4 G_\mathrm{cav}^4 }{\left(1- t^4 K G_\mathrm{cav}^4\right)}\right].
	\end{align}
	\item These equations can be solved numerically to give the theory line in Fig. \ref{fig:exercise}. One notices, amongst other things, that the width of the support has expanded in comparison to the Kesten-McKay law.
\end{itemize}
As we mentioned in the above discussion, one can derive the upper limit for the disorder $W_c$, in the case of a uniform distribution, using the forward approximation. Here, we delve into the details of this calculation (the outline of which is given in Ref. \cite{pietracaprina2016forward}).
\begin{itemize}
	\item Beginning with Eq.~(\ref{forwardapprox}) with $\epsilon_0 = 0$, show that one can write for the probability distribution of $x_L$
	\begin{align}
		P(x_L) &= \int \prod_{j = 1}^L [d\epsilon_j P(\epsilon_j)] \delta\left(x_L +2 \sum_{j=1}^L\ln \vert \epsilon_j/t \vert \right) \nonumber \\
		&=\int \frac{d\hat x}{2\pi} \prod_{j = 1}^L [d\epsilon_j P(\epsilon_j)] \exp\left[ i \hat x\left(x_L +2 \sum_{j=1}^L\ln \vert \epsilon_j/t \vert \right)\right].
	\end{align}
	\item Show that in the case of uniform disorder on the interval $[-W/2,W/2]$ one obtains
	\begin{align}
		P(x_L) &=\int \frac{d\hat x}{2\pi} \exp\left\{  L\left[ i\hat x\left(\frac{x_L}{L} +2  \log\left( \frac{W}{2t}\right)\right) -\log\left[t\left( 1+2 i \hat x\right)\right] \right] \right\} .
	\end{align}
	\item Assuming that $x_L\sim L$, and making the substitution $z_L = \frac{x_L}{2L} + \log\left( \frac{W}{2t}\right)$, perform a saddle point approximation to show that (ignoring the prefactor)
	\begin{align}
		P(x_L) &\sim \exp\left[-L\left(z_L - 1 - \ln z_L \right) \right].
	\end{align}
	\item Thus, using the rule of thumb that, out of $K^L$ samplings of $x_L$ from this distribution, the typical largest value of $x_L$ is given by $K^L P(x_L^\star) \approx 1$, show that
	\begin{align}
		\ln\left( K e z^\star_L e^{-z^\star_L} \right) = 0.
	\end{align}
	\item Hence, obtain Eq.~(\ref{upperbound}) by requiring $x_L^\star/L=0$ at the critical disorder.
\end{itemize}

\newpage

\section{Spectral theory of complex networks}\label{section:networks}

\begin{quotation}
	Networks are present everywhere. All we need is an eye for them. -- Albert-L\'aszl\'o Barab\'asi
\end{quotation}

\subsection{Complex networks, adjacency matrices, and Laplacian matrices}

The concept of a network \cite{newman2018networks} is a very simple one. We suppose that we have a set of nodes labelled $i = 1, \cdots, N$. Each node may be connected with a subset of the other nodes. If nodes $i$ and $j$ are connected, we say that there exists an `edge' between them. In principle, we may also associate `weights' with the edges. These weights could represent the rate of transport along the edge, for instance, or the strength of the interaction between the two entities that are represented by the nodes, depending on the system that the network describes. We thus see that networks are neat abstractions -- a kind of summary data structure. While they are simple to describe in their basic form, networks are complicated objects that are associated with many varied and rich phenomena. 

The study and use of networks has rather exploded in the past few decades. One could point to a few reasons for this, but undoubtedly the rise of the internet, and subsequently `big data', make networks an indispensable tool for gaining a mathematical foothold on technological problems in the modern age. The google algorithm \cite{beel2009google}, for example, explicitly uses network theory, and the huge datasets that describe social media networks provide a valuable wealth of information on human social phenomena \cite{ediger2010massive}. In fact, any situation in which there are many interacting components presents an opportunity to exploit the tools of network science. Such systems may include, for example, social networks \cite{o2008analysis}, epidemics \cite{pastor2015epidemic}, opinion dynamics \cite{starnini2025opinion}, ecological communities \cite{pilosof2017multilayer} and amorphous solids \cite{mungan2019networks}.

There exist two important mathematical representations of networks, which we will discuss here. One is the network adjacency matrix. This object, which we call $\underline{\underline{A}}$ and have met in the previous section, encodes which nodes are connected to which others in the network. That is, if nodes $i$ and $j$ are connected, we set $A_{ij} = 1$ and $0$ otherwise. One may also choose to add weights $J_{ij}$ to the edges, which we can represent via the weighted adjacency matrix with elements $W_{ij} = J_{ij} A_{ij}$. 

Typically, in mathematical models, the (weighted) adjacency matrix appears when the network in question describes interactions between different components. Instead, when the network is meant to describe diffusive transport (e.g. in the case of migration of organisms in an ecological model, or the spread of a disease amongst a social network), the Laplacian matrix is used. It is defined as $L_{ij} = J_{ij}A_{ij} - \delta_{ij} \sum_{k} J_{ik}A_{ik}$, with the first term capturing the `influx' into site $i$ from site $j$, while the second captures the total `outflux' from site $i$. One can also consider the normalised Laplacian (where one divides $L_{ij}$ by the degree of node $i$) as in Ref. \cite{akara2023random} for example, but we focus on the unnormalised case here.

Below, we will take two examples of random network models: the sparse Erd\H{o}s-R\'enyi (ER) graph and the configuration model. Importantly, the ER graph, similarly to the RRG in the previous section, is locally tree-like \cite{bordenave2015large}, which means that we can use cavity equations similar to Eqs.~(\ref{cavityanderson}) and (\ref{cavityanderson2}). For the configuration model, we will consider the case where all nodes have a large (but heterogeneous) number of neighbours, which will permit us to obtain closure of the cavity equations. As a result, we will develop a set of tools to handle a range of network architectures. 

\subsection{Sparse ER graph adjacency matrix}
\subsubsection{Cavity equations and solution strategy}
Let us first consider the case of the weighted adjacency matrix of the Erd\H{o}s-R\'enyi (ER) graph $\underline{\underline{W}}$. The ER graph is perhaps the simplest model of a complex network (aside from the RRG). To create an ER graph, we suppose that every possible edge between any pair of nodes may exist with probability $p/N$. This means that the degree distribution (i.e. the distribution of the number of neighbours per node) is binomial. For large $N$, we may approximate it with the Poisson distribution $P_k = e^{-p}p^k/k!$, such that the average degree of a node is $p$.

Since the ER graph is locally tree-like \cite{bordenave2015large}, just as the RRG was, we may once again use the cavity graph trick that we used in Section \ref{section:anderson} [see Eqs.~(\ref{cavityanderson}) and (\ref{cavityanderson2}) in particular]. The cavity equations are, in the present case,
\begin{align}
	G_{ii} &= \frac{1}{\omega  - \sum_{j\in \partial_i} J_{ij}^2 G_{jj}^{(i)}}, \nonumber \\
	G_{jj}^{(i)} &= \frac{1}{\omega  - \sum_{k\in\partial_j^{(i)}} J_{jk}^2 G_{kk}^{(j)}}. \label{cavityG}
\end{align}
In the following, we will take $J_{ij} = 1/\sqrt{p} = \mathrm{const.}$ as a simple example. One notes now, as was the case for the RRG, that the sums in the denominators of the above expressions contain a subextensive number of terms. That is, these sums do not necessarily concentrate if $p$ is small. If we took the limit $p \to \infty$, however, then the sums would concentrate, and we would obtain the semicircle law. We note that the cavity equations in Eq.~(\ref{cavityG}) can also be solved numerically using the population dynamics algorithm (see Section \ref{section:popdyn}).

To make progress therefore, we consider the case where $p$ is reasonably large, but much smaller than $N$, so that we are comfortably in the sparse regime. We then expand the cavity equations as a series in $1/p$, so that we obtain the sparse corrections to the semicircle law, as was done by Rodgers and Bray \cite{rodgers1988density} using replicas (see also Ref. \cite{kim1985density}). Our approach for doing this will be similar in spirit to that used in Section \ref{section:pertexpand}. We will subsequently discuss the non-perturbative contributions to the spectrum below, which are $\propto e^{-p}$, and are thus not captured by this expansion.

\subsubsection{Expansion in $1/p$ (inverse connectivity)}
Let us then expand Eqs.~(\ref{cavityG}) as a series in $1/p$, and therefore take into account the change in the eigenvalue density that arises from local fluctuations of the network connectivity. To do this, it turns out to be more convenient to write the cavity equations in the following form
\begin{align}
	G_{ii} &= \frac{1}{\omega  - \frac{1}{p}\sum_{j} A_{ij} G_{jj}^{(i)}}, \nonumber \\
	G_{jj}^{(i)} &= \frac{1}{\omega  -  \frac{1}{p}\sum_{k} A_{jk}^{(i)}   G_{kk}^{(j)}}, \label{cavityER}
\end{align}
where $\underline{\underline{A}}^{(i)}$ is the adjacency matrix of the network with node $(i)$ removed. Noting that the existence of every link in the graph exists independently, a useful observation is that $\langle (A_{ij})^r \rangle = \left\langle \left(A_{jk}^{(i)} \right)^r \right\rangle= p/N$, where $r$ is some arbitrary positive integer power.

Anticipating that the sum $ \frac{1}{p}\sum_{k} A_{jk}^{(i)}  G_{kk}^{(j)}$ will have small fluctuations about its mean when $p$ is large enough, we may expand as follows
\begin{align}
	G_{jj}^{(i)} &\approx \frac{1}{\omega  - G_\mathrm{cav}} + \frac{\left(\frac{1}{p}\sum_{k} A_{jk}^{(i)}  G_{kk}^{(j)}\right)-G_\mathrm{cav} }{(\omega  - G_\mathrm{cav})^2}  + \frac{\left[\left(\frac{1}{p}\sum_{k} A_{jk}^{(i)}  G_{kk}^{(j)}\right)-G_\mathrm{cav} \right]^2}{(\omega  - G_\mathrm{cav})^3} \nonumber \\
	&\approx \frac{1}{\omega  - G_\mathrm{cav}} + \frac{\left(\frac{1}{p}\sum_{k} A_{jk}^{(i)}  G_{kk}^{(j)}\right)-G_\mathrm{cav} }{(\omega  - G_\mathrm{cav})^2}  + \frac{\frac{1}{p^2}\sum_{k} A_{jk}^{(i)}  [G_{kk}^{(j)}]^2}{(\omega  - G_\mathrm{cav})^3} , \label{cavityGexpand} 
\end{align}
where we write $G_\mathrm{cav} = \left\langle G_{jj}^{(i)} \right\rangle$. To leading order in $1/p$, we thus find that the fluctuations of $G_{jj}^{(i)}$ go as $1/\sqrt{p}$
\begin{align}
	\left\langle \left(G_{jj}^{(i)} - G_\mathrm{cav}\right)^2 \right\rangle&\approx  \left\langle \frac{\left[\left(\frac{1}{p}\sum_{k} A_{jk}^{(i)}  G_{kk}^{(j)}\right)-G_\mathrm{cav} \right]^2}{(\omega  - G_\mathrm{cav})^4} \right\rangle \nonumber \\
	&\approx \frac{1}{p} G_\mathrm{cav}^6 +O\left( \frac{1}{p^2}\right) .
\end{align}
With this in mind, we see that we may use $\langle [G_{kk}^{(j)}]^2\rangle \approx G_\mathrm{cav}^2$ in the right-hand side of Eq.~(\ref{cavityGexpand}), and only incur an error of the order $O(1/p^2)$. We therefore obtain
\begin{align}
	G_\mathrm{cav} &= \frac{1}{\omega  - G_\mathrm{cav}} +  \frac{1}{p} G_\mathrm{cav}^5.
\end{align}
We may perform an analogous expansion of the equation for $G_{ii}$, and we obtain
\begin{align}
	G \equiv N^{-1}\sum_i G_{ii} &= \langle G_{ii}\rangle = \frac{1}{\omega  - G_\mathrm{cav}} +  \frac{1}{p} G_\mathrm{cav}^5 .
\end{align}
Hence, we see that $G \approx G_\mathrm{cav}$ to this accuracy in $1/p$. Finally, using this fact, we obtain a self-consistent series for $G$, similar to that in Eq.~(\ref{gapprox}) in the case of the RRG Anderson model,
\begin{align}
	G = \frac{1}{\omega - G} + \frac{1}{p} G^5 + O\left(\frac{1}{p^2}\right). \label{sparseerseries}
\end{align}
We may thus obtain the $1/p$ correction to the semicircle law by noting that $G^2 = \omega G -1 + O\left( \frac{1}{p}\right)$,
which allows us to write Eq.~(\ref{sparseerseries}) as a quadratic equation in $G$
\begin{align}
	G(\omega - G)  = 1 + \frac{1}{p} (\omega G - 1)^2+ O\left(\frac{1}{p^2}\right).
\end{align}
Finally, solving this quadratic, we find
\begin{align}
	G(\omega) = \frac{\omega(1+2/p) \pm \sqrt{\omega^2 - 4(1+1/p)}}{2(1 + \omega^2/p)} . \label{EMA}
\end{align}
Using the inverse Stieltjes transform in Eq.~(\ref{densefromres}) as usual, we obtain the following modified semicircle law
\begin{align}
	\rho(\omega) &=  \frac{2}{\pi\omega_c^2} \left\{ 1 + \frac{1}{p} \left[1-4 \frac{\omega^2}{\omega_c^2}\right]\right\} \sqrt{\omega_c^2-\omega^2}, \nonumber \\
	\omega_c^2 &= 4\left(1+\frac{1}{p}\right) .\label{semicirclesparse}
\end{align}
This result has previously been obtained in Refs. \cite{rodgers1988density, kim1985density}, and was subsequently rederived using the so-called `effective-medium' approximation \cite{semerjian2002sparse}, which happens to capture the leading $1/p$ correction, but is not systematic. We compare the expression in Eq.~(\ref{semicirclesparse}) to results from numerics in Fig. \ref{fig:SIR}.

One can extend this approach to obtain the higher-order corrections to the semicircle law, and indeed the corrections to the elliptic law in the case where $J_{ij} \neq J_{ji}$. This has been accomplished using the diagrammatic approach in Ref. \cite{baron2023pathintegralapproachsparse} (see Fig.~\ref{fig:popdyn}), but it could also easily be carried out using the cavity approach. More complicated spectral statistics, such as the eigenvalue correlations (which are non-trivial over large separations, but coincide with the GOE for small separations), have also been computed using the diagrammatic \cite{baron2025classes} and supersymmetric approaches \cite{mirlin1991universality}, but this is beyond the scope of our techniques here. 

\subsubsection{Outlier eigenvalue}
In the present case, since the elements $W_{ij} = A_{ij} J_{ij}$ have a finite mean $\langle W_{ij} \rangle = \sqrt{p}/N$, we expect there to be a possible outlier eigenvalue. Proceeding as we did in Section \ref{section:outlier}, we find that $G(\lambda_\mathrm{outlier}) = \frac{1}{\sqrt{p}}$. This then yields
\begin{align}
	\lambda_\mathrm{outlier} \approx \sqrt{p} + \frac{1}{\sqrt{p}} .
\end{align}
As was noted in Ref. \cite{baron2023pathintegralapproachsparse}, one should in principle take into account some off-diagonal contributions to the sum $N^{-1}\sum_{ij}G_{ij}$ in the course of this calculation, due to the sparsity of the matrix. However, since we ignore terms smaller than $O(1/p)$ in the above formula, then these contributions are not important here.

We therefore see that as long as $p\gtrapprox 4$, the outlier protrudes from the warped semicircle. We will see that this is important in our example below. 

\subsubsection{Single-defect approximation -- network hubs}
The preceding calculation allowed us to take into account the effect that fluctuations in the node degrees had on the eigenvalue spectrum. However, we assumed that these fluctuations were small so that we could obtain a series in $1/p$. How may we take into account atypical fluctuations? 

We turn once again to the single-defect approximation (SDA) that we introduced in Section \ref{section:SDA}. That is, we identify the anomalies of the network, and we attempt to understand the contribution that they make to the eigenvalue spectrum by assuming that all of their neighbours are `typical'. In the previous section, this meant that the anomalous node was associated with a large on-site energy $\epsilon_i$. Here, nodes that have anomalously large degree are those that constitute `defects' in the network, and therefore give rise to outlier eigenvalues and localised eigenvectors.

Mathematically, we find the outlier eigenvalues by searching for those poles of the first of Eqs.~(\ref{cavityG}) that lie outside the continuous bulk of the eigenvalue spectrum. Assuming that the degree of the node $i$ (which we denote $k_i$) is large, the sum in the denominator of Eq.~(\ref{cavityG}) now concentrates, and we may write 
\begin{align}
	G_{ii} \approx \frac{1}{\omega - \frac{k_i}{p} G}. \label{SDAER}
\end{align}
Here, we have assumed that each of the neighbours of $i$ are `typical', so that we may evaluate the average of the large sum with $\sum_j A_{ij} G_{jj}^{(i)} \approx k_i \langle G_{jj}^{(i)}\rangle$. We recall that we have established that $\langle G_{jj}^{(i)}\rangle \equiv G_\mathrm{cav} =G$ for such typical nodes, so that we may use the perturbative result in Eq.~(\ref{EMA}) for $G(\omega)$ in Eq.~(\ref{SDAER}). 

Importantly, we see now that we capture additional isolated poles of the resolvent at locations $\omega_k \approx k G(\omega_k)/p$ when $k$ is large. More precisely, a network `hub' of degree $k_\mathrm{hub}$ contributes an eigenvalue pair at
\begin{align}
	\lambda_\mathrm{hub} = \pm\sqrt{\frac{k_\mathrm{hub}}{p(1-p/k_\mathrm{hub})} + \frac{1}{k_\mathrm{hub}(1-p/k_\mathrm{hub})^3}}. \label{hubeigenvalue}
\end{align}
So, to contribute an eigenvalue that protrudes from the bulk region, the connectivity of the hub must be $k_\mathrm{hub} \gtrapprox 4 p$. Similar to Section \ref{section:SDA}, where we first introduced the SDA, the eigenvectors corresponding to these poles are localised around their corresponding network hubs \cite{semerjian2002sparse, biroli1999single, valigi2025eigenvalue}. We show this explicitly in the exercises. 

Hence, we may use the degree distribution $P_k = e^{-p}p^k/k!$ to arrive at an expression for the resolvent that is valid outside the semicircle-like bulk region
\begin{align}
	G_\mathrm{SDA}(\omega) = \sum_k e^{-p}\frac{p^k}{k!} \frac{1}{\omega -  \frac{kG}{p}} , \label{sdaeradjacency}
\end{align}
and the corresponding eigenvalue density is given by \cite{valigi2025eigenvalue}
\begin{align}
	\rho_\mathrm{SDA}(\omega) \sim \sqrt{\frac{p}{2\pi}} e^{-p} \left(\frac{e}{\omega^2} \right)^{p^2 \omega^2}.
\end{align}
Once again, we therefore see that the collection of additional isolated poles that the SDA describes constitutes a Lifshitz tail. This tail part of the spectrum extends indefinitely along the real axis for $N \to \infty$, but is suppressed by a factor $e^{-p}$. For this reason, the Lifshitz tails in this case disappear very quickly as we increase the connectivity of the graph. 

\subsubsection{Example: epidemic threshold in the SIR model}
Let us now consider the following simple SIR model \cite{tang2020review} of an epidemic to demonstrate how network structure can affect dynamics. We consider the numbers of susceptible $S_i$, infected $I_i$, and recovered $R_i$ individuals on various nodes $i$. The nodes of the network here could represent different towns or social cliques, for example. We take the network to be the weighted ER graph considered above. We also allow for the possible addition of a single artificial added network hub with degree $k_\mathrm{hub}$, which we randomly connect to the other nodes. 

The SIR equations are in this case
\begin{align}
	\frac{dS_i}{dt} &= - \beta S_i \left[I_i+ \frac{1}{\sqrt{p}}\sum_{j\neq i} A_{ij} I_j \right], \nonumber \\
	\frac{dI_i}{dt} &= \beta S_i \left[I_i+ \frac{1}{\sqrt{p}}\sum_{j\neq i} A_{ij} I_j \right] - \gamma I_i , \nonumber \\
	\frac{dR_i}{dt} &= \gamma I_i.
\end{align}
Here, we have defined the recovery rate $\gamma$, the on-site infection rate $\beta$, and the neighbouring site infection rate $\beta/\sqrt{p}$. The total population $S_i +I_i +R_i = 1$ is constant. We see immediately that there is a trivial fixed-point at $I_i = 0$, where there is no infection. We wish to understand the dynamical stability of this solution. This is determined by the largest eigenvalues of the matrix $\frac{\beta}{\sqrt{p}}\underline{\underline{A}}- (\gamma-\beta) \underline{\underline{\id}}$. 

From our considerations above, we see that, as long as the average degree $p$ is reasonably large and there is no network hub, the largest eigenvalue of the Jacobian is given by
\begin{align}
	\lambda_\mathrm{max} = \beta \left(1 + \sqrt{p} + \frac{1}{\sqrt{p}}\right) - \gamma . 
\end{align} 
In the presence of the network hub, we have instead
\begin{align}
	\lambda_\mathrm{max} = \beta \sqrt{\frac{k_\mathrm{hub}}{p(1-p/k_\mathrm{hub})} + \frac{1}{k_\mathrm{hub}(1-p/k_\mathrm{hub})^3}} - \gamma . 
\end{align}
The epidemic threshold is then simply given by $\lambda_\mathrm{max} = 0$. We demonstrate various possible scenarios below in Fig. \ref{fig:SIR}. 

\begin{figure}[h]
	\centering 
	\includegraphics[scale = 0.4]{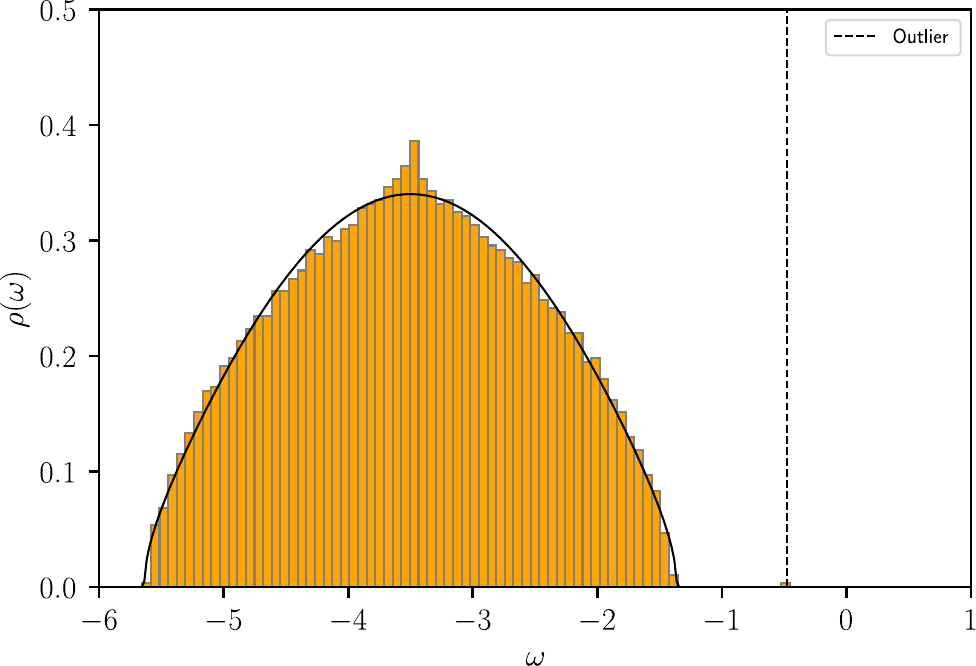}
	\includegraphics[scale = 0.4]{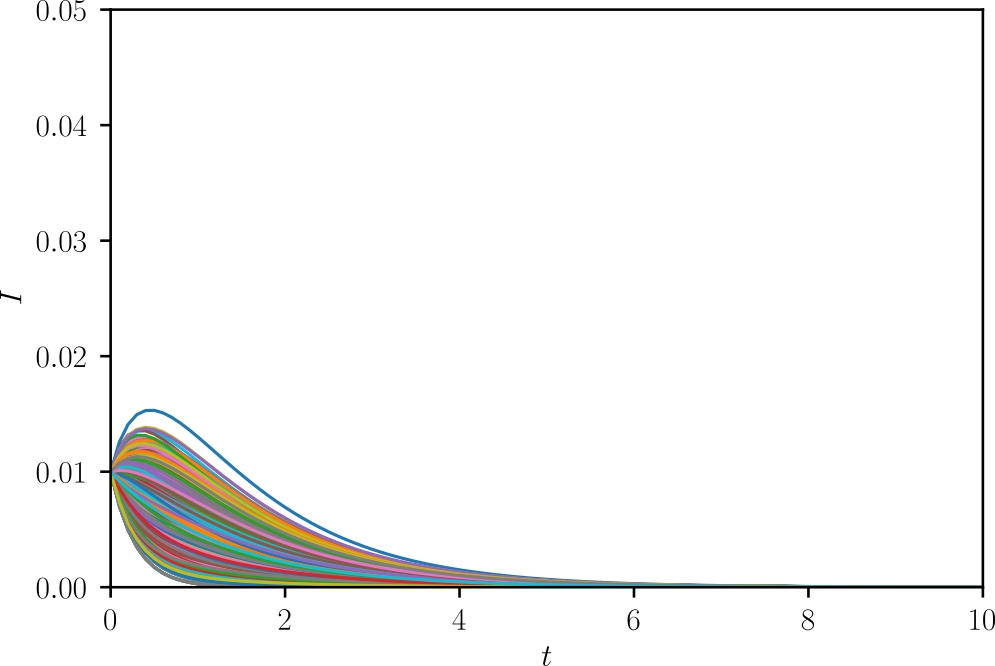}
	\includegraphics[scale = 0.4]{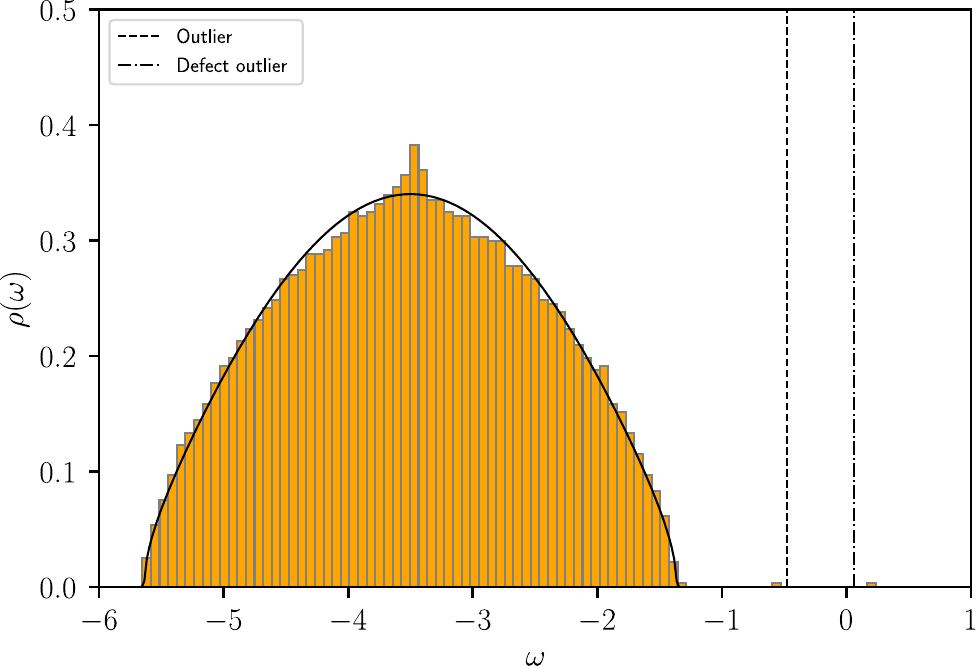}
	\includegraphics[scale = 0.4]{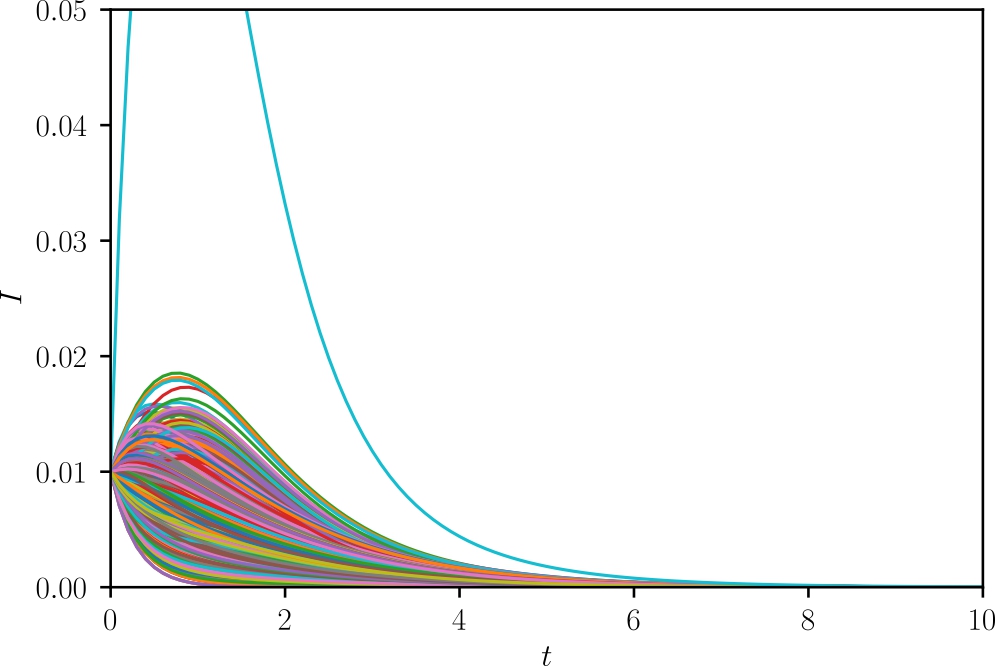}	
	\includegraphics[scale = 0.4]{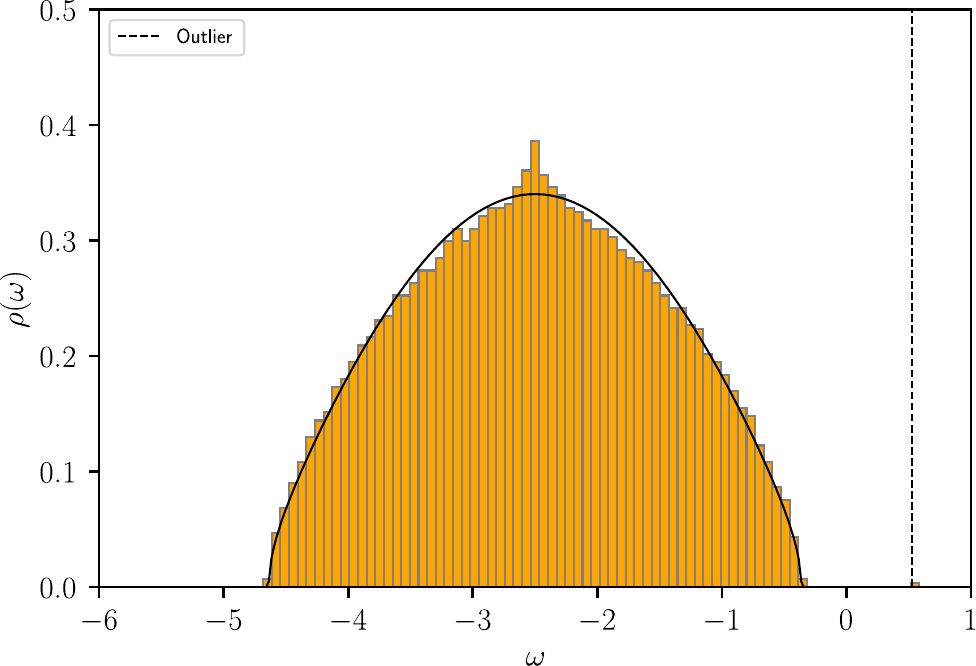}
	\includegraphics[scale = 0.4]{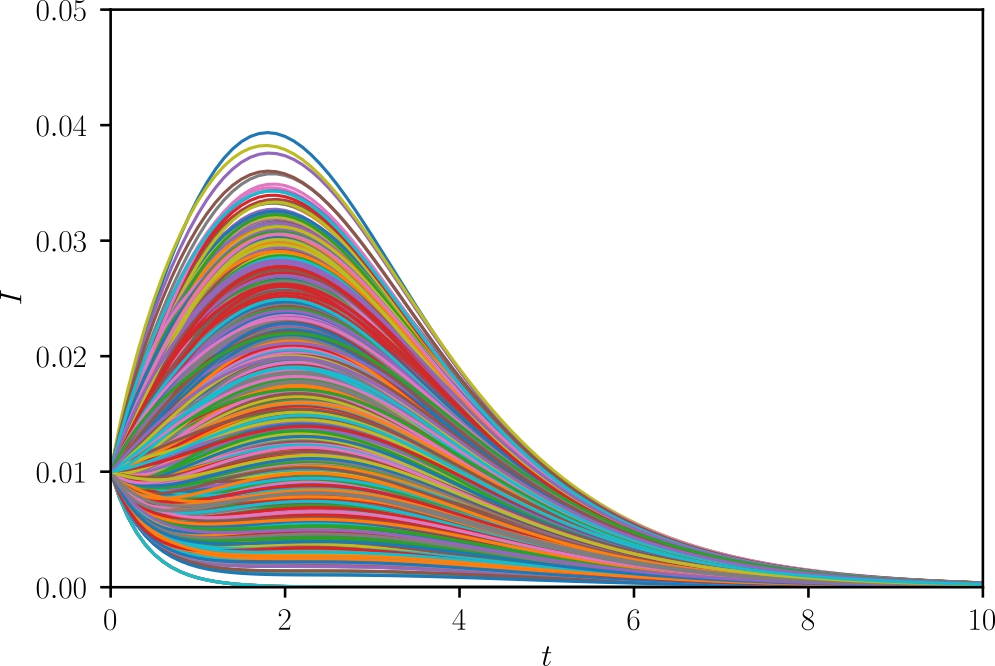}	
	\captionsetup{justification=raggedright,singlelinecheck=false}
	\caption{ SIR dynamics and the associated eigenvalue spectra, with and without a network hub. On the first row ($\gamma = 4.5$, without network hub), the $I = 0$ fixed point is stable, and the epidemic quickly goes extinct without many infections. In the second row, we show the dynamics with the network hub with degree $81$ (still, $\gamma = 4.5$). Here, the leading eigenvalue is greater than 0, and a localised eigenvector is unstable. Finally, in the third row, we set $\gamma = 3.5$ and we remove the network hub. In this case, the instability is associated with an extended eigenvector, and the epidemic is much more widespread. Other system parameters: $N = 4000$, $p = 7$, $\beta = 1$.  }\label{fig:SIR}
\end{figure} 

\subsection{Sparse ER graph Laplacian matrix}
\subsubsection{Modified cavity equations and solution strategy}
Having tackled the ER graph adjacency matrix, let us now consider the Laplacian matrix
\begin{align}
	L_{ij} = J_{ij} A_{ij} - \delta_{ij} \sum_{k} J_{ik} A_{ik},
\end{align}
which we will apply to a problem involving diffusion on networks below. The appropriate cavity equations for $\underline{\underline{L}}$ read
\begin{align}
	G_{ii} &= \frac{1}{\omega + \sum_{j\in\partial_i} J_{ij} - \sum_{j\in \partial_i} J_{ij}^2 G_{jj}^{(i)}}, \nonumber \\
	G_{jj}^{(i)} &= \frac{1}{\omega + \sum_{k\in\partial_j^{(i)}} J_{ij} - \sum_{k\in\partial_j^{(i)}} J_{jk}^2 G_{kk}^{(j)}},
\end{align}
where now $\underline{\underline{G}} = \left[\omega \underline{\underline{\id}} - \underline{\underline{L}} \right]^{-1}$ is the resolvent matrix of the Laplacian.

In principle, we may iterate these equations using the population dynamics algorithm described in Ref. \cite{kuhn2008spectra} or in Section \ref{section:popdyn} below, but here we pursue solutions by analytical means. Closed-form solutions are most readily available to us in the special case $J_{ij} = J = \mathrm{const.}$. We note that for the random regular graph, the spectrum is particularly simple in this case, and we obtain a shifted Kesten-McKay law. 

For the case of other graphs, like the ER graph, it is helpful to make the following substitution \cite{dean2002approximation, kuhn2008spectra}
\begin{align}
	G_{kk}^{(j)}(\omega) = \frac{K_{kk}^{(j)}(\omega)}{1+ J K_{kk}^{(j)}(\omega)},\hspace{1cm}
	G_{ii}(\omega) = \frac{K_{ii}(\omega)}{1+ J K_{ii}(\omega)},		
\end{align}
so that we obtain cavity equations in terms of the transformed resolvent 
\begin{align}
	K_{ii} = \frac{1}{\omega - J + \sum_{j\in\partial_i} \frac{J}{1 + J K_{jj}^{(i)}}}, \hspace{0.5cm}
	K_{jj}^{(i)} = \frac{1}{\omega - J + \sum_{k\in\partial_j^{(i)}} \frac{J}{1+ J K_{kk}^{(j)}}}.
\end{align}
Further, we notice the following 
\begin{align}
	K_{ii}(\omega) = \frac{1}{\omega - J + \sum_{j\in\partial_i} J - \sum_{j\in \partial_i} J^2 G_{jj}^{(i)}} = G_{ii}(\omega -J) .
\end{align}
Hence, we may make the further substitutions $H_{ii}(\omega) = K_{ii}(\omega+J)$ and $H_{jj}^{(i)}(\omega) = K_{jj}^{(i)}(\omega+J)$ to obtain
\begin{align}
	H_{ii} = \frac{1}{\omega  + \sum_{j\in\partial_i} \frac{J}{1 + J H_{jj}^{(i)}}}, \hspace{1cm} 
	H_{jj}^{(i)} = \frac{1}{\omega + \sum_{k\in\partial_j^{(i)}} \frac{J}{1+ J H_{kk}^{(j)}}}, \label{cavityH}
\end{align}
and, rather nicely, we may obtain the eigenvalue density in terms of the transformed resolvent
\begin{align}
	\rho(\omega) = \frac{1}{\pi N} \lim_{\epsilon \to 0} \mathrm{Im} \sum_{i}H_{ii}(\omega- i\epsilon). 
\end{align}
These substitutions have allowed us to do away with the varying diagonal elements of the Laplacian, in a manner of speaking, which present a difficulty over and above that presented by the adjacency matrix. We are thus in a position to make a systematic analysis once again.

\subsubsection{$1/p$ expansion}
Let us again expand Eqs.~(\ref{cavityH}) as a series in $1/p$, taking $J = 1/p$ so that the limiting spectrum is finite. We obtain
\begin{align}
	H = H_\mathrm{cav} = \frac{1}{\omega + \frac{1}{1+ H/p}} + \frac{1}{p} \frac{H^3}{(1+H/p)^2} \approx \frac{1}{\omega + \frac{1}{1+2 H/p}}. \label{hlaplace}
\end{align}
One notes that this expression has been found previously using the effective-medium approximation in Ref. \cite{baron2021persistent} (something similar was also found using the supersymmetric approach \cite{akara2023random}), where it was also shown that varying weights $J_{ij}$ could be taken into account. We may solve this quadratic to obtain the eigenvalue density, which takes the Mar\v{c}enko-Pastur form
\begin{align}
	\rho(\omega) = \frac{\sqrt{(\lambda_+- \omega)(\omega- \lambda_-)}}{4 \pi \omega /p} , \hspace{1cm}
	\lambda_\pm = -1 \pm 2\sqrt{\frac{2}{p}} - \frac{2}{p} .\label{pertlaplace}
\end{align}
\subsubsection{Single-defect approximation}
\begin{figure}[h]
	\centering 
	\includegraphics[scale = 0.6]{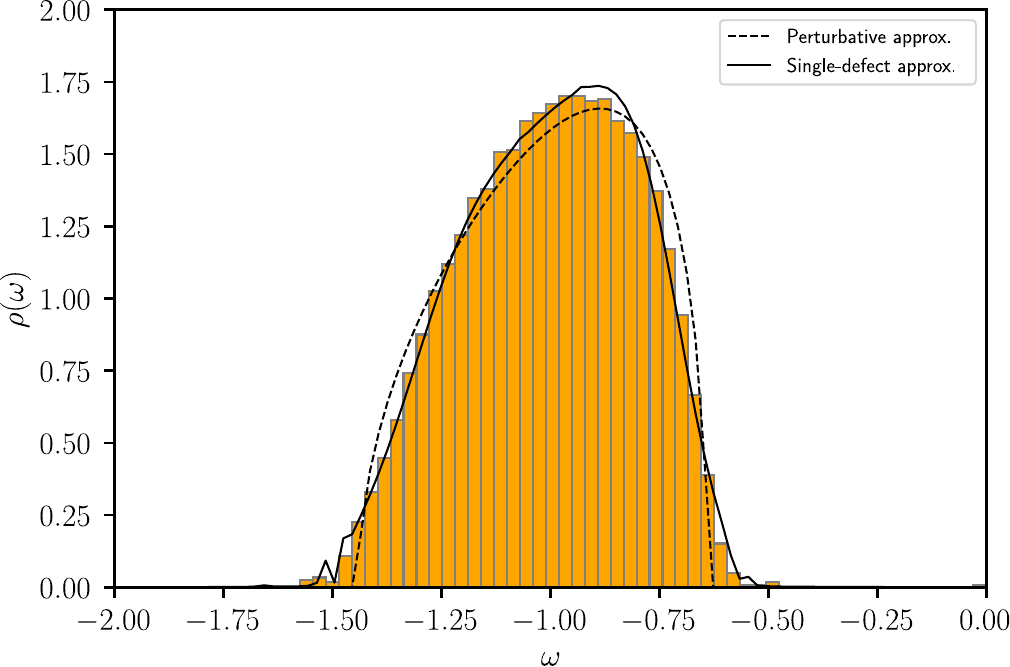}
	\captionsetup{justification=raggedright,singlelinecheck=false}
	\caption{ Eigenvalue density of the Laplacian matrix with $p = 50$. The theory lines are the perturbative result in Eq.~(\ref{pertlaplace}) and the SDA in Eq.~(\ref{sdalaplace}).}\label{fig:laplacian}
\end{figure}

Once again, we may also perform the single defect approximation, and we obtain an equation analogous to that in Eq.~(\ref{sdaeradjacency})
\begin{align}
	H_\mathrm{SDA} = \sum_k e^{-p} \frac{p^k}{k!} \frac{1}{\omega + \frac{k}{p + H(\omega)}},
\end{align}
from which we may extract the eigenvalue density via
\begin{align}
	\rho_\mathrm{SDA}(\omega) = \frac{1}{\pi} \lim_{\epsilon \to 0} \mathrm{Im}\left[\sum_k e^{-p} \frac{p^k}{k!} \frac{1}{\omega - i \epsilon + \frac{k}{p + H(\omega-i\epsilon)}}\right],\label{sdalaplace}
\end{align}
where we use the expression for $H(\omega)$ in Eq.~(\ref{hlaplace}) in the right-hand side of the above expression. We test both Eqs.~(\ref{pertlaplace}) and (\ref{sdalaplace}) in Fig. \ref{fig:laplacian}.

\subsubsection{Example: diffusion-induced instability and network topology}
To illustrate the utility of our knowledge of the Laplacian spectrum in context, let us consider a simple example of reaction-diffusion dynamics, namely, the Brusselator model \cite{prigogine1968symmetry}. In its original incarnation, this model was meant to capture the oscillatory and pattern-forming capabilities of the Belousov–Zhabotinsky reaction \cite{zhabotinsky2007belousov}, but we wish here only to use it to understand abstractly the combination of local reaction and diffusion on networks. We consider the concentration of two `chemicals' that may react and diffuse, and we consider two cases for the network on which they may diffuse. In one case, we consider a 1D ring of sites, and in the other, we consider the ER graph. 

What we will see is that, for certain parameter values, the system will tend to a homogenous fixed point solution in the absence of diffusion. However, for a threshold rate of diffusivity, this homogeneous solution becomes unstable, and spatial patterns form on the network. The diffusion-induced instability mechanism that we discuss below has applications in many models \cite{satnoianu2000turing}, particularly its original area of conception by Turing in the context of chemical morphogenesis \cite{am1952chemical}, and also in the study of ecosystem dynamics \cite{baron2020dispersal}.

The reaction-diffusion equations for the local concentrations of the two chemicals $x_i(t)$ and $y_i(t)$ are
\begin{align}
	\frac{dx_i}{dt} &= a + x_i^2 y_i - (1+b)x_i+ D_x\sum_{j} L_{ij} x_j , \nonumber \\
	\frac{dy_i}{dt} &= b x_i - x_i^2 y_i + D_y\sum_{j} L_{ij} y_j ,
\end{align}
where $a$ and $b$ are model constants, and $D_x$ and $D_y$ are diffusion coefficients. We see here that $x$ is an `activator' chemical, tending to lead to the greater production of itself and the other chemical, and $y$ is the `inhibitor' chemical. These equations have an homogeneous fixed point at $x_i^\star = a$ and $y_i^\star = b/a$.

Let us consider the stability of the homogeneous fixed point solution to small perturbations. Linearising about the fixed point by introducing $\delta x_i = x_i(t) - x_i^\star$ and $\delta y_i = y_i(t) - y_i^\star$, we obtain
\begin{align}
	\frac{d(\delta x_i)}{dt} &= (b-1)\delta x_i + a^2 \delta y_i+ D_x\sum_{j} L_{ij} \delta x_j, \nonumber \\
	\frac{d(\delta y_i)}{dt} &= -b \delta x_i - a^2 \delta y_i + D_y\sum_{j} L_{ij} \delta y_j .
\end{align}
The simplest way to understand stability in this case is to work in the eigen-basis of the Laplacian. If we decompose the perturbations into components in the directions of the eigenvectors of $\underline{\underline{L}}$ such that
\begin{align}
	\delta x_i(t) = \sum_\nu v_{i}^{(\nu)} c^{(\nu)}_x(t), \hspace{1cm} \delta y_i(t) = \sum_\nu v_{i}^{(\nu)} c^{(\nu)}_y(t),
\end{align} 
then we obtain
\begin{align}
	\frac{d c^{(\nu)}_x}{dt} &= (b-1)c^{(\nu)}_x + a^2 c^{(\nu)}_y+ \lambda_\nu D_x c^{(\nu)}_x, \nonumber \\
	\frac{dc^{(\nu)}_y}{dt} &= -b c^{(\nu)}_x - a^2 c^{(\nu)}_y + \lambda_\nu D_y c^{(\nu)}_y .
\end{align}
We thus see that in order to have a stable fixed point in the presence of diffusion, the eigenvalues of the following $2\times2$ Jacobian must have negative real parts for all possible values of $\lambda_\nu$
\begin{align}
	M_\nu = \begin{bmatrix}
		b-1 + \lambda_\nu D_x & a^2\\
		-b & -a^2 +D_y \lambda_\nu
	\end{bmatrix}
\end{align} 
Restating the above discussion more precisely, the Turing instability (otherwise known as the diffusion-induced instability) occurs when we have a stable homogeneous fixed point in the absence of diffusion $D_x = D_y = 0$, but the fixed point becomes unstable for some non-zero values of $D_x$ and $D_y$. Typically, the instability occurs when we have a slowly diffusing activator and a comparatively quickly diffusing inhibitor (i.e. small $D_x/D_y$). 

\begin{figure}[h]
	\centering 
	\includegraphics[scale = 0.32]{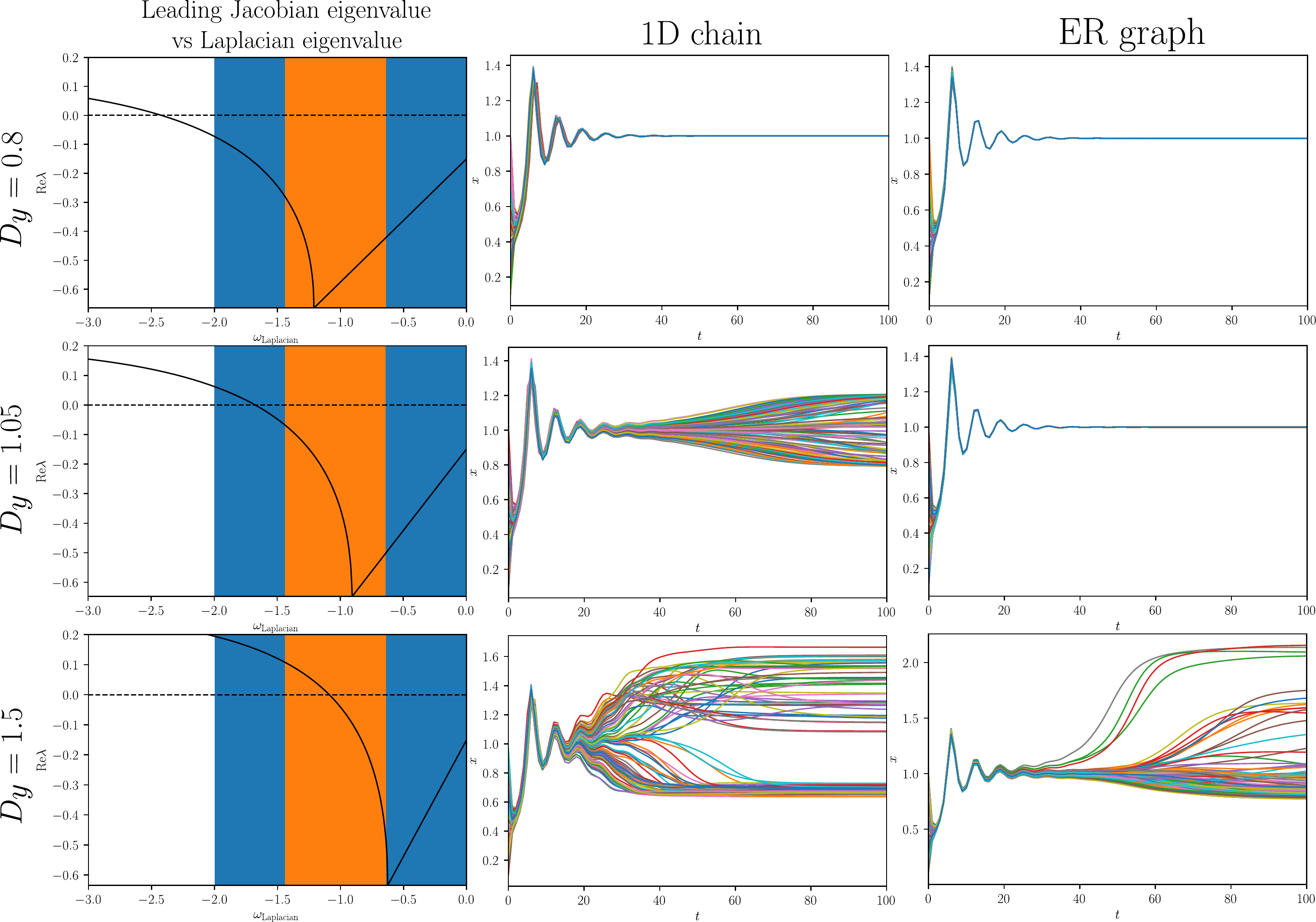}
	\captionsetup{justification=raggedright,singlelinecheck=false}
	\caption{Turing instability on the 1D chain and the ER graph with $p = 50$. In the leftmost column, we plot the leading eigenvalue of the Brusselator system linearised about its fixed point as a function of the Laplacian eigenvalue. The eigenvalue range of the ER graph laplacian (according excluding localised states) is highlighted in orange, and those of the 1D chain are highlighted in blue. In the second column we plot the dynamics of the $x$ components of all points on the 1D chain, and in the third, we plot the corresponding dynamics on the ER graph. The system parameters are $a = 1$, $b = 1.7$, $D_x = 0.05$. In the three rows, we show results for $D_y = 0.8$, $1.05$ and $1.5$. }\label{fig:turing}
\end{figure}

What we show now in Fig. \ref{fig:turing} is that since the range of $\lambda_\nu$ is different for different network topologies, the Turing instability occurs for different parameter sets for each network. For the 1D chain the range of $\lambda_\nu$ is $[-2, 0]$ (since the Fourier transform of the discrete Laplacian is $\cos(q)-1$). For the ER graph, we find from the our spectral analysis in Eq.~(\ref{pertlaplace}) that it is $[\left(-1 -2 \sqrt{\frac{2}{p}}- \frac{2}{p}\right), \left(-1 +2 \sqrt{\frac{2}{p}}- \frac{2}{p}\right)]$ (ignoring the Lifshitz tails, which fall off quickly outside this range). This means that some eigen-modes of the Laplacian become unstable in the case of the 1D chain while all those of the ER graph are stable for the same system parameters. We thus come to the conclusion that stability is dependent on graph architecture. 

\subsection{Dense configuration model networks}\label{section:denseconfiguration}
So far, we have considered the spectra of sparse ER graphs, and we utilised the $1/p$ expansion and the SDA to obtain analytical results. Now, we consider a slightly different network model -- the so-called configuration model \cite{newman2018networks}. In this case, we study a different limit, and we suppose that the network is well-connected, so that each node has many neighbours. Further, we suppose that the degree of each node $k_i$ is drawn from some arbitrary degree distribution $P_k$. This will allow us to contrast the effects of sparsity (small average degree) with those of degree heterogeneity.

According to the configuration model, we first assign each node a degree $k_i$. We attach to each node a set of edges equal to its assigned degree $k_i$, keeping the other ends of these edge loose/unassigned for the present. We then select pairs of `loose' edge ends at random to connect to each other until all the loose edges are paired. In the end, one can show that nodes with assigned degrees $k_i$ and $k_j$ are connected with probability \cite{rodgers2005eigenvalue, baron2022networks}
\begin{align}
	\langle A_{ij} \rangle =	1 - \left[ 1 - 2 \frac{k_i k_j}{(pN)^2}\right]^{pN/2} \approx \frac{k_i k_j}{pN}, \label{configprob}
\end{align}
where $p = \sum_k kp_k$ is the average degree of the network and the approximation holds when $\frac{k_i k_j}{(pN)}\ll 1$ for all $i$ and $j$. 

\begin{figure}[h]
	\centering 
	\includegraphics[scale = 0.6]{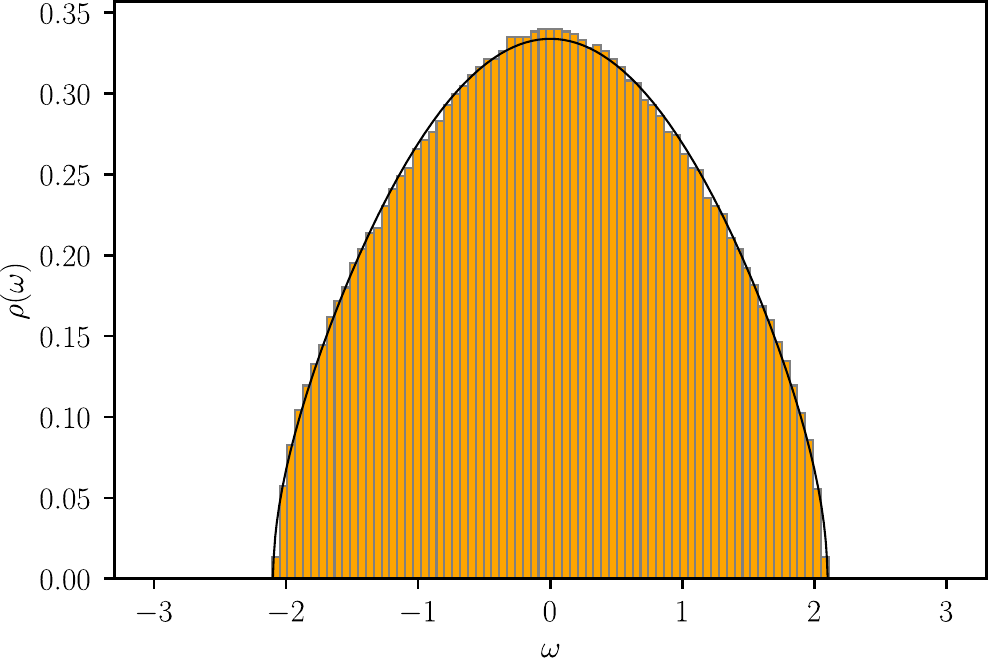}
	\captionsetup{justification=raggedright,singlelinecheck=false}
	\caption{ Eigenvalue density of the scaled adjacency matrix of a configuration model network. The degrees of the nodes are drawn from a uniform distribution in the range $[p- \sqrt{3} p s, p+ \sqrt{3} p s]$, where $p = 0.02 N$ and $N = 10^4$. The theory line is given by Eq.~(\ref{heterosemicircle}). }\label{fig:config}
\end{figure}

In this case, we can compute the spectrum of the weighted adjacency matrix relatively easily. Since the network is densely connected (i.e. $p \gg 1$), the cavity resolvent elements may be considered equal to the true resolvent elements in this case (with error being of order $1/p$). Considering for simplicity the weights to be $\pm 1/\sqrt{p}$ with equal probability, we obtain
\begin{align}
	G_{ii} &= \frac{1}{\omega - \frac{k_i}{p} A}, \nonumber \\
	A &= \frac{1}{N}\sum_{i} \frac{k_i}{p} G_{ii} = \frac{1}{N}\sum_{i} \frac{k_i/p}{\omega - k_i A/p} . \label{resolventconfiguration}
\end{align}
In principle, these are a closed set of equations for $A$ and $G = N^{-1} \sum_i G_{ii}$, which we could solve by averaging with respect to the degree distribution $P_k$. However, in the interest of generality, it is beneficial to consider the case where the degree heterogeneity $s^2 = \sum_k p_k \frac{k^2-p^2}{p^2}$ is small. We may handle this using a perturbative approach once again. One finds
\begin{align}
	A &\approx \frac{1}{\omega - A} + \frac{s^2 \omega A}{(\omega -A)^3} \approx  \frac{1}{\omega - A} + s^2 (A^3 + A^5), \nonumber \\
	G &\approx \frac{1}{\omega -A} + s^2 A^5. 
\end{align}
We thus see that to leading order in $s^2$, $A \approx G + s^2 A^3 \approx G + s^2 G^3$. Inserting this into the above expression for $G$, we arrive at 
\begin{align}
	G \approx \frac{1}{\omega - G} + 2 s^2 G^5. 
\end{align}
We can solve this in the same way as Eq.~(\ref{sparseerseries}), and we obtain
\begin{align}
	\rho(\omega) &= \frac{2}{\pi \omega_c^2} \left[1 + 2s^2 \left(1 - 4 \frac{\omega^2}{\omega_c^2} \right) \right]\sqrt{\omega_c^2 - \omega^2} , \nonumber \\
	\omega_c^2 &= 4(1+2 s^2). \label{heterosemicircle}
\end{align}
We thus see that degree heterogeneity broadens the eigenvalue spectrum, just as sparsity did. The above expression is tested against numerics in Fig. \ref{fig:config}.

Again like the sparse case, the above approach may be extended to Laplacian matrices \cite{da2025spectral} and to weights with asymmetric signs or varied magnitudes \cite{baron2022networks}. Further, the results obtained by considering configuration model networks may be extended to other networks with the same degree distribution \cite{baron2022networks}. We consider the non-hermitian case in the exercises.

\subsection{Exercises}
We will now derive the result in Eq.~(\ref{circularlawhetero}), which is a non-hermitian counterpart to the result in Eq.~(\ref{heterosemicircle}) for the eigenvalue density of the adjacency matrix of a heterogeneous network.

We imagine that the adjacency matrix is constructed according to the configuration model as defined in the discussion around Eq.~(\ref{configprob}). However, in this instance, we imagine that the weights $J_{ij}$ are drawn from a distribution with statistics
\begin{align}
	\langle J_{ij} \rangle = 0,\,\,\, \langle J_{ij}^2 \rangle  = \sigma^2_0, \,\,\, \langle J_{ij} J_{ji} \rangle  = 0 .
\end{align}
\begin{itemize}
	\item Show that the relevant Hermitised cavity equations in this case [analogous to Eqs.~(\ref{cavitynonherm})] are 
	\begin{align}
		\mathcal{H}_{ii} = \left( \mathcal{Z} - \sum_{j\in \partial_i} \mathcal{J}_{ij} \mathcal{H}^{(i)}_{jj}\mathcal{J}_{ji}\right)^{-1} ,
	\end{align}
	so that
	\begin{align}
		\sum_{j\in \partial_i} \mathcal{J}_{ij} \mathcal{H}^{(i)}_{jj}\mathcal{J}_{ji} = \begin{bmatrix}
			\sum_{j\in \partial_i} J_{ij}^2 D_{jj}^{(i)} & \sum_{j\in \partial_i} J_{ij} C_{jj}^{(i)} J_{ji}\\
			\sum_{j\in \partial_i} J_{ij} B_{jj}^{(i)} J_{ji} & \sum_{j\in \partial_i}  A_{jj}^{(i)} J_{ji}^2
		\end{bmatrix} \label{myblockex}
	\end{align}
	\item We now suppose that, since the sums are over are a large number of components, they can be replaced with their ensemble averages, \textit{conditioned} on the connectivity of the node $i$. That is, given the expression for the probability of connection in Eq.~(\ref{configprob}), show that
	\begin{align}
		\sum_{j\in \partial_i} J_{ij}^2 D_{jj}^{(i)} \approx \frac{\sigma^2_0k_i}{pN} \sum_{j} k_j D_{jj},
	\end{align}
	and find the corresponding expressions for the other components of Eq.~(\ref{myblockex}). 
	\item Hence, demonstrate that
	\begin{align}
		A_{ii} &= \frac{1}{\Delta_i} \left( i\eta - \frac{\sigma^2_0k_i}{pN} \sum_{j} k_j A_{jj} \right) , \nonumber \\
		\Delta_i &= \left(i\eta -\frac{\sigma_0^2 k_i}{pN} \sum_{j}  k_j A_{jj}\right)\left(i\eta -\frac{\sigma_0^2 k_i}{pN} \sum_{j}  k_j D_{jj}\right) - \vert z \vert^2 . 
	\end{align}
	and find the corresponding expressions for the other components.
	\item By evaluating $\sum_i k_i A_{ii}$, demonstrate that the two solutions to the sets of equations are given by $\frac{\sigma_0^2}{pN} \sum_i \frac{k_i^2}{\Delta_i} = -1$ and $\sum_i k_i A_{ii} = 0$. Show that on the boundary of the eigenvalue spectrum, where both conditions are satisfied, we have
	\begin{align}
		\sum_i \frac{\sigma_0^2 k_i^2}{pN} = \vert z \vert^2.
	\end{align}
	\item Hence, derive Eq.~(\ref{circularlawhetero}). 
\end{itemize}
We claimed in the text that the `outlier' eigenvalues that correspond to hubs of the network [see Eq.~(\ref{hubeigenvalue})] have eigenvectors that are localised about these hubs. Following similar steps to Section \ref{section:localisation}, we may indeed demonstrate that such eigenvectors are localised. 
\begin{itemize}
	\item Consider Eq.~(\ref{sdaeradjacency}), and consider a hub of the network with degree $k_\mathrm{hub}$. Supposing that $k_\mathrm{hub} \gg p$ for simplicity, so that $G_\mathrm{cav}(\lambda_\mathrm{hub}) \approx 1/\lambda_\mathrm{hub}$, show that this hub gives rise to an eigenvalue pair at
	\begin{align}
		\lambda_\mathrm{hub} \approx \pm \sqrt{\frac{k_\mathrm{hub}}{p}} .
	\end{align}
	Supposing the hub is at a site $i$, show that the local resolvent close to the eigenvalue is
	\begin{align}
		G_{ii}(\omega) \approx \frac{1}{2} \frac{1}{\omega- \lambda_\mathrm{hub}} .
	\end{align}
	\item Since this eigenvalue pair corresponds to an isolated pole of the local resolvent, show that (using the reasoning of Section \ref{section:localisation})
	\begin{align}
		[G_{ii}(\lambda_\mathrm{hub})]^{-1} = 0, \Rightarrow [G_{ii}^{(j)}(\lambda_\mathrm{hub})]^{-1} = \frac{1}{p} G_{jj}^{(i)}(\lambda_\mathrm{hub}),
	\end{align}
	where $j$ is a neighbour of $i$.
	\item Using again that we expect $\lambda_\mathrm{hub}$ to be large, show that
	\begin{align}
		G_{jj}(\lambda_\mathrm{hub}+\delta) = \frac{G_{jj}^{(i)}}{G_{ii}^{(j)}} G_{ii}(\lambda_\mathrm{hub}+\delta) \approx \frac{1}{p \lambda_\mathrm{hub}^2} G_{ii}(\lambda_\mathrm{hub}+\delta) , 
	\end{align}
	where $\delta$ is a small quantity. 
	\item Hence, argue that the eigenvectors are exponentially localised  about the hub, and show that the IPR is approximately (ignoring fluctuations in the degrees of the surrounding nodes)
	\begin{align}
		q^{-1} \approx \frac{1}{4}\left[1 + \frac{1}{k_\mathrm{hub}}\frac{1}{1-\frac{p}{k_\mathrm{hub}^2}} \right].
	\end{align}
	
\end{itemize}

\newpage

\part{Additional analytical tools}\label{part:additionaltools}

\section*{Overview}
\begin{quotation}
	It is better to solve one problem five different ways, than to solve five problems one way. -- George P\'olya
\end{quotation}

In the first part of these notes, we derived a number of the most famous results in random matrix theory. We presented the Wigner semicircle law, the Girko elliptic law, the Mar\v{c}enko-Pastur law, the Kesten-McKay law, and we also discussed dynamic mean-field theory and illustrated the phenomenon of localisation, amongst many other things. All of this was accomplished with one method -- the cavity method. So, one might naturally wonder why we should need anything else. 

Firstly, the cavity method is not always the simplest method to derive results. In fact, in some cases, it has not yet been able to derive certain results that have been deduced by other approaches. Secondly, the cavity method is not the only one that is used in the literature. Indeed, different authors may believe, quite justifiably, that their pet method is the more straightforward, palatable or flexible. In which case, one must be familiar with a broad set of methodologies if one is to fully appreciate the wider literature. Finally, there is the psychological aspect. Alternative analytical methods often inspire us to think about problems in different ways, or even ask questions that might otherwise not have been natural using a different analytical framework. 

Therefore, in this second part, we exhibit some further analytical tools, which can be of great use in both random matrix theory calculations and disordered system more generally. In the present case, the focus is purely on the analytical methodology, and we therefore take the simplest examples of calculations using each method to demonstrate the procedures. This means that, in large part, we rederive results like the Wigner semicircle law that are already given in Part 1, with a few exceptions where we demonstrate some possible extensions, which are mainly discussed in the exercises.

First, we present the method that was originally used by Wigner to derive his semicircle law. Aside from being of historical interest, it serves to illustrate how combinatorics enter into random matrix theory. Although Wigner did not use them, we show how Feynman diagrams can be useful for understanding the combinatorics. We then extend this approach to the case of the generalised Mar\v{c}enko-Pastur equation. Afterwards, we discuss free probability theory, which has some deep connections with the combinatorial approach, and it also provides us with some extremely helpful tricks for handling random matrix sums and products.

Secondly, we discuss a class of methods that we will term `auxiliary field' approaches. These are the replica, supersymmetric, and path-integral approaches. Broadly speaking, these methods seek to write the resolvent matrix elements as exponential integrals over some additional `fields'. These methods largely take their inspiration from physics, specifically field theory. By writing the resolvent elements in this form, we facilitate taking the disorder average (the average over the random matrix elements). Although such approaches are a bit superfluous for deriving the average eigenvalue density in the classic cases that are the main focus of this document, they drastically simplify more complicated computations. In each case, we discuss precisely in which contexts each method really excels, which will often be for computations outside the scope of these notes. It is the hope that as the reader progresses to such harder and more modern problems, Part 2 of these notes provides a helpful stepping stone.

\newpage

\section{Wigner's combinatorial approach to the semicircle law}\label{section:wignerapproach}
Working in the 1950s, much of the methodology in Part 1 of these notes was not available to Wigner. As such, it was necessary for him to be somewhat more resourceful when deriving his eponymous semicircle law. Wigner's insight \cite{wigner1955characteristic, wigner1993characteristic} was that we can relate the moments of the eigenvalue density $\rho(\omega)$ to the moments of the matrix $\underline{\underline{J}}$. For example, we have from elementary linear algebra that $N^{-1}\mathrm{Tr} \underline{\underline{J}} = N^{-1}\sum_\nu \lambda_\nu$. More generally, we have 
\begin{align}
	N^{-1} \mathrm{Tr} \underline{\underline{J}}^r = N^{-1}\sum_{\nu} \lambda_{\nu}^r. \label{wignerobs}
\end{align}
Furthermore, when the matrix elements obey the statistics in Eq.~(\ref{statisticswigner}), one finds that the objects $N^{-1}\mathrm{Tr} \underline{\underline{J}}^r = \sum_{i_1,i_2,\cdots, i_r} J_{i_1 i_2}J_{i_2 i_3}\cdots J_{i_r i_1}$ concentrate on their average by the law of large numbers, so that $N^{-1}\mathrm{Tr} \underline{\underline{J}}^r  \to N^{-1} \langle  \mathrm{Tr} \underline{\underline{J}}^r\rangle$ for $N \to \infty$. That is, we may write
\begin{align}
	\int d\omega \,\omega^r \rho(\omega) = N^{-1} \langle \mathrm{Tr} \underline{\underline{J}}^r \rangle.  \label{momenttrace}
\end{align}
We have thus reduced the problem to that of computing the traces of products of the random matrix $\underline{\underline{J}}$. Our strategy will now be to find a recursive relation for these moments, and we will thus have uniquely determined the form of $\rho(\omega)$. 

Let us consider the first few values of $r$ in Eq.~(\ref{momenttrace}). For $r = 1$, one finds rather simply
\begin{align}
	\int d\omega \, \omega \rho(\omega) = N^{-1} \left\langle \sum_{i} J_{ii} \right\rangle = N^{-1}  \sum_{i}\left\langle J_{ii} \right\rangle = 0,
\end{align}
where we have used the statistics in Eq.~(\ref{statisticswigner}). In fact, more generally, we can see that the odd moments vanish since the traces for odd $r$ involve averages of odd numbers of the $J_{ij}$ multiplied together. Even if we consider instances where some of the indices take the same values, we always end up with an average of an odd number of $J_{ij}$s, which evaluates to zero. To be more clear, let us take the example $r = 3$. One has 
\begin{align}
	&N^{-1} \left\langle \sum_{ijk} J_{ij}J_{jk}J_{ki} \right\rangle = N^{-1}  \sum_{ijk}\left\langle J_{ij} J_{jk}J_{ki} \right\rangle \nonumber \\
	&= \sum_{(i,j,k)} \left\langle J_{ij} \right\rangle \left\langle J_{jk} \right\rangle \left\langle J_{ki} \right\rangle +3 \sum_{(i,j)}\left\langle J_{ij}  J_{ji} \right\rangle \left\langle J_{ii} \right\rangle +  \sum_{i} \left\langle J_{ii}^3 \right\rangle = 0 ,
\end{align}
where $(i,j,k)$ indicates combinations where none of the indices are equal. Let us then turn our attention to the cases where $r$ is even. In the case $r=2$, we have trivially [noting the symmetry $J_{ij} = J_{ji}$ and the variance given in Eq,~(\ref{statisticswigner})]
\begin{align}
	N^{-1}  \sum_{ij}\left\langle J_{ij} J_{ji} \right\rangle = 1,
\end{align}
where we ignore the subleading contribution due to the diagonal elements. For $r=4$, we have something more non-trivial. 

The following observation makes our lives easier as we proceed. In order for a combination of indices $(i,j)$ [of a factor $J_{ij}$] to contribute to the sum, it must be repeated. That is, $\langle J_{12} J_{23} J_{32} J_{21}\rangle = \langle J_{12}^2 \rangle \langle J_{23}^2\rangle= 1/N^2$ contributes to the sum, but $\langle J_{12} J_{24} J_{45} J_{51}\rangle =\langle J_{12} \rangle \langle J_{24} \rangle \langle J_{45}\rangle \langle J_{51}\rangle = 0$ does not. However, if indices are repeated an odd number of times, the term vanishes -- e.g. $\langle J_{12}^3 J_{23} \rangle = 0$. We therefore have the following remaining terms in the case $r = 4$
\begin{align}
	N^{-1}  \sum_{ijkl}\left\langle J_{ij} J_{jk}J_{kl}J_{li} \right\rangle= N^{-1}\sum_{(i,j, k)} \left[\left\langle J_{ij}^2 \right\rangle  \left\langle J_{ik}^2 \right\rangle + \left\langle J_{ij}^2 \right\rangle  \left\langle J_{jk}^2 \right\rangle \right]  +N^{-1} \sum_{(i,j)}\left\langle J_{ij}^4 \right\rangle . \label{4pairings}
\end{align}
In each of these terms, we have constrained different pairs of the matrix elements $J_{ij}$, $J_{jk}$, $J_{kl}$ and $J_{li}$ to have the same indices. In the first, we have paired $J_{ij}$ with $J_{jk}$ and $J_{kl}$ with $J_{li}$. In the second, we instead pair $J_{ij}$ with $J_{li}$ and $J_{kl}$ with $J_{jk}$. Finally, in the last term, we pair $J_{ij}$ with $J_{kl}$ and $J_{jk}$ with $J_{li}$, and in so doing, we require either $j= l$ and $i=k$. Since the latter average vanishes identically, we are left with $\langle J_{ij}^4 \rangle$. 

Evaluating Eq.~(\ref{4pairings}), one arrives at 
\begin{align}
	N^{-1}  \sum_{ijkl}\left\langle J_{ij} J_{jk}J_{kl}J_{li} \right\rangle = 2 + 1/N. 
\end{align}
We thus see that one of these terms vanishes in the thermodynamic limit. As we go to higher orders, it becomes inconvenient to continue the evaluation of each term by hand. It is better if we can spot a pattern and thus evaluate successive traces systematically. Wigner's approach was to reason abstractly about the `non-crossing paths' to be taken through the set of indices to make a non-vanishing pairing. Similar to Ref. \cite{bray1979evidence}\footnote{We note that by making a diagrammatic series expansion of the resolvent itself, Bray and Moore \cite{bray1979evidence} provide perhaps one of the most succinct derivations of the semicircle law.}, we will be more explicit here, and make Feynman diagrams to keep track of the pairings. For example, 

\begin{figure}[h]
	\centering 
	\includegraphics[scale = 0.3]{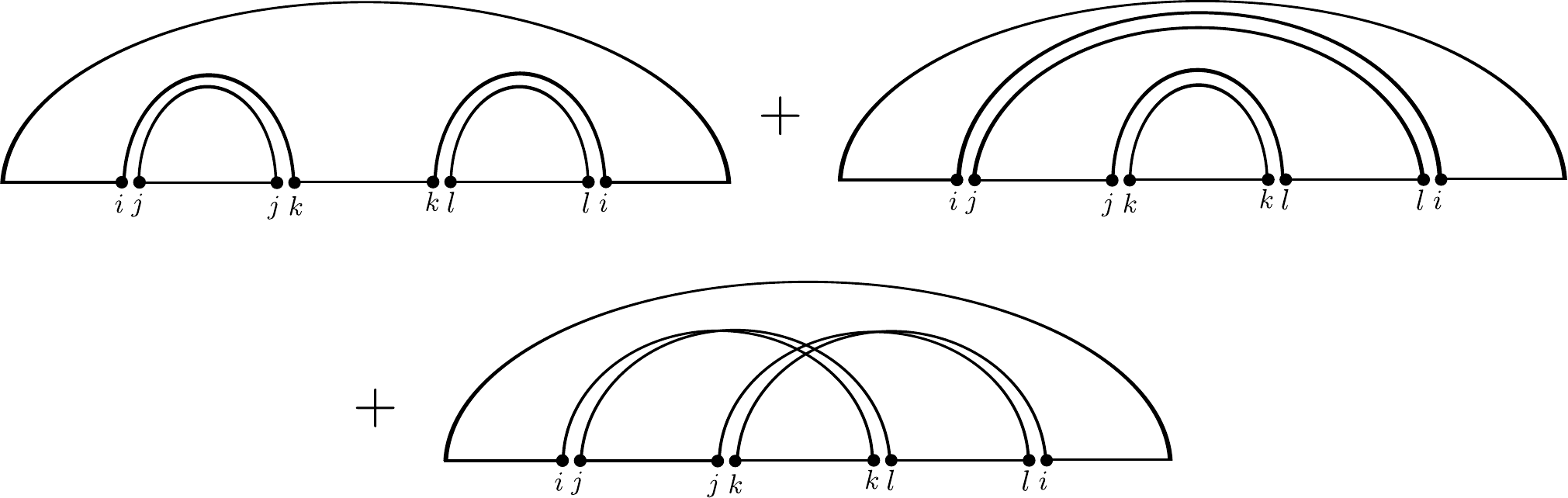}
	\captionsetup{justification=raggedright,singlelinecheck=false}
	\captionsetup{labelformat=empty}
\end{figure}

In these diagrams, we have represented the indices of each matrix with a pair of dots. Matrices that are multiplied together in the product $N^{-1}\langle\mathrm{Tr}\underline{\underline{J}}^4\rangle$ are represented by neighbouring pairs of dots in the diagram, and the dots that share an index (owing to their position in the sum) are connected with a horizontal line. We represent the different `pairings' of the indices by joining the indices that are constrained to be the same with arcs. The terms are in the same order as they are in Eq.~(\ref{4pairings}). These are known as rainbow diagrams, and they first appeared in the literature surrounding QCD \cite{brezin1978planar}. We note that they can also be derived using the path integral method [see Section \ref{section:pathintegral}].

We thus see that the non-vanishing terms in the trace for $r = 4$ correspond to the diagrams where the arcs do not cross. This is a generic property as we go to higher values of $r$. In general, diagrams are proportional to $N^{E-A-1}$, where $E$ is the number of disconnected (by arcs) sets of horizontal edges, and $A$ is the number of double arcs. This is because each disconnected set of horizontal edges evaluate to $\propto\sum_{i}\delta_{ii}$, and each double arc indicates a factor of $\langle J^2\rangle \propto 1/N$. When arcs cross, this connects sets of horizontal edges that would otherwise not be connected, and thus reduces the power of $N$ of the diagram. Wigner \cite{wigner1955characteristic} also noted that it is only the non-crossing pairings that contribute in the thermodynamic limit. As a passing aside, we point out that it is precisely this non-crossing combinatorics that connects random matrix theory with free probability theory \cite{mingo2017free} (see Section \ref{section:freeprob}).

The non-crossing diagrams corresponding to $N^{-1}\mathrm{Tr}\underline{\underline{J}}^6$ are
\begin{figure}[h]
	\centering 
	\includegraphics[scale = 0.3]{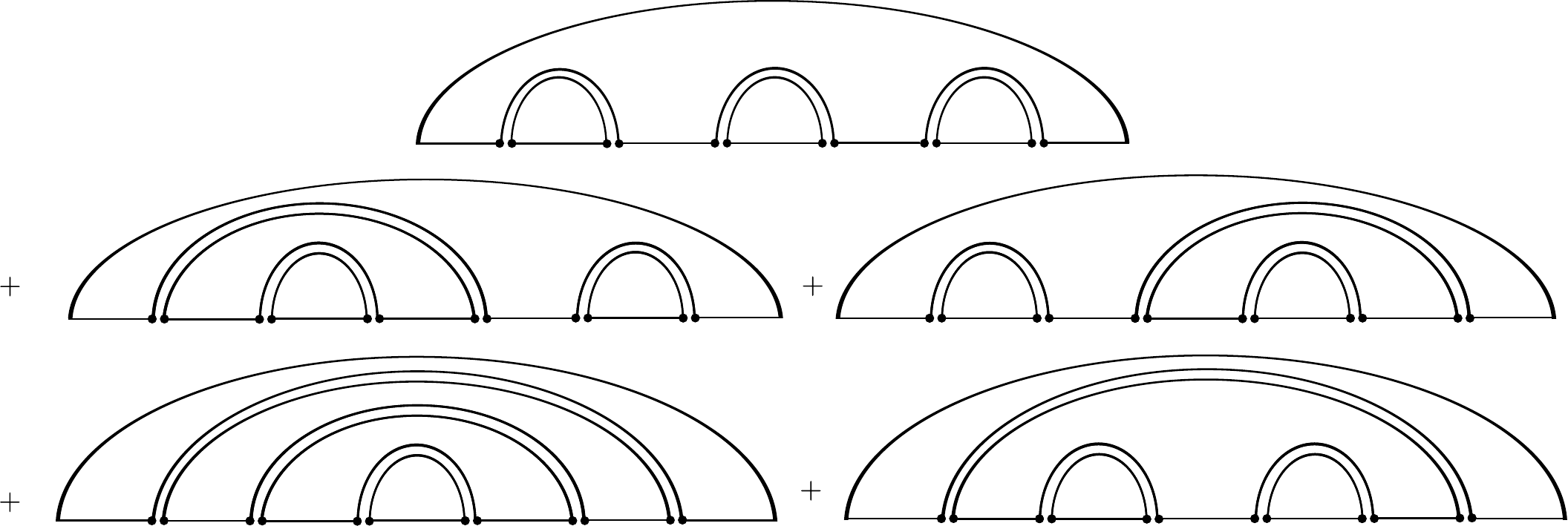}
	\captionsetup{justification=raggedright,singlelinecheck=false}
	\captionsetup{labelformat=empty}
\end{figure}

So, we have now translated the problem of finding the moments of the eigenvalue density $\rho(\omega)$ into a combinatorial one. That is, the moment $\int d\omega\, \omega^r\rho(\omega)$ is equal (in the limit $N\to \infty$) to the number of non-crossing pairings of $r$ pairs of dots. Let us denote the number of non-crossing diagrams associated with a particular value of $r$ as $t_r = N^{-1} \langle\mathrm{Tr}\underline{\underline{J}}^r\rangle$ (where this equality holds for $N \to \infty$). The question remains -- what are the numbers $t_r$? 

Following Wigner, we approach the problem recursively. We may split each set of diagrams for fixed $r$ into different groups according to the length of the first arc. If the first arc (i.e. the arc whose leftmost point begins the first pair of dots) skips over $r-2$ pairs of dots to the last available pair of dots, then underneath this arc, we may insert all of the diagrams associated with the $r-2$ term (i.e. $t_{r-2}$ diagrams). One can see that the last two diagrams in the $r=6$ case above indeed enclose the diagrams of the $r= 4$ case beneath one outer arc. If the arc instead skips over only $2k$ pairs of dots, then we may fit $t_{2k}$ arrangements of arcs underneath the first arc. Since only $r- 2k - 2$ pairs of dots remain, we then have $t_{r- 2(k-1)}$ arrangements of arcs that can occupy the remaining sites. That is, writing $r = 2 m$, we have the recursion relation for $m>1$
\begin{align}
	t_{2m} = \sum_{k = 1}^{m-1} t_{2(m-k)} t_{2(k-1)} ,
\end{align}
where $t_{2}=1$ and $t_0 = 1$. To evaluate $t_r$ in general, we utilise the generating function
\begin{align}
	t(x) = \sum_{m=0}^\infty x^{2m} t_{2m}.
\end{align}
Using the recursion relation, we then have 
\begin{align}
	t(x) = 1+ x^2 +\sum_{m=2}^\infty x^{2m}  \sum_{k = 1}^{m-1} t_{2(m-k)} t_{2k} = 1 + x^2 [t(x)]^2 . 
\end{align}
We take a moment to note that $t(x)$ is related to the trace of the resolvent via $G(z) = z^{-1} t(1/z)$. This can be seen using a series expansion of the resolvent [see Eq.~(\ref{resseries})]. Indeed, one can also resum a diagrammatic series for the resolvent directly (see Fig. \ref{fig:rainbowdiagrams}). Given our discussion in Section \ref{section:semicircle}, we could now simply use the inverse Stieltjes transform of Eq.~(\ref{densefromres}), and obtain the semicircle law. That is, we have in some sense already finished the calculation.

However, for historical interest, we proceed with Wigner's own derivation, which did not exploit the inverse Stieltjes transform. We do this partially to show just how useful the inverse Stieltjes transform really is! Solving for $t(x)$, we obtain
\begin{align}
	t(x) = \frac{1 - \sqrt{1-4x^2}}{2 x^2}. 
\end{align}
Expanding this as a series in $x$, we finally find that
\begin{align}
	t_{2m} = \frac{1}{m+1} \binom{2m}{m},
\end{align}
which are exactly the Catalan numbers. Now, we must find the function $\rho(\omega)$ that satisfies 
\begin{align}
	\int d\omega\, \omega^{2m}\rho(\omega) = \frac{1}{m+1} \binom{2m}{m}.
\end{align}
Wigner noticed that if we write $\frac{ds}{d\omega}=\omega\rho(\omega)$, we can integrate by parts to show that
\begin{align}
	\int d\omega s \omega^{2m} = -\frac{2}{m+2} \frac{(2m)!}{m!(m+1)!}.
\end{align}
We may also show that 
\begin{align}
	4\int d\omega \omega^{2m-1} \frac{ds}{d\omega} -  	\int d\omega \omega^{2m+1} \frac{ds}{d\omega} &= \frac{(2m)!}{m!(m+1)!}\left[4- \frac{2(2m-1)}{m+2} \right]\nonumber \\
	&= \frac{6}{m+2} \frac{(2m)!}{m!(m+1)!}.
\end{align}
We therefore arrive at the differential equation
\begin{align}
	\frac{1}{6}\left(\frac{4}{\omega}- \omega \right) \frac{ds}{d\omega} = -\frac{1}{2} s ,
\end{align}
which we may integrate to obtain 
\begin{align}
	s = C (4 -\omega^2)^{3/2},
\end{align}
where $C$ is a constant of integration. Using the relation $\rho(\omega) = \omega^{-1}\frac{ds}{d\omega}$, we obtain the semicircle law, up to the normalisation constant, which can be determined by simple integration.

We thus see that a good amount of ingenuity was required by Wigner to deduce $\rho(\omega)$ from its moments without using the inverse Stieltjes transform. With this being said, the combinatorial approach used by Wigner, when combined with the inverse Stieltjes transform and Feynman diagrams, is a very powerful and adaptable tool. We demonstrate its flexibility now in Section \ref{section:mpgeneral} and also in Section \ref{section:pathintegral}.

\section{Combinatorial approach to the Mar\v{c}enko-Pastur equation}\label{section:mpgeneral}
As was stated in Section \ref{section:mplaw}, it is impractical to derive the general Mar\v{c}enko-Pastur relation in Eq.~(\ref{generalmplaw}) using the cavity approach. This owes to the fact that we rely on a diagonal (or close to diagonal) structure of the population covariance matrix when we use this approach. If we wish to be more general, it is more convenient to use a different approach. 

One elegant method is to exploit the S-transform of free probability theory, which allows us to relate the resolvent of the product of two (possibly random) matrices to the resolvents of the two matrices individually. This will be discussed later in Section \ref{section:freeprob}. Here, we will demonstrate the flexibility of the diagrammatic approach that we discussed in the previous section by extending it to the general Mar\v{c}enko-Pastur problem. Later, in Section \ref{section:pathintegral}, we will also discuss how diagrammatic techniques can also be used to handle a range of other ensembles in the context of the path-integral formalism.

We wish to evaluate the trace of the following resolvent
\begin{align}
	\underline{\underline{A}} = \left[z \underline{\underline{\id}}_N - \frac{1}{n}\underline{\underline{X}}\underline{\underline{X}}^T \right]^{-1}. 
\end{align}
It will be helpful also to consider the resolvent
\begin{align}
	\underline{\underline{D}} = z\left[z \underline{\underline{\id}}_n - \frac{1}{n}\underline{\underline{X}}^T\underline{\underline{X}} \right]^{-1}. 
\end{align} 
Recalling $A = N^{-1}\mathrm{Tr}\underline{\underline{A}}$ and $D = n^{-1}\mathrm{Tr}\underline{\underline{D}}$, we will also use the general relation
\begin{align}
	D = 1-\gamma + \gamma z A . \label{dintermsofa}
\end{align}
This relation follows from the fact that $\underline{\underline{X}}\underline{\underline{X}}^T$ and $\underline{\underline{X}}^T\underline{\underline{X}}$  have the same non-zero eigenvalues. However, if $n<N$, then $\underline{\underline{X}}^T$ has a non-zero kernel, and $\underline{\underline{A}}$ has precisely $N-n$ zero eigenvalues, meaning that $\underline{\underline{A}}$ has a pole $(1-\gamma^{-1})/z$, which is not possessed by $D/z$. Conversely, if $N<n$, then $D/z$ posses a pole $(1-\gamma)/z$ that is not possessed by $A$. 

We suppose that the random variables have a generic correlator
\begin{align}
	\langle X_{i}^\alpha X_{j}^\beta \rangle = \delta_{\alpha \beta} \Sigma_{ij},
\end{align}
for an arbitrary $\Sigma_{ij}$, so that each realisation of the vector $\underline{X}^\alpha$ is independent. Further, we suppose that the matrix $\underline{\underline{\Sigma}}$ has an empirical eigenvalue density $\rho_\Sigma(\omega)$. For this to be a bounded spectrum, we require the diagonal elements of $\underline{\underline{\Sigma}}$ to be finite, and the off diagonal elements scale as $1/\sqrt{N}$. 

Let us then follow a similar line of reasoning to Wigner and consider a series expansion of $A(z) = N^{-1} \sum_i A_{ii}$. This reads
\begin{align}
	A =& \frac{1}{z} + \frac{1}{z^2 N n} \sum_{\alpha;i} X_i^\alpha X_i^\alpha +  \frac{1}{z^3 N n^2} \sum_{\alpha\beta;ij} X_i^\alpha X_j^\alpha X_j^\beta X_i^\beta \nonumber \\
	&+ \frac{1}{z^4 N n^3} \sum_{\alpha\beta\gamma;i j k} X_i^\alpha X_j^\alpha X_j^\beta X_k^\beta X_k^\gamma X_i^\gamma + \cdots . \label{resseriesmp}
\end{align}
Supposing that each of the terms in the sum each individually concentrates on its mean (one can convince oneself of this fairly easily by examining the first few terms), we may then consider the ensemble average $A \approx \langle A \rangle$ when $N\to\infty$.

For ease of writing, we will assume in the following argument that the variables $X_{i}^\alpha$ are Gaussian random variables, which enables us to use Wick's/Isserlis' theorem directly. This simplifies the argument considerably. To generalise what follows to non-Gaussian matrices, one must make a scaling assumption of the higher-order cumulants of $X_i^\alpha$. One can then deduce from this scaling that one may effectively use Wick's theorem, understanding that the discarded terms are subleading in $N$. This is in some sense what we did in the case of Wigner's semicircle above.

Under the assumption that we may use Wick's theorem, let us evaluate some of the terms in Eq.~(\ref{resseriesmp}). Taking the term proportional to $1/z^3$ for example, we have
\begin{align}
	\frac{1}{n^2N}\sum_{\alpha\beta;ij} \left\langle X_i^\alpha X_j^\alpha X_j^\beta X_i^\beta \right\rangle &= \frac{1}{n^2N}\sum_{\alpha\beta;ij}\Bigg[ \left\langle X_i^\alpha X_j^\alpha\right\rangle \left\langle X_j^\beta X_i^\beta \right\rangle  \nonumber \\
	&+\left\langle X_i^\alpha X_i^\beta  \right\rangle \left\langle X_j^\alpha X_j^\beta \right\rangle+  \left\langle X_i^\alpha X_j^\beta\right\rangle \left\langle  X_j^\alpha X_i^\beta \right\rangle \Bigg] \nonumber \\
	&= \frac{1}{n^2N}\sum_{\alpha\beta;ij} \left[\Sigma_{ij} \Sigma_{ji} + \delta_{\alpha\beta} \Sigma_{ii}\Sigma_{jj} + \delta_{\alpha\beta} \Sigma_{ij}\Sigma_{ji}  \right] .
\end{align}
We see that the first two of these terms are of the order $O(N^0)$, whereas the third scales as $1/N$ (assuming that $n$ and $N$ are of the same order of magnitude). We may represent these Wick pairings diagrammatically as follows
\begin{figure}[H]
	\centering 
	\includegraphics[scale = 0.3]{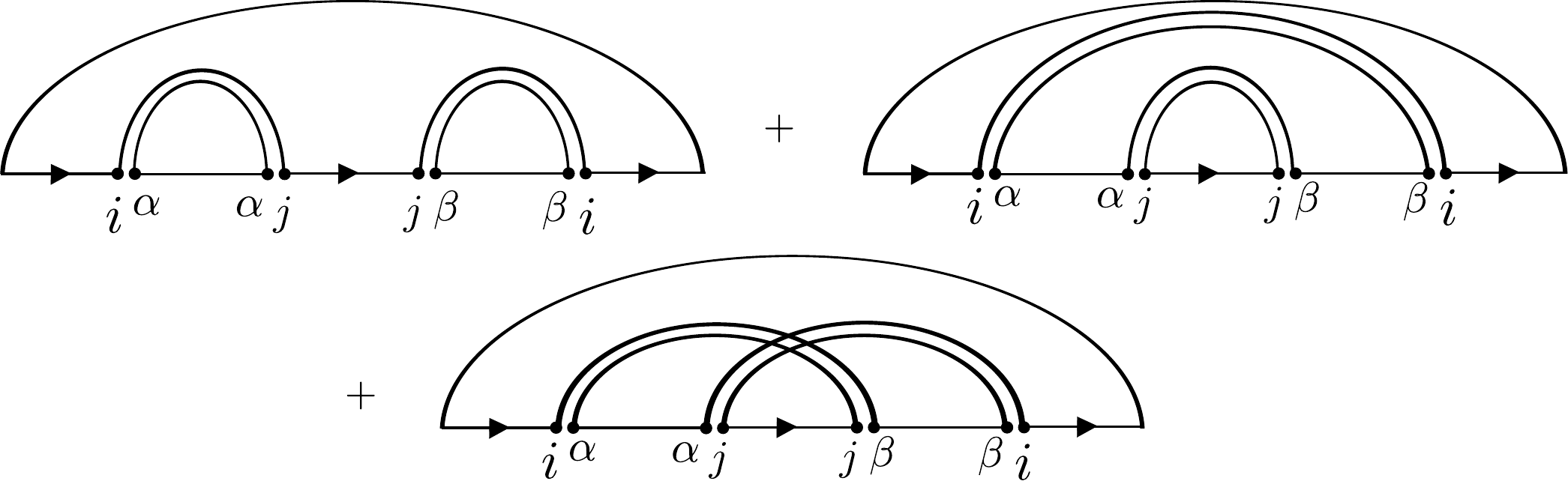}
	\captionsetup{justification=raggedright,singlelinecheck=false}
	\captionsetup{labelformat=empty}
\end{figure}
Here, we represent each $X_{i}^\alpha$ by a pair of points positioned together, similar to the Wigner case. Now, we connect lower indices that are constrained to be the same (by virtue of their position in the matrix product) by a directed horizontal edge, whereas pairs of upper indices that are constrained to be the same are connected by a plain edge. Crucially, we again connect the nodes corresponding to factors of $X_i^\alpha$ that are Wick-paired together with double arcs. Once again, we find that only the planar (non-crossing) diagrams survive in the limit $N \to \infty$. 

We see that the diagrams here are very similar to those in Section \ref{section:wignerapproach}, with the exception that edges alternate between being directed and undirected. To find $A$, we must sum the full set of planar diagrams that begin and end with directed edges. It will help us also to consider the diagrammatic series of $D$, which corresponds to the set of diagrams that begin and end with undirected edges. More precisely, we have
\begin{align}
	D =& 1 + \frac{1}{z n^2} \sum_{\alpha;i} X_i^\alpha X_i^\alpha +  \frac{1}{z^2 n^3} \sum_{\alpha\beta;ij} X_i^\alpha X_i^\beta X_j^\beta X_j^\alpha \nonumber \\
	&+ \frac{1}{z^3 n^4} \sum_{\alpha\beta\gamma;i j k} X_i^\alpha X_i^\beta X_j^\beta X_j^\gamma X_k^\gamma X_k^\alpha + \cdots . \label{resseriesmpD}
\end{align}
Taking the term proportional to $1/z^2$ for example, we have
\begin{align}
	\frac{1}{n^3}\sum_{\alpha\beta;ij} \left\langle X_i^\alpha X_i^\beta X_j^\beta X_j^\alpha  \right\rangle &= \frac{1}{n^3}\sum_{\alpha\beta;ij}\bigg[ \left\langle  X_i^\alpha X_i^\beta\right\rangle \left\langle X_j^\beta X_j^\alpha  \right\rangle \nonumber \\
	&+\left\langle X_i^\alpha X_j^\alpha  \right\rangle \left\langle   X_i^\beta X_j^\beta\right\rangle +  \left\langle  X_i^\alpha X_j^\beta\right\rangle \left\langle  X_i^\beta X_j^\alpha  \right\rangle \bigg] \nonumber \\
	&= \frac{1}{n^3}\sum_{\alpha\beta;ij} \left[\delta_{\alpha\beta}\Sigma_{ii} \Sigma_{jj} +  \Sigma_{ij}\Sigma_{ji} + \delta_{\alpha\beta} \Sigma_{ij}\Sigma_{ji}  \right] .
\end{align}
The diagrammatic representations in this case are given by
\begin{figure}[H]
	\centering 
	\includegraphics[scale = 0.3]{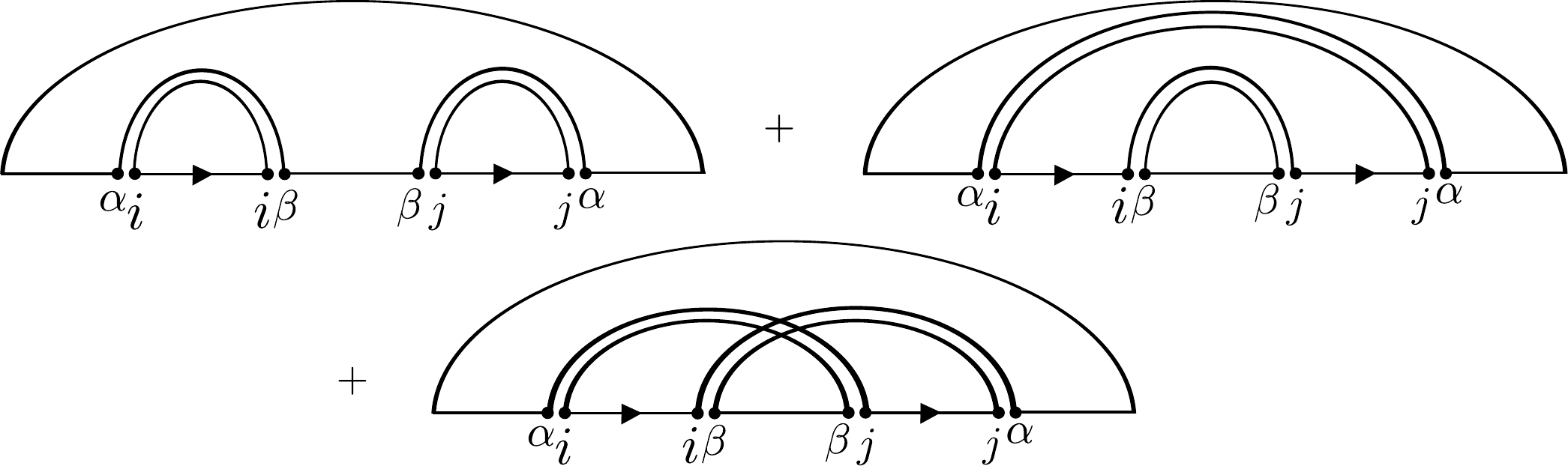}
	\captionsetup{justification=raggedright,singlelinecheck=false}
	\captionsetup{labelformat=empty}
\end{figure}
The trace $D$ is then given by the sum of all such diagrams, i.e., all those that are planar (non-crossing) and start and end with undirected edges.

Let us now then consider the full set of diagrams for $A$, i.e. the full set of non-crossing diagrams beginning and ending in directed edges. Somewhat differently to Wigner, we do not attempt to find a recursion relation, but instead resum the series directly. To do this, we split the full series into sub-series as follows
\begin{figure}[H]
	\centering 
	\includegraphics[scale = 0.26]{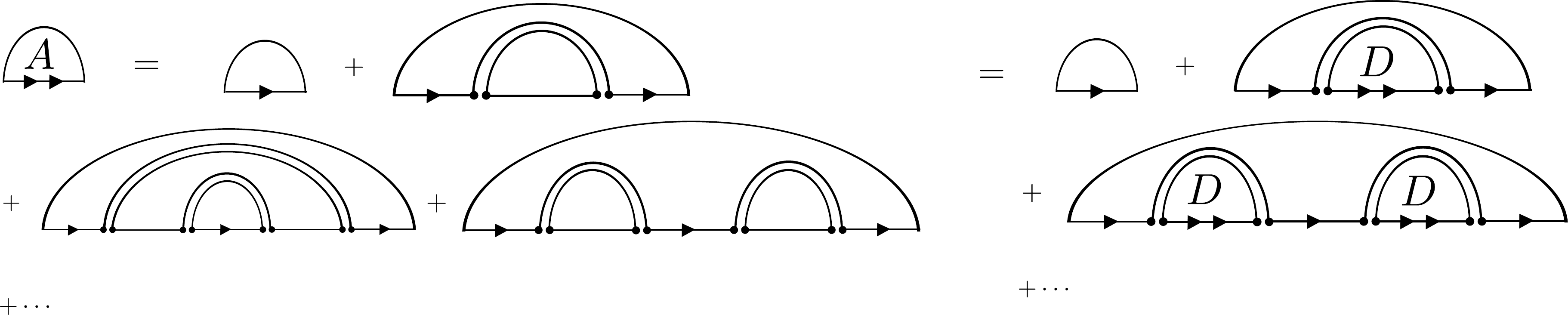}
\end{figure}
Here, we have identified sub-series containing all the diagrams that possess the same number of `external arcs'. That is, all of the possible diagrams whose first two nodes are connected to the last two have 1 external arc. If the first pair of nodes are connected to any other pair than the last, and then the next pair of nodes directly after this is connected to the last, we say that such a diagram has two external arcs, and so on. For example, in the left-hand series, we would say that the second and third diagram have a single external arc, while the fourth has two and the first has zero.

We recognise that to produce, for example, the sub-series with a single external arc, we can take the full series for $D$ (i.e. diagrams that begin and end with undirected edges) and put a single pair of new nodes at either end, connected by an arc. Similar statements can be made about the series with two eternal arcs, and so on. This is what is represented by the right-hand diagrammatic series above. We have represented the full series of diagrams for the trace $D$ as an edge with a double arrow.

Being careful to keep track of the factors of $1/z$ and $1/n$, we therefore arrive at the geometric series
\begin{align}
	A = \frac{1}{z} + \frac{D}{z^2 N} \sum_{i} \Sigma_{ii} + \frac{D^2}{z^3 N} \sum_{ij} \Sigma_{ij}\Sigma_{ji} + \frac{D^3}{z^4 N} \sum_{ijk} \Sigma_{ij}\Sigma_{jk}\Sigma_{ki} + \cdots .
\end{align}
Now, we may use the fact that each term involves traces of powers of $\underline{\underline{\Sigma}}$. Using Wigner's observation in Eq.~(\ref{wignerobs}), we find
\begin{align}
	A = \int d\nu \rho_\Sigma(\nu) \left[  \frac{1}{z} + \frac{D \nu}{z^2}  + \frac{D^2\nu^2}{z^3 }  + \frac{D^3\nu^3}{z^4 }  + \cdots \right] = \int d\nu \frac{ \rho_\Sigma(\nu)}{z - \nu D}. 
\end{align}
Finally, using the relation in Eq.~(\ref{dintermsofa}), we arrive at the general Mar\v{c}enko-Pastur equation in Eq.~(\ref{generalmplaw}). For the sake of deriving the result in Eq.~(\ref{ledoitpeche}), it is helpful instead to note the following series, which can also be obtained from diagrams by using $\frac{1}{N}\mathrm{Tr}\left(\underline{\underline{A}}\,\underline{\underline{\Sigma}}\right) = \frac{1}{N}\left\langle\mathrm{Tr}\left(\underline{\underline{A}}\,\underline{\underline{\Sigma}}\right) \right\rangle$ for $N \to \infty$
\begin{align}
	\frac{1}{N}\sum_{ij}A_{ij}\Sigma_{ji} = \frac{1}{z N}\sum_{ij}\Sigma_{ji} + \frac{D}{z^2 N } \sum_{ij}\Sigma_{ij}\Sigma_{ji} + \frac{D^2}{z^3 N} \sum_{ijk} \Sigma_{ik}\Sigma_{kj}\Sigma_{ji} + \frac{D^3}{z^4 N} \sum_{ijkl} \Sigma_{ik}\Sigma_{kl}\Sigma_{lj}\Sigma_{ji} + \cdots .
\end{align}
We therefore obtain [which ultimately leads to Eq.~(\ref{generalmplaw})]
\begin{align}
	\sum_\mu\lambda^\Sigma_\mu \, (\underline{v}^{(\mu)})^T \underline{\underline{A}}(z) \underline{v}^{(\mu)} &= \frac{1}{N}\mathrm{Tr}\left(\underline{\underline{A}}\,\underline{\underline{\Sigma}}\right) = \sum_\mu\lambda^\Sigma_\mu \left[ \frac{1}{z}+ \frac{D}{z^2 } \lambda^\Sigma_\mu + \frac{D^2}{z^3 }(\lambda^\Sigma_\mu)^2  + \frac{D^3}{z^4 } (\lambda^\Sigma_\mu)^3 + \cdots \right] \nonumber \\
	&=\int d\nu \frac{\nu \rho_\Sigma(\nu)}{z - \nu \left[ 1-\gamma + \gamma z A(z)\right]} . \label{helpfullp} 
\end{align}
We thus see that diagrammatics permits us not only to compute the resolvent for a variety of ensembles, but it also allows us to evaluate more exotic quantities. Indeed, diagrammatic series have been used to evaluate such quantities as the 2-point eigenvalue density correlations \cite{brezin1994correlation} and the left-right eigenvector overlap in non-hermitian ensembles \cite{janik1997non, nowak2018probing}. We also show in Section \ref{section:pathintegral} how diagrammatics can be used in the context of the path integral formalism to perform perturbative expansions in model parameters other than $1/N$.

\newpage

\section{RMT and free probability theory}\label{section:freeprob}
Free probability theory aims to extend the idea of statistical independence to random quantities that are non-commutative. This theory, which was developed originally by Dan Voiculescu \cite{voiculescu1983symmetries} in an abstract pure mathematical context, clearly should have great relevance for random matrices. In fact, the connection between free probability and random matrix theory is deep. In some sense, Wigner's semicircle law can be regarded as a central limit theorem for free random variables \cite{voiculescu1991limit}. It is also no accident that non-crossing diagrams emerged in our combinatorial derivation of the semicircle and Mar\v{c}enko-Pastur laws above. Non-crossing partitions are an integral aspect of free probability theory, being necessary to define the so-called `free cumulants'.  

In this Section, we provide a recipe, using Voiculescu's $R$- and $S$-transforms \cite{voiculescu1986addition, voiculescu1987multiplication, voiculescu1991limit}, to compute the eigenvalue density of sums and products of free random matrices, whose resolvents we already know separately. We do not prove any of the results used here. For this, the reader is better directed to pedagogical introductions to free probability theory, in particular Refs. \cite{mingo2017free, nica2006lectures}, and to Ref. \cite{potters2020first}, where many more examples of the application of free probability theory in RMT are given. What we provide here is merely a brief user's guide, and some selected examples to highlight the remarkable utility of the tools of free probability.

We note that while we only discuss Hermitian matrices here, efforts have also been made to extend the concepts of free probability to non-Hermitian matrices \cite{burda2015quaternionic, janik1997non2, burda2011multiplication}.

\subsection{When are two matrices `asymptotically free'?}

To be considered asymptotically free with respect to each other, and therefore for the theorems stated below to apply, two square matrices $\underline{\underline{A}}$ and $\underline{\underline{B}}$, each of dimension $N$, must satisfy
\begin{align}
	\lim_{N\to \infty}\left\langle \frac{1}{N}\mathrm{Tr}\left[P_1\left(\underline{\underline{A}} \right) P_2\left(\underline{\underline{B}} \right)P_3\left(\underline{\underline{A}} \right) P_4\left(\underline{\underline{B}} \right) \cdots  \right] \right\rangle  = 0
\end{align}
where the product may end in $P_k(\underline{\underline{A}})$ or $P_k(\underline{\underline{B}})$, and $P_{r}(\cdot)$ are polynomials satisfying
\begin{align}
	\lim_{N\to\infty}\left\langle \frac{1}{N}\mathrm{Tr}\left[P_k\left(\underline{\underline{A}} \right) \right] \right\rangle = \lim_{N\to\infty}\left\langle \frac{1}{N}\mathrm{Tr}\left[P_r\left(\underline{\underline{B}} \right)  \right] \right\rangle = 0.
\end{align}
That is, in words, the alternating product of centred polynomials of mutually free  matrices is centred. As a general rule, if two matrices have independent entries, and at least one is unitarily invariant, then they will satisfy the requirement for mutual freeness. However, we note that independence alone will not suffice as a condition for freeness. For example, suppose $\underline{\underline{A}}$ and $\underline{\underline{B}}$ are two independent diagonal random matrices with centred diagonal elements, i.e.
\begin{align}
	A_{ij} = a_i \delta_{ij}, \hspace{1cm} \langle a_i \rangle = 0,
\end{align}
and similarly for $\underline{\underline{B}}$. In this case, we may compute, e.g.
\begin{align}
	\lim_{N\to\infty}\left\langle \frac{1}{N}\mathrm{Tr}\left[\underline{\underline{A}}\, \underline{\underline{B}} \,\underline{\underline{A}} \, \underline{\underline{B}}  \right] \right\rangle  = \lim_{N \to \infty}\frac{1}{N} \sum_{i} \langle a_i^2 \rangle \langle b_i^2 \rangle \neq 0.
\end{align}
However, if on the other hand $\underline{\underline{A}}$ and $\underline{\underline{B}}$ are two independent GUE matrices, for example, we would instead find
\begin{align}
	\lim_{N\to\infty}\left\langle \frac{1}{N}\mathrm{Tr}\left[\underline{\underline{A}}\, \underline{\underline{B}} \,\underline{\underline{A}} \, \underline{\underline{B}}  \right] \right\rangle  &= \lim_{N \to \infty}\frac{1}{N} \sum_{ijkl} \langle A_{ij}A_{kl}\rangle  \langle B_{jk}B_{li} \rangle \nonumber \\
	&= \lim_{N \to \infty}\frac{1}{N^3} \sum_{ijkl} \delta_{il}  \delta_{jk} \delta_{jl} \delta_{kl} = \lim_{N \to \infty} \frac{1}{N^2} = 0 .
\end{align}
Indeed, if $\underline{\underline{B}}$ is diagonal, and $\underline{\underline{A}}$ is GUE, we have
\begin{align}
	\lim_{N\to\infty}\left\langle \frac{1}{N}\mathrm{Tr}\left[\underline{\underline{A}}\, \underline{\underline{B}} \,\underline{\underline{A}} \, \underline{\underline{B}}  \right] \right\rangle  &= \lim_{N \to \infty}\frac{1}{N} \sum_{ij} \langle A_{ij}A_{ji}\rangle  \langle B_{j}B_{i} \rangle \nonumber \\
	&= \lim_{N \to \infty}\frac{1}{N^2} \sum_{ij} \delta_{ij} b_i^2 = \lim_{N \to \infty} \frac{1}{N} = 0 .
\end{align}
So, assuming that the freeness property holds, for the two matrices $\underline{\underline{A}}$ and $\underline{\underline{B}}$, we will now describe how one may compute the eigenvalue density of their sum and their product. 

\subsection{$R$-transform: sums of free random matrices}
Let us first understand the $R$-transform \cite{voiculescu1986addition}, which permits us to compute the eigenvalue density of the sum of two free random matrices. We first define the `Blue's function' $B(z)$, which is the functional inverse of the Green's function 
\begin{align}
	B(G(z)) = z.
\end{align}
The $R$-transform is then defined as
\begin{align}
	R(z) = B(z) - \frac{1}{z}. 
\end{align}
Substituting $z = G(w)$, we then see that
\begin{align}
	R(G) = w(G) - \frac{1}{G}.
\end{align}
This has the following meaning. To find the $R$-transform, we must first have an expression for $G(w)$, which is obtained via our usual means (e.g. the cavity method). We may then invert this expression to obtain $w(G)$, and thus the $R$-transform. 

The key property of the $R$-transform is how it relates to matrix sums. Consider the $R$-transform of $\underline{\underline{A}}+\underline{\underline{B}}$, which we denote $R_{A+B}$. The crucial property is, as long as $\underline{\underline{A}}$ and $\underline{\underline{B}}$ are free,
\begin{align}
	R_{A+B}(z) = R_A(z) + R_B(z). 
\end{align}
So, to obtain the resolvent of the sum $\underline{\underline{A}}+\underline{\underline{B}}$, we follow these steps. First, find $G_A(w)$ and $G_B(w)$. Invert these expressions to find $w_A(G)$ and $w_B(G)$. Then, compute
\begin{align}
	R_{A+B}(G) = w_A(G) + w_B(G) - \frac{2}{G}. 
\end{align}
To find the resolvent of the sum $G_{A+B}(z)$ we then use the original definition of the $R$-transform to write
\begin{align}
	G_{A+B}(z) = \frac{1}{z - R_{A+B}(G_{A+B}(z))}, \label{rtransforminversion}
\end{align}
which we solve for $G_{A+B}(z)$.

\begin{figure}[H]
	\centering 
	\includegraphics[scale = 0.6]{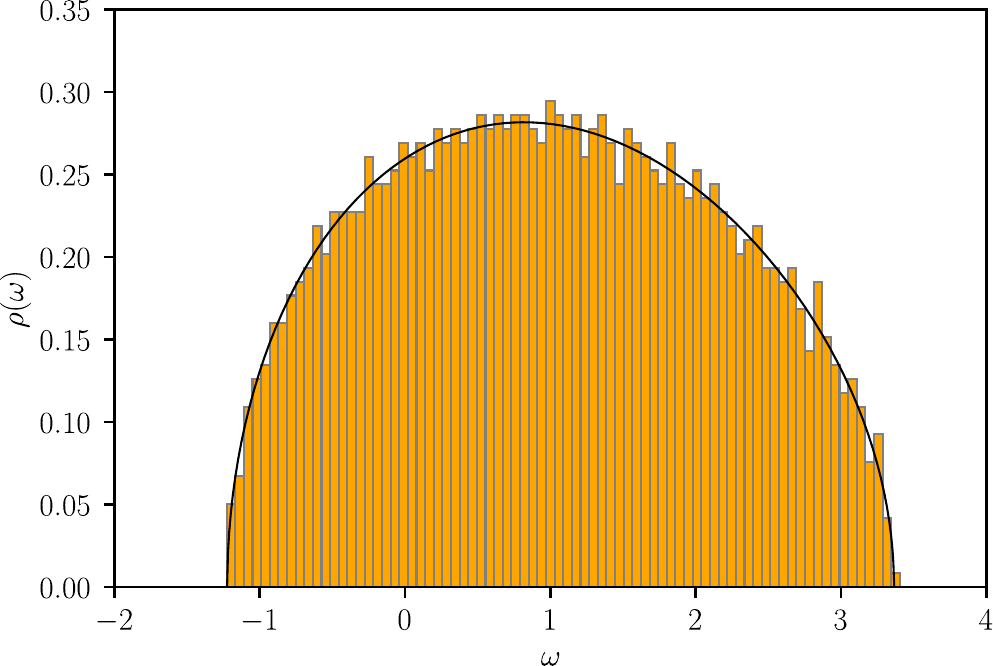}
	\captionsetup{justification=raggedright,singlelinecheck=false}
	\caption{ Eigenvalue density of the sum of a GOE matrix and a standard Wischart matrix. The matrix sizes are $N = 2000$. The theory line is given by Eq.~(\ref{sumWischartGOE}). }\label{fig:free_sum}
\end{figure}

\subsubsection*{Example: GOE $+$ diagonal matrix }
In the exercises, we consider the sum of a Wishart matrix and a GOE matrix (see Fig. \ref{fig:free_sum}). First, however, we take a somewhat trivial example, which we can verify by other means, to demonstrate the method. 

Let us take the example of the sum of a GOE matrix $\underline{\underline{A}}$ and a diagonal matrix $\underline{\underline{B}}$. First, we must compute the resolvents of both matrices. We are extremely familiar with the resolvent of the GOE, which yields the semicircle law [see e.g. Eq.~(\ref{ressemicircle})]
\begin{align}
	G_A(z) = \frac{1}{z - G_A(z)} .
\end{align}
We can write the resolvent of the diagonal matrix $\underline{\underline{B}}$ simply in terms of the distribution of diagonal elements $\rho_B(D)$
\begin{align}
	G_B(z) = \int dD \frac{\rho_B(D)}{z - D}. 
\end{align} 
We can  solve easily for $w_A(G) = G + G^{-1}$ and $w_B(G)$ is the implicit solution to $G = \int dD \rho_B(D)/[w_B(G) - D]$. For the sum, we thus see that 
\begin{align}
	R_{A+B}(G) = w_B(G) -\frac{1}{G} + G.
\end{align}
Inverting the $R$-transform using Eq.~(\ref{rtransforminversion}), we thus find
\begin{align}
	w_B(G_{A+B}(z)) = z- G_{A+B}(z).
\end{align}
Therefore, we finally obtain 
\begin{align}
	G_{A+B}(z) = \int dD \frac{\rho_B(D)}{z - D - G_{A+B}(z)} .
\end{align} 
This is exactly what one would from studying the cavity equations in Eq.~(\ref{cavity}), taking $J_{ii}$ to be drawn from $\rho_B(\cdot)$. 

\subsection{$S$-transform: products of free random matrices}
We now instead consider the $S$-transform \cite{voiculescu1987multiplication}, which allows us to handle products of random matrices. We begin by defining the moment generating function of the matrix $\underline{\underline{A}}$
\begin{align}
	M(z) = \frac{1}{N}\mathrm{Tr}\left[\sum_{r = 1}^\infty z^{r} \underline{\underline{A}}^{r}\right].
\end{align}
This is related to the resolvent via
\begin{align}
	G(z) = \frac{1}{z} + \frac{1}{z} M\left( \frac{1}{z}\right). \label{gandm}
\end{align}
We then define the functional inverse of $M(z)$ 
\begin{align}
	\chi\left( M(z) \right) = z.
\end{align}
The $S$-transform is then given by
\begin{align}
	S(z) = \frac{1+z}{z} \chi\left(z\right).
\end{align}
Assigning $z = M(w)$, we thus see that
\begin{align}
	S\left(M\right) = \frac{1 + M}{M} w(M).
\end{align}
This has the following meaning. To find the $S$-transform, we first find $M(w)$, which can be obtained from Eq.~(\ref{gandm}). One then inverts this to find $w(M)$, which then yields the $S$-transform.

Crucially, the $S$-transform of the product of two mutually free matrices is the product of their $S$-transforms, i.e.
\begin{align}
	S_{AB}(z) = S_A(z) S_B(z). 
\end{align}
Thus, to find the eigenvalue density of the product of two free matrices, we first find $w_A(M)$ and $w_B(M)$, and then we evaluate
\begin{align}
	S_{AB}(M) = \left( \frac{1+M}{M}\right)^2 w_A(M) w_B(M). 
\end{align}
One then solves the following for $M_{AB}(z)$
\begin{align}
	\frac{1+M_{AB}(z)}{M_{AB}(z)} z = S_{AB}\left(M_{AB}(z) \right). \label{sinversion}
\end{align}
Finally, the resolvent is computed via Eq.~(\ref{gandm}), i.e. $G_{AB}(z) = z^{-1} + z^{-1}M_{AB}(z)$.

\begin{figure}[H]
	\centering 
	\includegraphics[scale = 0.6]{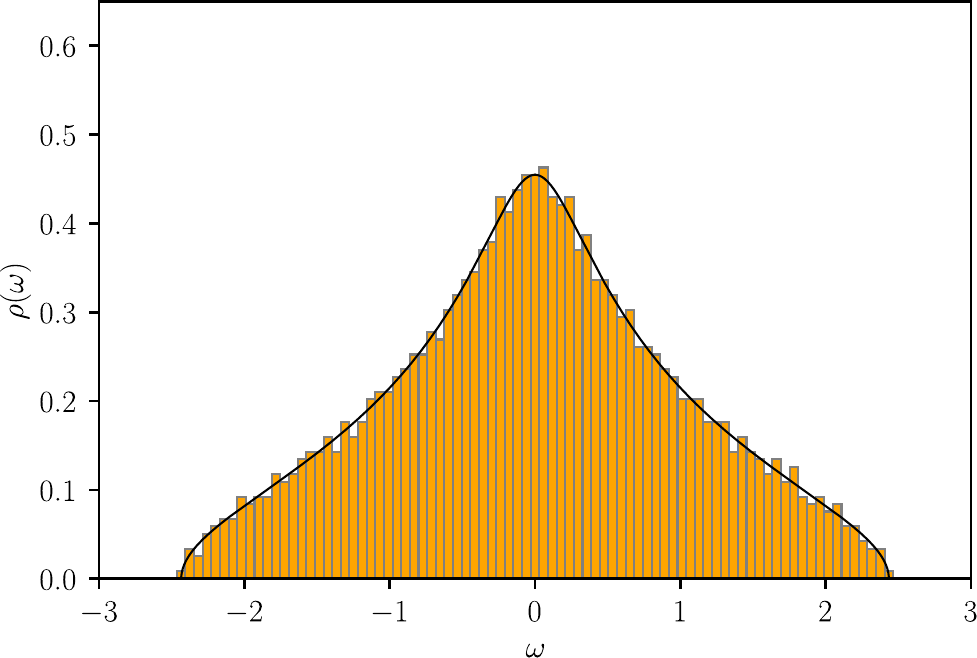}
	\captionsetup{justification=raggedright,singlelinecheck=false}
	\caption{ Eigenvalue density of the product of a standard Wischart and a GOE matrix. The matrix size is $N = 2000$. The theory line is given by Eq.~(\ref{productGOEWischart}). }\label{fig:free_product}
\end{figure}

\subsubsection*{Example: Generalised Mar\v{c}enko-Pastur equation }
Here, we present what is perhaps a more straightforward derivation of the generalised Mar\v{c}enko-Pastur equation in Eq.~(\ref{generalmplaw}). An alternative example, the product of a Wischart matrix and a GOE matrix, is given in the exercises. 

The `trick' here is to note that a sample covariance matrix with a general population covariance matrix $\underline{\underline{\Sigma}}$ can be written in terms of a standard Wischart matrix as follows. We define the Wishart matrix via its elements
\begin{align}
	W_{ij} = \frac{1}{n} \sum_{\alpha} x_i^\alpha x_{j}^\alpha, \label{wischart}
\end{align}
where $x_{i}^\alpha$ are centred Gaussian random variables with the correlator
\begin{align}
	\langle x_{i}^\alpha x_j^\beta \rangle = \delta_{\alpha\beta} \delta_{ij}. 
\end{align}
That is, a Wischart matrix is a sample covariance matrix with a corresponding population covariance matrix that is given by the identity matrix. To obtain sample covariance matrices for a general population covariance matrix, we then construct
\begin{align}
	\underline{\underline{C}} = \sqrt{\underline{\underline{\Sigma}}}\, \underline{\underline{W}}\sqrt{\underline{\underline{\Sigma}}} .
\end{align}
One can then verify easily that $\langle C_{ij} \rangle = \Sigma_{ij}$. Since we know that the $S$-transform of a product of free matrices is the product of their $S$-transforms, we have that $S_C = S_\Sigma S_W$. 

Let us now compute the $S$-transforms of both matrices. Supposing we know the eigenvalue density of $\underline{\underline{\Sigma}}$, we have using $G_\Sigma(z) = \int d\mu \rho_\Sigma(\mu)/[z-\mu]$,
\begin{align}
	1 + M_\Sigma(z) = \int d\mu \frac{\rho_\Sigma(\mu)}{1 - z \mu} .
\end{align}
For the Wischart matrix, we have the resolvent from Eq.~(\ref{aandd}). This then leads to the following expression for its $S$-transform
\begin{align}
	S_W(M) = \frac{1}{1+ \gamma M} . \label{StransformWischart}
\end{align}
The $S$-transform of the product is thus 
\begin{align}
	S_C(M) = w_\Sigma(M) \frac{1+M}{M} \frac{1}{1+ \gamma M} ,
\end{align}
where $w_\Sigma(M)$ is the implicit solution to $1 + M = \int d\mu \frac{\rho_\Sigma(\mu)}{1 - w_\Sigma(M) \mu} $. We may thus use the inversion formula in Eq.~(\ref{sinversion}) to find
\begin{align}
	w_\Sigma(M_C(z)) = [1+\gamma M_C(z)]z .
\end{align}
Using the implicit definition of $w_\Sigma(M_C(z))$, we then arrive at
\begin{align}
	1 + M_C(z) = \int d\mu \frac{\rho_\Sigma(\mu)}{1 - [1+\gamma M_C(z)] z \mu} .
\end{align}
Finally, using the relation between the generating function and the resolvent in Eq.~(\ref{gandm}), we recover Eq.~(\ref{generalmplaw}).

\subsection{Exercises}
In the following, to demonstrate the versatility of the $R$- and $S$-transforms, we will now derive the eigenvalue density for two example ensembles that would otherwise be very inconvenient to handle using any of the techniques that we have so far discussed. 

Let us first consider the sum of a GOE matrix and a Wischart matrix.
\begin{itemize}
	\item Using the resolvent of a Wischart matrix given in Eq.~(\ref{wischart}), derive its $R$-transform. The $R$-transform of the GOE matrix is given by $R(z) = z$, as we found earlier.
	\item Let us now consider the sum of a GOE matrix and a Wischart matrix. Use the $R$-transform to show that the resolvent of this sum is given by 
	\begin{align}
		G(z) = \frac{1}{z - G - \frac{1}{1-\gamma G}} . \label{sumWischartGOE}
	\end{align}
	\item Show that for small $\gamma$, we can expand to expand to leading order in $\gamma$ to obtain
	\begin{align}
		G (z - 1 - G) = 1 + \gamma G^2, 
	\end{align}
	and the eigenvalue density is thus approximated by 
	\begin{align}
		\rho(\omega) \approx \frac{\sqrt{(\lambda_+ -\omega)(\omega- \lambda_-)}}{2\pi(1+\gamma)} , \hspace{1cm} \lambda_{\pm} = 1 \pm 2 \sqrt{1+\gamma} .
	\end{align}
	\item The numerical solution of Eq.~(\ref{sumWischartGOE}) is used in conjunction with the inverse Stieltjes transform in Eq.~(\ref{densefromres}) [i.e. $\rho(\omega) = \pi^{-1} \lim_{\epsilon\to 0}\mathrm{Im}G(\omega-i \epsilon)$] to produce the solid line in Fig. \ref{fig:free_sum}. 
\end{itemize}
Let us now consider the product of the same two matrices. 
\begin{itemize}
	\item First, we must compute the $S$-transform of both matrices. The $S$-transform of the Wischart matrix is given in Eq.~(\ref{StransformWischart}). Show that the $S$-transform of the GOE matrix is given by 
	\begin{align}
		S(M) = \pm \frac{1}{\sqrt{M}}. 
	\end{align}
	\item Therefore, show using Eq.~(\ref{gandm}) that the resolvent of the product is given by
	\begin{align}
		\frac{G(1+\gamma (z G -1))}{\sqrt{z G -1}} = 1. \label{productGOEWischart}
	\end{align}
	\item Expand this expression for small $\gamma$. Replace any terms that are higher-order than quadratic in $G$ using the zeroth-order approximation, and solve the resulting quadratic expression for $G$ to show that
	\begin{align}
		\rho(\omega) \approx \frac{1}{2\pi} \frac{\sqrt{4 + 8 \gamma - \omega^2}}{1 + 2 \gamma \omega^2} .
	\end{align}

\end{itemize}

\newpage

\section{Replica method}\label{section:replicas}
We now move on to describe a series of three methods that are ubiquitous in the theory of disordered systems, and thus appear time and time again in the RMT literature. The three methods largely employ the same strategy. The strategy is to write the resolvent matrix elements, which are at the centre of our analysis, as integrals over additional `fields' where the disorder (i.e. the random matrix entries) appears only in the exponent of the integrand. We term these approaches `auxiliary field' methods. The motivation for introducing the auxiliary fields is that this expedites taking the average over the disorder. Further, generalisations such as the introduction of block structure to the random matrix, non-Hermiticity, or non-Gaussianity are (arguably) simpler to handle than with the cavity method. Although we will not discuss this in depth here, these methods also facilitate the calculation of more complicated quantities than the average eigenvalue density, such as eigenvalue correlation functions \cite{efetov1983supersymmetry, kamenev1999level,baron2025classes}, which would be far more difficult to compute with the cavity method.

Roughly speaking, the replica, supersymmetric and path integral methods that we now discuss all accomplish the same feat. Namely, they handle the inconvenient determinant factor that appears when we write the resolvent matrix elements using an exponential integral representation. More precisely, let us consider the problem of deriving Wigner's semi-circle law. We wish to compute the resolvent matrix elements as defined in Eq.~(\ref{resdef}). We may write the following standard Fresnel integral (following Edwards and Jones \cite{edwards1976eigenvalue})
\begin{align}
	G_{kl}(\omega) &= \int \left(\prod_{j=1}^N \frac{e^{i\pi/4} d\phi_j}{\sqrt{2\pi}}\right) \frac{i\phi_k \phi_l}{\mathcal{Z}}\exp\left[ -\frac{i}{2} \sum_{i,j}\phi_i [(\omega- i\epsilon)\delta_{ij} - J_{ij}] \phi_j \right], \nonumber \\
	\mathcal{Z} &= \int \left(\prod_{j=1}^N \frac{e^{i\pi/4} d\phi_j}{\sqrt{2\pi}}\right) \exp\left[ -\frac{i}{2} \sum_{i,j}\phi_i [(\omega- i\epsilon)\delta_{ij} - J_{ij}] \phi_j \right] ,\label{fresnel}
\end{align}
where we note that the positive regulariser $\epsilon$ ensures convergence. Noting that the object of primary interest to us is the trace of the resolvent $G(\omega)  = N^{-1} \sum_i G_{ii}(\omega)$, we may also write
\begin{align}
	G(\omega) &= -\frac{2}{N}\frac{\partial \ln \mathcal{Z}}{\partial \omega} . \label{gfromlogz}
\end{align}
What we would like to do now is to find the ensemble average $\langle G(\omega) \rangle = N^{-1} \sum_i \langle G_{ii}(\omega) \rangle$. However, the determinant factor $\mathcal{Z}$ in the first of Eqs.~(\ref{fresnel}) is currently preventing us from doing this. Each of the auxilliary field methods presents a different way, essentially, of handling the errant factor of $\mathcal{Z}$. 

With the present case of the replica method, one tackles this by using the representation in Eq.~(\ref{gfromlogz}), which requires us to average the \textit{logarithm} of $ \mathcal{Z}$. We will now discus how the replica method allows us to proceed along these lines. 

\subsection{The replica trick and statistical physics}\label{section:statphys}

Frequently in statistical physics, one is tasked with calculating the so-called `partition function', which is usually labelled $Z$ after the German word \textit{Zustandssumme} (it is no accident that we have used similar notation $\mathcal{Z}$ above). The free energy, which enables one to understand the thermodynamics of the system, is then given by $F = - k_B T \ln Z$. Ordinarily, the partition function is calculated by integrating/summing the Boltzmann weight for each microstate of the system. 

Let us take the Ising model for example. In this system, we have a set of magnetic `spins' $\sigma_i \in \{-1,+1\}$ that are arranged on a lattice. We encapsulate the lattice and the coupling between spins with the weighted adjacency matrix $\underline{\underline{J}}$, where here $J_{ij} = J$ if $i$ and $j$ are neighbours on the lattice, and $J_{ij} = 0$ otherwise. The Hamiltonian (or `energy function', loosely speaking) of the system when it is in a given spin configuration $\{\sigma\}$ is then
\begin{align}
	H(\{\sigma\}) = - \sum_{ij} J_{ij} \sigma_i \sigma_j . \label{hamiltonian}
\end{align}
The partition function for the Ising model with this Hamiltonian is
\begin{align}
	Z = \sum_{\{\sigma\}} e^{-\beta H(\{\sigma\})}, 
\end{align}
where here we sum over all possible spin configurations and we write $\beta = (k_B T)^{-1}$. The probability of the system being in any one spin state is then, according to Boltzmann,
\begin{align}
	P(\{\sigma\}) = \frac{1}{Z}e^{-\beta H(\{\sigma\})}.
\end{align}
The reason that we may observe any spin state with finite probability is due to thermal fluctuation. Importantly, we note that in calculating the partition function, we only sum over the degrees of freedom that fluctuate due to thermal noise. 

In the study of \textit{disordered} physical systems, we deal with models where the couplings $J_{ij}$ are drawn randomly from some distribution, just as in random matrix theory. However, these quantities are \textit{fixed} for any one realisation of the system (so-called \textit{quenched disorder}). This means that they do not fluctuate due to thermal noise. For the Ising model with quenched random couplings, the Hamiltonian is exactly the same as in Eq.~(\ref{hamiltonian}), except $\underline{\underline{J}}$ is now a random matrix. These random couplings can represent the disordered interactions between magnetic domains in peculiar `spin-glass' materials, such as dilute solutions of Mn in Cu \cite{edwards1975theory} (i.e. small magnetic domains separated by non-magnetic host material). The original model of such materials by Edwards and Anderson \cite{edwards1975theory} involved classical magnetic dipoles. The `solvable' model of a spin glass using Ising spins was formulated by Sherrington and Kirkpatrick soon after \cite{sherrington202550}.

To proceed with a thermodynamic analysis, it is necessary to evaluate the free energy. To do this, we assume that the free energy is `self-averaging'. That is, the free energy of any one particular large sample is equal to the average over all possible realisations of $\underline{\underline{J}}$. We therefore compute
\begin{align}
	F = -k_B T \langle \ln Z \rangle . \label{freeenergy}
\end{align}
We note that there is an important difference here between assuming  a self-averaging free energy $F$ and a self-averaging partition function $Z$. Averaging $Z$ \textit{before} taking the logarithm to find $F$ would be akin to treating the couplings as fluctuating due thermal effects. This is known as the \textit{annealed average}. The case in which one is typically interested instead involves averaging $\ln Z$ as above, which is known as the \textit{quenched average}. The quenched average is also precisely the problem with which we are presently faced in our RMT calculation. 

A fixture in the methods of disordered systems theory is the so-called replica method (or $n\to 0$ method), which was first developed by Edwards and Anderson \cite{edwards1975theory, sherrington202550}, and later used for random matrices by Edwards and Jones \cite{edwards1976eigenvalue}. This trick enables us to perform the average of the logarithm. At its core, the replica trick simply exploits the identity 
\begin{align}
	\ln Z = \lim_{n \to 0}\frac{Z^n -1}{n} . \label{replicatrick}
\end{align}
That is, we need only compute $\langle Z^n \rangle$ to perform the average of the logarithm. If we take $n$ to be an integer, we see that the average of the logarithm can be replaced by an average of an $n$-fold `replicated' system, which we can compute easily. The difficulty arises in taking the smooth limit $n \to 0$. In the Physics literature, one fragrantly disregards this difficulty and extends the results that are derived for integer $n$ to continuous $n$. In the end, the results that one obtains are usually correct (but not always \cite{verbaarschot1985critique,zirnbauer1999another}). The Parisi solution \cite{parisi1979infinite, parisi1980order,parisi1983order} to the Sherrington-Kirkpatrick model that was obtained with replicas has been proved correct rigorously by Talagrand \cite{talagrand2006parisi}. Further, the solution for the spherical $p$-spin model has also been obtained using the cavity approach \cite{gradenigo2020solving}. So, we at least know that the replica approach provides the correct results for the Sherrington-Kirkpatrick and spherical $p$-spin models \cite{franz2006note}, and we will show here that it also produces correct results in the random matrix context.

We take a brief moment to comment on so-called `replica symmetry', which is a notion that invariably accompanies the use the replica trick. In the course of computing the average $\langle Z^n\rangle$, one will usually introduce order parameters such as the `overlap' $Q_{ab} = N^{-1}\sum_i\langle\langle \sigma^a_i \sigma^b_i \rangle\rangle$, where the indices $a$ and $b$ indicate possibly different replicas, and the double brackets here indicate an average over the couplings \textit{and} the thermal noise that may arise in physics models. In the replica symmetric case, the solution for $Q_{ab}$ is such that $Q_{ab} = q_d$ when $a = b$ and $Q_{ab} = q_0$ otherwise. When replica symmetry is broken, $Q_{ab}$ ($a\neq b$) takes heterogeneous values depending on $a$ and $b$. 

Replica symmetry happens to be obeyed for our simple calculation of the eigenvalue density. In fact, for our case, the replicas decouple for large $N$, and the quenched and annealed averages are the same (as we will show). In more complicated systems such as the Sherrington-Kirkpatrick or spherical $p$-spin models mentioned above, the quenched and annealed averages are not necessarily equivalent, and one also observes the phenomenon of replica-symmetry breaking for certain system parameters \cite{thouless1977solution,de1978stability}. This replica-symmetry breaking indicates a transition of the free energy landscape from a simple single minimum to a more complex structure with many minima. This is known as the spin-glass transition \cite{mezard1987spin}, and it entails many interesting glassy phenomena such as ergodicity-breaking and ageing. It has also been found that for more complicated calculations in random matrix theory, such as the eigenvalue density correlations, considerations of replica symmetry breaking are also relevant \cite{kamenev1999level}. We also illustrate briefly why replica symmetry breaking becomes necessary in the context of the spherical $p$-spin model below, since it is in this context that the replica method really becomes indispensable.

\subsection{Ensemble-averaged replicated partition function: recovering the semicircle}
Let us now use the replica trick in Eq.~(\ref{replicatrick}) to compute $\langle G(\omega) \rangle$ from Eqs.~(\ref{fresnel}) and (\ref{gfromlogz}), with a view to obtaining the Wigner semicircle. When the couplings $J_{ij} = J_{ji}$ are each i.i.d random variables with zero mean and variance $\langle J_{ij}^2 \rangle = \sigma^2/N$, the `partition function' in Eq.~(\ref{fresnel}) can be interpreted as being associated with a simple all-to-all connected spin model, where each spin can take values all along the real axis. While this spin model is somewhat trivial, generalisations such as the spherical $p$-spin model with $p>2$ exhibit replica symmetry breaking \cite{crisanti1992spherical}, as we will discuss.

We can write the replicated partition function [c.f. Eq.~(\ref{fresnel})] as 
\begin{align}
	\mathcal{Z}^n = \int \left(\prod_{a=1}^n\prod_{j=1}^N \frac{e^{i\pi/4}d\phi^a_j}{\sqrt{2\pi}}\right) \exp\left[ -\frac{i}{2} \sum_{a}\sum_{i,j}\phi^a_i [(\omega- i\epsilon)\delta_{ij} - J_{ij}] \phi^a_j \right] . 
\end{align}
Let us now compute the average $\langle \mathcal{Z}^n \rangle$. Using the independence of the couplings $J_{ij}$, we need only compute
\begin{align}
	\left \langle \exp\left[ i \sum_{a}\phi^a_i J_{ij}\phi^a_j \right]  \right\rangle \approx 1 - \frac{\sigma^2}{2N} \sum_{a,b} \phi_i^a \phi_i^b \phi_j^a \phi_j^b \approx \exp\left[ - \frac{\sigma^2}{2N} \sum_{a,b} \phi_i^a \phi_i^b \phi_j^a \phi_j^b\right],
\end{align}
where we have used that $N$ is large, and we have neglected terms proportional to higher moments of $J_{ij}$, which assume to be negligible in comparison. We then have
\begin{align}
	\langle \mathcal{Z}^n \rangle = \int \left(\prod_{a=1}^n\prod_{j=1}^N \frac{e^{i\pi/4} d\phi^a_j}{\sqrt{2\pi}}\right) &\exp\left[ -\frac{i}{2} \sum_{a}\sum_{i}(\phi^a_i)^2 (\omega- i\epsilon)  - \frac{\sigma^2}{4N} \sum_{i,j}\sum_{a,b} \phi_i^a \phi_i^b \phi_j^a \phi_j^b \right] . \label{replicaaveraged}
\end{align}
We have thus exchanged what was initially a Gaussian integral (or, more precisely, a Fresnel integral) containing the disorder in Eq.~(\ref{fresnel}) for one in which we now have no disorder and a quartic `interaction term' in the exponent. Such an integral cannot be evaluated exactly, but as usual, we can exploit the large $N$ limit to arrive at a result. 

In principle, we could proceed under the assumption of replica symmetry. We could introduce an overlap order parameter $Q_{ab}$ as described above (this approach is explored in the exercises below), make a replica symmetric ansatz (see e.g. Refs. \cite{edwards1976eigenvalue,rodgers1988density} ), and solve for $q_d$ and $q_0$. However, it turns out that the replicas in fact decouple -- this is a stronger statement than there merely being replica symmetry. This means that $q_0$ (the overlap between different replicas) is zero, and the annealed and quenched averages of the partition function $\mathcal{Z}$ are equivalent. We presently demonstrate the replica decoupling using combinatorics. 

\subsection{Replica decoupling}\label{section:replicadecoupling}
We now show that the replicas decouple in this case, and thus that the quenched and annealed averages are equivalent. This is something of a technical aside, which is often glossed over in calculations. The reader who is satisfied with simply believing that $\langle \mathcal{Z}^n\rangle = \langle \mathcal{Z} \rangle^n$ may skip to Eq.~(\ref{decoupled}). We largely follow the reasoning of Edwards and Warner below \cite{edwards1980effect}.

One way to proceed from Eq.~(\ref{replicaaveraged}) would be to expand the exponential with the quartic exponent as a series. One could then perform the integrals over $\phi_i^a$, since these are simply Gaussian integrals, and one could subsequently resum the series. Let us examine the first few terms. The first term in the series is the trivial
\begin{align}
	I_0= \int \left(\prod_{a=1}^n\prod_{j=1}^N \frac{e^{i\pi/4} d\phi^a_j}{\sqrt{2\pi}}\right) \exp\left[ -\frac{i}{2} \sum_{a}\sum_{i}(\phi^a_i)^2 z \right] = \frac{1}{z^{Nn/2}} ,
\end{align}
where we write $z = \omega - i\epsilon$. Now, we introduce the notation
\begin{align}
	\left\langle \cdots\right\rangle_\mathcal{Z} = \int \left(\prod_{a=1}^n\prod_{j=1}^N \frac{e^{i\pi/4} d\phi^a_j}{\sqrt{2\pi/ z}}\right) \left( \cdots \right)\exp\left[ -\frac{i}{2} \sum_{a}\sum_{i}(\phi^a_i)^2 (\omega- i\epsilon) \right] .
\end{align}
The subsequent term is more non-trivial. We have
\begin{align}
	z^{Nn/2} I_{1} &=  -\frac{\sigma^2}{4N} \left\langle \sum_{i,j}\sum_{a,b} \phi_i^a \phi_i^b \phi_j^a \phi_j^b \right\rangle_\mathcal{Z} \nonumber \\
	&=  -\frac{\sigma^2}{4N} \sum_{i,j}\sum_{a,b} \bigg[ \left\langle \phi_i^a \phi_i^b \right\rangle_\mathcal{Z} \left\langle \phi_j^a \phi_j^b \right\rangle_\mathcal{Z} + \left\langle \phi_i^a \phi_j^a  \right\rangle_\mathcal{Z} \left\langle  \phi_i^b\phi_j^b \right\rangle_\mathcal{Z}+ \left\langle \phi_i^a \phi_j^b  \right\rangle_\mathcal{Z} \left\langle  \phi_i^b\phi_j^a \right\rangle_\mathcal{Z} \bigg] \nonumber \\
	&=  \frac{\sigma^2}{4N} \frac{1}{z^2} \sum_{i,j}\sum_{a,b} \left[ \delta_{ab} + \delta_{ij} + \delta_{ij}\delta_{ab} \right] = \frac{\sigma^2}{4} \frac{1}{z^2} \left[ n N + n^2 + n\right],\label{s1}
\end{align}
where we have multiplied by the appropriate normalising factor so that we may use Wick's (Isserlis') theorem.

\begin{figure}[H]
	\centering 
	\includegraphics[scale = 0.3]{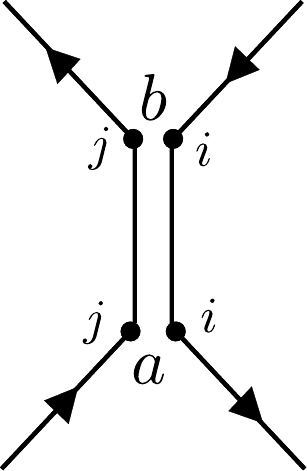} 
	\hspace{3cm}
	\includegraphics[scale = 0.3]{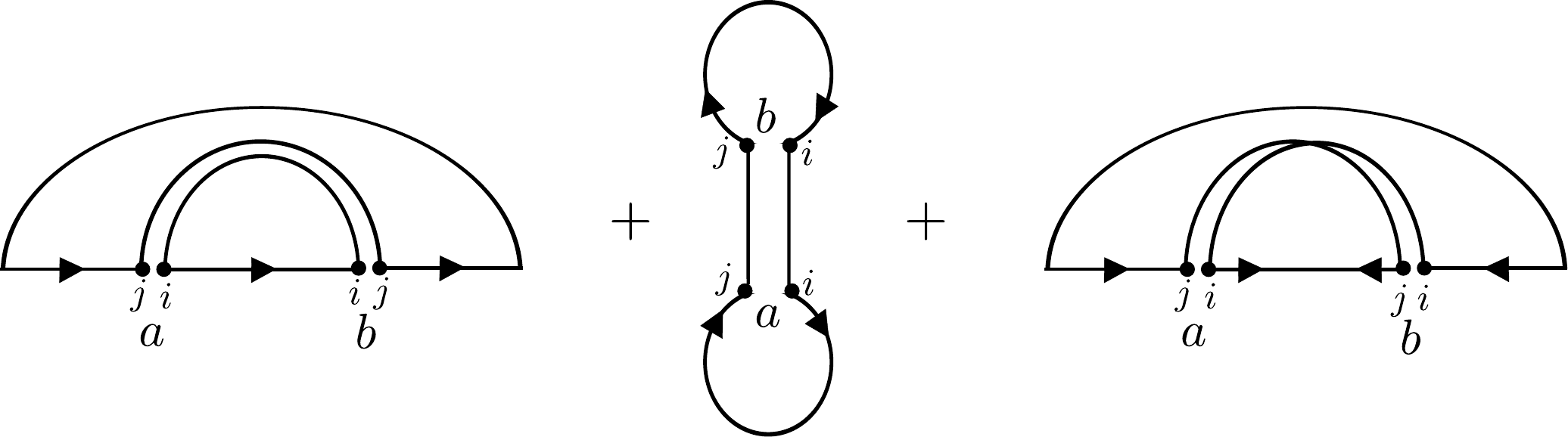}
	\captionsetup{justification=raggedright,singlelinecheck=false}
	\caption{(Top) The diagrammatic representation of the quartic term before pairing. (Bottom) The possible pairings of the `loose ends' of left diagram when calculating the average  $\left\langle \sum_{i,j}\sum_{a,b} \phi_i^a \phi_i^b \phi_j^a \phi_j^b \right\rangle_\mathcal{Z}$. }\label{fig:replicadiagrams}
\end{figure}

Of the terms in Eq.~(\ref{s1}), we see that only the pairing $\left\langle \phi_i^a \phi_i^b \right\rangle_\mathcal{Z} \left\langle \phi_j^a \phi_j^b \right\rangle_\mathcal{Z}$ is relevant in the thermodynamic limit. Following Edwards and Warner \cite{edwards1980effect}, we can represent the terms in Eq.~(\ref{s1}) with diagrams. This is shown in Fig. \ref{fig:replicadiagrams}.

As is shown in the left diagram, we represent each of the four $\phi_i^a$, $\phi_i^b$, $\phi_j^a$, and $\phi_j^b$ with a vertex. To distinguish the two vertices with the same lower (site) index $i$ or $j$ from each other, we give a different directionality to the `free' edges that will later be paired. Vertices with the same upper (replica) index are arranged as a pair. Vertices with the same lower index are connected by an undirected edge. 

The three Wick pairings in the third line of Eq.~(\ref{s1}) are then represented by the right-hand panel of Fig. \ref{fig:replicadiagrams}. If two of $\phi$s are averaged together, then we connect their directed edges. We then obtain diagrams that are very reminiscent of those in Section \ref{section:wignerapproach}. As was the case there, the dominant contributions are represented by planar rainbow diagrams. This is because these diagrams have the greatest number of disconnected parts, which corresponds to the greatest power of $N$.

\begin{figure}[H]
	\centering 
	\includegraphics[scale = 0.3]{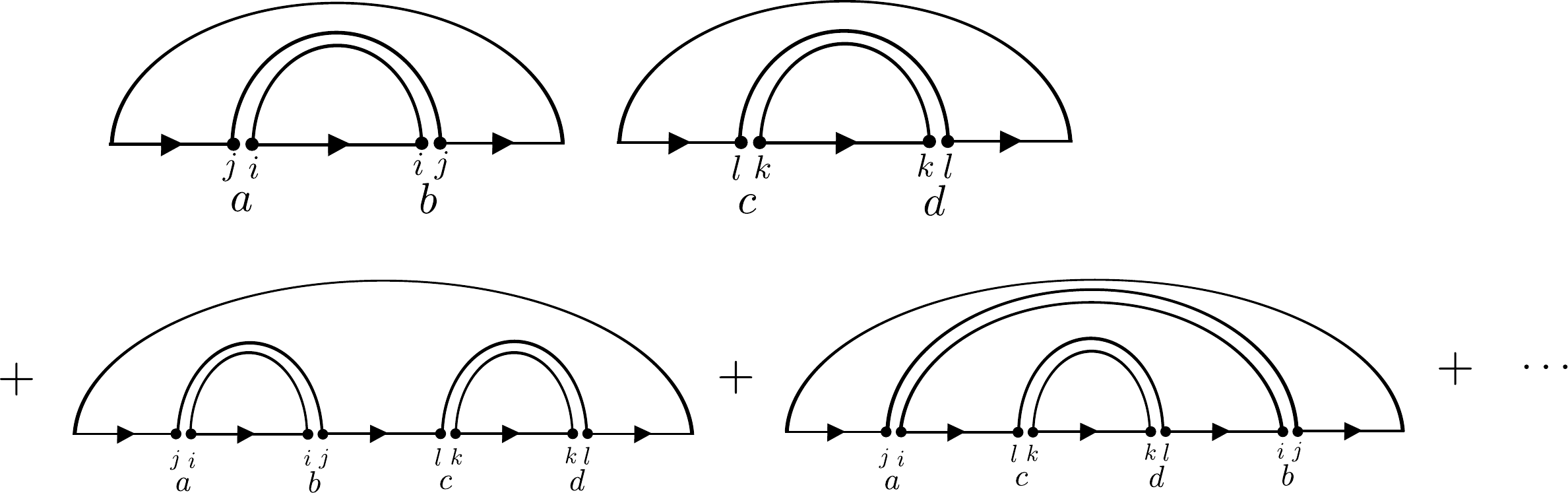}
	\captionsetup{justification=raggedright,singlelinecheck=false}
	\caption{Possible planar Wick pairings for the second term in the expansion of $\langle \mathcal{Z}^n\rangle$.}\label{fig:replicadiagrams2}
\end{figure}

Let us now consider the second non-trivial term in the expansion. In this case there are $7!!$ Wick pairings of the variables $\phi_i^a$. However, there is again only one dominant pairing. We can see this from the diagrams in Fig. \ref{fig:replicadiagrams2}. 
The diagram that has the highest order in $N$ is the diagram that is the square of the leading term of $I_1$. The `connected' diagrams that mix indices cannot have a greater (or equal) number of disconnected pieces, and therefore are lower order in $N$. In other words, terms that Wick-pair $\phi$s with differing lower indices (e.g. $\langle \phi_i^a \phi_j^b\rangle_\mathcal{Z} \propto \delta_{ij}\delta_{ab}$) give rise to Kronecker deltas, which reduce the order of magnitude of the term in $N$.

Therefore, we have 
\begin{align}
	z^{Nn/2} I_{2} &\approx \frac{\sigma^4}{32 N^2} \sum_{i,j,k,l} \sum_{a,b,c,d}  \left\langle \phi_i^a \phi_i^b \right\rangle_\mathcal{Z} \left\langle \phi_j^a \phi_j^b \right\rangle_\mathcal{Z} \left\langle \phi_k^c \phi_k^d \right\rangle_\mathcal{Z} \left\langle \phi_l^c \phi_l^d \right\rangle_\mathcal{Z}   \nonumber \\
	&= \frac{\sigma^4 N^2}{32  z^4}  \sum_{a,b,c,d} \delta_{ab} \delta_{cd}  = \frac{\sigma^4 N^2 n^2}{32  z^4} .
\end{align}
We note that these dominant Wick pairings always constrain the upper (replica) indices of $\phi$s with the same lower index to be the same. This pattern continues to higher orders. We thus see that it is possible to replace the factor $ - \frac{\sigma^2}{4N} \sum_{i,j}\sum_{a,b} \phi_i^a \phi_i^b \phi_j^a \phi_j^b $ in the exponent of Eq.~(\ref{replicaaveraged}) with $ - \frac{\sigma^2}{4N} \sum_{i,j}\sum_{a} \phi_i^a \phi_i^a \phi_j^a \phi_j^a $ when $N$ is large. We therefore obtain
\begin{align}
	\langle \mathcal{Z}^n \rangle &= \int \left(\prod_{a=1}^n\prod_{j=1}^N \frac{e^{i\pi/4} d\phi^a_j}{\sqrt{2\pi}}\right) \exp\left[ -\frac{i}{2} \sum_{a}\sum_{i}(\phi^a_i)^2 (\omega- i\epsilon) - \frac{\sigma^2}{4N} \sum_{i,j}\sum_{a} \phi_i^a \phi_i^a \phi_j^a \phi_j^a \right]= \langle \mathcal{Z} \rangle^n . 
\end{align}
That is, the replicas decouple. For this reason, we see that the logarithm and the disorder average commute in the computation of $\langle G(\omega) \rangle$ from Eq.~(\ref{gfromlogz}). That is, we may write (without replica indices)
\begin{align}
	\langle G(\omega) \rangle &= 2 N^{-1}\frac{\partial \ln \langle \mathcal{Z} \rangle}{\partial \omega}\nonumber \\
	&= \frac{1}{\langle \mathcal{Z}\rangle} \int \left(\prod_{j=1}^N \frac{e^{\frac{i\pi}{4}} d\phi_j}{\sqrt{2\pi}}\right)\left( \sum_i \frac{i\phi_i^2}{N} \right) \exp\left[ -\frac{i z}{2} \sum_{i}\phi_i^2  - \frac{\sigma^2}{4N} \sum_{i,j} \phi_i^2  \phi_j^2 \right] ,
\end{align}
where we have
\begin{align}
	\langle \mathcal{Z}\rangle &= \int \left(\prod_{j=1}^N \frac{e^{i\pi/4} d\phi_j}{\sqrt{2\pi}}\right) \exp\left[ -\frac{i z}{2} \sum_{i}\phi_i^2  - \frac{\sigma^2}{4N} \sum_{i,j} \phi_i^2  \phi_j^2 \right]. \label{decoupled}
\end{align}
With this simplification, we may now more easily perform a saddle-point approximation of $\langle\mathcal{Z}\rangle$. We note that another way to show this is by introducing overlap matrix and showing that the off-diagonal elements of this matrix are nil. This is explored more in the exercises below. 

\subsection{Saddle-point approximation and Wigner semicircle law}\label{section:spareplica}
Into Eq.~(\ref{decoupled}) we now introduce the `order parameter' $q = N^{-1}\sum_i \phi_i^2$ via the complex exponential representation of the delta function
\begin{align}
	1= \int dq\, N\, \delta\left(N q - \sum_i \phi_i^2\right) = \int \frac{d\hat q \,dq}{2\pi} \; N \exp\left[ i N \hat q q - i \hat q \sum_i \phi_i^2 \right].
\end{align}
We thus obtain
\begin{align}
	\langle G(\omega)\rangle &= \langle \mathcal{Z}\rangle^{-1}\int\frac{d\hat q \,dq}{2\pi} \; \int \left(\prod_{j=1}^N \frac{e^{i\pi/4} d\phi_j}{\sqrt{2\pi}}\right) \left( i q\right)\exp\left[ -N \left( \frac{i z}{2} q- i\hat q q + \frac{\sigma^2}{4} q^2\right)   - i \hat q \sum_i \phi_i^2 \right] \nonumber \\
	&=  \langle \mathcal{Z}\rangle^{-1}\int\frac{d\hat q \,dq}{2\pi} \;  (iq) \exp\left[ -N \left( \frac{i z}{2} q- i\hat q q + \frac{\sigma^2}{4} q^2 + \frac{1}{2}\ln\left(2 \hat q\right)\right)  \right] \nonumber \\
	&\equiv  \langle \mathcal{Z}\rangle^{-1}\int\frac{d\hat q \,dq}{2\pi} \;  (iq) \exp\left[ -N \Omega(\hat q, q) \right] ,
\end{align}
and we also have
\begin{align}
	\langle \mathcal{Z}\rangle = \int\frac{d\hat q \,dq}{2\pi} \;   \exp\left[ -N \Omega(\hat q, q) \right] .
\end{align}
By introducing the order parameter, we were able to factorise the integrals over $\phi_i$ so that each could be carried out separately in a trivial way. 

We have thus obtained integrals that can be evaluated by exploiting the large value of $N$ and the saddle-point technique. That is, since $\langle G(\omega)\rangle$ involves a ratio of integrals with integrands of the form $f(q,\hat q) \exp[-N \Omega(q, \hat q)]$ and $N$ is large, each integral will be dominated by the `saddle point'\footnote{The saddle point is so called because any stationary point of an analytic function is a saddle as a consequence of the Cauchy-Riemann equations.} $(q^\star, \hat q^\star)$ at which $\Omega(q, \hat q)$ is stationary. Even if this corresponds to complex values of $q$ and $\hat q$, we can deform the contour to go through the saddle point in such a way that the remaining integrals can be performed using Laplace's method \cite{bender1999advanced}. As such, and noting that the integral for $\langle \mathcal{Z}\rangle$ shares the same saddle point $(q^\star, \hat q^\star)$ as that of $\langle G(\omega)\rangle \langle \mathcal{Z}\rangle$, we find that $\langle G(\omega)\rangle = i q^\star$ for $N \to \infty$.

More precisely, we obtain for the saddle point of both $\langle \mathcal{Z} \rangle $ and $\langle G(\omega)\rangle\langle \mathcal{Z}\rangle$
\begin{align}
	\partial_q \Omega &= \frac{i z}{2} - i \hat q_\star + \frac{\sigma^2}{2} q_\star = 0, \nonumber \\
	\partial_{\hat q} \Omega &= - i q_\star + \frac{1}{2\hat q_\star} = 0,
\end{align}
and thus 
\begin{align}
	\sigma^2 (iq_\star)^2 - z (i q_\star)  + 1  = 0,\hspace{2cm}
	\hat q_\star = \frac{1}{2 i q_\star} .
\end{align}
We can therefore write 
\begin{align}
	\langle G(\omega)\rangle = i q_\star = \frac{z - \mathrm{sign}[\mathrm{Re}(z)]\sqrt{z^2 - 4 \sigma^2}}{2 \sigma^2}. 
\end{align}
Using Eq.~(\ref{densefromres}), we therefore finally arrive once again at the Wigner semicircle law. 

\subsubsection{Summary of the calculation}
To summarise, our calculation was comprised of the following steps:
\begin{itemize}
	\item First, we observed that to compute the resolvent of our random matrix, we had to compute the disorder average of the logarithm of a partition function $\langle \ln\mathcal{Z}\rangle$. To this end, we rewrote the logarithm using the replica trick. We thus had to compute the disorder average of an $n$-fold replicated partition function $\langle \mathcal{Z}^n\rangle$.
	\item Secondly, we carried out the disorder average, obtaining an integral expression over $n\times N$ `fields' $\{\phi^a_i\}$, which had a quartic contribution in the exponent. 
	\item We then showed explicitly that the replicas decoupled: $\langle \mathcal{Z}^n\rangle=\langle \mathcal{Z}\rangle^n$. This is often just assumed to be the case in calculations of the eigenvalue density.
	\item We then introduced an `order parameter' $q$. This allowed us to decouple the (no-longer replicated) fields $\{\phi_i\}$, and replace the quartic term with a quadratic one, which did not mix the site indices. That is, we decoupled the sites. 
	\item After integrating out $\{\phi_i\}$, we then carried out a saddle-point approximation of the remaining integral over $q$ and its conjugate. We then arrived at the result.
\end{itemize}
Even more succinctly, the procedure is: disorder average, order parameter introduction, saddle-point approximation. Broadly speaking, the auxiliary field methods all follow this recipe (unless we are using a diagrammatic/series expansion approach).

\subsection{Replica symmetry breaking: spherical $p$-spin}\label{section:rsb}

In calculating the resolvent by using the replica approach, we found that the replicas decoupled, and we therefore found that we could replace the quenched average with an annealed average [i.e. we showed that $\langle \ln \mathcal{Z} \rangle = \ln \langle \mathcal{Z} \rangle$]. However, this certainly isn't always the case. For the spherical $p$-spin model, of which we investigated the relaxational dynamics in Section \ref{section:pspin}, one must employ a `replica symmetry breaking' (RSB) scheme in order to properly evaluate $\langle \ln Z \rangle$ in the limit $N \to \infty$\footnote{This statement is true for $p\geq 3$. For $p = 2$, the model is entirely solvable using the results of random matrix theory. See exercises.}. We briefly explore what this means mathematically, without delving into the physical significance in any great detail. The reader is directed to the pedagogical overviews of the spherical $p$-spin model in Refs. \cite{castellani2005spin, barrat1997p, zamponi2010mean}, and the classic Refs. \cite{mezard1987spin, crisanti1992spherical, crisanti1993spherical} for a deeper understanding of the physical significance of RSB for glassy systems. 

\subsubsection{Equilibrium (static) treatment of the $p$-spin model}
Let us return to the spherical $p$-spin model discussed in Section \ref{section:pspin}. Instead of studying the dynamics now, we will compute the free energy, as discussed in Section \ref{section:statphys}, which allows us to understand the thermodynamic equilibrium properties of the system. In particular, we can extract the static order parameters, and detect any phase transitions. 

Specifically, we use the replica trick to compute the free energy per spin [see Eq.~(\ref{freeenergy})] for the $p$-spin system, whose Hamiltonian is given in Eq.~(\ref{pspinhamiltonian}). To this end, we compute (setting $k_B = 1$, and writing $\beta = 1/T$)
\begin{align}
	\langle Z^n \rangle &= \int D\sigma \left\langle \exp\left[ \beta \sum_a\left(\sum_{i_1<\cdots < i_p} J_{i_1,\cdots,i_p}\sigma_{i_1}^a\cdots \sigma_{i_p}^a +\sum_{i} h_i \sigma_i^a \right)\right] \right\rangle_J \prod_a \delta\left(\sum_i (\sigma_i^a)^2 -N\right) \nonumber \\
	&= \int D\sigma \exp\left[ N\frac{\beta^2}{4} \sum_{ab}\left( \frac{1}{N} \sum_i \sigma_i^a\sigma_i^b\right)^p+\beta\sum_{i;a} h_i \sigma_i^a\right]  \prod_a \delta\left(\sum_i (\sigma_i^a)^2 -N\right), \label{pspinreplicatedpartition}
\end{align}
where we use the shorthand $D \sigma = \prod_{i a}( d\sigma_i^a/\sqrt{2\pi})$, and the distribution of the coupling constants obeys Eq.~(\ref{pspinstats}) with $J = 1$. We see that the random matrix partition function corresponds to the case $h = 0$, $\beta = 1$ and $p = 2$, without a spherical constraint. From now on, we will set $h_i = 0$ for the sake of simplicity.

In general, the replicas do not decouple in this case. Instead, to perform the saddle-point computation, we introduce the $n \times n$ `overlap' order parameter matrix
\begin{align}
	Q_{ab} = \frac{1}{N}\sum_{i} \sigma_i^a \sigma_i^b ,
\end{align}
where the spherical constraint is satisfied by setting $Q_{aa} = 1$. Imposing this using the complex exponential representations of delta functions, we have
\begin{align}
	\langle Z^n \rangle &= \int DQ D\hat q D\sigma \exp\left[ N \sum_{ab}\left(\frac{\beta^2}{4}  Q_{ab}^p + i\hat q_{ab} Q_{ab} - \frac{1}{N}\sum_i i \hat q_{ab} \sigma_i^a \sigma_i^b  \right)\right]  \nonumber \\
	&=\int DQ D\hat q \exp\left[ N \sum_{ab}\left(\frac{\beta^2}{4}  Q_{ab}^p + i\hat q_{ab} Q_{ab} \right)  - \frac{N}{2}\ln \det (2 i \underline{\underline{\hat q}})\right] ,  
\end{align}
Let us now carry out the integrals over the variables $Q_{ab}$ and $\hat q_{ab}$ using a saddle point approximation. We will find it most convenient to carry out the integral over $\hat q_{ab}$ first. That is, we wish to carry out the integrals $\int D\hat q \exp\left[ -NS_\lambda \right]$ with
\begin{align}
	S_\lambda = -iN \sum_{ab}\hat q_{ab} Q_{ab}   + \frac{N}{2}\ln \det (2 i \underline{\underline{\hat q}}) .
\end{align}
To do this, we make use of the identity 
\begin{align}
	\frac{\partial }{\partial M_{ab}} \ln \det \underline{\underline{M}} = \left( \underline{\underline{M}}\right)^{-1}_{ba} .
\end{align}
The saddle-point equation is thus (letting $\lambda_{ab} = i \hat q_{ab}$)
\begin{align}
	\frac{1}{2}(\underline{\underline{\lambda}}^{-1})_{ab} &= Q_{ab},  .
\end{align}
For this to be a valid saddle-point, we require the following Hessian matrix to have only positive eigenvalues\footnote{We note that we may find the derivative $\partial (\lambda^{-1})_{ab}/\partial \lambda_{cd}$ by differentiating $\sum_{b} (\lambda^{-1})_{ab} \lambda_{bc} = \delta_{ac}$.} 
\begin{align}
	\frac{\partial^2 S_\lambda}{\partial \lambda_{ab} \partial \lambda_{cd}} = \frac{1}{2}(\underline{\underline{\lambda}}^{-1})_{ac} (\underline{\underline{\lambda}}^{-1})_{bd} = 2 Q_{ac}Q_{bd},
\end{align}
which we see must be true because $\underline{\underline{Q}}$ is positive definite by definition. We thus obtain 
\begin{align}
	\langle Z^n \rangle &\propto \int DQ  \exp\left[ \frac{N}{2} \left( \frac{\beta^2}{2}\sum_{ab} Q_{ab}^p + \ln \det \underline{\underline{Q}}\right)\right] . 
\end{align}
Finally, carrying out the remaining saddle-point integration over $Q_{ab}$, we may write for the free energy density and the accompanying saddle-point equation for the overlap matrix $\underline{\underline{Q}}$
\begin{align}
	F &= \lim_{n\to 0} \left[ - \frac{1}{2\beta n} \left( \frac{\beta^2}{2}\sum_{ab} Q_{ab}^p + \ln \det \underline{\underline{Q}}\right)\right], \nonumber \\
	0 &=\frac{\beta^2 p}{2} Q_{ab}^{p-1} + (\underline{\underline{Q}}^{-1})_{ab}, \nonumber \\
	Q_{aa} &= 1. \label{fsaddle}
\end{align}
We note that in order for the saddle-point approximation made here to be valid, we should also investigate the Hessian matrix
\begin{align}
	\beta n\frac{\partial^2 F}{\partial Q_{ab} \partial Q_{cd}} = - \left[ \frac{\beta^2 p(p-1)}{2} Q_{ab}^{p-2} \delta_{ac}\delta_{bd} - (Q^{-1})_{ac} (Q^{-1})_{bd}\right],
\end{align}
and show that all of its eigenvalues are positive. We will demonstrate how this requirement can help us to rule out incorrect ansatz solutions for $Q_{ab}$. 

Our objective now is to solve the saddle-point equations for $Q_{ab}$ in Eq.~(\ref{fsaddle}). However, this is no trivial task. It requires us to make an ansatz for the form of the matrix $\underline{\underline{Q}}$, and to interpret carefully the limit $n \to 0$. We note that, importantly, the order parameters $Q_{ab}$ have a physical interpretation. Although they must be defined in the replicated system, their $n \to 0$ limits quantify the overlaps of the original $p$-spin system. Specifically, we have (see Ref. \cite{castellani2005spin} for an explanation of this)
\begin{align}
	\frac{1}{N}\sum_i \langle \langle \sigma_i\rangle_T^2 \rangle_J = \lim_{n \to 0} \frac{2}{n(n-1)} \sum_{a>b} Q_{ab}, \label{overlapinterpretation}
\end{align}
where $Q_{ab}$ is understood to be the saddle point value, and $\langle \cdot \rangle_T$ and $\langle \cdot \rangle_J$ indicate an average over realisations of the thermal noise and the disorder, respectively. We thus see that a non-zero overlap $Q_{ab}$ indicates that the spins have some preferred direction, which varies from realisation to realisation, depending on the disorder. This peculiar behaviour is indicative of the transition to the \textit{spin glass phase}, and corresponds to the free energy landscape developing many distinct minima. This will be important as we use the overlap matrix to investigate the presence of a phase transition below.

\subsubsection{Replica-symmetric (RS) solution}
The simplest possible ansatz is the replica-symmetric (RS) one. That is, let us suppose that at the saddle point we have
\begin{align}
	Q_{ab} &= q_0 +(1-q_0)\delta_{ab} , \nonumber \\
	\Rightarrow (Q^{-1})_{ab} &= \frac{1}{1-q_0} \delta_{ab} - \frac{q_0}{(1-q_0)[1+(n-1)q_0]}(1-\delta_{ab}). 
\end{align}
Substituting this into Eq.~(\ref{fsaddle}) we obtain for the free energy and the saddle point condition for $q_0$ respectively\footnote{One evaluates $\det \underline{\underline{Q}}$ by using the eigenvalues of $ \underline{\underline{Q}}$. These are $\lambda_1 = 1-q_0$, with degeneracy $n-1$, and $\lambda_2 = 1+(n-1)q_0$ with degeneracy 1. One must expand the logarithm to first order in $n$ to get the correct $n \to 0$ limit.}
\begin{align}
	F &= \lim_{n\to 0}\bigg\{- \frac{1}{2\beta n} \Big[\frac{\beta^2}{2}\left( n +n(n-1) q_0^p \right) + (n-1)\ln (1-q_0)  + \ln(1+(n-1)q_0)\Big]\bigg\},\nonumber \\
	\frac{\beta^2 p}{2} q_0^{p-1} &= \frac{q_0}{(1-q_0)[1+(n-1)q_0]} .
\end{align}
After taking the limit $n\to 0$, we hen have 
\begin{align}
	F &= - \frac{1}{2\beta} \left[\frac{\beta^2}{2}\left( 1 - q_0^p \right) + \ln (1-q_0) + \frac{q_0}{1-q_0}\right],\nonumber \\
	q_0^{p-1} (1-q_0)^2 &= \frac{2}{p} T^2 q_0.
\end{align}
The latter equation here has two solutions. The first is the trivial `paramagnetic' solution, for which $q_0 = 0$. In this case, we obtain for the free energy $F = -1/(4T)$. However, beyond a critical temperature $T^\star$, another solution emerges such that $q_0 \neq 0$. By inspection, this solution is the dominant contribution in the saddle-point computation, and has a lower associated free energy, before we take the limit $n\to 0$ and $n$ is still an integer. We highlight a crucial subtlety here; after the saddle point computation has been performed and we take $n \to 0$, the free energy in the paramagnetic case diverges to $-\infty$ and the non-trivial solution has finite (i.e. higher) free energy. This is an expected, albeit peculiar, feature of the replica method that we have to live with. The important thing to recognise is that the saddle point computation is carried out for integer $n$, and we \textit{subsequently} take the limit $n \to 0$ once we have the result. The dominant contribution is therefore the one for which the free energy (and equivalently the exponent of the saddle-point integral) is least \textit{before} we promote $n$ to a real variable and take $n\to 0$. Let us now analyse the validity of the two solutions to the saddle point equations. 

The eigenvalues of the Hessian are found by solving
\begin{align}
	&-\bigg[\left(\frac{\beta^2 p(p-1)}{2} q_0^{p-2}- \frac{1}{(1-q_0)^2} \right)\delta Q_{ab} + \frac{q_0}{(1-q_0)^2[1+ (n-1)q_0]} \sum_{d} \left(\delta Q_{ad} + \delta Q_{bd}\right) \nonumber \\
	&\hspace{1cm}- \frac{q_0^2}{(1-q_0)^2[1+ (n-1)q_0]^2}\sum_{ab} \delta Q_{ab}\bigg] = \Lambda \delta Q_{ab},
\end{align}
from which we obtain\footnote{These are found by setting $\sum_{d} \delta Q_{ad}= 0$, $\sum_{ad} \delta Q_{ad}= 0$, and $ \delta Q_{ad}= 1$ respectively.}
\begin{align}
	\Lambda_1 &= \frac{1}{(1-q_0)^2} - \frac{\beta^2 p(p-1)}{2}q_0^{p-2}, \nonumber \\
	\Lambda_2 &= \Lambda_1 - (n-2) \frac{q_0}{(1-q_0)^2[1+ (n-1)q_0]}, \nonumber \\
	\Lambda_3 &= \Lambda_2 - n \frac{q_0}{(1-q_0)^2[1+ (n-1)q_0]}+ n (n-1)\frac{q_0^2}{(1-q_0)^2[1+ (n-1)q_0]^2} .
\end{align}
These eigenvalues have degeneracies $n(n-3)/2$, $n-1$ and $1$ respectively. We see that the eigenvalues are always positive for the paramagnetic solution. In the non-trivial case however, we have $\beta^2 p q_0^{p-2}/2 = (1-q_0)^{-2}$, and thus $\Lambda_1 = (2-p)/(1-q_0)^2$, which is always negative for $p>2$, and we see that the non-trivial RS solution cannot contribute. Since the paramagnetic solution predicts an infinitely negative energy density for $T \to 0$, which obviously cannot be correct, we see that something has gone wrong here. We must therefore try a more general ansatz in the hopes of finding another contribution to the saddle-point computation.

\subsubsection{One-step replica symmetry breaking (1RSB)}
To go beyond the RS ansatz, we must assume that the overlaps between certain pairs of replicas are different to others. The simplest way to do this is to split to replicas into groups. The overlaps between replicas within the same group take the value $q_1$, whereas overlaps between replicas in different groups are $q_0$. Importantly, we introduce the so-called symmetry-breaking parameter $m$, which is the number of replicas that are in the same group. We note that the symmetry breaking that occurs here is not in the sense that any one replica is special -- of course this cannot be. It is the symmetry of the replicas with respect to \textit{one another} that is broken.

The one-step replica symmetry breaking ansatz is instead (for $m = 3$ and $n = 6$)
\begin{align}
	\underline{\underline{Q}} = \begin{bmatrix}
		1 & q_1 & q_1 & q_0 & q_0 & q_0 \\
		q_1 & 1 & q_1 & q_0 & q_0 & q_0 \\
		q_1 & q_1 & 1 & q_0 & q_0 & q_0 \\
		q_0 & q_0 & q_0 & 1 & q_1 & q_1  \\
		q_0 & q_0  & q_0 & q_1 & 1 & q_1 \\
		q_0 & q_0 & q_0 & q_1 & q_1 & 1 
	\end{bmatrix} .
\end{align}
Inserting this into Eq.~(\ref{fsaddle}), we have instead\footnote{In this case, the eigenvalues of $ \underline{\underline{Q}}$ are $\lambda_1 = 1-q_1$, with degeneracy $n-n/m$, $\lambda_2 = m(q_1-q_0)+1-q_1$ with degeneracy $n/m-1$, and $\lambda_3 = n q_0 + m(q_1-q_0)+ 1-q_1$ with degeneracy 1.}
\begin{align}
	-2 \beta F =& \frac{\beta^2}{2} \left[1 + (m-1)q_1^p - m q_0^p \right]+ \frac{m-1}{m} \ln(1-q_1) \nonumber \\
	&+ \frac{1}{m}\ln\left[m (q_1-q_0) + 1-q_1 \right] + \frac{q_0}{m(q_1-q_0)+1-q_1},
\end{align}
where we note that one must expand the logarithm to leading order in $n$ to obtain the correct $n \to 0$ limit. Optimising this with respect to $q_0$, $q_1$ and $m$, we find
\begin{align}
	q_0 &= 0, \nonumber \\
	(1-m)\left[\frac{\beta^2 p}{2}q_1^{p-1} - \frac{q_1}{(1-q_1)[(m-1)q_1+1]}  \right] &= 0, \nonumber \\
	\frac{\beta^2}{2}q_1^p + \frac{1}{m^2} \ln\left( \frac{1-q_1}{1-(1-m)q_1}\right) + \frac{q_1}{m[1-(1-m)q_1]} &= 0. 
\end{align}
We note that in the $n\to 0$ limit, we need no longer expect $m$ to be an integer either. Instead, we promote it to a real number, which turns out to satisfy $0\leq m \leq1$ \cite{castellani2005spin}. The simple solution here is $m = 1$ so that
\begin{align}
	\frac{\beta^2}{2}q_1^p + \ln\left(1-q_1\right) + q_1 &= 0,
\end{align}
of which we see $q_1 = 0$ is a solution. We thus recover the paramagnetic solution that we saw for the RS case, which we noted was a valid solution. 

\begin{figure}[h]
	\centering 
	\includegraphics[scale = 0.45]{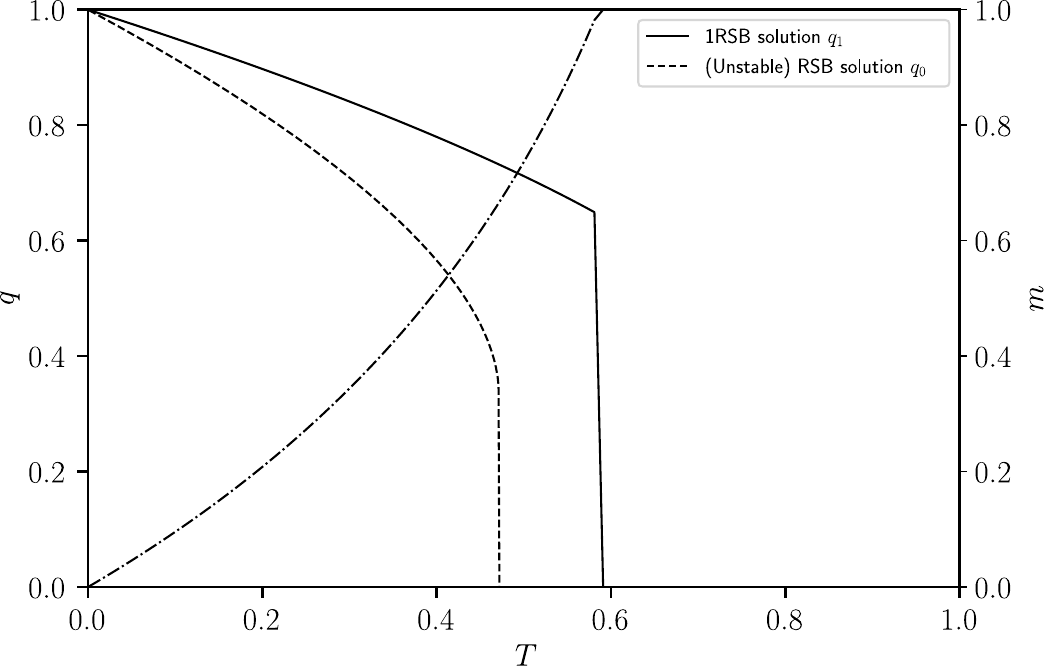}
	\includegraphics[scale = 0.45]{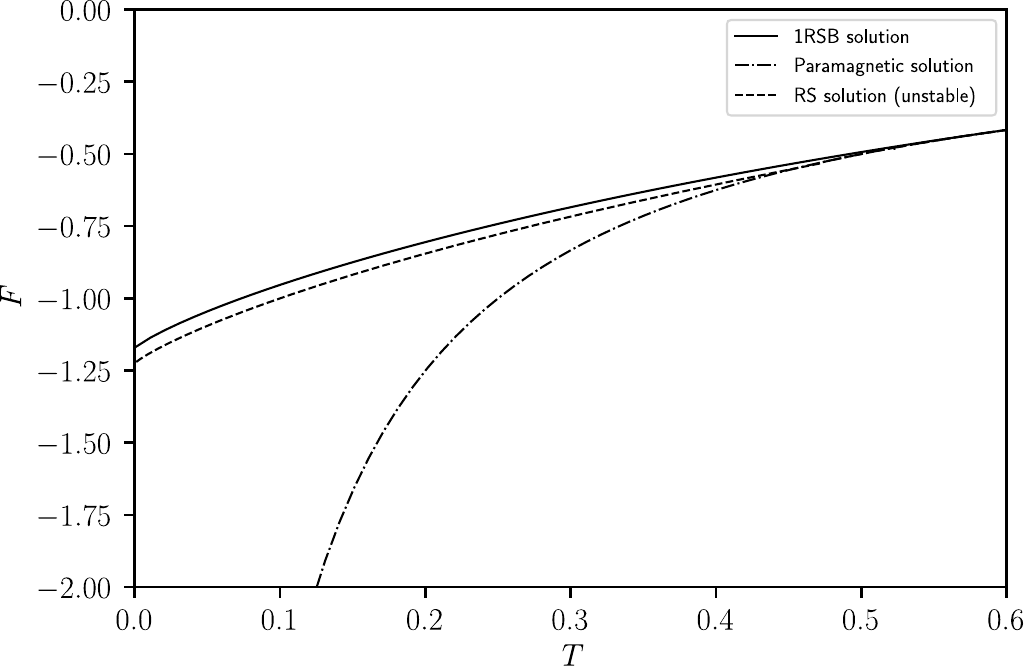}
	\captionsetup{justification=raggedright,singlelinecheck=false}
	\caption{(Left) The order parameters $q$ and $m$ for $p = 3$. As we reduce the temperature, a non-zero overlap order parameter emerges at $T_c \approx 0.586$, and the symmetry breaking parameter becomes $m<1$. The RSB solution underestimates $T_c$ and the overlap. (Right) The free energy has a discontinuous derivative at the transition. While the paramagnetic solution has a diverging free energy at zero, the 1RSB solution tends to a finite value. The RS solution is again inaccurate. }\label{fig:rsborderparameters}
\end{figure}

Let us then see under what circumstances we obtain a non-trivial solution. If we plot the function $f(q_1) =\frac{\beta^2}{2}q_1^p + \ln\left(1-q_1\right) + q_1$, we see that a maximum develops as we decrease $T$, and this maximum moves upwards. When the maximum hits $f = 0$, we obtain a new non-zero solution for $q_1$. The critical temperature at which the transition occurs is \cite{barrat1997p}
\begin{align}
	T_c = y \sqrt{\frac{p}{2y}}(1-y)^{\frac{p}{2}-1}, \hspace{1cm} \frac{2}{p} = -2y \frac{1-y + \ln y}{(1-y)^2} . \label{Tcstatic}
\end{align}
Decreasing $T$ below $T_c$, $m$ becomes less than 1, and $q_1$ increases towards 1. From Eq.~(\ref{overlapinterpretation}), we thus see that $T_c$ marks the onset of the spin-glass transition. At this temperature, the free energy has a discontinuous derivative, and the observable $\frac{1}{N}\sum_i \langle \langle \sigma_i\rangle_T^2 \rangle_J$ takes on a non-zero value. The order parameters and the free energy per spin are plotted as a function of temperature in Fig. \ref{fig:rsborderparameters}. 

One can show that this solution is `stable' in the sense that the Hessian has all real eigenvalues, of which there are nine for the 1RSB ansatz \cite{crisanti1992spherical}. In fact, the 1RSB solution is all we need for all temperatures. This is in contrast to the Sherrington-Kirkpatrick model, for example, which requires Parisi's ultrametric ansatz \cite{parisi1979infinite, parisi1980order, parisi1983order}.

\begin{figure}[H]
	\centering 
	\includegraphics[scale = 0.5]{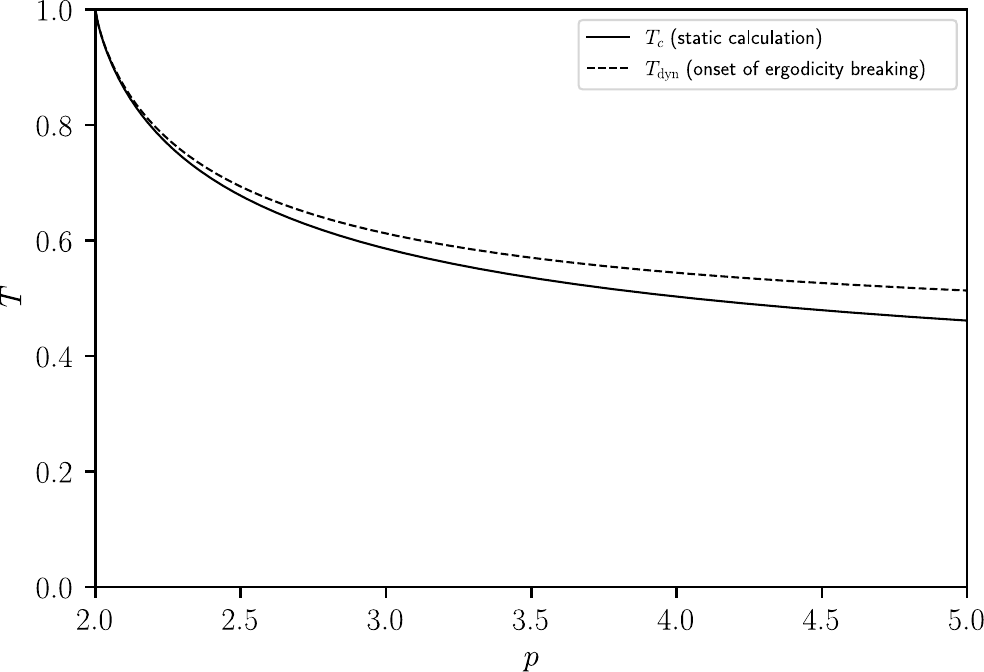}
	\captionsetup{justification=raggedright,singlelinecheck=false}
	\caption{Critical temperature as a function of $p$. The onset of ergodicity breaking occurs at a higher temperature than the thermodynamic transition. However, the critical temperatures from the dynamic and static approaches coincide for $p = 2$, where there is no non-trivial glassy behaviour.}\label{fig:rsbtc}
\end{figure}

We also plot, in Fig. \ref{fig:rsbtc}, the critical temperature as a function of $p$ (which we plot as a continuous variable with the understanding that only integer values are really possible). This is compared with the critical temperature as understood from the dynamical approach in Eq.~(\ref{tc}). We notice that the dynamical transition, at which ergodicity breaks, occurs at a higher temperature. This is due to the fact that the system gets `stuck' in metastable states with large energy barriers \cite{castellani2005spin, crisanti1993spherical}, and is thus unable to equilibrate. That is, as we decrease temperature, ergodicity breaks \textit{before} the thermodynamic transition occurs. 

\subsection{Non-Hermitian random matrices: elliptic law}\label{section:replicanonherm}

While we have so far discussed Hermitian matrices, we can indeed extend the replica procedure to the non-hermitian case. Below, we show how the replica approach can be used to rederive the elliptic law. This is accomplished using the so-called `eigenvalue potential' construction [see also  Eq.~(\ref{potential})]
\begin{align}
	\Phi(z,z^\star) &= -\frac{1}{N}\ln\mathrm{det}\left[\eta^2 \underline{\underline{\id}} +  (z^\star\underline{\underline{\id}} -\underline{\underline{J}}^T) (z\underline{\underline{\id}} -	\underline{\underline{J}}) \right]. \label{phicandrho}
\end{align}
The correctly regularised resolvent for the non-Hermitian case [see Eq.~(\ref{cdef})] and the 2D eigenvalue density are then extracted via
\begin{align}
	 C(z,z^\star) &= \frac{1}{N} \mathrm{Tr} \left\{ (z^\star \underline{\underline{\id}} -\underline{\underline{J}}^T ) \left[(z \underline{\underline{\id}} -\underline{\underline{J}} )(z^\star \underline{\underline{\id}} -\underline{\underline{J}}^T ) + \eta^2 \underline{\underline{\id}}  \right]^{-1}\right\} =-\frac{\partial}{\partial z } \Phi(z,z^\star), \nonumber \\
	\rho^{(2)}(z) &= \frac{1}{\pi}  \lim_{\eta\to 0}\frac{\partial}{ \partial z^\star} C(z,z^\star) =-\frac{1}{\pi}  \lim_{\eta\to 0}\frac{\partial^2}{\partial z \partial z^\star} \Phi(z,z^\star) .
\end{align}
We thus see that to find the disorder-averaged eigenvalue density, which can be extracted from the disorder-averaged $\Phi(z,z^\star)$, we are once again charged with computing the average of a logarithm. What is particularly nice about the construction $\Phi(z,z^\star)$, which is close to Girko's \cite{girko1985circular,girko1986elliptic}, is that it involves the determinant of an Hermitian matrix. That is, we have Hermitised the problem. This allows us to proceed with far less difficulty. Using complex Gaussian integration, we may rewrite the determinant as an area integral over the complex plane. This takes somewhat more effort than in the hermitian random matrix case, but we remark that the exponential integral representation of $\Phi(z,z^\star)$ that we derive in Eq.~(\ref{generalexpint}) applies for any matrix $\underline{\underline{J}}$, so one needn't be careful to adapt this derivation when generalising this calculation.

\begin{tcolorbox}[colback=blue!10!white,colframe=blue!90!black,title=Lemma: Determinants of Hermitian matrices as Gaussian integrals]
	Suppose we have an Hermitian matrix $\underline{\underline{H}} =  \underline{\underline{X}} + i \underline{\underline{Y}}$, where $\underline{\underline{X}}$ is real and symmetric and $\underline{\underline{Y}}$ is real and anti-symmetric. Since $\underline{\underline{H}} $ is hermitian, we have $(\mathrm{det}\underline{\underline{H}})^2 = \mathrm{det}(\underline{\underline{X}}^2+\underline{\underline{Y}}^2)$. We can therefore also write the determinant of $\underline{\underline{H}}$ as the root of the determinant of a block matrix
	\begin{align}
		\mathrm{det}\underline{\underline{H}} = \sqrt{\mathrm{det} \begin{bmatrix}
				\underline{\underline{X}} & -\underline{\underline{Y}} \\
				\underline{\underline{Y}} & \underline{\underline{X}}
		\end{bmatrix}} .
	\end{align}
	We can thus write this a Gaussian integral over real variables
	\begin{align}
		(\mathrm{det}\underline{\underline{H}})^{-1} = \int \left(\prod_i\frac{du^{(x)}_i du^{(y)}_i}{\pi}\right) \exp\Bigg[-S(\underline{u})  \Bigg],
	\end{align}
	where 
	\begin{align}
		S=\begin{bmatrix}
			(\underline{u}^{(x)})^T &(\underline{u}^{(y)})^T 
		\end{bmatrix} \begin{bmatrix}
			\underline{\underline{X}} & -\underline{\underline{Y}} \\
			\underline{\underline{Y}} & \underline{\underline{X}}
		\end{bmatrix}\begin{bmatrix}
			\underline{u}^{(x)}  \\
			\underline{u}^{(y)} 
		\end{bmatrix}
	\end{align}
	Writing $u_i = u^{(x)}_i + i u^{(y)}_i$, we can finally rewrite this more compactly as
	\begin{align}
		(\mathrm{det}\underline{\underline{H}})^{-1} = \int \left(\prod_i\frac{d^2u_i}{\pi}\right) \exp\left[- \sum_{ij} u_i^\star H_{ij} u_j \right].\label{deth}
	\end{align}
\end{tcolorbox}

\subsubsection{Exponential integral representation}
We therefore may write the eigenvalue potential as  
\begin{align}
	\exp[N\Phi(z,z^\star)] = \int \left(\frac{d^2u_i}{\pi} \right) \exp&\bigg[- \eta^2 \sum_i \vert u_i \vert^2 - \sum_{i,j,k} u^\star_i (z^\star \delta_{ik}- J_{ik}^T)(z \delta_{kj}- J_{kj})u_j  \bigg] .\label{withouths}
\end{align}
However, we still have an integrand whose exponent is quadratically dependent on the disorder. This makes taking the disorder average unnecessarily difficult. To decouple the disordered terms, we exploit an extremely useful identity in field theory computations. This is the Hubbard-Stratonovich (HS) transformation. \begin{tcolorbox}[colback=blue!10!white,colframe=blue!90!black,title=Lemma: Hubbard-Stratonovich (HS) transformation]
	The HS transformation can take various similar forms, but in this case we use
	\begin{align}
		\exp\left[ -\sum_i x_i^2\right] = \int \left(\prod_i \frac{d\phi_i}{\sqrt{\pi}} \right)\exp\left[ - \sum_i \phi_i^2 - 2i \sum_i x_i \phi_i\right].\label{hstrans}
	\end{align}
	This can be understood easily as the Fourier relation for a product of uncoupled Gaussians. We take advantage of similar identities in the context of the SUSY approach later in Section \ref{section:susy}.
\end{tcolorbox}

Now, suppose that the matrix $\underline{\underline{H}}$ in Eq.~(\ref{deth}) can be written as a product $\underline{\underline{H}} =\underline{\underline{M}}^\dagger\underline{\underline{M}}$, where $\underline{\underline{M}} = \underline{\underline{M}}^{(x)} +i \underline{\underline{M}}^{(y)}$ is not necessarily Hermitian. We may thus apply the Hubbard-Stratonovich transformation to Eq.~(\ref{deth}) by writing
\begin{align}
	\begin{bmatrix}
		\underline{\underline{X}} & -\underline{\underline{Y}} \\
		\underline{\underline{Y}} & \underline{\underline{X}}
	\end{bmatrix} = \begin{bmatrix}
		\underline{\underline{M}}^{(x)} &\underline{\underline{M}}^{(y)} \\
		-\underline{\underline{M}}^{(y)} & \underline{\underline{M}}^{(x)}
	\end{bmatrix}\begin{bmatrix}
		\underline{\underline{M}}^{(x)} &-\underline{\underline{M}}^{(y)} \\
		\underline{\underline{M}}^{(y)} & \underline{\underline{M}}^{(x)}
	\end{bmatrix}.
\end{align}
Identifying the vectors $\underline{\phi}$ and $\underline{x}$ in Eq.~(\ref{hstrans}) as
\begin{align}
	\underline{\phi} =  \begin{bmatrix}
		\underline{v}^{(x)}  \\
		\underline{v}^{(y)} 
	\end{bmatrix}, \hspace{1cm}
	\underline{x} = \begin{bmatrix}
		\underline{\underline{M}}^{(x)} &-\underline{\underline{M}}^{(y)} \\
		\underline{\underline{M}}^{(y)} & \underline{\underline{M}}^{(x)}
	\end{bmatrix} \begin{bmatrix}
		\underline{u}^{(x)}  \\
		\underline{u}^{(y)} 
	\end{bmatrix}, 
\end{align}
we therefore obtain
\begin{align}
	(\mathrm{det}\underline{\underline{M}}^\dagger\underline{\underline{M}})^{-1} = \int \left(\prod_i\frac{du^{(x)}_i du^{(y)}_i dv^{(x)}_i dv^{(y)}_i}{\pi^2}\right) \exp\Bigg[-S(\underline{u},\underline{v} )  \Bigg] ,
\end{align}
where 
\begin{align}
	S(\underline{u},\underline{v} )=& -\sum_i  (v^{(x)}_i)^2+ (v^{(y)}_i)^2\nonumber \\
	&+2i\sum_{ij} \bigg(v^{(x)}_i M^{(x)}_{ij} u^{(x)}_j +v^{(y)}_i M^{(x)}_{ij} u^{(y)}_j +v^{(y)}_iM^{(y)}_{ij}u^{(x)}_j- v^{(x)}_i M^{(y)}_{ij} u^{(y)}_j \bigg), \nonumber \\
	=&-\begin{bmatrix}
		(\underline{v}^{(x)})^T &(\underline{v}^{(y)})^T 
	\end{bmatrix}\begin{bmatrix}
		\underline{v}^{(x)}  \\
		\underline{v}^{(y)} 
	\end{bmatrix}+2i\begin{bmatrix}
		(\underline{v}^{(x)})^T &(\underline{v}^{(y)})^T 
	\end{bmatrix} \begin{bmatrix}
		\underline{\underline{M}}^{(x)} &-\underline{\underline{M}}^{(y)} \\
		\underline{\underline{M}}^{(y)} & \underline{\underline{M}}^{(x)}
	\end{bmatrix} \begin{bmatrix}
		\underline{u}^{(x)}  \\
		\underline{u}^{(y)} 
	\end{bmatrix}.
\end{align}
More succinctly, letting $v_i = v_i^{(x)} + iv_i^{(y)}$, we have the general relation
\begin{align}
	(\mathrm{det}\underline{\underline{M}}^\dagger\underline{\underline{M}})^{-1} =& \int \left(\prod_i\frac{du^2_i dv^2_i}{\pi^2}\right) \exp\Bigg[-\sum_i \vert v_i \vert^2 +i\sum_{ij}\left( u^\star_i M^\dagger_{ij} v_j + v_i^\star M_{ij} u_j \right)  \Bigg] .
\end{align}
We thus obtain for our specific case $\underline{\underline{M}} =  z\underline{\underline{\id}} - \underline{\underline{J}}$
\begin{empheq}[box={\fboxsep=5pt\fbox}]{align}
	\exp[N\Phi(z,z^\star)] =  \int &\left(\frac{d^2u_i d^2v_i}{\pi^2} \right) \exp\Bigg[-\sum_i(\vert v_i \vert^2+ \eta^2  \vert u_i \vert^2) \nonumber \\
	&+ i\sum_{i,j} [u^\star_i (z^\star \delta_{ij}- J_{ij}^T)v_j +v_i^\star(z \delta_{ij}- J_{ij})u_j ] \Bigg] .\label{generalexpint}
\end{empheq}
Although we have had to put some effort in to get to this point, this representation not only forms the starting point of our calculation presently, but it is also applies generally (see e.g. \cite{haake1992statistics, baron2020dispersal, baron2022eigenvalues,patil2024spectral}), since the matrix $\underline{\underline{J}}$ has been left entirely generic up to this point. 
\subsubsection{Disorder average and elliptic law} 
We are now in a position to take the average over the disorder. That is, we now wish to find $\langle \Phi(z,z^\star) \rangle$, which will yield the disorder-averaged resolvent $\langle C(z,z^\star)\rangle$, and thus the average eigenvalue density via Eqs.~(\ref{phicandrho}). This can be accomplished using replicas, and we proceed in a way analogous to the Hermitian case, but we do not reproduce the steps in detail here.

We first introduce the replicated disorder-averaged eigenvalue potential $\langle \Phi^n \rangle$. After averaging, one can show again that the replicas decouple \cite{sommers1988spectrum, edwards1976eigenvalue, baron2022eigenvalues} by performing an expansion of the quartic term in the exponent as we did in Section \ref{section:replicadecoupling}. We may thus consider the annealed average, i.e. $\exp\left[N\langle \Phi(z,z^\star) \rangle\right] = \left\langle\exp\left[N \Phi(z,z^\star) \right]\right\rangle$. Using similar reasoning, one can also show that it is possible to neglect the terms involving $N^{-1}\sum_i u_i u_i$, $N^{-1}\sum_i v_i v_i$, and $N^{-1}\sum_i v_i u_i$ in the exponent of the integrand.

One finally arrives at
\begin{align}
	\exp\left[N\langle \Phi(z,z^\star) \rangle\right] = & \int \left(\frac{d^2u_i d^2v_i}{\pi^2} \right) \exp\Bigg[-\sum_i\vert v_i \vert^2+ \eta^2  \vert u_i \vert^2 -i\sum_i (u_i^\star v_i z^\star +u_i v_i^\star z ) \nonumber \\
	&-\frac{\sigma^2}{2N}\sum_{i,j} \left[2\vert u_i\vert^2 \vert v_j\vert^2 + \Gamma (u_i^\star v_i u_j^\star v_j + u_i v_i^\star u_j v_j^\star) \right] \Bigg].
\end{align}
Proceeding analogously to the Hermitian case, we now introduce `order parameters' and subsequently perform a saddle point approximation. Due to the complex quantities $w\equiv N^{-1}\sum_i u_i^\star v_i$ and its complex conjugate, we therefore require a complex exponential representation of the 2D Dirac delta function. The real quantities $q_u \equiv N^{-1} \sum_{i} \vert u_i\vert^2$ and $q_v \equiv N^{-1} \sum_{i} \vert v_i\vert^2$ may be introduced via ordinary 1D Dirac delta functions in their complex exponential representation. 
\begin{tcolorbox}[colback=blue!10!white,colframe=blue!90!black,title=Lemma: Complex exponential representation of the Dirac delta in the complex plane]
	We wish to write a complex exponential expression for 
	\begin{align}
		\delta^{(2)}(f - f_0) = \delta(f^{(x)}-f^{(x)}_0) \delta(f^{(y)}-f^{(y)}_0) . 
	\end{align}
	where $f = f^{(x)} + i f^{(y)}$ and $f_0 = f_0^{(x)} + i f_0^{(y)}$. We may write simply
	\begin{align}
		\delta^{(2)}(f - f_0) &= \int \frac{d\hat f^{(x)} d\hat f^{(y)}}{(2\pi)^2} \exp\big[ i \hat f^{(x)}\left(f^{(x)} - f_0^{(x)} \right) \nonumber \\
		&\hspace{1cm} +  i \hat f^{(y)}\left(f^{(y)} - f_0^{(y)} \right)\big] \nonumber \\
		&= \int \frac{d^2\hat F}{\pi^2} \exp\left[ i \hat F^\star\left(f - f_0 \right) +  i \hat F\left(f^\star - f_0^\star \right)\right]
	\end{align}
	where we substitute $\hat F =  \hat f/2$. 
\end{tcolorbox}
Armed with this, we thus obtain 
\begin{align}
	\exp\left[N\langle \Phi(z,z^\star) \rangle\right] = \int \mathcal{D}[\cdots] \exp\left[ N(\Psi + \Theta + \Omega)\right], \label{withops}
\end{align}
where $\mathcal{D}[\cdots] $ is short-hand for the integration over the order parameters $q_u$, $q_v$ and $w$, their associated `hatted' conjugate variables, and factors of $2$ and $\pi$, and we have
\begin{align}
	\Psi &= i \hat q_u q_u + i \hat q_v q_v + i (\hat w w^\star + \hat w^\star w), \nonumber \\
	\Theta &=  -\eta^2q_u - q_v - \sigma^2 q_u q_v - i (w z^\star +w^\star z) - \frac{\Gamma\sigma^2}{2}(w^2 + (w^\star)^2), \nonumber \\
	\Omega &= \ln\left\{\int \left(\frac{d^2u d^2v}{\pi^2} \right) \exp\left[- i\left(\hat q_u u^\star u + \hat q_v v^\star v + \hat w^\star u^\star v +\hat w u v^\star   \right)\right]\right\} \nonumber \\
	&= - \ln\vert\hat w \hat w^\star - \hat q_u \hat q_v\vert + \mathrm{const}. ,
\end{align}
where we have used the standard Fresnel integral \cite{edwards1976eigenvalue} to evaluate $\Omega$. We have thus succeeded in integrating out the fields $u_i$ and $v_i$, and we are left with a far smaller number of integration variables.

At this stage, it is helpful to note the following. Using the definition in Eq.~(\ref{phicandrho}), we see that we may write the resolvent
\begin{align}
	\langle C(z,z^\star) \rangle = \exp\left[-N\langle \Phi(z,z^\star) \rangle\right] \int \mathcal{D}[\cdots] (i w^\star)\exp\left[ N(\psi + \Theta + \Omega)\right] . \label{cfromphi}
\end{align}
Hence, we may evaluate Eq.~(\ref{cfromphi}) using the saddle point method once again, and $C(z, z^\star)$ will simply be given by the value of $iw^\star$ at the saddle point. The equations that determine the saddle point are
\begin{align}
	i \hat q_u =  \eta^2 + \sigma^2 q_v, &\hspace{1cm} i \hat q_v = 1+ \sigma^2 q_u,  \nonumber \\
	i \hat w = i z + \Gamma \sigma^2 w^\star, &\hspace{1cm} i \hat w^\star = i z^\star + \Gamma \sigma^2 w , \nonumber \\
	i q_u = -\frac{\hat q_v}{\hat w \hat w^\star - \hat q_u \hat q_v}, &\hspace{1cm} i q_v =  -\frac{\hat q_u}{\hat w \hat w^\star - \hat q_u \hat q_v}, \nonumber \\
	iw^\star= \frac{\hat w^\star}{\hat w \hat w^\star - \hat q_u \hat q_v}, &\hspace{1cm} iw= \frac{\hat w}{\hat w \hat w^\star - \hat q_u \hat q_v} .
\end{align}
We first note that 
\begin{align}
	q_u q_v - w w^\star = \frac{1}{\hat w \hat w^\star - \hat q_u \hat q_v} \equiv \frac{1}{\Delta} .
\end{align}
Eliminating $\hat q_u$, we therefore see that there are two solutions for $\eta \to 0$: one where $q_v = 0$, and one where $\Delta = \sigma^2$. In the first instance, we see that by eliminating $i \hat w$ that
\begin{align}
	\Gamma \sigma^2 (iw^\star)^2 -z(iw^\star) +1 = 0.
\end{align}
This corresponds to Eq.~(\ref{csolanalytic}). For the second solution, we have 
\begin{align}
	i w^\star &= \frac{\hat w^\star}{\sigma^2} = \frac{z^\star - \Gamma \sigma^2 (iw)}{\sigma^2}, \nonumber \\
	i w &= \frac{\hat w}{\sigma^2} = \frac{z - \Gamma \sigma^2 (i w^\star)}{\sigma^2} ,
\end{align}
which corresponds to Eq.~(\ref{csolnonanalytic}). We have thus succeeded in obtaining the same two solutions for the resolvent as in Section \ref{section:ellipse}, from which we obtained the elliptic law.

As a quick aside, we note that although the fact that at the saddle-point $iw = (iw^\star)^\star$ may at first seem troubling, one should note the following. Let $w = w^{(x)} + iw^{(y)}$ and $w^\star =  w^{(x)} + iw^{(y)}$. The (real) variables over which we integrate in Eq. (\ref{cfromphi}), in order to impose the delta function $\delta(w - N^{-1} \sum_{i}u^\star_i v_i)$, are $w^{(x)} $ and $ w^{(y)} $. To perform the saddle point computation, we have to use analytical continuation to deform the integrals, which are initially over the real axis, to contours in the complex plane. As such, the ostensibly real variables $w^{(x)} $ and $ w^{(y)} $ take on complex values at the saddle point. Hence, $w$ and $w^\star$ are no longer necessarily complex conjugates when evaluated at the saddle point.

\subsection{Advantages of using replicas}
Although the above calculations may seem somewhat protracted compared to the cavity method, the replica procedure becomes much more user-friendly after repeated exposure, particularly when computing the average eigenvalue density. In particular, if we guess ahead of time that the replicas will decouple, we can simply begin with the formula in Eq.(\ref{decoupled}) or its equivalent for the ensemble of interest. In practice therefore, one may often neglect to use replicas entirely, and simply assume that the quenched and annealed averages for $\langle G(\omega)\rangle$ are equivalent. 

With this in mind, the replica method then becomes a highly flexible and convenient tool for computing the average eigenvalue density. Indeed, it is straightforward to extend the replica method to block-structured matrices \cite{baron2020dispersal, patil2024spectral}, matrix products \cite{agliari2019marchenko}, etc. Indeed, one can also use the same approach as in Section \ref{section:outlier} to calculate the outlier eigenvalue. 

Clearly, the downside of the replica method is its lack of rigour. We had to assume that we could carry over the results for integer $n$ to continuous $n$, which is nothing short of divination. And yet, the results that we obtain work! However, one must be careful. Some examples of more sophisticated calculations have been found where a na\"ive replica treatment falls short, e.g. in the calculation of the microscopic eigenvalue density correlations \cite{verbaarschot1985critique, kamenev1999wigner, zirnbauer1999another}.

With this being said, the replica method is really most illuminating for systems in which the annealed and quenched averages are not the same. Phenomena such as replica symmetry breaking reveal changes in the behaviour of the system that signify glassy physics, as we have seen. The replica method is thus indispensable in this context as a framework for understanding glassy systems \cite{mezard1987spin,charbonneau2023spin}.

\subsection{Exercises}
We derived the Wigner semicircle law once again using the replica method above. In doing so, we showed explicitly that the replicas decoupled. An alternative way to proceed, which we now explore, is to introduce the overlap order parameters $Q_{ab}$.
\begin{itemize}
	\item Beginning with Eq.~(\ref{replicaaveraged}), show that we may introduce the overlap matrix as we did in our study of the $p$-spin model to obtain (with $z = \omega - i\epsilon$, and taking $\sigma= 1$ for simplicity)
	\begin{align}
		\langle \mathcal{Z}^n \rangle = \int DQ D\hat q \exp\bigg[ -N\bigg( \frac{i z }{2}\sum_a Q_{aa} &+ \sum_{ab}\left( \frac{Q_{ab}^2}{4} + i \hat q_{ab}Q_{ab}\right) + \frac{1}{2} \ln \det (2i\underline{\underline{\hat q}})\bigg)\bigg] . 
	\end{align}
	\item Carry out the saddle-point approximation for the integral over the variables $\{\hat q_{ab}\}$, and show that the saddle-point is given by $2Q_{ab} = (\underline{\underline{\lambda}}^{-1})_{ab}$, where $i \hat q_{ab} = \lambda_{ab}$. 
	\item Carry out the subsequent saddle-point approximation for the integral over $Q_{ab}$ to obtain the condition 
	\begin{align}
		-\frac{iz}{2} \delta_{ab}- \frac{1}{2}Q_{ab} + \frac{1}{2}\left(\underline{\underline{Q}}^{-1}\right)_{ab} = 0.
	\end{align} 
	\item Using the replica-symmetric ansatz $Q_{ab} = q_d \delta_{ab} + q_0(1-\delta_{ab})$, show that
	\begin{align}
		q_d &= \frac{q_d +(n-2)q_0}{(q_d-q_0)[q_d + (n-1)q_0]}-iz \nonumber \\
		q_0 &= \frac{q_0}{(q_d-q_0)[q_d + (n-1)q_0]} .
	\end{align}
	\item Taking the $q_0 = 0$ solution, evaluate $\langle G \rangle = -\frac{2}{N} \frac{\partial}{\partial \omega} \lim_{n \to 0} [\langle \mathcal{Z}^n \rangle -1]/n$ to derive the semicircle law once more. We note that the off-diagonal elements of the overlap matrix evaluate to zero here, indicating the decoupling of the replicas. 
\end{itemize}
We also studied the $p$-spin model using the replica method above. To get the correct saddle-point solution, it was necessary for us to use an ansatz for the overlap matrix that broke replica symmetry. Here, we show that this isn't at all necessary for the case $p = 2$, for which we can use results that we have previously obtained using random matrix theory. In the $p=2$ case, the partition function is self-averaging, and thus there is no need for replicas, or indeed replica symmetry breaking.
\begin{itemize}
	\item In the case $p =2$, show that we may write the partition function (without taking the disorder average) in terms of the eigenvalues of $\underline{\underline{J}}$
	\begin{align}
		Z &= \int D\sigma d\hat q \exp\left[\frac{\beta}{2}  \sum_{ij}J_{ij}\sigma_i \sigma_j + i \hat q\left(\sum_i \sigma_i^2 - N\right)\right]\nonumber \\
		&= \int D\phi d\hat q \exp\left[\frac{\beta}{2}  \sum_{\nu}\lambda_\nu(\phi_\nu)^2  + i \hat q\left(\sum_\nu \phi_\nu^2 - N\right)\right],
	\end{align}
	where we write $\sigma_i = \sum_\nu v_i^{(\nu)} \phi_\nu$, where $\underline{v}^{(\nu)}$ is the eigenvector of $\underline{\underline{J}}$ with eigenvalue $\lambda_\nu$. 
	\item Assuming that for the largest eigenvalue we have $\mathrm{Re}(\beta\lambda_\mathrm{max}/2 + i \hat q)<0$, carry out the integration over $\{\phi_\nu\}$ to show that
	\begin{align}
		Z = \int  d\hat q \exp\left[  - iN \hat q - \frac{1}{2} \sum_\nu \ln\left[-(2i\hat q + \beta \lambda_\nu)\right] \right]. 
	\end{align}
	\item Carry out a saddle point integration over $\hat q$, and show that the saddle-point is given by 
	\begin{align}
		G(z) = \beta,
	\end{align}
	where $z = -2i\hat q/\beta$, and $G(z)$ is the resolvent of the random matrix $\underline{\underline{J}}$, which is given by Eq.~(\ref{ressemicircle}) [i.e. it is the resolvent that yields the semicircle law]. Thus, argue that this equation only has a solution for $\beta<1$. 
	\item In this case, show that $z = \beta + 1/\beta$, and hence that indeed the assumption we made earlier for the convergence of the integrals is satisfied. 
	\item Show that the thermal averages of the spin eigenmodes are thus (as long as $\beta<1$) 
	\begin{align}
		\langle \phi_\nu^2\rangle &= \frac{1}{Z}\int D\phi d\hat q \, \phi_\nu^2 \,\exp\left[\frac{\beta}{2}  \sum_{\nu}\lambda_\nu(\phi_\nu)^2  + i \hat q\left(\sum_\nu \phi_\nu^2 - N\right)\right]\nonumber \\
		&= \frac{1}{\beta^2 +1- \beta \lambda_\nu}.
	\end{align}
	\item In the case $\beta>1$, we must be more careful when finding the saddle-point value of $\hat q$. Reason that, for $\lambda_\mathrm{max} = 2$ being the edge of the Wigner semicircle where the eigenvalue density is vanishingly small, we must have for $\beta >1$
	\begin{align}
		\frac{1}{N} \frac{1}{z-\lambda_\mathrm{max}} \approx \beta.
	\end{align}
	Hence, show that for $\beta>1$, the eigenmode $\phi_\mathrm{max}$ corresponding to the eigenvalue $\lambda_\mathrm{max}$ takes the vast majority of the weight in the sum $\sum_\nu \phi_\nu^2 = N$, and hence that $T = 1$ is the critical temperature at which a transition in the order parameter $N^{-1}\sum_i \langle \langle \sigma_i\rangle_T^2\rangle_J$ occurs.
\end{itemize}
Let us also study the relaxational dynamics in the $p = 2$ case. We will see that the critical temperature obtained from dynamics coincides with that of the thermodynamic computation in this case. 
\begin{itemize}
	\item Beginning with the dynamics
	\begin{align}
		\dot s_i = - r(t) s_i + \sum_j J_{ij}s_j + \xi_i(t), \hspace{1cm} \langle \xi_i(t) \xi_j(t') \rangle = 2T\delta_{ij} \delta(t-t'), 
	\end{align}
	show that, using It\^o's lemma, along similar lines to Eq.~(\ref{itoed}), we have
	\begin{align}
		r(t) = \frac{1}{N} \sum_{ij} s_i J_{ij} s_{j} + T,
	\end{align}
	\item  Writing $v_i^{(\nu)}$ for the $i$-th component of the eigenvector corresponding to the eigenvalue $\lambda_\nu$ of $\underline{\underline{J}}$, we define $\phi_\nu(t)$ via $s_i(t) = \sum_\nu \phi_\nu(t) v^{(\nu)}_i$. Thus, write dynamic equations for $\phi_\nu(t)$, and write $r(t)$ in terms of $\phi_\nu(t)$.
	\item Let us assume that $r(t) \to r = \mathrm{const.}$. Using It\^o's lemma again, show that 
	\begin{align}
		\frac{d}{dt}\left( \langle \phi_\nu^2\rangle\right) = -2(r-\lambda_\nu)\langle \phi_\nu^2\rangle +2T. 
	\end{align}
	Hence, show that for $r>2$, we have $\langle \phi_\nu^2\rangle = \frac{T}{r-\lambda_\nu}\left[1-e^{-2(r-\lambda_\nu)t}\right]$. 
	\item Show that we may write the constraint for $t \to \infty$ as (with $\beta = 1/T$)
	\begin{align}
		\frac{1}{N}\sum_\nu \frac{1}{r - \lambda_\nu} = \beta.
	\end{align}
	Hence, show that we obtain the same behaviour from the dynamic approach as we do from statics.
\end{itemize}

\newpage

\section{Supersymmetric (SUSY) approach}\label{section:susy}
As we discussed in Section \ref{section:replicas}, the replica method permitted us to proceed with evaluation of the resolvent in Eq.~(\ref{fresnel}). This was accomplished by writing $\langle G(\omega) \rangle = -2/N \partial_\omega \langle \ln \mathcal{Z}\rangle$, and using the identity $\ln x= \lim_{n\to 0}(x^n-1)/n$. Of course, this method is fairly suspect, since we must take $n$ to be an integer in order to simplify the calculation. 

Happily, there is an alternative (and less offensive to those interested in mathematical rigour) approach that also begins with Eq.~(\ref{fresnel}). This is the supersymmetric (SUSY) approach. The downside of this method in comparison to the replica approach is that the analysis is somewhat more algebraically involved; a greater number of auxiliary fields are introduced, which can mean that the requisite manipulations are more laborious.

With that being said, the sequence of steps for the SUSY approach is more-or-less identical to that used in conjunction with the replica method. That is, once we have our integral expression for the resolvent, we perform the disorder average, we introduce `order parameters' to decouple the various sites, we integrate out the superfluous fields, and subsequently perform a saddle-point approximation \cite{bender1999advanced} to evaluate the remaining integrals over the `order parameters', exactly as with the replica method.

We note that our discussion here will not demonstrate the true strengths of the SUSY approach, which really shines when performing calculations that are non-perturbative in $1/N$. For example, SUSY provides an elegant means of computing the microscopic (i.e. on a separation scale of $1/N$) eigenvalue density correlations \cite{efetov1983supersymmetry, mirlin1991universality}, the eigenvalue density in the `almost hermitian' regime \cite{fyodorov1997almost,fyodorov1997almost2} (where $\langle J_{ij}J_{ji}\rangle/\langle J_{ij}^2\rangle -1\sim N^{-1}$), or even the distribution of resolvent elements \cite{mirlin1994distribution,mirlin1994statistical}. There are works that provide a pedagogical introduction to performing these kinds of more complicated calculations \cite{efetov1983supersymmetry,verbaarschot2004supersymmetric,zuk1994introduction, guhr2010supersymmetry}. Here, we set our sights a little lower, and simply rederive the semicircle law using the SUSY approach. The hope is that we may present, in as transparent a manner as possible, the essence of the method without having to deal with any additional difficulties. In the exercises, we also derive an exact expression (valid for finite $N$) for the eigenvalue density in the case of the GUE (see Fig. \ref{fig:GUE_n10}).

The trick that we employ to avoid the use of replicas is to represent the determinant factor $\mathcal{Z}^{-2} = \det\left(z\underline{\underline{\id}} - \underline{\underline{J}} \right)$ as an integral over anti-commuting Grassmann variables. First, let us discuss what Grassmann variables are, and how they help us to accomplish this.

\subsection{Grassmann variables, Berezin's integration rules and the determinant}
Let us introduce the Grassmann variables $\{\xi_i\}$. These objects are anti-commuting such that
\begin{align}
	\xi_i \xi_j = -\xi_j \xi_i , 
\end{align}
meaning that $\xi_i^2 = 0$. However, the Grassmann variables commute with ordinary variables, and are commuting when it comes to addition. That is, if $a$ is an ordinary complex number
\begin{align}
	a \xi_i = \xi_i a, \hspace{1cm} a + \xi_i = \xi_i +a, \hspace{1cm} \xi_i + \xi_j =  \xi_j + \xi_i  .
\end{align}
We then introduce the integration rules due to Berezin \cite{berazin2012method}
\begin{align}
	\int d \xi_i \xi_i = 1, \hspace{1cm} \int d \xi_i  = 0.
\end{align}
We note that the order of integration matters here. For $i \neq j$, we write
\begin{align}
	\int d \xi_i d\xi_j \,\xi_j \xi_i = 1, \hspace{1cm} \int d \xi_j d\xi_i \, \xi_j \xi_i = -1 .
\end{align}
These rules permit us to make a rather useful representation of the determinant. We recall that the determinant of an $N \times N$ matrix $\underline{\underline{M}}$ can be written 
\begin{align}
	\det \underline{\underline{M}} = \sum_{i_1, i_2, \cdots, i_N} \epsilon_{i_1, i_2, \cdots, i_N} M_{1, i_1} M_{2, i_2} \cdots M_{N, i_N},  
\end{align} 
where $\epsilon_{i_1, i_2, \cdots, i_N} $ is the Levi-Civita symbol. Its entries are $1$ if $(i_1, i_2, \cdots, i_N)$ is an even permutation of $(1,2,\cdots ,N)$ [i.e. it can be obtained by swapping adjacent entries of $(1,2,\cdots ,N)$ an even number of times], $-1$ if it is an odd permutation, and 0 if any of the indices are equal. We can immediately see the connection with the anticommuting Grassmann variables. 

To write the Levi-Civita symbol in terms of Grassmann variables, we introduce a second set $\{\eta_i\}$, which have the same rules as $\{\xi_i\}$
\begin{align}
	\eta_j \eta_i = -\eta_i \eta_j , \hspace{1cm} \eta_j \xi_i = -\xi_i \eta_j.
\end{align}
The Levi-Civita symbol can therefore be written in terms of the Grassmann variables as follows
\begin{align}
	\epsilon_{i_1, i_2, \cdots, i_N} = \int d\xi_N d\eta_N \cdots  d\xi_1  d\eta_1\, \eta_{1}\xi_{i_1}  \cdots   \eta_{N}\xi_{i_N} .
\end{align}
As one does for functions of complex variables, we can define functions of Grassmann variables via their Taylor series. A convenient form in which to write the determinant is therefore
\begin{align}
	\det \underline{\underline{M}} = \int \left( \prod_i d\xi_i d\eta_i \right) \exp\left[\sum_{i,j} \eta_i M_{ij} \xi_j\right]. \label{detgrassmann}
\end{align}
One can see this as follows. Due to the integration rules $\int d\xi_i = \int d\eta_i = 0$ and $\xi_i^2 = 0$, we see that when we expand the exponential we only retain the $N$-th order term
\begin{align}
	&\int \left( \prod_i d\xi_i d\eta_i \right) \exp\left[\sum_{i,j} \eta_i M_{ij} \xi_j\right] = \int \left( \prod_i d\xi_i d\eta_i \right) \frac{1}{N!}\left[\sum_{i,j} \eta_i M_{ij} \xi_j\right]^N ,
\end{align}
where it is understood that the product is ordered such that $\prod_i d\xi_i d\eta_i = d \xi_N d\eta_N \cdots d\xi_1 d\eta_1$. When we now expand the bracket, we only retain those terms in which we have no repeats of the variables $\eta_i$ and $\xi_j$. Let us consider a particular term in the expansion $\eta_{i_1} \xi_{j_1} \cdots \eta_{i_N} \xi_{j_N}$. Since pairs of Grassmann variables commute with each other, i.e. $\eta_1 \xi_1 \eta_2 \xi_2 = \eta_2 \xi_2 \eta_1 \xi_1 $, we see that every such term can be reordered to give 
\begin{align}
	\eta_{i_1} \xi_{j_1} \cdots \eta_{i_N} \xi_{j_N} = \eta_{1} \xi_{k_1} \cdots \eta_{N} \xi_{k_N} .
\end{align}
We thus see that every permutation $(k_1, \cdots, k_N)$ appears $N!$ times, cancelling the prefactor. Summing over the different permutations of $\{k_i\}$, one arrives at
\begin{align}
	\int \left( \prod_i d\xi_i d\eta_i \right) \exp\left[\sum_{i,j} \eta_i M_{ij} \xi_j\right] &= \int \left( \prod_i d\xi_i d\eta_i \right) \sum_{k_1, \cdots ,k_N} \eta_1 M_{1,i_i}\xi_{i_1} \cdots  \eta_N  M_{N,i_N}\xi_{i_N} \nonumber \\
	&= \sum_{i_1, \cdots, i_N} \epsilon_{i_1, \cdots, i_N}M_{1,i_1}\cdots M_{N,i_N} ,
\end{align}
which is precisely the determinant.

\subsection{Supersymmetric representation of the resolvent and disorder average}
Let us now return to the expression for th resolvent in Eq.~(\ref{fresnel}). By using an additional set of `bosonic' variables (ordinary commuting numbers)
we may write 
\begin{align}
	&G_{kk}(z) =  \left[\det\left(z\underline{\underline{\id}} - \underline{\underline{J}} \right)\right]^{1/2} \int \left(\prod_{j=1}^N \frac{e^{i\pi/2} d\phi_j}{\sqrt{2\pi}}\right) i \phi_k \phi_k\exp\left[ -\frac{i}{2} \sum_{i,j}\phi_i [z\delta_{ij} - J_{ij}] \phi_j \right] \nonumber \\
	&= \det\left(z\underline{\underline{\id}} - \underline{\underline{J}} \right)\int \left(\prod_{j=1}^N \frac{e^{i\pi/2} d\phi_j d\chi_j}{2\pi}\right) i \phi_k \phi_k  \exp\left[ -\frac{i}{2} \sum_{i,j} [z\delta_{ij} - J_{ij}] \left(\phi_i\phi_j + \chi_i \chi_j\right) \right]. 
\end{align}
We may then finally use the expression in Eq.~(\ref{detgrassmann}) to introduce the `fermionic' component to our supersymmetric expression for the resolvent, and we thus write
\begin{align}
	G_{kk}(\omega) =&  \int \left(\prod_{j=1}^N \frac{e^{i\pi/2} d\phi_j d\chi_j d\xi_j d\eta_j}{2\pi}\right) i \phi_k \phi_k \exp\left[  \sum_{i,j} \left(z\delta_{ij} - J_{ij}\right)\left[ \eta_i\xi_j-\frac{i}{2}\left(\phi_i\phi_j + \chi_i \chi_j\right) \right] \right]. 
\end{align}
We have thus succeeded in writing the resolvent in a form that involves the disorder only in the exponent, as we desire. Now that we have this, we are in a position to carry out the disorder average. Alternatively, we may define the partition function
\begin{align}
	Z[h] = \int &\left(\prod_{j=1}^N \frac{e^{i\pi/2} d\phi_j d\chi_j d\xi_j d\eta_j}{2\pi}\right) \exp\bigg[  \sum_{i,j} \left(z\delta_{ij} - J_{ij}\right) \eta_i\xi_j \nonumber \\
	&-\frac{i}{2} \sum_{i,j} \left((z-h)\delta_{ij} - J_{ij}\right) \left(\phi_i\phi_j + \chi_i \chi_j\right) \bigg]. 
\end{align}
from which we find $\langle G \rangle = N^{-1}\partial_h Z\vert_{h=0}$. Importantly, the partition function is normalised to unity such that $Z[0] = 1$.

Let us now take the example of the GOE, with $\langle J_{ij}\rangle = 0$ and $\langle J_{ij}^2 \rangle$. Proceeding along very similar lines to the replica approach, we use the independence of the random matrix elements to write
\begin{align}
	&\left\langle \exp\left\{  - J_{ij}\left[ \eta_i\xi_j + \eta_j\xi_i-i\left(\phi_i\phi_j + \chi_i \chi_j\right) \right] \right\} \right\rangle\nonumber \\
	 &= 1 +\frac{1}{2N}\left[ \eta_i\xi_j + \eta_j\xi_i-i\left(\phi_i\phi_j + \chi_i \chi_j\right) \right]^2 + O\left(\frac{1}{N^2}\right) \nonumber \\
	&\approx \exp\bigg\{\frac{1}{N}\bigg[\eta_i\xi_j\eta_j\xi_i - \frac{1}{2}\left(\phi_i\phi_j + \chi_i \chi_j\right)^2 - i \left(\eta_i\xi_j + \eta_j\xi_i \right)\left(\phi_i\phi_j + \chi_i \chi_j\right)  \bigg]  \bigg\}. 
\end{align}
We thus have
\begin{align}
	\langle Z[h]\rangle &= \int \left(\prod_{j=1}^N \frac{e^{\frac{i\pi}{2}} d\phi_j d\chi_j d\xi_j d\eta_j}{2\pi}\right) \exp\left\{  z\sum_{i} \left[ \eta_i\xi_i-\frac{i}{2}\left(\phi_i\phi_i + \chi_i \chi_i\right) \right] \right\} \nonumber \\
	&\times \exp\bigg\{-\frac{1}{2N} \sum_{ij}\bigg[\eta_i\xi_i\eta_j\xi_j + \frac{1}{2}\left(\phi_i\phi_j + \chi_i \chi_j\right)^2 + i \left(\eta_i\xi_j + \eta_j\xi_i \right)\left(\phi_i\phi_j + \chi_i \chi_j\right)  \bigg]  \bigg\} . \label{susyaveraged}
\end{align}
Once again, we have traded the Gaussian integral with disorder for a quartic action after the disorder average. We note that this expression is exact if $J_{ij}$ are drawn from the GOE. Similarly to how we proceeded in the context of the replica method [see the discussion around Eq.~(\ref{hstrans})], we now use a Hubbard-Stratonovich transformation to handle the quartic Grassmannian term. 

\subsection{Hubbard-Stratonovich transformation}
Consider the integral
\begin{align}
	\int \frac{dq}{\sqrt{2\pi/N}} \exp\left[ -N\frac{q^2}{2} - iq \sum_i \eta_i \xi_i\right]= \int \frac{dq}{\sqrt{2\pi}} \exp\left[ -N\frac{q^2}{2}\right]\sum_r \frac{1}{r!}\left( - iq \sum_i \eta_i \xi_i\right)^r .
\end{align}
For even values of $r=2k$, we have $\int \frac{dq}{\sqrt{2\pi}} \exp\left[ -N\frac{q^2}{2} \right] \frac{q^{2k}}{(2k)!} = \frac{(2k-1)!!}{(2k)!N} = \frac{1}{2^k k! N}$, and for odd values of $r$, we have simply instead $\int \frac{dq}{\sqrt{2\pi}} \exp\left[ -N\frac{q^2}{2} \right] \frac{q^{r}}{r!} = 0$. Therefore, we have the following Hubbard-Stratonovich transformation (that we see is valid for Grassmann variables)
\begin{align}
	\exp\left\{-\frac{1}{2N} \sum_{ij}\eta_i\xi_i\eta_j\xi_j  \right\} = \int \frac{dq}{\sqrt{2\pi/N}} \exp\left[ -N\frac{q^2}{2} - iq \sum_i \eta_i \xi_i\right]  ,
\end{align}
and thus
\begin{align}
	&\langle Z[h]\rangle= \int \frac{dq}{\sqrt{2\pi}}\int \left(\prod_{j=1}^N \frac{e^{i\pi/2} d\phi_j d\chi_j d\xi_j d\eta_j}{2\pi}\right) \exp\left\{ -N\frac{q^2}{2} -\frac{i(z-h)}{2}\sum_{i}  \left(\phi_i\phi_i + \chi_i \chi_i\right)  \right\} \nonumber \\
	&\times \exp\bigg\{-\frac{1}{2} \sum_{ij}\bigg[ \frac{1}{2N}\left(\phi_i\phi_j + \chi_i \chi_j\right)^2 + i \left(\eta_i\xi_j + \eta_j\xi_i \right)\left(\frac{1}{N}\left(\phi_i\phi_j + \chi_i \chi_j\right) + \left(iz  + q\right) \delta_{ij}\right)  \bigg]  \bigg\}  .
\end{align}
We are now in a position to carry out the integrals over the Grassmann variables. Defining the $N \times 2$ matrix $\underline{\underline{X}} = (\underline{\phi}, \underline{\chi})$, we see that
\begin{align}
	\underline{\phi}\,\underline{\phi}^T + \underline{\chi}\,\underline{\chi}^T = \underline{\underline{X}} \,\underline{\underline{X}}^T , 
\end{align}
and thus we obtain by once again exploiting the formula for the determinant
\begin{align}
	\int \left(\prod_{j=1}^N   d\xi_j d\eta_j\right) &
	\exp\left\{-i\sum_{ij}  \eta_i\xi_j \left[\frac{1}{N}\left(\phi_i\phi_j + \chi_i \chi_j\right) + \left(iz  + q\right) \delta_{ij}\right]  \right\} = \det\left[ \left(z - iq \right)\underline{\underline{\id}}_N - \frac{i}{N} \underline{\underline{X}} \,\underline{\underline{X}}^T \right] \nonumber \\
	&= \left(z - iq \right)^{N-2}\det\left[ \left(z - iq \right)\underline{\underline{\id}}_2 - \frac{i}{N}  \underline{\underline{X}}^T\underline{\underline{X}} \right] ,
\end{align}
where we have used Sylvester's determinant identity [see Eq.~(\ref{sylvester})] for the second equality. Finally, we arrive at an expression entirely in terms of the commuting variables
\begin{align}
	&\langle Z[h]\rangle= \int \frac{dq}{\sqrt{2\pi/N}}\int \left(\prod_{j=1}^N \frac{e^{i\pi/2} d\phi_j d\chi_j }{2\pi}\right) \det\left[ \left(z - iq \right)\underline{\underline{\id}}_2 - \frac{i}{N}  \underline{\underline{X}}^T\underline{\underline{X}} \right] \nonumber \\
	&\times \exp\bigg\{-N\frac{q^2}{2} + (N-2) \ln\left(z-iq\right)  -\frac{1}{4N} \sum_{ij}\left(\phi_i\phi_j + \chi_i \chi_j\right)^2  -\frac{i(z-h)}{2}\sum_{i}  \left(\phi_i\phi_i + \chi_i \chi_i\right)  \bigg\}  .
\end{align}
\subsection{Introduction of `order parameters', saddle-point approximation and Wigner's semicircle}
Now that we have rid ourselves of the Grassmann variables, we are in a position to introduce order parameters, and subsequently perform a saddle point approximation, in a similar way to the calculation in Section \ref{section:spareplica}.

We introduce the following quantities via Dirac delta functions in their complex exponential representation
\begin{align}
	C_{\phi} = \frac{1}{N}\sum_i \phi_i^2, \hspace{1cm}  C_{\phi\chi} = \frac{1}{N}\sum_i \phi_i \chi_i, \hspace{1cm} C_{\chi} = \frac{1}{N}\sum_i \chi_i^2,
\end{align}
after which we carry out the integrals over $\phi_i$ and $\chi_i$ to obtain
\begin{align}
	\langle Z[h]\rangle =& \int \frac{dq dC_\phi d\hat C_\phi dC_\chi d\hat C_\chi dC_{\phi\chi} d\hat C_{\phi\chi}}{(2\pi/N)^{7/2}}   \det\left[ \left(z - iq \right)\underline{\underline{\id}}_2 - i \begin{bmatrix}
		C_\phi & C_{\phi\chi}\\
		C_{\phi\chi} & C_{\chi}
	\end{bmatrix} \right]\nonumber \\
	&\times \exp\bigg\{-N\frac{q^2}{2} + (N-2) \ln\left(z-iq\right)-\frac{N}{4}\left(C_\phi^2 + 2 C_{\phi\chi}^2 + C_\chi^2 \right)   -N\frac{i(z-h)}{2}  \left(C_\phi + C_\chi\right)  \bigg\} \nonumber \\
	&\times \exp\left\{N\left( C_\phi \hat C_\phi + C_\chi \hat C_\chi+ C_{\phi\chi} \hat C_{\phi\chi}\right) -  \frac{N}{2} \ln (\hat C_{\phi\chi}^2-4 \hat C_{\phi} \hat C_{\chi})\right\} .
\end{align}
We notice that the determinant prefactor is not relevant for determining the saddle point. For this reason, we may safely ignore it, and we may therefore carry out the integral over $q$ separately to the other integrals. The saddle point for the variable $q$ satisfies
\begin{align}
	q^\star = \frac{1}{q^\star +iz} , \label{qsaddle}
\end{align}
and thus we have
\begin{align}
	&\langle Z[h]\rangle = \int \frac{dC_\phi d\hat C_\phi dC_\chi d\hat C_\chi dC_{\phi\chi} d\hat C_{\phi\chi}}{(2\pi/N)^{3}}   \frac{\det\left[ \left(z - iq \right)\underline{\underline{\id}}_2 - i \begin{bmatrix}
		C_\phi & C_{\phi\chi}\\
		C_{\phi\chi} & C_{\chi}
	\end{bmatrix} \right]}{\sqrt{1-\frac{1}{(z-i q^\star)^2}}}\nonumber \\
	&\times \exp\bigg\{-N\frac{(q^\star)^2}{2} + (N-2) \ln\left(z-iq^\star\right)-\frac{N}{4}\left(C_\phi^2 + 2 C_{\phi\chi}^2 + C_\chi^2 \right)   -N\frac{i(z-h)}{2}  \left(C_\phi + C_\chi\right)  \bigg\} \nonumber \\
	&\times \exp\left\{N\left( C_\phi \hat C_\phi + C_\chi \hat C_\chi+ C_{\phi\chi} \hat C_{\phi\chi}\right) -  \frac{N}{2} \ln (\hat C_{\phi\chi}^2-4 \hat C_{\phi} \hat C_{\chi})\right\} .
\end{align}
Finally, we once again use the saddle-point approximation to carry out the integrals over the remaining order parameters. We obtain the following saddle-point equations
\begin{align}
	C_\phi =  \frac{-2 \hat C_\chi}{(\hat C_{\phi\chi}^2-4 \hat C_{\phi} \hat C_{\chi})}, &\hspace{1cm} \hat C_\phi = \frac{1}{2} (C_\phi + i z), \nonumber \\
	C_{\phi\chi} =  \frac{\hat  C_{\phi\chi}}{(\hat C_{\phi\chi}^2-4 \hat C_{\phi} \hat C_{\chi})}, &\hspace{1cm} \hat C_{\phi\chi} = C_{\phi\chi}, \nonumber \\
	C_{\chi} =  \frac{-2 \hat C_\phi}{(\hat C_{\phi\chi}^2-4 \hat C_{\phi} \hat C_{\chi})}, &\hspace{1cm} \hat C_{\chi} = \frac{1}{2} (C_\chi + iz).
\end{align}
There are two solutions to these equations. Either $\hat C_{\phi\chi} =0$, in which case $C_\phi = 1/(2\hat C_\phi)$ and 
\begin{align}
	C_\phi = \frac{1}{C_\phi + iz}, \hspace{1cm} C_\chi = \frac{1}{C_\chi + iz}. \label{saddlepointcphi} 
\end{align}
On the other hand, we may also have $\hat C_{\phi\chi}^2-4 \hat C_{\phi} \hat C_{\chi} = 1$. This gives rise to a set of degenerate solutions. However, in the case of this set of solutions, we obtain a non-vanishing exponent at the saddle-point, so that $\langle G_{kk} \rangle \propto e^{\alpha N}$, which cannot be the case. 

Considering instead the first saddle point solution however, we see that the terms in the exponent cancel at the saddle point to yield something independent of $N$. We define 
\begin{align}
	-N S =& -N\frac{(q^\star)^2}{2} + (N-2) \ln\left(z-iq^\star\right)-\frac{N}{4}\left(C_\phi^2 + 2 C_{\phi\chi}^2 + C_\chi^2 \right)   -N\frac{i(z-h)}{2}  \left(C_\phi + C_\chi\right) \nonumber \\
	&+N\left( C_\phi \hat C_\phi + C_\chi \hat C_\chi+ C_{\phi\chi} \hat C_{\phi\chi}\right) -  \frac{N}{2} \ln (\hat C_{\phi\chi}^2-4 \hat C_{\phi} \hat C_{\chi}) .
\end{align}
Since $q^\star = C_\phi = C_\chi$, $\hat C_{\phi\chi} = C_{\phi\chi} = 0$ and $\hat C_\phi = \hat  C_\chi = 1/(2C_\phi)$, we thus have at the saddle point [using also Eq.~(\ref{qsaddle})]
\begin{align}
	-NS &=iNq^\star h - 2 \ln\left(z-iq^\star\right) . 
\end{align}
Since we expect $Z[0] = 1$ to hold after the saddle-point approximation, then we must have $Z[h] = e^{i N q^\star h}$. Indeed, one can check this explicitly by carefully computing the appropriate Hessian and carrying out the Gaussian integral about the saddle point. We finally obtain
\begin{align}
	\langle G \rangle = \frac{1}{N} \partial_h Z\vert_{h = 0} = i q^\star ,
\end{align}
where $q^\star$ satisfies Eq.~(\ref{qsaddle}). Using the inverse Stieltjes transform in Eq.~(\ref{densefromres}), we thus recover the Wigner semicircle law once again.

\subsection{Generalisations and alternative routes}
Although we have not had to deal with replicas, and as a result we have not had to make any replica symmetry ansatz or otherwise motivate replica decoupling, the procedure above is comparatively protracted. Indeed, there are ways to significantly simplify what we have done, as we will now discuss briefly. The reason for presenting the calculation as above was to make the manipulations as transparent as possible. This allowed us to see clearly the correspondence with the sequence of manipulations made in the replica computation.

An alternative way to proceed, which is more amenable to generalisation, is to use the so-called Mirlin-Fyodorov approach \cite{fyodorov1991density,mirlin1991universality}. Instead of using the Hubbard-Stratonovich transformation, this approach uses a more general functional analogue, which enables one to perform calculations for many kinds of random matrix ensemble (see Ref.\cite{akara2023random} for a pedagogical introduction). 

We also explore an alternative Hubbard-Stratonovich transformation in the exercises, which permits us to decouple all of the bosonic and fermionic fields simultaneously. As a result, following Ref. \cite{verbaarschot2004supersymmetric}, we are able to derive an exact (for any $N$) expression for the eigenvalue density in the case of the GUE. Such finite $N$ calculations are expedited by the SUSY approach. One notes that for the GOE, the Hubbard-Stratonovich would require 16 new variables \cite{verbaarschot1985grassmann, guhr2010supersymmetry}, as opposed to the 4 in the GUE case, which is why it made sense to proceed as we did above.

\subsection{Advantages of SUSY}

Although what we have presented may seem like a very drawn-out way to derive the semicircle law, the `supersymmetric' (SUSY) formalism is extremely powerful, and has been used in many contexts in random matrix theory to carry out calculations that may otherwise be daunting. It is particularly useful in those calculations for which perturbative series (in $1/N$, or some other system parameter) cannot be utilised. For example, the computation of the microscopic 2-point eigenvalue correlations can be obtained relatively straightforwardly via this approach \cite{verbaarschot1985grassmann,zuk1994introduction,efetov1983supersymmetry}, as can the eigenvalue density in the so-called `weakly non-Hermitian regime' \cite{fyodorov1997almost,fyodorov1997almost2}, and the left-right eigenvector overlaps of non-Hermitian matrices \cite{wurfel2024mean,chalker1998eigenvector}. The distribution of the local density of states can also be computed \cite{mirlin1994distribution,mirlin1994statistical}, as can the eigenvalue density of more complicated random matrix ensembles such as sparse graph adjacency and Laplacian matrices \cite{akara2023random}.

With that being said, the integration over the saddle-point manifold that must be carried out in some of the aforementioned calculations is not for the faint-of-heart, and requires an in-depth discussion. For this reason, we have contented ourselves here with simply demonstrating the basics of the SUSY formalism via the semicircle law. For further reading, and a more complete pedagogical introduction to the SUSY formalism and its application to RMT (particularly the 2-point correlations) the reader is directed to Refs. \cite{efetov1983supersymmetry,verbaarschot2004supersymmetric,zuk1994introduction,guhr2010supersymmetry}.

\begin{figure}[h]
	\centering 
	\includegraphics[scale = 0.6]{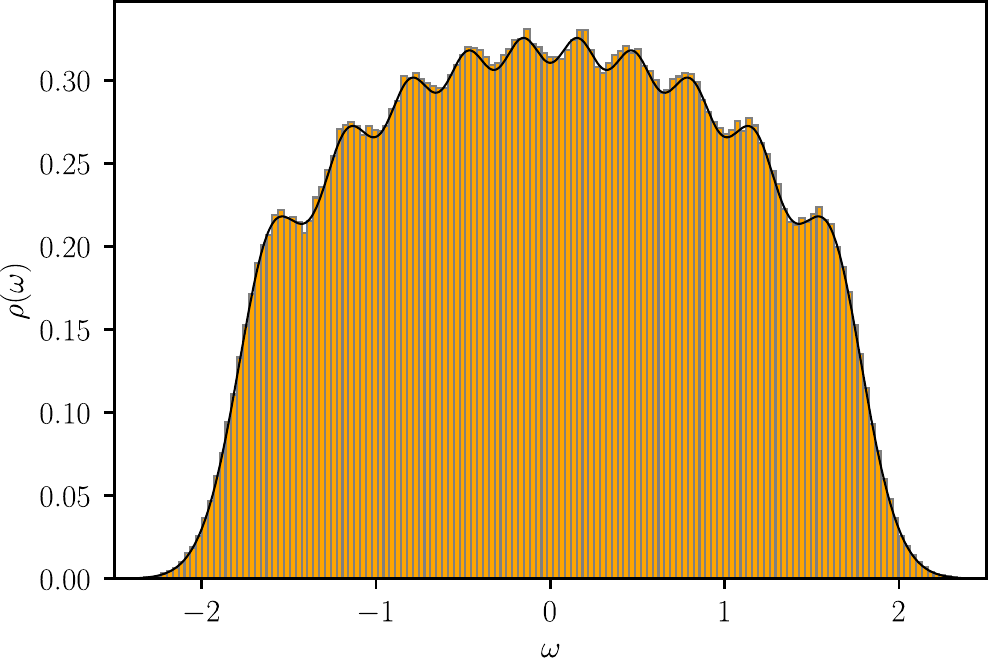}
	\captionsetup{justification=raggedright,singlelinecheck=false}
	\caption{Eigenvalue density of the GUE ensemble, with $N = 10$, averaged over $10^5$ realisations. The theory line is given by Eq.~(\ref{GUEallN}).}\label{fig:GUE_n10}
\end{figure} 

\newpage

\subsection{Exercises}
As is often the case, there are a variety of ways of proceeding with these calculations. We will now explore an alternative method, which exploits a combined bosonic-fermionic version of the Hubbard-Stratonovich (HS) transformation. To make things easier, we also use the more compact expression for $(\det \underline{\underline{H}})^{-1}$ in Eq.~(\ref{deth}), and we consider the GUE, rather than the GOE, for which the relevant HS transformation is simpler. Here, we mostly follow the notes by Verbaarschot \cite{verbaarschot2004supersymmetric}, without using the graded vector/matrix notation.
\begin{itemize}
	\item Show that we may write
	\begin{align}
		Z[h] = \int \left(\prod_{j=1}^N \frac{ d^2\phi_j d\xi_j d\eta_j}{-\pi}\right) \exp\bigg[-  i\sum_{i,j}& \left(z\delta_{ij} - J_{ij}\right) \eta_i\xi_j -i\sum_{i,j} \phi^\star_i\left((z-h)\delta_{ij} - J_{ij}\right) \phi_j  \bigg].
	\end{align}
	Hence, by taking the ensemble average, show that for the GUE with $\langle J_{ij}^2 \rangle = 0$ and $\langle J_{ij}J_{ji} \rangle = \frac{1}{N}$ we have
	\begin{align}
		\langle Z[h] \rangle = \int \left(\prod_{j=1}^N \frac{d^2\phi_j d\xi_j d\eta_j}{-\pi}\right) \exp\bigg\{ -i \sum_{i}\left[z\eta_i\xi_i +(z-h) \phi^\star_i \phi_i \right]\bigg\}\nonumber \\
		\exp\bigg\{ - \frac{1}{2N}\sum_{ij}\left[\phi_i^\star \phi_j + \eta_i\xi_j \right]\left[\phi_j^\star \phi_i + \eta_j\xi_i \right]\bigg\}.
	\end{align}
	\item Now, in order to handle the quartic term, we perform a HS transformation. Letting $\sigma_{BB}$ and $\sigma_{FF}$ be ordinary bosonic variables, and $\sigma_{BF}$ and $\sigma_{FB}$ be Grassmann variables, show by integrating over these new variables that
	\begin{align}
		\langle Z[h] \rangle =& \int d\sigma \left(\prod_{j=1}^N \frac{d^2\phi_j d\xi_j d\eta_j}{-\pi}\right) \exp\bigg\{ -\frac{N}{2} \left(\sigma_{BB}^2 + \sigma_{FF}^2 - 2 \sigma_{BF}\sigma_{FB} \right)\bigg\}\nonumber \\
		&\times\exp\bigg\{ - i\sum_{j}\bigg[\phi_j^\star (\sigma_{BB} + z - h) \phi_j + \phi_j^\star \sigma_{BF} \xi_j + \eta_j \sigma_{FB} \phi_j + \eta_j (i\sigma_{FF} + z) \xi_j \bigg]\bigg\},
	\end{align}
	where $d\sigma = d\sigma_{BB}d\sigma_{FF}d\sigma_{FB}d\sigma_{BF}/(2\pi)$. 
	\item Now, integrate out the variables $\phi_j$, $\phi^\star_j$, $\xi_j$ and $\eta_j$ to derive the `$\sigma$-model'
	\begin{align}
		\langle Z[h] \rangle = \int d\sigma  \bigg\{\frac{(i\sigma_{FF}+z)}{(z+\sigma_{BB}-h)}\left[ 1 + \frac{\sigma_{BF}\sigma_{FB}}{(i\sigma_{FF}+z)(z+\sigma_{BB}-h)}\right]\bigg\}^N\nonumber \\
		\times\exp\bigg\{ -\frac{N}{2} \left(\sigma_{BB}^2 + \sigma_{FF}^2 - 2 \sigma_{BF}\sigma_{FB} \right)\bigg\} .
	\end{align}
	Hint: It is easier to first integrate out the Grassmann variables.
	\item Finally, integrate out the Grassmann variables $\sigma_{BF}$ and $\sigma_{FB}$, and make the shifts $\sigma_{BB} \to \sigma_{BB} + h - z$ and $\sigma_{FF} \to \sigma_{FF} + i z$ to obtain
	\begin{align}
		\langle Z[h] \rangle = N \int \frac{d\sigma_{FF} d\sigma_{BB}}{2\pi}&  \left[ \frac{(i\sigma_{FF})^{N}}{(\sigma_{BB})^{N}} + \frac{(i\sigma_{FF})^{N-1}}{(\sigma_{BB})^{N+1}}\right]\exp\bigg\{ -\frac{N}{2} \left[(\sigma_{BB}+h-z)^2 + (\sigma_{FF}+iz)^2 \right]\bigg\} ,
	\end{align}
	and evaluate this using the saddle-point method to obtain the semicircle law once again.
\end{itemize}
Remarkably, we have reduced our initial integral over $4N$ variables to one involving only 2. What is truly impressive however is that in this case we may evaluate the remaining integrals exactly to obtain an expression that is valid for any value of $N$. One notes that the same result can be obtained using the orthogonal polynomial method \cite{mehta2004random}. We begin by deriving two helpful identities involving the Hermite ploynomials.
\begin{itemize}
	\item The definition of the Physicist's Hermite polynomials, according to the Rodrigues formula, is
	\begin{align}
		H_n(z) =(-1)^n e^{z^2} \frac{d^n}{dz^n} e^{-z^2} . 
	\end{align}
	Using this, demonstrate that
	\begin{align}
		\int_{-\infty}^{\infty} dx (ix)^n e^{-(x+iz)^2} &= \int_{-\infty}^{\infty} dx \frac{e^{z^2}}{(-2)^n} \frac{d^n}{dz^n} e^{-(x^2 + 2ix z)} \nonumber \\
		&= \frac{\sqrt{\pi}}{2^n} H_n(z), 
	\end{align}
	and
	\begin{align}
		\int_{-\infty}^\infty dy y^{-n} e^{-(y-z)^2}  &=  \frac{1}{(n-1)!} \int dy' \frac{(-1)^{n-1}}{z-y'} \partial_{y'}^{n-1} e^{-(y')^2} \nonumber \\
		&= \frac{1}{(n-1)!} \int dy' \frac{1}{z-y'} e^{-(y')^2} H_{n-1}(y').
	\end{align}
	\item Hence, evaluate the partition function to obtain
	\begin{align}
		&\langle Z[h] \rangle = \sqrt{\frac{N}{2}}\frac{N \sqrt{\pi}}{2^{N+1} N!} \int dy e^{-\frac{Ny^2}{2}}\frac{\left[H_N\left(\sqrt{\frac{N}{2}} z \right)H_{N-1}\left(\sqrt{\frac{N}{2}} y \right)-H_{N-1}\left(\sqrt{\frac{N}{2}} z \right)H_{N}\left(\sqrt{\frac{N}{2}} y \right) \right]}{z-y-h} .
	\end{align}
	\item Therefore, using that $\langle G \rangle = 2 \partial_h \langle Z[h]\rangle \vert_{h= 0} /N$ and $\rho(\omega) =\lim_{\epsilon \to 0} \mathrm{Im} G(\omega - i\epsilon)/\pi$, exploit the Christoffel-Darboux formula
	\begin{align}
		\sum_{k = 0}^n\frac{H_k(x)H_k(y)}{k! 2^k} = \frac{1}{2^{n+1}n!} \frac{H_n(y) H_{n+1}(x) - H_n(x)H_{n+1}(y)}{x-y} ,
	\end{align}
	to obtain the final expression (which is valid for all $N$)
	\begin{align}
		\rho(\omega) = \frac{e^{-\frac{Nz^2}{2}}}{\sqrt{2\pi N}}\sum_{k = 0}^{N-1}\frac{\left[H_k\left(\sqrt{\frac{N}{2}}\omega\right)\right]^2}{2^k k!} . \label{GUEallN}
	\end{align}
\end{itemize}
We test the formula in Eq.~(\ref{GUEallN}) in Fig. \ref{fig:GUE_n10}. We see that the averaged eigenvalue density exhibits peculiar oscillations for smaller values of $N$.

\newpage

\section{MSRJD path-integral formalism}\label{section:pathintegral}
The path integral formalism was most famously pioneered by Richard Feynman in his quest to produce a Lagrangian formulation of quantum mechanics \cite{feynman2005principle,feynman1948space}. Feynman, building on the work of Dirac, Wheeler and others, demonstrated that the quantum mechanical probability amplitude could be considered as the sum of contributions from all possible `paths' that a particle could take. This approach has been shown to be equivalent to the more familiar formulation of Schr\"odinger. Feynman would later go on to apply the path-integral formalism to quantum electrodynamics \cite{feynman1949space}, and would for this work later win the Nobel prize along with Tomonaga and Schwinger in 1965. 

Since the Schr\"odinger equation and the Fokker-Planck equation are related by a relatively straightforward change of variables (involving switching to an imaginary time coordinate) \cite{altlandsimons}, it is also natural to construct a path-integral representation of stochastic processes. In particular, Onsager and Machlup formulated the path integral measure for Gaussian random processes \cite{onsager1953fluctuations}, while Doi and Peliti formulated an operator and path-integral formalism for individual-based dynamics \cite{doi1976second, peliti1985path}. The precise formalism that we shall use here is referred to as the MSRJD path integral formalism after Martin, Siggia, Rose \cite{martin1973statistical}, Janssen \cite{janssen1976lagrangean} and de Dominicis \cite{dedominicis1978dynamics}.

The path-integral approach has many uses. In stochastic systems, it is well-suited to handling non-Markovian phenomena such as delayed reactions \cite{brett2013stochastic} or subdiffusion \cite{baron2019stochastic}, and within this formalism one may perform Van Kampen system-size expansions and WKB/eikonal approximations (i.e. finding the most likely paths of least action) \cite{altlandsimons}. For disordered systems and RMT, the path integral method permits one to easily take the disorder average, providing a helpful alternative to replicas (or indeed SUSY), as pointed out by de Dominicis \cite{dedominicis1978dynamics}. 

In this Section, we will discuss two broad methodologies that one may use in conjunction with the path-integral approach to compute quantities of interest for disordered dynamical systems. First, we demonstrate how the DMFT effective process may be derived using a saddle-point approach. Secondly, we show how Feynman diagrammatic series allow us to compute the response/correlation (or other) functions of the dynamical system, and how such quantities allow us to deduce spectral properties of random matrices. 

In particular, we show how the MSRJD path integral provides an alternative auxiliary field method to replicas and Grassmann variables. We mean this in the following sense. Replicas/SUSY allow one to construct an exponential integral representation of the resolvent matrix in which the disorder appears only as a linear contribution to the exponent. This makes taking the disorder average much easier. We show here how the path integral accomplishes the same feat, arguably in a more simple way compared to the SUSY approach, and in a more rigorous manner compared to replicas. 

For other pedagogical texts that highlight various applications of the path-integral approach, in disordered systems and beyond, the reader is directed to Refs. \cite{galla2024generating, hertz2016path, altlandsimons}.   

\subsection{Path integral of an SDE with arbitrary noise}
We begin by formulating the MSRJD path integral for a simple single-component system with noise. We will subsequently generalise our findings to the many-component disordered systems that are our true interest. Suppose we have a generic single-component SDE
\begin{align}
	\dot x = f(x,t) + \xi(t) + h(t),\label{stochproc}
\end{align}
where $f(x,t)$ is a non-random function of $x(t)$ and $t$, $h(t)$ is an arbitrary external field, and $\xi(t)$ is a centred coloured Gaussian noise with correlator
\begin{align}
	\langle \xi(t) \xi(t') \rangle_\xi = F(t,t') .
\end{align}
Here $\langle \cdots \rangle_\xi$ indicates an average over realisations of the noise. We wish to write a path-integral expression for the so-called `generating functional' of this process. The generating functional is to a random process what the generating function is to a random variable. It contains the complete information about the statistics of the process, and one may extract these statistics by differentiating with respect to the `source fields'. 

As a brief reminder, consider the generating function of two correlated random variables $x$ and $y$ with joint distribution $P(x,y)$. The generating function is then
\begin{align}
	Z[\psi] = \int dx dy e^{i x \psi_x + i y \psi_y} P(x,y).
\end{align}
We obtain the statistics of $x$ and $y$ by differentiation with respect to $\psi_x$ and $\psi_y$. For example, $\langle x y^2 \rangle =i \partial ^3 Z/\partial \psi_y^2 \partial \psi_x$.

The generating functional for the process in Eq.~(\ref{stochproc}) can be thought of as a high-dimensional generating function for the set of values of $x$ along a trajectory: $\{x(t)\}$. As such, we now introduce a source field $\psi(t)$ for every point on the trajectory. 

With this in mind, we construct the generating functional by first discretising Eq.~(\ref{dynamicalsystem}) (we will later take the continuous time limit). The discretised dynamics is given by
\begin{align}
	x_i(t+\Delta) = x_i(t) +\Delta f(x(t), t) +  \Delta \xi(t) + \Delta h(t).
\end{align}
We can thus write for the joint probability distribution of the Gaussian noise variables
\begin{align}
	P_\xi(\underline{\xi})=\frac{1}{\mathcal{N}}\exp\left[- \frac{1}{2} \underline{\xi}^T \underline{\underline{F}}^{-1} \underline{\xi}\right] ,
\end{align}
where $\mathcal{N}$ is a normalisation constant. We now construct the generating functional using a series of Dirac delta functions to enforce this dynamics. For example, if the probability distribution of $x(0)$ (the initial condition) is $P_0(x(0))$, then the joint probability of seeing a particular pair of $x(\Delta)$ and $x(0)$ is
\begin{align}
	P(x(0), x(\Delta)) =& \int d\xi(0)\, P_\xi(\xi(0))\, P_0(x(0))\nonumber \\
	&\times \delta\left( x(\Delta) - x(0) -\Delta\left[f(x(0), 0) + h(0)+  \xi(0)\right] \right).
\end{align}
That is, for a particular starting point $x(0)$, the value at the next time step $x(\Delta)$ is fully determined by the value of $\xi(0)$. Integrating over all possible values of the random noise, we obtain the probability of the single-step `trajectory'.

Generalising this reasoning for many steps, the probability of observing a particular trajectory $\{ x(0), x(\Delta), \cdots , x(M\Delta) \}$ is thus
\begin{align}
	P(\{ x(0), x(\Delta), &\cdots , x(M\Delta) \}) = \int \left(\prod_{m= 0}^{M-1} d\xi(m\Delta) \right) \nonumber \\
	\times  \prod_{m = 0}^{M-1} \delta( x((m+1)\Delta) &- x(m\Delta) -\Delta\left[f(x(m\Delta), m\Delta) + h(m\Delta)+  \xi(m\Delta)\right] ) \nonumber \\
	&\times \frac{1}{\mathcal{N}}\exp\left[- \frac{1}{2} \underline{\xi}^T \underline{\underline{F}}^{-1} \underline{\xi}\right] P_0(x(0)) .\label{trajectoryprob}
\end{align}
We are now in a position to compute the generating functional, which we define as (in analogy to the generating function)
\begin{align}
	&\langle Z(\left\{ \underline{\psi}, \underline{h} \right\})\rangle_\xi= \int \left( \prod_{m = 0}^{M} dx(m\Delta) \right) P(\{ x(0), x(\Delta), \cdots , x(M\Delta) \}) \nonumber \\
	&\hspace{2cm}\times\exp\left[i \Delta  \sum_{m = 0}^{M} \psi(m\Delta) x(m\Delta) \right].
\end{align}
Using the expression for the trajectory probability in Eq.~(\ref{trajectoryprob}), and writing each of the delta functions as a complex exponential integral, we have
\begin{align}
	&\langle Z(\left\{ \underline{\psi}, \underline{h} \right\})\rangle_\xi= \int \left( dx(M\Delta)\prod_{m = 0}^{M-1} dx(m\Delta)d\hat x(m\Delta) d\xi(m\Delta) \right)\frac{1}{\mathcal{N}}\exp\left[- \frac{1}{2} \underline{\xi}^T \underline{\underline{F}}^{-1} \underline{\xi}\right] P_0(x(0)) \nonumber \\
	&\times\exp\left[i \Delta  \sum_{m = 0}^{M-1} \psi(m\Delta) x(m\Delta) \right] \nonumber \\
	&\times\prod_{m = 0}^{M-1} \exp\Bigg\{ i \hat x(m\Delta)\big[x((m+1)\Delta) - x(m\Delta) -\Delta\left[f(x(m\Delta), m\Delta) + h(m\Delta)+  \xi(m\Delta)\right] \big]\Bigg\}.
\end{align}
Finally, if we carry out the Gaussian integrals over the noise variables $\xi$, and subsequently take the continuous-time limit $\Delta \to 0$, the generating functional is then given by the more elegant form
\begin{align}
	&\langle Z[\psi, h ] \rangle_\xi =\int D[x, \hat x]\exp\left[i \int dt \psi(t) x(t) \right] \nonumber \\
	&\times \exp\left[ i\int dt \, \hat x \left(\dot x - f(x,t) - h(t)\right)-\frac{1}{2} \int dt \int dt' \hat x(t) \hat x(t') F(t,t') \right] ,\label{sdegenfunct}
\end{align}
where $D[x, \hat x]$ indicates integration with respect to all possible `paths' describing both the coordinate $\{ x(t)\}$ and the conjugate `momentum' $\{ \hat x(t) \}$, absorbing some constant factors and the initial condition. Importantly, we have the normalisation condition 
\begin{align}
	\langle Z[\psi = 0, h ] \rangle_\xi = 1.
\end{align}
We have thus succeeded in constructing a path-integral expression that contains all of the statistical information of the process in Eq.~(\ref{stochproc}). In principle, we could now compute quantities such as $\langle x(t) x(t') \rangle_\xi$ by taking derivatives of Eq.~(\ref{sdegenfunct}). Indeed, one can also calculate response functions by using $R(t,t') =\langle \frac{\delta x(t)}{\delta h(t')} \rangle_\xi = -i\langle x(t) \hat x(t')  \rangle_Z$ where we write
\begin{align}
	&\langle \cdots \rangle_Z = \int D[x, \hat x]\left[ \cdots \right]\exp\left[i\int dt \, \hat x \left(\dot x - f(x,t) - h(t)\right)-\frac{1}{2} \int dt \int dt' \hat x(t) \hat x(t') F(t,t') \right] .
\end{align}
However, the requisite integrals are not always simple to compute if $f(x,t)$ is non-linear in $x$, for example. Indeed, perturbative methods using Feynman diagrams exist for precisely this purpose \cite{hertz2016path}. We will use similar tactics below when calculating the response functions of a disordered dynamical system.

\subsection{Relation between the resolvent and the response functions of a linear dynamical system}
Below, we will generalise the reasoning of the last subsection to the case of many-component systems. Our reasons for doing so are two-fold. Firstly, we will be able to use the path integral to derive dynamic mean field theories. Secondly, we will use it rederive results for the spectra of random matrices. For this latter objective, we require first to make some helpful observations.

Consider a real symmetric random matrix $\underline{\underline{J}}$. Following Ref. \cite{baron2022eigenvalues}, suppose we construct the following linear dynamical system
\begin{align}
	\dot x_i = -\omega x_i + \sum_{j}J_{ij}x_j +\xi_i(t)+ h_i(t),  \label{dynamicalsystem}
\end{align}
where $h_i(t)$ is an arbitrary function of time and $\xi_i(t)$ is a centred Gaussian white noise with correlator $\langle \xi_i(t)\xi_i(t')\rangle = \sigma_T^2 \delta(t-t')$. We again imagine that the coupling constants $J_{ij}$ have statistics
\begin{align}
	\langle J_{ij} \rangle = \frac{\mu}{N},\,\,\, \langle J_{ij}^2 \rangle - \langle J_{ij} \rangle ^2 = \frac{\sigma^2}{N}, \,\,\, \langle J_{ij} J_{ji} \rangle- \langle J_{ij} \rangle \langle J_{ji} \rangle  = \frac{\Gamma\sigma^2}{N} .\label{ellipsestatspath}
\end{align}
Let us define the response function as the functional derivative $R_{ik}(t,T) = \delta x_i(t)/\delta h_k(T) \vert_{h = 0}$. One obtains
\begin{align}
	\partial_t R_{ik}(t,T) = - \omega R_{ik}(t,T) + \sum_j J_{ij} R_{jk}(t,T) + \delta_{ij}\delta(t-T).
\end{align}
We then see that the response functions are related to the resolvent of the random matrix [see Eq.~(\ref{resdef})] via the Laplace transform
\begin{empheq}[box={\fboxsep=6pt\fbox}]{align}
	\mathcal{L}_t\left[ R_{ij}(t,0)\right](-i\epsilon) =  \left[\left((\omega-i\epsilon)\underline{\underline{\id}}  - \underline{\underline{J}} \right)^{-1}\right]_{ij},\label{laplaceresponse}
\end{empheq}
where $\mathcal{L}_t[f(t)](\eta) = \int_0^t dt f(t) e^{-\eta t}$ is the Laplace transform. Thus, if we can find the response functions of the system in Eq.~(\ref{dynamicalsystem}), we obtain the resolvent. In the symmetric case $\Gamma = 1$, we can therefore also find the eigenvalue spectrum of $\underline{\underline{J}}$ via Eq.~(\ref{densefromres}) (i.e. as long as $\underline{\underline{J}}$ is Hermitian). 

\subsection{Path-integral construction for a disordered system}
We now extend our MSRJD generating functional approach to the system in Eq.~(\ref{dynamicalsystem}). For a given realisation of the disorder, the generating functional is constructed (once again) by first discretising Eq.~(\ref{dynamicalsystem}) and enforcing the relationships between the variables at subsequent time-steps through Dirac delta functions. The discretised dynamics is given by
\begin{align}
	x_i(t+\Delta) = x_i(t) +\Delta\left[-\omega x_i(t) + \sum_{j}J_{ij}x_j(t) + h_i(t)\right] + \sqrt{\Delta \sigma^2_T } \epsilon_i(t), \label{linearsystemdiscrete}
\end{align}
where now $\epsilon(t)$ are independent standard Gaussian random variables. Following identical steps to the single-component process, we eventually end up with, after taking the continuous-time limit $\Delta \to 0$,
\begin{align}
	&\langle Z[\psi, h \vert \{J_{ij}\}]\rangle_\xi=\int D[x, \hat x] \exp\left[i \sum_{i}\int dt \psi_i x_i - \frac{\sigma_T^2}{2}\sum_i\int dt \hat x_i^2(t) \right]\nonumber \\
	&\times\exp\left[ i\sum_{i} \int dt \, \hat x_i \left(\dot x_i + \omega x_i- \sum_{ j}J_{ij} x_j - h_i \right)\right].
\end{align}
where again $D[x, \hat x]$ indicates integration with respect to all possible trajectories of the variables $\{ x_i(t)\}$ and their conjugate `momenta' $\{ \hat x_i(t) \}$, absorbing some constant factors and the initial condition. Once more, normalisation entails
\begin{align}
	\langle Z[\psi = 0, h \vert \{J_{ij}\}] \rangle_\xi = 1.
\end{align}
We note that the object $\langle Z[\psi, h \vert \{J_{ij}\}] \rangle$ is a merely a Gaussian integral over many variables. Below, we will show that while have begun with a simple Gaussian process, after averaging over $J_{ij}$, we arrive at an effective stochastic dynamics that is non-Gaussian. The challenge is then to handle this non-Gaussianity.

Averaging over realisations of the matrix $\underline{\underline{J}}$, we write
\begin{align}
	\langle Z[\psi, h] \rangle =& \int D[x, \hat x] \Bigg\langle\exp\left[- \frac{\sigma_T^2}{2}\sum_i \int dt \hat x_i^2(t ) + i \sum_{i}\int dt \psi_i x_i\right]\nonumber \\
	&\hspace{1cm}\times\exp\left[  i\sum_{i} \int dt \, \hat x_i \left(\dot x_i + \omega x_i- \sum_{ j}J_{ij} x_j - h_i \right)\right]\Bigg\rangle. \label{genfunct}
\end{align}
The disorder-averaged generating function is the object at the centre of our analysis. What makes it particularly useful is the fact that we can obtain the statistics of the dynamics via functional differentiation. For example, we have
\begin{align}
	\langle x_i(t) \rangle_Z = -i\frac{\delta \langle Z\rangle}{\delta \psi_i(t)} \bigg\vert_{\psi = 0}, &  \hspace{0.5cm} \langle x_i(t) x_j(t') \rangle_Z = -\frac{\delta \langle Z\rangle }{\delta \psi_i(t) \delta \psi_j(t')} \bigg\vert_{\psi = 0}, \nonumber \\
	\langle \hat x_i(t) \rangle_Z =& i\frac{\delta \langle Z\rangle}{\delta h_i(t)} \bigg\vert_{\psi = 0} = 0, \label{genfunctderivataives}
\end{align}
where we now use the shorthand
\begin{align}
	\langle \cdots \rangle_Z = \int D[x, \hat x] \left[ \cdots\right]\Bigg\langle\exp\left[ i\sum_{i} \int dt \, \hat x_i \left(\dot x_i + \omega x_i- \sum_{ j}J_{ij} x_j  \right)\right]\Bigg\rangle ,
\end{align}
and we note that by normalisation we still have $\langle Z[\psi = 0, h] \rangle = 1$, which is why averages of the `hatted' conjugate momentum variables on their own are nil. Particularly relevant for us are the disorder-averaged response functions, which can be extracted from the generating functional as
\begin{empheq}[box={\fboxsep=6pt\fbox}]{align}
	\langle R_{ij}(t,0) \rangle = -i\left\langle \frac{\delta^2 Z}{ \delta \psi_i(t) \delta h_j(0)} \right\rangle \bigg\vert_{\psi = 0, h = 0} = -i\langle x_i(t) \hat x_j(0) \rangle_Z .\label{responsefunctions}
\end{empheq}
In writing this formula for the response functions, we have arrived at what we require to compute the eigenvalue density. That is, we now have an expression for the resolvent (essentially) in terms of an integral over exponential functions in which the disorder appears linearly in the exponent. This is also what we strove for when we used replicas to compute the resolvent in Section \ref{section:replicas}. In this sense, as was observed by de Dominicis \cite{dedominicis1978dynamics}, we have used time as a substitute for replicas (or indeed Grassmann variables as in Section \ref{section:susy}).  

We note that we deal here with the disorder-averaged response functions (and other order parameters), rather than the response functions for a particular realisation of the matrix $\underline{\underline{J}}$. Now, taking the disorder average plays the role of `smoothing' the set of Dirac delta peaks that constitutes the empirical eigenvalue spectrum of a single finite-$N$ realisation. That is, we implicitly assume here that the fluctuations of the response functions are negligible when calculating the average, although it can also be shown by the path integral method that these are indeed negligible \cite{baron2025classes}.

\subsection{DMFT via a saddle-point computation and Wigner's semicircle}\label{section:dmftpath}
Now, beginning with Eq.~(\ref{genfunct}), we wish to understand the statistics of $x_i(t)$. In particular, we would like to extract the response functions $\langle R_{ij}(t,0) \rangle$. We will do this by first developing a dynamic mean-field theory using the path integral. 

Specifically, our strategy will now be as follows (following similar steps to the replica and SUSY computations). First, we will carry out the disorder average explicitly. Secondly, we introduce `order parameters' via Dirac delta functions, which allow us to decouple the different components. Finally, by making a saddle-point approximation \cite{bender1999advanced}, we arrive at an expression for $\langle Z[\psi,h]\rangle$ that represents $N$ decoupled stochastic processes. This then enables us to read off an `effective process' for each component that reproduces the correct statistics of the ensemble. This effective process is the dynamic mean-field theory (DMFT) that we seek, from which we can deduce the order parameters, including the response functions. In fact, we will recover the same DMFT that was derived in Section \ref{section:dmft} using the cavity method.

Examining the disordered term in Eq.~(\ref{genfunct}) and taking the disorder average, we find [using the statistics in Eq.~(\ref{ellipsestatspath})]
\begin{align}
	&\Bigg\langle\exp\left[-i \int dt \,  \sum_{i<j} \left(J_{ij}\hat x_i   x_j + J_{ji}\hat x_j   x_i \right) \right]\Bigg\rangle \nonumber \\
	&\approx \prod_{i<j}\bigg\{ 1-i \frac{\mu}{N} \int dt \left[\hat x_i x_j + \hat x_j x_i \right] - \frac{\sigma^2}{2N} \int dt dt' \left[\hat x_i(t) x_j(t) \hat x_i(t') x_j(t') + \Gamma\hat x_i(t) x_j(t) \hat x_j(t') x_i(t')  \right]\bigg\} \nonumber \\
	&\approx \exp\bigg\{ -i \frac{\mu}{N} \int dt \sum_{ij}\hat x_i x_j - \frac{\sigma^2}{2N} \int dt dt' \sum_{ij} \left[\hat x_i(t) x_j(t) \hat x_i(t') x_j(t') + \Gamma\hat x_i(t) x_j(t) \hat x_j(t') x_i(t')  \right] \bigg\},
\end{align}
where we neglect terms that are subleading in $N$. We therefore have
\begin{align}
	\label{genfunctav}&\langle Z[\psi, h] \rangle = \int D[x, \hat x] \exp\left[i \sum_{i}\int dt \psi_i x_i \right]\exp\left[ - \frac{\sigma_T^2}{2}\sum_i \int dt \hat x_i^2(t )\right] \\
	&\times \exp\left[ i\sum_{i} \int dt \, \hat x_i \left(\dot x_i + \omega x_i- h_i \right) -i \frac{\mu}{N} \int dt \sum_{ij}\hat x_i x_j\right]\nonumber \\
	&\times\exp\left\{ - \frac{\sigma^2}{2N} \int dt dt' \sum_{ij} \left[\hat x_i(t) x_j(t) \hat x_i(t') x_j(t') + \Gamma\hat x_i(t) x_j(t) \hat x_j(t') x_i(t')  \right] \right\}. \nonumber 
\end{align}
We now introduce the order parameters
\begin{align}
	M(t) \equiv  \frac{1}{N} \sum_{j} x_j(t), &\hspace{1cm} C(t,t') \equiv  \frac{1}{N} \sum_{j} x_j(t) x_j(t'), \nonumber \\
	 Q(t,t') &\equiv \frac{1}{N} \sum_j  x_j(t) \hat x_j(t') ,
\end{align}
and we thus obtain
\begin{align}
	\langle Z[\psi, h] \rangle = \int D[\cdots] \exp[N(\Psi+\Omega)],
\end{align}
where $D[\cdots]$ represents a functional integral measure over $M(t)$, $C(t,t')$, $Q(t,t')$ and their conjugate `hatted' variables at all times $t$ and $t'$, as well as some constant factors, and we have
\begin{align}
	\Psi &= i \int dt \hat M(t) M(t) + i \int dtdt'\left[\hat C(t,t') C(t,t') + \hat Q(t,t') Q(t,t')\right], \nonumber \\
	\Omega &=\frac{1}{N} \sum_i\ln \left\{ \int D[x_i, \hat x_i] e^{S_i[x_i, \hat x_i]}\right\} ,
\end{align}
where we define the single-component action
\begin{align}
	&S_i[x_i, \hat x_i] =  i \int dt \psi_i x_i + i\int dt \, \hat x_i \left(\dot x_i + \omega x_i - h_i \right) - \frac{\sigma_T^2}{2}\sum_i \int dt \hat x_i^2(t ) \\
	& -i\int dt \hat M(t) x_i(t) -i\int dt dt' [\hat C(t,t') x_i(t) x_i(t') +\hat Q(t, t') x_i(t) \hat x_i(t')] \nonumber \\
	& - i  \mu\int dt  \hat x_i(t) M(t) - \frac{\sigma^2}{2} \int dtdt' \left[\hat x_i(t) \hat x_i(t') C(t,t') + \Gamma \hat x_i(t) x_i(t')  Q(t,t') \right] \nonumber .
\end{align}
Since for $\langle Z[\psi,h]\rangle$ we have an integral with a large factor in the exponent of the integrand, we now perform the saddle-point computation for the integrals over the order-parameters and their hatted conjugates. We expect the saddle-point computation to be increasingly accurate as $N\to \infty$.

We find the following saddle-point equations
\begin{align}
	\hat M(t) = \frac{\mu}{N} \sum_i \langle \hat x_i(t) \rangle_i, &\hspace{0.5cm} M(t) =  \frac{1}{N} \sum_i\langle x_i(t) \rangle_i , \nonumber \\
	\hat C(t,t') =- \frac{i\sigma^2}{2N} \sum_i \langle \hat x_i (t) \hat x_i(t') \rangle_i, &\hspace{0.5cm} C(t,t') =  \frac{1}{N} \sum_i\langle x_i(t) x_i(t') \rangle_i, \nonumber \\	
	\hat Q(t,t') = -\frac{i\Gamma \sigma^2}{2N} \sum_i  \langle \hat x_i(t) x_i(t') \rangle_i ,&\hspace{0.5cm} Q(t,t') =  \frac{1}{N}\sum_i\langle x_i(t) \hat x_i(t') \rangle_i, \label{saddlepoint}
\end{align}
where we write
\begin{align}
	\langle \cdots \rangle_i &=  \frac{1}{Z_i[\psi_i,h_i]}\int D[x_i, \hat x_i] [\cdots]e^{S_i[x_i, \hat x_i]},
\end{align}
and we define the single-component generating functional
\begin{align}
	&Z_i[\psi_i, h_i] = \int D[x_i, \hat x_i] e^{S_i[x_i, \hat x_i]}.
\end{align}
Now arrives the subtle point. One notes that each of the averages in Eq.~(\ref{saddlepoint}) $\langle x_i(t)\rangle_i$ (and so on) are independent of their index $i$ (if the external fields are the same across components). We thus see that, after the saddle point integration is carried out, the full generating functional $\langle Z[\psi,h]\rangle$ can be written as 
\begin{align}
	\langle Z[\psi, h] \rangle =\prod_i Z_i[\psi_i, h_i],
\end{align}
and each of the order parameters $M(t)$, $C(t,t')$ may be computed by considering an average with the respect to the statistics of a single component, i.e.
\begin{align}
	M(t) = \frac{1}{N}\sum_i \langle x_i(t)\rangle_i = \langle x_1(t) \rangle_1.
\end{align}
We therefore see that the system can be considered as a set of independent single-component processes (since the generating functional factorises). Further, since each single process is representative of every other component in the system (they are statistically equivalent), it suffices to consider averages over realisations of a single-component stochastic process to reproduce the statistics of the ensemble. In this sense, the components \textit{decouple} in the limit $N \to \infty$, and we see that a mean-field approximation is exact in this limit.

Making the minor observation that averages of only hatted variables are nil, and thus $\hat M(t) = \hat C(t,t') = 0$, and defining $R(t,t') = -iQ(t,t')$, one finally arrives at
\begin{align}
	&Z_1[\psi_1, h_1] = \int D[x_1, \hat x_1]\exp\bigg[ i \int dt \, x_1\,\psi_1 - \frac{\sigma_T^2}{2}\int dt \hat x_1^2(t )\nonumber \\
	&+ i\int dt \, \hat x_1 \left(\dot x_1 + \omega x_1 - h_1 \right)   - i  \mu\int dt  \hat x_1(t) M(t) \nonumber \\
	&- \frac{\sigma^2}{2} \int dtdt' \hat x_1(t) \hat x_1(t') C(t,t')- i\Gamma\sigma^2 \int dtdt'  \hat x_1(t) x_1(t')  R(t,t') \bigg].\label{singlecomponentZ}
\end{align}
The objects $M(t)$, $R(t,t')$ and $C(t,t')$ are quantities that are to be determined self-consistently according to
\begin{align}
	M(t) = \langle x_1(t) \rangle_1 ,&\hspace{0.5cm} C(t,t') = \langle x_1(t) x_1(t') \rangle_1 , \nonumber \\
	R(t,t') = -i Q(t,t') &=  -i\langle x_1(t) \hat x_1(t') \rangle_1 = \frac{\delta \langle x_1(t) \rangle_1}{\delta h_1(t')} .
\end{align}  
Finally, we may read off the single-component SDE to which this generating functional corresponds by comparing with Eq.~(\ref{sdegenfunct})
\begin{empheq}[box={\fboxsep=6pt\fbox}]{align}
	\dot x = -\omega x(t) + \mu M(t)+ \Gamma \sigma^2 \int dt' R(t,t') x(t') + \eta(t) +\xi(t)+h(t),\label{effproc}
\end{empheq}
where the centred Gaussian noise $\eta(t)$ has correlator
\begin{align}
	\langle \eta(t) \eta(t') \rangle_\eta = \sigma^2 C(t,t') = \sigma^2 \langle x(t) x(t') \rangle_\eta,
\end{align}
and the other order parameters can also be obtained by averaging over realisations of the noise
\begin{align}
	M(t) = \langle x(t) \rangle_\eta , \hspace{1cm} R(t,t') = \left\langle \frac{\delta x(t)}{\delta h(t')} \right\rangle_\eta .
\end{align}
We have therefore succeeded in re-deriving the dynamic mean-field as in Section\ref{section:dmft}. We note that while the `decoupling' of the different components was an assumption that had to be made in Section \ref{section:dmft}, here it emerges naturally as a consequence of the limit $N \to \infty$.

We may thus obtain the response functions via functional differentiation
\begin{align}
	\partial_t R(t,t') = - \omega R(t,t') + \Gamma \sigma^2 \int dt'' R(t,t'') R(t'',t') + \delta(t,t') .
\end{align}
Recalling the relationship between the Laplace transform of the response functions and the resolvent in Eq.~(\ref{laplaceresponse}), we thus obtain an expression for the resolvent of the random matrix $\underline{\underline{J}}$ of our original stochastic process in Eq.~(\ref{dynamicalsystem}) [assuming time translational invariance $R(t,t') = R(t-t')$]
\begin{align}
	\Gamma \sigma^2 G^2-(\omega-i\epsilon)G + 1= 0.
\end{align}
We therefore recover the semicircle law once again in the case $\Gamma = 1$. 

For $\Gamma \neq1$, we note that we obtain the same expression as for the resolvent outside the bulk region of the eigenvalue spectrum in the case of the elliptic law [see Eq.~(\ref{csolanalytic})]. To obtain the non-analytic expression for the resolvent, it is necessary to use a different path integral construction \cite{baron2023pathintegralapproachsparse}, just as it was necessary to use Girko's block-matrix construction in Section \ref{section:ellipse}. 

\subsection{Fluctuation-dissipation theorem (FDT)}\label{section:fdt}
Due to its ubiquity in the study of statistical physics, and our use of it in the context of the spherical $p$-spin model in Section \ref{section:pspin}, we briefly use the path integral to derive the fluctuation dissipation theorem (FDT) in Eq.~(\ref{fdt}). This applies for any system in thermodynamic equilibrium with time-reversal symmetry. That is, we must assume that the initial state of the system is drawn from Boltzmann's distribution 
\begin{align}
	P(\{s_i\}) = \frac{1}{\mathcal{Z}}e^{-\frac{H(\{s_i\})}{T}}. 
\end{align}
Given that it contains all statistical information about the dynamical system at hand, the MSRJD generating function is ideal for understanding general symmetries and properties of stochastic systems. The final state at $t = t_f$ of the system is given by (assuming that the system is in a Boltzmann distributed initial state at $t = -t_f$) the following MSRJD path integral
\begin{align}
	& Z[\psi,h]  = \int D[x, \hat x] \frac{1}{\mathcal{Z}}e^{-\frac{H(\{x_i(-t_f)\})}{T}} \nonumber \\
	&\times\exp\left[- T\sum_i \int_{-t_f}^{t_f} dt \hat x_i^2(t )+  i\sum_{i} \int_{-t_f}^{t_f} dt \, \hat x_i \left(\dot x_i + \frac{\delta H(t)}{\delta x_i(t)}  \right)\right]\nonumber \\
	&\times\exp\left[ i \int_{-t_f}^{t_f} dt \, x_i\,\psi_i \right] . 
\end{align}
Now, we examine the action under the following change of variables
\begin{align}
	t \to -t, \hspace{1cm} x_i(t) \to x_i(-t), \hspace{1cm} i\hat x_i(t) \to i\hat x_i(-t) -\frac{1}{T} \dot x_i(-t) . \label{transform}
\end{align}
We obtain
\begin{align}
	S =& - T\sum_i \int_{-t_f}^{t_f} dt \hat x_i^2(t )+  i\sum_{i} \int_{-t_f}^{t_f} dt \, \hat x_i(t) \left(\dot x_i(t) + \frac{\delta H(t)}{\delta x_i(t)}  \right) \nonumber \\
	\to& - T\sum_i \int_{-t_f}^{t_f} dt \hat x_i^2(-t )+  i\sum_{i} \int_{-t_f}^{t_f} dt \, \hat x_i(-t) \left(-\dot x_i(-t) + \frac{\delta H(-t)}{\delta x_i(-t)}  \right) \nonumber \\
	&+ 2i \sum_i \int_{-t_f}^{t_f} dt \dot x_i(-t) \hat x_i(-t) - \frac{1}{T}\sum_i\int_{-t_f}^{t_f}dt \dot x_i \frac{\delta H}{\delta x_i} .
\end{align}
We notice that the first of these two additional terms compensates for the minus sign that has been picked up by the $\dot x_i$ term of the original action, while the latter is a total derivative. That is, we see that the action is invariant under the transformation as long as there is equilibrium. That is, we obtain the time-reversed generating functional (with the starting point now being the Boltzmann distribution at $t_f$ and traversing back in time)
\begin{align}
	& Z[\psi,h]  = \int D[x, \hat x] \frac{1}{\mathcal{Z}}e^{-\frac{H(\{x_i(t_f)\})}{T}} \exp\left[- T\sum_i \int_{-t_f}^{t_f} dt \hat x_i^2(-t ) \right] \\
	&\times\exp\left[ i\sum_{i} \int_{-t_f}^{t_f} dt \, \hat x_i(-t) \left(\dot x_i(-t) + \frac{\delta H(-t)}{\delta x_i(-t)}  \right)+ i \int_{-t_f}^{t_f} dt \, x_i(-t)\,\psi_i(-t) \right] \nonumber . 
\end{align}
To obtain the fluctuation dissipation theorem, we exploit the transformation in Eq.~(\ref{transform}). Quite simply, one immediately recovers the FDT
\begin{align}
	R_{ii}(t-t') &= -i\langle x_i(t) \hat x_i(t') \rangle_Z = -i\langle x_i(-t) \hat x_i(-t') \rangle_Z + \frac{1}{T}\langle x_i(-t) \dot x_i(-t') \rangle_Z \nonumber \\
	&= \frac{1}{T}\partial_{t'}C(t-t'),
\end{align} 
where we have used the time translational invariance and causality presuming that $t>t'$. 

\subsection{Feynman diagrams via the path integral}
An alternative way to evaluate the disorder-averaged response functions is to construct a series expansion of the path integral. By appropriately collecting terms, we arrive at a $1/N$ expansion of the response functions. The terms in this series can be kept track of more easily with the use of Feynman diagrams, in a very similar way to Sections \ref{section:wignerapproach} and \ref{section:mpgeneral}.

This approach has some advantages over the saddle-point computation performed above in the context of more sophisticated calculations. For example, the saddle-point approach neglects correlations between components by construction. Such correlations can be computed using a Feynman diagrammatic approach in relatively straightforward manner \cite{baron2025classes}, whereas an integration over the saddle-point manifold would be necessary when using the saddle-point approach (this has been done \cite{altland2000wigner} using the Keldysh dynamic formalism \cite{kamenev1999electron,kamenev2009keldysh}). It is also possible to make a perturbative expansion of the response functions in model parameters other than $1/N$ using the diagrammatic approach \cite{baron2023pathintegralapproachsparse, baron2025classes} (see also the exercises). 

Here, we use the computation of the semicircle law to demonstrate the diagrammatic approach. Needless to say, the diagrammatic approach is very much `overkill' in this case -- we merely use the semicircle law as a simple example to exhibit the method.  

We note that the same diagrammatic series can also be used to derive the elliptic law (and other non-hermitian ensembles) \cite{kuczala2016eigenvalue, baron2023pathintegralapproachsparse}. We also note that a diagrammatic approach can also be used to evaluate quantities other than the response functions, such as the correlation function $C(t,t')$ or the mean $M(t)$ \cite{hertz2016path}.

\subsubsection{Evaluation using Wick's theorem}
We begin with Eq.~(\ref{responsefunctions}). Writing the path integral explicitly, and consulting Eq.~(\ref{genfunctav}), we may write the disorder-averaged response functions (in the case $\mu = 0$ and $\Gamma = 1$) as
\begin{align}
	&\left\langle \frac{1}{N} \sum_k R_{kk}(T,0) \right\rangle  = \int D[x, \hat x] \left(-\frac{i}{N} \sum_k x_k(T) \hat x_k(0) \right)\exp\left[ S_0 + S_\mathrm{int}\right],
\end{align}
where we identify the so-called `bare' action and the interaction term respectively
\begin{align}
	S_0 &= i\sum_{i} \int dt \, \hat x_i \left(\dot x_i + \omega x_i \right) - \frac{\sigma_T^2}{2}\int dt \hat x_i^2(t ), \nonumber \\
	S_\mathrm{int} &= - \frac{\sigma^2}{2N} \int dt dt' \sum_{ij} \left[\hat x_i(t) x_j(t) \hat x_i(t') x_j(t') + \hat x_i(t) x_j(t) \hat x_j(t') x_i(t')  \right].
\end{align}
To evaluate the response functions that we desire, we now expand the interaction term
\begin{align}
	N^{-1}\sum_{k} \left\langle R_{kk}(T,0) \right \rangle= -i 	N^{-1}\sum_{k}\sum_{r} \left\langle \frac{S_\mathrm{int}^r}{r!} x_k(T) \hat x_k(0) \right\rangle_0, \label{expsum}
\end{align}
where $\langle \cdot \rangle_0$ indicates an average with respect to the bare action, i.e.
\begin{align}
	\langle O \rangle_0 = \int D[x,\hat x, y, \hat y] \, O \, e^{S_0} . 
\end{align}
Each of the terms in the sum in Eq.~(\ref{expsum}) are simple complex Gaussian integrals, which may be evaluated. To make our evaluation of the terms more systematic, we use Wick's theorem. Wick's theorem states that, for an action that is quadratic in the dynamic variables (as the bare action above is), the average of an even number of the dynamic variables is given by the sum of all possible combinations of the variables averaged in pairs. 

Let us take an example of the application of Wick's theorem. The average of four dynamic variables with respect to the bare action simplifies as follows
\begin{align}
	&\langle x_k(t)\hat x_k(t') x_i(T) \hat x_{i}(T')\rangle_0 = \langle x_k(t)\hat x_k(t') \rangle_0 \langle x_i(T) \hat x_{i}(T')\rangle_0 \nonumber \\
	&  + \langle x_k(t)\hat x_i(T') \rangle_0 \langle x_i(T) \hat x_{k}(t')\rangle_0+ \langle x_k(t) x_i(T) \rangle_0 \langle \hat x_i(t') \hat x_{k}(T')\rangle_0  \nonumber \\
	&= \langle x_k(t)\hat x_k(t') \rangle_0 \langle x_i(T) \hat x_{i}(T')\rangle_0 + \langle x_k(t)\hat x_i(T') \rangle_0 \langle x_i(T) \hat x_{k}(t')\rangle_0 ,
\end{align}
where we used that $\langle x_k(t) x_i(T) \rangle_0 \langle \hat x_i(t') \hat x_{k}(T')\rangle_0 = 0$, since averages of only hatted variables evaluate to nil [as we saw in the discussion surrounding Eq.~(\ref{genfunctderivataives})]. 

Crucially for our calculation below, we have the following relation for the bare response function
\begin{align}
	-i\langle x_i(t) \hat x_{j}(T) \rangle_0 = R^{(0)}_{ij} (t,T) = \delta_{ij} e^{- \omega(t-T)} \Theta(t-T) , \label{bareresponse}
\end{align}
where $\Theta(\cdot)$ is the Heaviside function. This means that we can evaluate averages with respect to the bare action explicitly. The expression in Eq.~(\ref{bareresponse}) can be obtained simply by computing the response functions of the dynamical system corresponding to the bare action [i.e. Eq.~(\ref{dynamicalsystem}) with $J_{ij} = 0$].

\subsubsection{Feynman diagrams as a combinatorial tool}
Keeping track of the huge variety of `Wick pairings' in the sum in Eq.~(\ref{expsum}) is a daunting task. A useful strategy is to use Feynman diagrams, which help to identify the terms that are (non-)vanishing in the limit $N\to \infty$, as well as being a convenient bookkeeping tool. 

We have already seen that Wick pairings can vanish if one of the pairs is solely composed of `hatted' conjugate variables. Aside from this, terms can also vanish due to the time ordering of the dynamic variables $x_i(t)$ and $\hat x_i(t')$ since $R^{(0)}_{ij}(t, t') = 0$ for $t<t'$ due to causality. They can also vanish in the thermodynamic limit $N \to \infty$. This is because $R^{(0)}_{ij}(t,t') \propto \delta_{ij}$, and therefore some Wick pairings will be subleading in $1/N$ once we carry out the sums over the indices $i,j, \cdots$. In the last case in particular, Feynman diagrams are particularly helpful. This is because only diagrams with a planar topology \cite{brezin1978planar, t1993planar} (with no crossing arcs) survive. These happen to be the same rainbow diagrams that appeared when we considered the combinatorics in Wigner's approach to the semicircle law in Section \ref{section:wignerapproach}.

As simple examples, let us consider the Feynman diagrams that arise from the $r = 1$ term in Eq.~(\ref{expsum}). Since there are 6 dynamic variables in this term, we could in principle have $5!!= 15$ diagrams to evaluate. However, most of these cancel identically, and we end up with only four non-zero Wick pairings. These are
\begin{align}
	&-iN^{-1} \sum_{k} \frac{1}{1!} \langle x_k(T)	\hat x_k(0) S_\mathrm{int}\rangle_0  = \frac{\sigma^2(-i)^3}{1!2N^2} \sum_{i,j,k}   \int dt  dt' \Bigg[ \langle x_k(T) \hat x_j(t) \rangle_0 \langle x_i(t) \hat x_i(t') \rangle_0 \langle x_j(t') \hat x_k(0) \rangle_0 \nonumber \\
	&+\langle x_k(T)  \hat x_i(t')\rangle_0 \langle x_j(t') \hat x_j(t) \rangle_0 \langle x_i(t)\hat x_k(0) \rangle_0+ \langle x_k(T) \hat x_i(t')  \rangle_0 \langle  x_j(t') \hat x_i(t) \rangle_0 \langle x_j(t) \hat x_k(0) \rangle_0\nonumber \\
	&\hspace{4cm}+ \langle x_k(T) \hat x_i(t) \rangle_0 \langle x_j(t) \hat x_i(t') \rangle_0 \langle x_j(t') \hat x_k(0) \rangle_0 \Bigg]\nonumber \\
	& = \frac{\sigma^2}{N^2} \sum_{i,j,k} \frac{1}{1!}  (-i)^3 \int dt \int dt' \Bigg[ \langle x_k(T) \hat x_j(t) \rangle_0 \langle x_i(t) \hat x_i(t') \rangle_0 \langle x_j(t') \hat x_k(0) \rangle_0 \nonumber \\
	&\hspace{4cm}+ \langle x_k(T) \hat x_i(t) \rangle_0 \langle x_j(t) \hat x_i(t') \rangle_0 \langle x_j(t') \hat x_k(0) \rangle_0 \Bigg]\nonumber \\
	&= \frac{\sigma^2}{N^2}  \int^T_0 dt \int_{0}^t dt' e^{-\omega(T-t)} e^{-\omega(t-t')} e^{-\omega t'}\sum_{ijk}  \left[ \delta_{kj} \delta_{ii} \delta_{jk} + \delta_{ki} \delta_{ji} \delta_{jk} \right]\nonumber \\
	&=\sigma^2 \int^T_0 dt \int_{0}^t dt' e^{-\omega(T-t)} e^{-\omega(t-t')} e^{-\omega t'} \left( 1 + \frac{1}{N}\right)  , \label{firstorderexample}
\end{align}
where we note that a combinatorial factor $1/2$ has cancelled due to the symmetry $(i\leftrightarrow j,t \leftrightarrow t')$.

Examples of terms that have been neglected involve pairings such as \begin{align}
	\langle x_k(T) \hat x_k(0)  \rangle_0 \langle x_i(t) \hat x_i(t') \rangle_0 \langle x_j(t') \hat x_j(t)  \rangle_0 = 0,
\end{align}
which evaluates to nil by causality (i.e. either $t$ or $t'$ must come first), and
\begin{align}
	\langle x_k(T)  x_j(t') \rangle_0 \langle x_i(t) \hat x_i(t') \rangle_0 \langle \hat x_j(t) \hat x_k(0) \rangle_0  = 0,
\end{align}
which involves a bare average of only hatted variables.

We note that the final integral in Eq.~(\ref{firstorderexample}) is a convolution, and so the Laplace transform is evaluated easily
\begin{align}
	\mathcal{L}_T\left\{-iN^{-1} \sum_{k} \frac{1}{1!} \langle x_k(T)	\hat x_k(0) S_\mathrm{int}\rangle_0 \right\}(\eta) = \frac{\sigma^2}{(\omega+\eta)^3} \left( 1 + \frac{1}{N}\right)  . 
\end{align}

Clearly, it would be inconvenient to proceed purely algebraically. We therefore now present the diagrammatic representation of the above Wick pairings. We adopt the convention of only listing those digrams that have different topological structure, and thus we write (noting that we swap the indices $j$ and $i$ to obtain the diagrammatic representations of the other non-zero Wick pairings) 
\begin{figure}[H]
	\centering 
	\includegraphics[scale = 0.4]{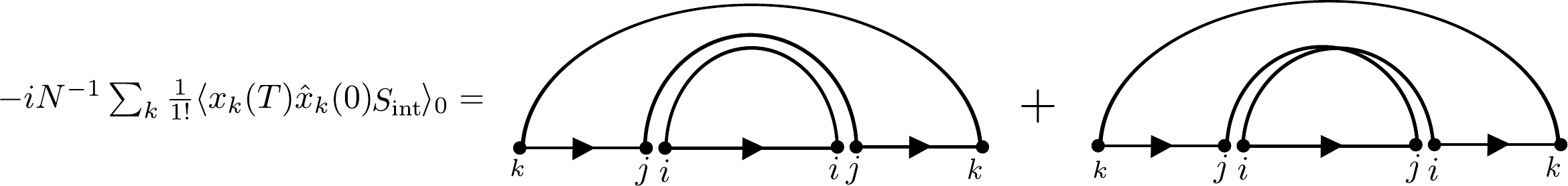}
	\captionsetup{labelformat=empty}
\end{figure}
The above diagrams should be interpreted as follows (see also Ref. \cite{baron2022eigenvalues}):  A dot on the left-hand end of a directed edge represents an $\hat x$-type variable, and a dot on the right-hand end of a directed edge represents an $x$-type variable. Pairs of dots positioned together have the same time coordinate. The $x$ and $\hat x$ variables connected by an arc are constrained to have the same index. Double arcs carry a multiplicative factor of $\sigma^2/N$. Points connected by horizontal edges are Wick-paired together (averaged with respect to the bare action), and thus evaluate to the bare response function. Because $R^{(0)}_{ij}(t,t') = 0$ for $t<t'$, the time coordinate of an $x$-type variable must always be greater than that of an $\hat x$-type variable, hence the directionality of the edges. Finally, all internal times (i.e. not corresponding to the nodes at either end of the diagram) and all indices are summed/integrated over. 

We therefore see that the second diagram, with the twisted arc, contributing to $-iN^{-1} \sum_{k} \frac{1}{1!} \langle x_k(T)	\hat x_k(0) S_\mathrm{int}\rangle_0$  is negligible in the thermodynamic limit. In general, diagrams are proportional to $N^{E - A -1}$, where $E$ is the number of disconnected (by arcs) sets of directed horizontal edges, and $A$ is the number of double arcs (i.e. excluding the one connecting the end points). This is because each disconnected set of horizontal edges evaluate to $\propto\sum_{i}\delta_{ii}$, and each double arc indicates a factor of $\sigma^2/N$. When arcs cross, this connects sets of edges that would otherwise not be connected, and thus reduces the power of $N$ of the diagram. This means that the right-hand diagram with the twisted double arc is $O(N^{-1})$, and therefore negligible in the thermodynamic limit. 

Let us take the additional example of the set of diagrams that arise from the $r= 2$ term in Eq.~(\ref{expsum}) to further elucidate why we focus only on planar diagrams in the limit $N\to \infty$. At $r = 2$, one has the following surviving terms (where we now ignore twisted double arcs)
\begin{align}
	&-\frac{i}{2! } N^{-1} \sum_k\langle x_k(T) \hat x_k(0) S_{\mathrm{int}}^2\rangle_0 \nonumber \\
	&= \frac{\sigma^4}{N^3}\int dt_1 dt'_1 dt_2 dt'_2 \sum_{k,i_1, j_1,i_2, j_2} \Bigg[ R^{(0)}_{ki_1}(T,t_1)  R^{(0)}_{j_1j_1}(t_1, t_1')   R^{(0)}_{i_1i_2}(t_1', t_2)  R^{(0)}_{j_2 j_2}(t_2,t_2')   R^{(0)}_{i_2 k}(t_2', 0) \nonumber \\
	&\hspace{1cm}+  R^{(0)}_{ki_1}(T,t_1) R^{(0)}_{j_1 i_2}(t_1, t_2)  R^{(0)}_{j_2j_2}(t_2 t_2')  R^{(0)}_{i_2 j_1}(t_2', t_1')  R^{(0)}_{i_1 k}(t_1', 0) \nonumber \\
	&\hspace{1cm}+ R^{(0)}_{ki_1}(T,t_1)  R^{(0)}_{j_1 i_2}(t_1,t_2)  R^{(0)}_{j_2 j_1}(t_2 ,t_1')   R^{(0)}_{i_1j_2}(t_1', t_2') R^{(0)}_{i_2 k}(t_2',0)
	\Bigg]   , \label{secondordersurviving}
\end{align}
and we have used that there is a symmetry between the times labelled $1$ and $2$ (which has cancelled the factor of $2!$ from Eq.~(\ref{expsum})) and symmetry between dashed and undashed times (which has cancelled a factor of $(2)^2$ from $S_\mathrm{int}^2$). We note that due to this kind of symmetry, the specific labelling of the vertices in the diagrams is again irrelevant. The number of ways of ordering the times always cancels the appropriate multiplicative factor, and so the only salient feature of a diagram is its topology \cite{hertz2016path}. 

The Wick pairings in Eq.~(\ref{secondordersurviving}) can be represented diagrammatically as
\begin{figure}[H]
	\centering 
	\includegraphics[scale = 0.45]{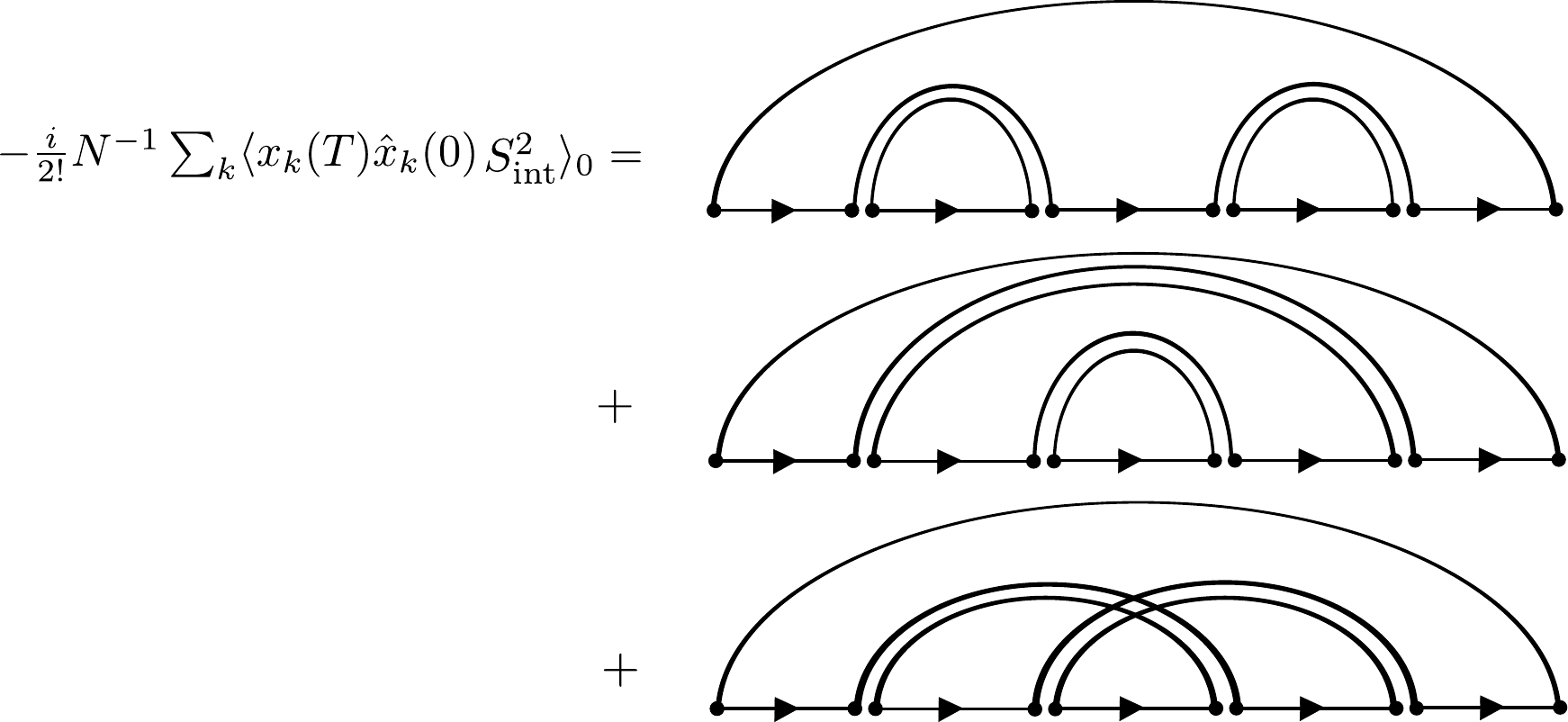}
	\captionsetup{labelformat=empty}
\end{figure}
These first two digrams each have three disconnected sets of directed edges, which corresponds to three factors of $\sum_l \delta_{ll}$. This cancels the factor of $N^{-3}$. In contrast, the third term in Eq.~(\ref{secondordersurviving}) is represented by a diagram whose directed edges are all connected by arcs. This means that one obtains only a single factor of $\sum_{l} \delta_{ll}$ after summing over all other indices. One thus finds that this diagram is an $O(N^{-2})$ contribution. Thus, only the first two of these Wick pairings survives in the thermodynamic limit. Explicitly, we have
\begin{align}
	&\frac{1}{N}\sum_{k} \lim_{\eta\to 0}\mathcal{L}_T\left\{-\frac{i}{2! }\langle x_k(T) \hat x_k(0) S_{\mathrm{int}}^2\rangle_0\right\}(\eta) = \frac{\sigma^4}{(\omega+\eta)^5N^3}\sum_{k, i_1, j_1, i_2, j_2}\Bigg[   \delta_{k,i_1} \delta_{j_1,j_1} \delta_{i_1,i_2} \delta_{j_2,j_2} \delta_{j_2,k}	 \nonumber \\
	&\hspace{4cm}+  \delta_{k,i_1} \delta_{j_1,i_2} \delta_{j_2,j_2} \delta_{i_2,j_1}\delta_{i_1,k}  +   \delta_{k,i_1} \delta_{j_1,i_2} \delta_{j_2,j_1} \delta_{i_1,j_2} \delta_{i_2,k} \Bigg]	,
\end{align} 
where we see that the first two products of Kronecker deltas evaluate to $1$, whereas the final set gives $1/N^2$. In the limit $N\to \infty$, we therefore have
\begin{align}
	&\frac{1}{N}\sum_{k} \mathcal{L}_T\left\{-\frac{i}{2! }\langle x_k(T) \hat x_k(0) S_{\mathrm{int}}^2\rangle_0\right\}(\eta) = \frac{2\sigma^4}{(\omega +\eta)^5}	,
\end{align} 
We thus see that the `non-planar' diagram (with crossing arcs) again gives a contribution that vanishes in the limit $N\to \infty$ and only the planar diagrams survive. We emphasise again that a factor of $1/(2! (2)^2)$ cancelled due to time ordering, which enabled us to consider only diagrams that differed by topology. 

To summarise, we have so far argued that the following simplifying rules apply generally: 
\begin{enumerate}
	\item The only Wick pairings that we need to consider pair solely hatted and unhatted dynamic variables.
	\item The only non-vanishing Wick pairings for $N\to \infty$ correspond to planar diagrams with non-crossing and non-twisted arcs.
	\item The number of combinations of Wick pairings that are equivalent up to time ordering always exactly cancels a prefactor, allowing us to discard the labelling of the internal nodes in the Feynman diagrams.
\end{enumerate}

One therefore sees that the sum in Eq.~(\ref{expsum}) can be evaluated in the thermodynamic limit by considering the set of all planar rainbow diagrams. As a final example, we find the following non-vanishing diagrams for the third-order term $r = 3$
\begin{figure}[H]
	\centering 
	\includegraphics[scale = 0.45]{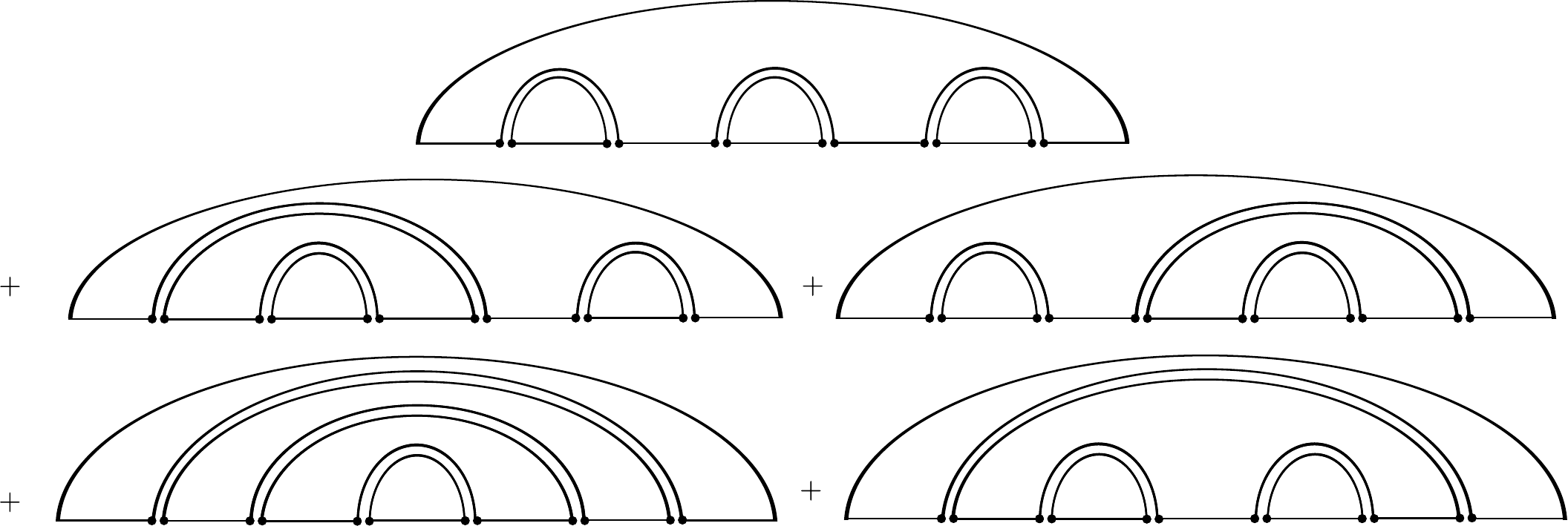}
	\captionsetup{labelformat=empty}
\end{figure}
These diagrams give us 
\begin{align}
	&\frac{1}{N} \sum_k \mathcal{L}_T \left\{-\frac{i}{3!}\left\langle x_k(T) \hat x_k(0) S_\mathrm{int}^3 \right\rangle_0 \right\}(\eta) = \frac{5\sigma^6}{(\omega+\eta)^7}.
\end{align}
We thus see how the formidable task of evaluating the series in Eq.~(\ref{expsum}) simplifies to summing a series of planar diagrams, each of which can be evaluated in terms of elementary functions. 

\subsubsection{Resummation of the diagrammatic series}
We employ one additional diagrammatic convention to simplify the notation when we perform sums over many diagrams. We denote a sum of planar diagrams by an edge with a double arrow, accompanied by a label for identification purposes. For example, let us take the surviving planar diagrams for the second-order term above $-\frac{i}{2!}  \langle x_k(t) \hat x_l S_\mathrm{int}^2\rangle_0 \equiv (O_2)_{kl}$, for which we write
\begin{figure}[H]
	\centering 
	\includegraphics[scale = 0.45]{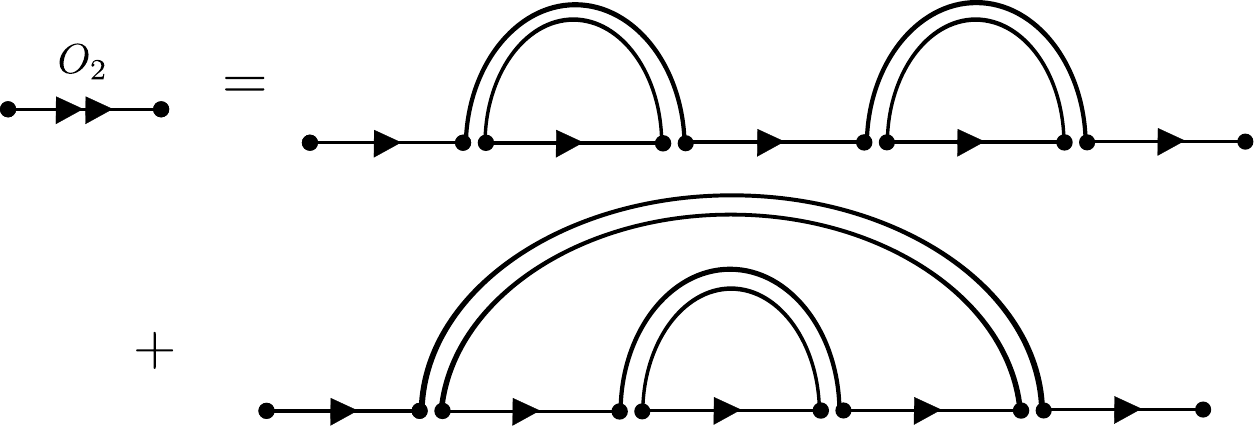}
	\captionsetup{labelformat=empty}
\end{figure}
When we draw an arc over a double-arrowed edge, this is also to be interpreted as a sum of diagrams. Precisely, for the example above we have
\begin{figure}[H]
	\centering 
	\includegraphics[scale = 0.45]{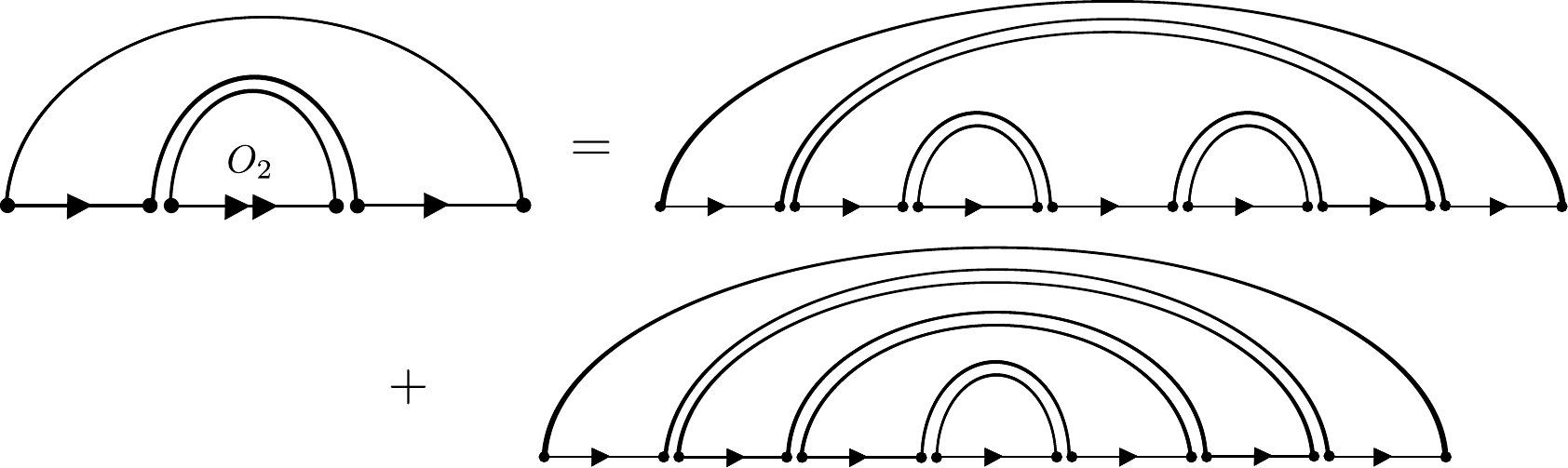}
	\captionsetup{labelformat=empty}
\end{figure}

\begin{figure}[H]
	\centering 
	\includegraphics[scale = 0.22]{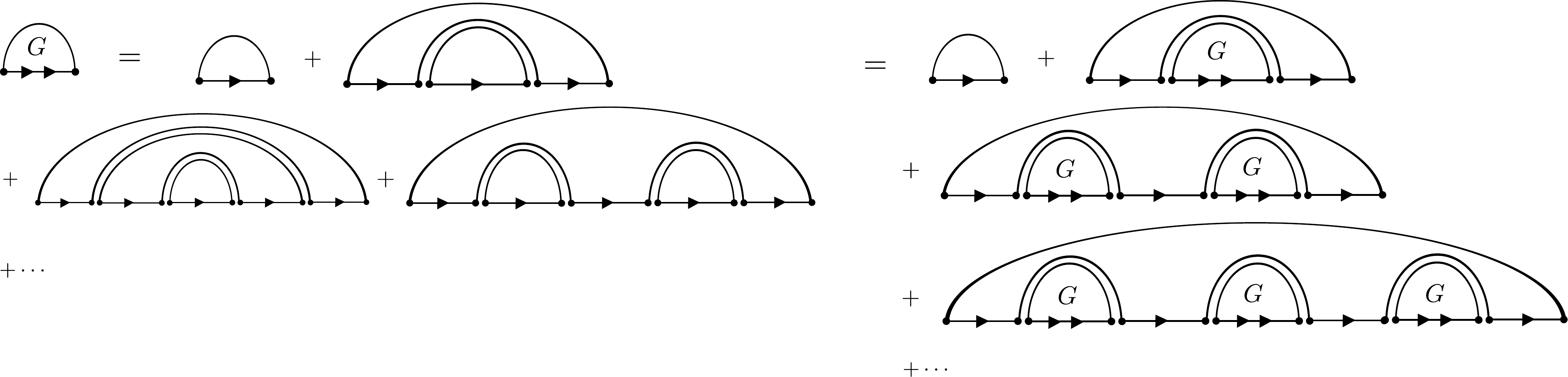}
	\caption{The sum over all possible diagrams. Recognising the self-similarity of the series, this can be rewritten as a geometric series. }\label{fig:rainbowdiagrams}
\end{figure}

We use this convention in Fig. \ref{fig:rainbowdiagrams} below to sum the full series of planar diagrams. We summarise the argument briefly here as to why the two series in Fig. \ref{fig:rainbowdiagrams} are same. 

Let us say that a diagram has $r_e$ `external arcs' if, by following a completely connected path of vertices from the leftmost vertex to the rightmost, we traverse $r_e$ arcs. We can categorise a general planar rainbow diagram by the number of external arcs that it has, since no arcs intersect. The full collection of diagrams with a single external arc, for example, can then be found by taking every planar diagram in the series, placing each of them inside a single arc, and attaching two directed edges to either side. Similar statements apply for diagrams with any number of external arcs. The complete series of planar diagrams can therefore be generated by summing together all of the sets of diagrams with $r_e = 1, 2, 3, \cdots$ external arcs, where under each `external' arc is the sum of all diagrams in the series. In this statement, we have thus identified a self-similarity quality of the series, which allows us to perform the resummation.

Because of this argument, we thus see why the full series of rainbow diagrams can be represented by the simpler series involving $G(\omega)$ (the dressed resolvent) in Fig. \ref{fig:rainbowdiagrams}. This simpler series is recognised to be geometric and is given by
\begin{align}
	G(\omega +\eta) &=  N^{-1} \sum_k\mathcal{L}_T\left[-i\langle x_k(T) \hat x_k(0)  \rangle_S \right](\eta)\nonumber \\
	&= \frac{1}{\omega+\eta} + \frac{\sigma^2}{(\omega+\eta)^2} G(\omega +\eta) + \frac{\sigma^4}{(\omega+\eta)^3} G^2(\omega+\eta) + \cdots . \label{selfconsistentseriesexplicit}
\end{align}
This series can be resummed to give
\begin{align}
	G(\omega+\eta) = \frac{1}{\omega+\eta - \sigma^2G(\omega+\eta)}. 
\end{align}
Finally, using Eq.~(\ref{densefromres}), we recover the Wigner semi-circle law
\begin{align}
	\rho(\omega) = \frac{1}{2 \pi \sigma^2}\sqrt{4 \sigma^2 - \omega^2} .
\end{align}
We note that in resumming the diagrammatic series this way, we have avoided a great deal of the difficulty that Wigner faced. This resummation, spotting the self-similarity of the series, was facilitated by the diagrammatics.

\subsection{Advantages of the path integral approach}
As a tool for random matrix theory and beyond, the path integral approach has certain advantages over the other methods that we have presented. Firstly, one has no need to invoke Grassmann variables or replicas. In addition to the spectral information, which we may extract from the response functions of a linear system, we also get additional information about the dynamics `for free', such as the correlation and the response functions. These observables can be used for understanding ageing phenomena, ergodicity breaking, or other dynamical transitions. 

The dynamic approach is also particularly well-suited to producing diagrammatic series, which are useful for performing perturbative analyses (see exercises), and one is also easily able to take into account additional correlations or other novel disorder statistics \cite{baron2022eigenvalues} (see exercises). Further, it is also possible to compute the 2-point eigenvalue density correlation functions \cite{baron2025classes, altland2000wigner}. We also saw that we were able to recover the fluctuation-dissipation theorem from the MSRJD path integral. One is also able to handle non-linearities in the system of interest perturbatively \cite{hertz2016path}. 

\subsection{Exercises}
Let us take now an example that highlights how the path integral approach can be used to perform perturbative computations \cite{baron2023pathintegralapproachsparse, baron2025classes}. We consider a random matrix ensemble with non-vanishing higher-order moments such that $J_{ij} = J_{ji}$ and 
\begin{align}
	\langle J_{ij}^2 \rangle = \frac{\sigma^2}{N}, \hspace{1cm} \langle J_{ij}^4 \rangle = \frac{\sigma^4 \alpha}{N} ,
\end{align}
and we assume $\alpha$ is small. We will thus ignore the higher moments and seek a perturbtive series that is accurate to first order in $\alpha$. An example of such an ensemble would be the ER adjacency matrix. That is, if we suppose that $P(a_{ij}) = (1-p/N) \delta(a_{ij}) + p/(2N) [\delta(a_{ij}-p^{-1/2})+ \delta(a_{ij}+p^{-1/2})]$, then we see that $\sigma = 1$ and $\alpha = 1/p$. 
\begin{itemize}
	\item Show that in this case we obtain the following expression for the disorder-averaged generating functional of the system in Eq.~(\ref{linearsystemdiscrete}) [analogous to Eq.~(\ref{genfunctav})]
	\begin{align}
		&\langle Z[\psi, h] \rangle \approx \int D[x, \hat x] \exp\left[i \sum_{i}\int dt \psi_i x_i - \frac{\sigma_T^2}{2}\sum_i \int dt \hat x_i^2(t )+ i\sum_{i} \int dt \, \hat x_i \left(\dot x_i + \omega x_i- h_i \right) \right]\nonumber \\
		&\times\exp\Bigg\{ - \frac{1}{N}\sum_{ij} \bigg\{  \frac{\sigma^2}{2!}\left[\int dt   \left(\hat x_i(t) x_j(t) + \hat x_j(t) x_i(t) \right) \right]^2- \frac{\sigma^4 \alpha}{4!}\left[\int dt\left(   \hat x_i(t) x_j(t) + \hat x_j(t) x_i(t)  \right)\right]^4\bigg\} \Bigg\},
	\end{align}
	where we only keep terms that are first order in $\alpha$ in the exponent.
	\item We now consider the additional Feynman diagrams that come about due to the term in the action proportional to $\alpha$, which we label
	\begin{align}
		S_\alpha = \frac{\sigma^4 \alpha}{4!N}\sum_{ij}  \left[\int dt\left(   \hat x_i(t) x_j(t) + \hat x_j(t) x_i(t)  \right)\right]^4 .
	\end{align}
	To lowest order in $\alpha$ and $\sigma$, we have the following contribution to the response functions 
	\begin{align}
		-iN^{-1} \sum_{k} \frac{1}{1!} \langle x_k(T)	\hat x_k(0) S_\mathrm{\alpha}\rangle_0. 
	\end{align}
	Show that the only non-vanishing diagram associated with this term can be represented as
	\begin{figure}[H]
		\centering 
		\includegraphics[scale = 0.3]{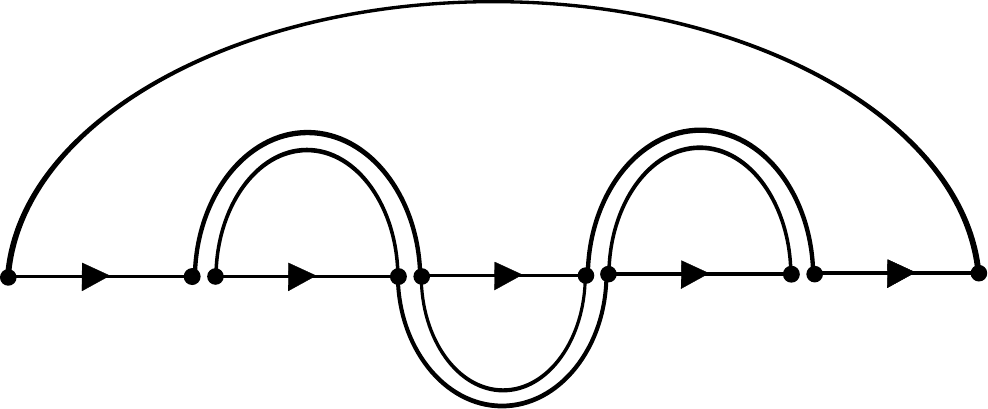}
		\captionsetup{labelformat=empty}
	\end{figure}
	\item Show that this diagram evaluates to
	\begin{align}
		\mathcal{L}_T\left\{-iN^{-1} \sum_{k} \frac{1}{1!} \langle x_k(T)	\hat x_k(0) S_\mathrm{\alpha}\rangle_0 \right\}(\eta) = \frac{\sigma^4 \alpha}{(\omega+\eta)^5}  . 
	\end{align}
	\item The full series of diagrams that are of order $\alpha$ can be represented by 
	\begin{figure}[H]
		\centering 
		\includegraphics[scale = 0.3]{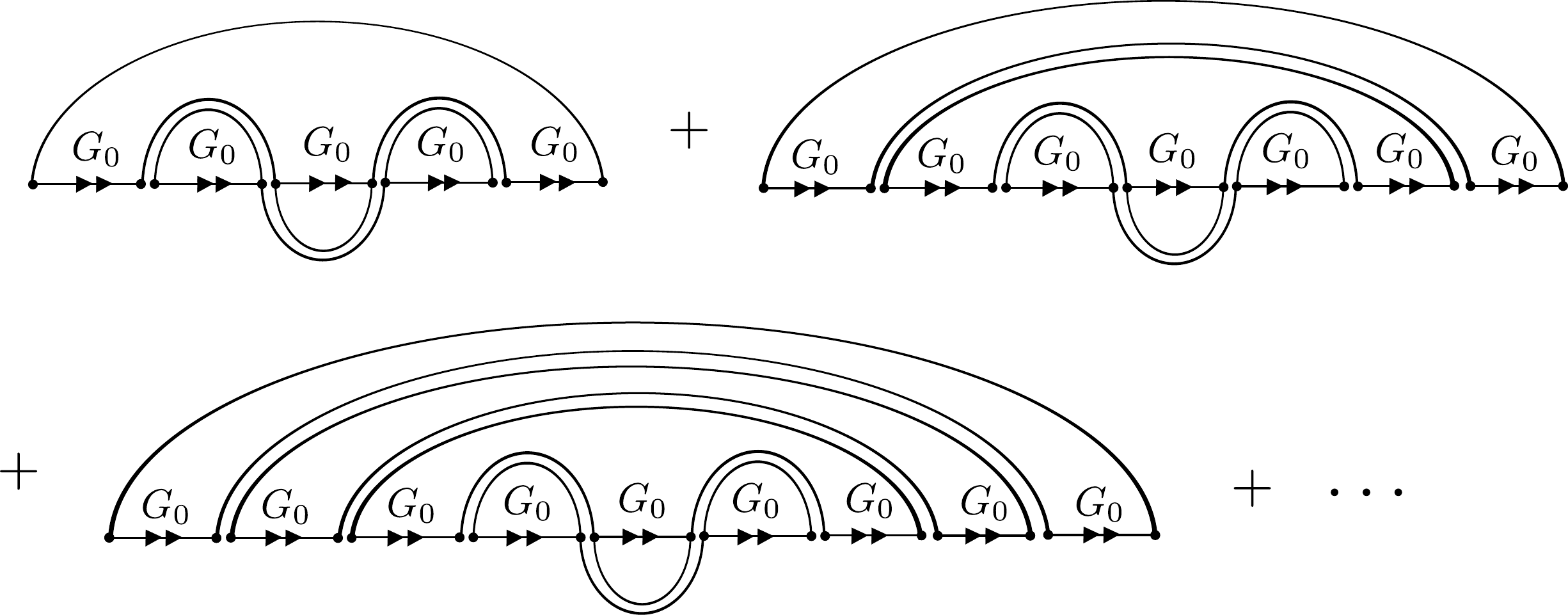}
		\captionsetup{labelformat=empty}
	\end{figure}
	where the double arrow labelled $G_0$ indicates the full series of rainbow diagrams (associated with the semicircle law) that appear in Fig. \ref{fig:rainbowdiagrams}. Hence, show that the resolvent can be approximated to leading order in $\alpha$ as
	\begin{align}
		G(\omega+\eta) \approx \frac{1}{\omega +\eta- \sigma^2 G_0} + \alpha \sigma^4  \frac{G_0^5}{1- \sigma^2 G_0^2},
	\end{align}
	where the zeroth order resolvent satisfies $G_0 = 1/(\omega + \eta - \sigma^2 G_0)$.
	\item Hence, obtain a self-consistent expression involving only $G$ (not $G_0$) that is accurate to first order in $\alpha$, similar to Eq.~(\ref{sparseerseries}).
\end{itemize}
Let us now consider the case where the matrix $\underline{\underline{J}}$ has heterogeneous statistics, such as might come about in a dense configuration model network [see Section \ref{section:denseconfiguration}]. That is, let us suppose that $J_{ij} = J_{ji}$, $\langle J_{ij} \rangle = 0$ and
\begin{align}
	\langle J_{ij}^2 \rangle = \frac{k_i k_j \sigma^2}{pN},
\end{align}
where $\{k_i\}$ are discrete variables drawn from a `degree distribution' $P_k$, and $p = \sum_k k P_k$ is the average degree. The higher order moments of $J_{ij}$ are subleading in $1/N$
\begin{itemize}
	\item Considering $k_i$ to be fixed quantities, show that in this case the disorder-averaged generating function is given by
	\begin{align}
		&\langle Z[\psi, h] \rangle \approx \int D[x, \hat x] \exp\left[i \sum_{i}\int dt \psi_i x_i - \frac{\sigma_T^2}{2}\sum_i \int dt \hat x_i^2(t )\right]\nonumber \\
		&\times \exp\left[ i\sum_{i} \int dt \, \hat x_i \left(\dot x_i + \omega x_i- h_i \right) \right] \nonumber \\
		&\times\exp\left\{ - \frac{1}{N}\sum_{ij}   \frac{\sigma^2 k_ik_j}{2!p}\left[\int dt   \left(\hat x_i(t) x_j(t) + \hat x_j(t) x_i(t) \right) \right]^2  \right\}.
	\end{align}
	\item Following the steps in Section \ref{section:dmftpath}, we introduce the modified order parameters 
	\begin{align}
		Q(t,t') = \frac{1}{N} \sum_{j} k_j x_j(t) \hat x_j(t') , \hspace{1cm} D(t,t') = \frac{1}{N} \sum_{j} k_j x_j(t)  x_j(t') .
	\end{align}
	Show that, after performing the saddle-point approximation, the single-component partition function [analogous to Eq.~(\ref{singlecomponentZ})] is given by
	\begin{align}
		&Z_1[\psi_1, h_1] = \int D[x_1, \hat x_1]\exp\bigg[ i \int dt \, x_1\,\psi_1 - \frac{\sigma_T^2}{2}\int dt \hat x_1^2(t )+i\int dt \, \hat x_1 \left(\dot x_1 + \omega x_1 - h_1 \right)\nonumber \\
		& \hspace{1cm}  - \frac{\sigma^2 k_1}{2 p} \int dtdt' \hat x_1(t) \hat x_1(t') D(t,t') - i\sigma^2 k_1 \int dtdt'  \hat x_1(t) x_1(t')  A(t,t') \bigg],
	\end{align}
	where $D(t,t') = \langle k_1 x_1(t) x_1(t')\rangle_1$ and $A(t,t') = - i \langle k _1 x_1(t) \hat x_1(t') \rangle_1 = \frac{\delta \langle x_1(t) \rangle_1}{\delta h_1(t')}$, and where in taking these averages, we also average over the variable $k_1$. Thus read off the effective process, similar to Eq.~(\ref{effproc})
	\begin{align}
		&\dot x_1 = - \omega x_1 + \sigma^2 \frac{k_1}{p} \int_0^t dt' A(t,t') x_1(t') + \sqrt{k_1/p}\eta(t) + h_1(t), \nonumber \\
		&\langle \eta(t) \eta(t') \rangle = \sigma^2 \langle k_1 x_1(t) x_1(t') \rangle_1.
	\end{align}
	This effective process, which is dependent on the quenched (fixed) random variable $k_1$, is known as a heterogeneous mean field theory \cite{aguirre2024heterogeneous,park2024incorporating,poley2025interaction}.
	\item By functionally differentiating the effective process, show that the response function $R_1(t,t') = \langle \frac{\delta x_1(t)}{\delta h(t')} \rangle$ satisfies
	\begin{align}
		\partial_t R_1(t,t') = - \omega R_1(t,t') + \frac{k_1}{p} \sigma^2 \int dt'' A(t,t'') R_1(t'',t') + \delta(t,t') .
	\end{align}
	Hence, show that the resolvent of the matrix $\underline{\underline{J}}$ obeys Eq.~(\ref{resolventconfiguration}).
	
\end{itemize}

\section{Population dynamics method for evaluating the cavity equations}\label{section:popdyn}

In Sections \ref{section:anderson} and \ref{section:networks}, we approximated the solution of the cavity equations on sparse networks, and we were thus able to compute the eigenvalue density of the adjacency matrices (and Laplacian matrices) of complex networks. Specifically, we examined the random regular graph (RRG) in Section \ref{section:anderson} and the Erd\H{o}s-R\'enyi (ER) graph in Section \ref{section:networks}. 

We now describe a simple numerical method, the population dynamics algorithm \cite{kuhn2008spectra, dean2002approximation, rogers2010new, livan2018introduction}, that will allow us to obtain an approximation of the eigenvalue density in the limit $N \to \infty$, without any high-connectivity approximations. Although we will only describe this method for the case of the ER graph adjacency and Laplacian matrices, it can easily be generalised for non-Hermitian matrices \cite{metz2019spectral, mambuca2022dynamical}, networks with heterogeneous weights \cite{da2025spectral}, network degree correlations \cite{rogers2010spectral}, and the case of heterogeneous diagonal elements \cite{semerjian2002sparse}.

The general principle behind the approach is the following. Recall the cavity equations for the ER graph adjacency matrix in Eq.~(\ref{cavityG}). With these equations, we first find the cavity resolvent elements at every point in the network, and subsequently use these to compute the true resolvent. When the network is infinite in size, an equivalent formulation is to consider the \textit{distribution} of cavity resolvent elements. The trace of the full resolvent is then obtained by taking an average over this distribution. That is, we have for the ER graph adjacency matrix
\begin{align}
	G' \stackrel{d}{=} \frac{1}{\omega - i \epsilon - \sum_{i = 1}^{k'} G^{(\mathrm{cav})}_i }, \hspace{1cm} G^{(\mathrm{cav})} \stackrel{d}{=} \frac{1}{\omega - i \epsilon - \sum_{i = 1}^{k-1} G^{(\mathrm{cav})}_i },
\end{align}
where $G'$, $G^{(\mathrm{cav})}$, $k'$ and $k$ are random variables and $\stackrel{d}{=}$ indicates an equivalence in distribution. The full averaged resolvent is obtained via $G(\omega) = \langle G'\rangle$. 

We note the following important properties of the random variables on the right-hand side of the above equations. The cavity resolvent elements $G^{(\mathrm{cav})}_i$ can each be considered as being independently drawn from a distribution $P_\mathrm{cav}(G^{(\mathrm{cav})})$. The degree of each node on the graph $k'$ is drawn from a Poisson degree distribution $P_{k'} = \frac{p^{k'}}{k'!}e^{-p}$ (in the large $N$ limit). Finally, let us now consider the cavity graph, with a specific node removed. The distribution of the degrees $k$ of the neighbours of this removed node is $k P_k/p$. 

\begin{figure}[t]
	\centering 
	\includegraphics[scale = 0.47]{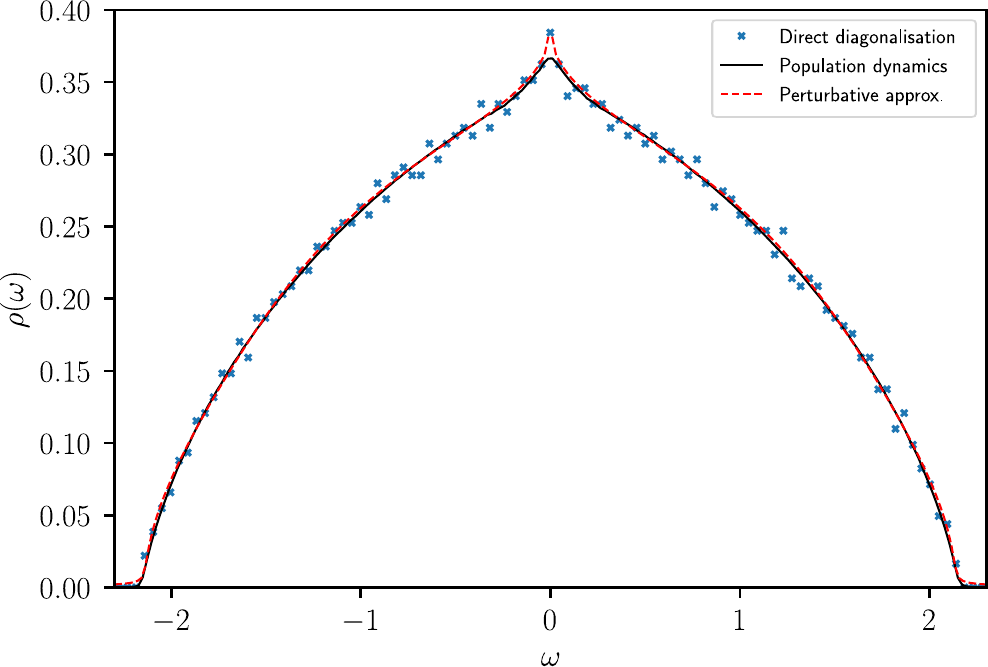}
	\includegraphics[scale = 0.47]{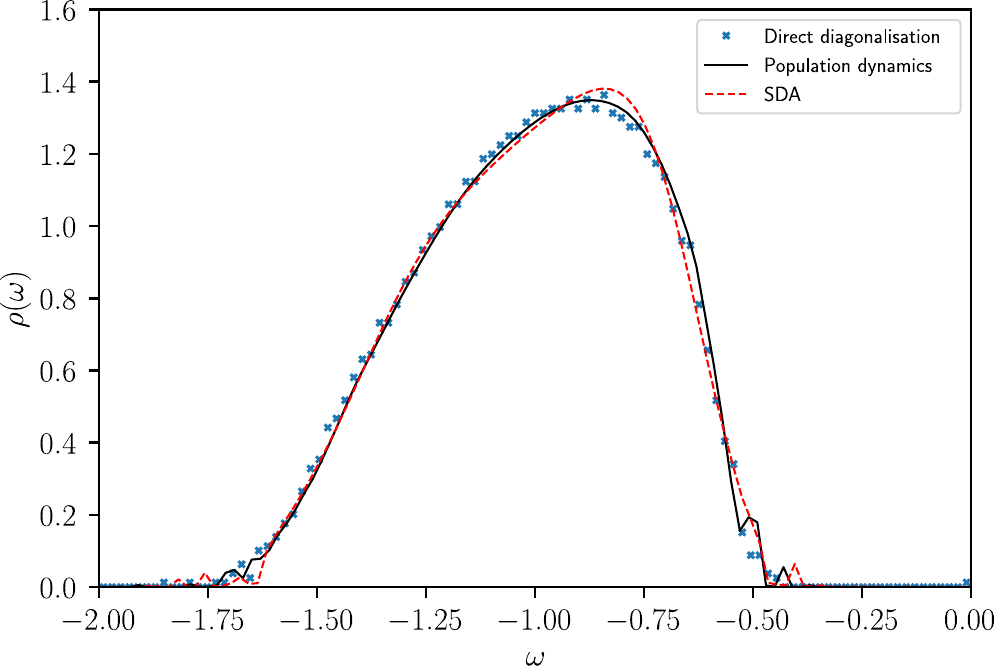}
	\captionsetup{justification=raggedright,singlelinecheck=false}
	\caption{ Eigenvalue density of the scaled adjacency matrices (left) with $p = 7$ and Laplacian matrices (right) with $p = 30$ of the ER graph. Direct diagonalisation results are performed for $N = 4000$. Population dynamics results are performed with a population size of $10^4$ and $\epsilon = 10^{-5}$. The theory lines are given by the numerical solution to $G = (\omega - G)^{-1} + \frac{1}{p}G^5 + \frac{1}{p^{2}}(G^7 + G^9)$ (left) [this is the second-order perturbative version of Eq.~(\ref{sparseerseries}), which is derived in Ref. \cite{baron2023pathintegralapproachsparse}], and the SDA in Eq.~(\ref{sdalaplace}) (right).}\label{fig:popdyn}
\end{figure}

We therefore obtain a self-consistency condition on the distribution $P_\mathrm{cav}(G^{(\mathrm{cav})})$. That is, the following self-consistent equations must satisfied
\begin{align}
	P_\mathrm{cav}(g) =& \sum_k \frac{k}{p} \frac{p^k}{k!}e^{-p} \int dg \int \prod_{i = 1}^{k-1}\left[d G_i^{(\mathrm{cav})} \, P_\mathrm{cav}(G_i^{(\mathrm{cav})})\right] \delta \left( g - \frac{1}{\omega - i\epsilon - \frac{1}{p}\sum_{i= 1}^{k-1} G_i^{(\mathrm{cav})}}\right) . \label{cavitypop}
\end{align}
One may then evaluate the normalised trace of the resolvent by computing
\begin{align}
	G(\omega) =& \sum_{k'} \frac{p^{k'}}{k'!}e^{-p} \int dG' \,G' \int \prod_{i = 1}^{k'}\left[d G_i^{(\mathrm{cav})} \, P_\mathrm{cav}(G_i^{(\mathrm{cav})})\right] \delta \left( G' - \frac{1}{\omega - i\epsilon - \frac{1}{p}\sum_{i= 1}^{k'} G_i^{(\mathrm{cav})}}\right) .
\end{align}
The remaining challenge is to find $P_\mathrm{cav}(G^{(\mathrm{cav})})$. This is primarily what is accomplished by the population dynamics algorithm, which is implemented as follows.
\begin{enumerate}
	\item Initialise a population $G_i^{(\mathrm{cav})}$ with random complex values and $i = 1, \cdots, N_\mathrm{pop}$
	\item For each $i$, draw a value of $k_i$ from the distribution $k P_k/p$.
	\item For each $i$, select at random a subset $S_i$ of size $k_i -1$ from the range $j = 1, \cdots, N_\mathrm{pop}$, and calculate $G'_i =  \frac{1}{\omega - i\epsilon - \frac{1}{p}\sum_{j \in S_i} G_j^{(\mathrm{cav})}}$. 
	\item Set $G_i^{(\mathrm{cav})} = G_i'$. 
	\item Repeat from Step 2 until the quantity $G_\mathrm{cav} = N_\mathrm{pop}^{-1}\sum_i G_i^{(\mathrm{cav})}$ converges.
	\item Once convergence has occurred, we begin the sampling phase. Perform Steps 2 to 4 once more. 
	\item Draw a value $k_j$ from the distribution $P_k$. Select a random subset $S_j$ of size $k_j$ from the range $l = 1, \cdots , N_\mathrm{pop}$. Compute $G_j = \frac{1}{\omega - i \eta - \frac{1}{p} \sum_{l \in S_j} G_l^{(\mathrm{cav})}}$. 
	\item Repeat Step 7 a number $N_\mathrm{samp}$ times to obtain a set $\{G_j\}$. Compute the quantity $G_\alpha = N_\mathrm{samp}^{-1} \sum_{j=1}^{N_\mathrm{samp}} G_j$.
	\item Repeat Steps 6 to 8 a number $N_\mathrm{iter}$ times to obtain a set $\{G_\alpha\}$. Compute $G(\omega) = N_\mathrm{iter}^{-1} \sum_{\alpha=1}^{N_\mathrm{iter}} G_\alpha$. This is our final estimate of the resolvent for a fixed value $\omega$.
	\item Repeat Steps 1 to 9 for all desired values of $\omega$. The eigenvalue density is then obtained via $\rho(\omega) = \pi^{-1}\mathrm{Im} G(\omega)$.
\end{enumerate}
The results of performing this procedure are show in Fig. \ref{fig:popdyn}. To adapt the method for the Laplacian matrix, we instead use the update rule $G'_i =  \frac{1}{\omega - i\epsilon + \frac{1}{p}\sum_{j \in S_i} \frac{1}{1+G_j^{(\mathrm{cav})}/p}}$.

Let us discuss briefly the rationale behind this algorithm. For a fixed $\omega$, Steps 1 to 5 above produce a population of $N_\mathrm{pop}$ complex values $G_i^{(\mathrm{cav})}$. This population can be thought of as having been sampled from the distribution $P_\mathrm{cav}$ defined in Eq.~(\ref{cavitypop}). Steps 6 to 10 are somewhat more subtle in their motivation. In principle, we could stop at Step 5, compute a set of $G_{j}$ as defined in Step 7 but with size $N_\mathrm{pop}$, and finally compute $G(\omega) = N_\mathrm{pop}^{-1}\sum_{j=1}^{N_\mathrm{pop}}G_j$. However, since $N_\mathrm{pop}$ is a finite number, one finds that the distribution $P_\mathrm{cav}$ is not always well represented by a single sample, especially near any spectral singularities. Instead, it is better to continue to iterate the dynamics of the cavity resolvent population $G_j$, and average over `time' as well as each realisation of the population. 

One notes further that we allow for the use of two different complex regularisers $\epsilon$ and $\eta$. This allows one to `smooth' the spectrum by choosing a slightly larger value of $\eta$, for example, if one so chooses. Depending on the ensemble, it can be useful to tune these parameters if there are spectral singularities.

\end{document}